%% file: thesis_template.tex
\documentclass[twoside,onecolumn,12pt,a4paper]{book}
\usepackage{anuthesis}
\usepackage{fancyhdr}
\usepackage{graphicx}

\usepackage{geometry} 

\usepackage{sidecap}
\usepackage{subfigure}
\usepackage{natbib}
\usepackage{amsmath}
\usepackage{tensor}
\newcommand{\Lagr}{\mathcal{L}}
\newcommand{\Mani}{\mathcal{M}}
\newcommand{\Hilb}{\mathcal{H}}
\begin{document}

\pagenumbering{roman}  
\bibliographystyle{apa-good}
\begin{titlepage}
    \begin{center}
        \vspace*{1cm}
        
        \huge
        \textbf{Appearing Out of Nowhere:\\
        The Emergence of Spacetime in Quantum Gravity}
        
        
        \vspace{1.5cm}
        \Large
        \textbf{Karen Crowther}
        
        \vfill
        
        Thesis submitted for the degree of\\
        Doctor of Philosophy
        
        \vspace{0.8cm}
        
        \includegraphics[width=0.4\textwidth]{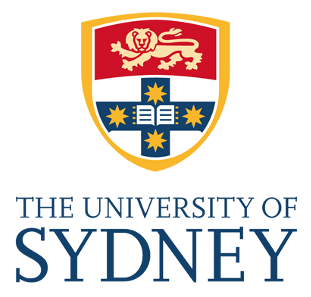}
        
        \Large
        Centre for Time\\
        Department of Philosophy\\
        University of Sydney\\
        Australia\\
        January 2014
        
    \end{center}
\end{titlepage}

\newpage
\mbox{}

\include{abstract}  

\include{thanks}  

\tableofcontents

\pagenumbering{arabic} 
\setcounter{page}{1}  

\include{Intro}

\include{Emerg}

\include{Eff}

\include{Univ}

\include{analogue}

\include{discrete}

\include{moo}

\include{concl}

\include{MyBibliography}  
\bibliography{thesis}
\end{document}

%% file: abstract.tex
\begin{center}
%

%
    \textbf{Abstract}
\end{center}
\small
Quantum gravity is understood as a theory that, in some sense, unifies general relativity (GR) and quantum theory, and is supposed to replace GR at extremely small distances (high-energies). It may be that quantum gravity represents the breakdown of spacetime geometry described by GR. The relationship between quantum gravity and spacetime has been deemed ``emergence'', and the aim of this thesis is to investigate and explicate this relation. After finding traditional philosophical accounts of emergence to be inappropriate, I develop a new conception of emergence by considering physical case studies including condensed matter physics, hydrodynamics, critical phenomena and quantum field theory understood as effective field theory. 

\paragraph*{}
This new conception of emergence is unconcerned with the ideas of reduction and derivation (i.e. it holds that we may have emergence with reduction or without it). Instead, a low-energy theory (or model) is understood as emergent from a high-energy theory if it is novel and autonomous compared to the high-energy theory, and the low-energy physics is dependent \textit{in a particular, minimal sense} on the high-energy physics (this dependence is revealed by the techniques of effective field theory and the renormalisation group). While novelty is construed in a broad sense, the autonomy comes essentially from the underdetermination of the high-energy theory by the low-energy theory, which reflects the minimal way in which the emergent, low-energy theory depends on the high-energy one. It results from the scaling behaviour of the theories and the limiting relations between them, and is demonstrated by the renormalisation group and effective field theory techniques, the idea of universality, and the phenomenon of symmetry-breaking.

\paragraph*{}
These ideas are important in exploring the relationship between quantum gravity and GR, where GR is understood as an effective, low-energy theory of quantum gravity. Without experimental data or a theory of quantum gravity, we rely on principles and techniques from other areas of physics to guide the way. As well as considering the idea of emergence appropriate to treating GR as an effective field theory, I investigate the emergence of spacetime (and other aspects of GR) in several concrete approaches to quantum gravity, including examples of the condensed matter approaches, the ``discrete approaches'' (causal set theory, causal dynamical triangulations, quantum causal histories and quantum graphity) and loop quantum gravity.

\normalsize

%% file: thanks.tex
\begin{center}
    \Large
    \textbf{Acknowledgements}
\end{center}

An enormous thank you to Dean Rickles and Huw Price for being the best supervisors I ever could have wished for. I am grateful for having benefited not only from their expertise, but their continual support and encouragement. 

\noindent Thank you!
\\ \\
This thesis has been improved by the corrections recommended by the amazing Christian W\"{u}thrich, Eleanor Knox, and Jonathan Bain. I am very appreciative of their thoughtful comments and suggestions.
\\ \\
I'd also like to say thanks to my Centre for Time comrades, Karim Th\'{e}bault, Pete Evans and Sam Baron, as well as the awesome Cambridge ``phil phys gang'', particularly Jeremy Butterfield, Arianne Shahvisi and Lena Zuchowski. I am thankful, too, for the hospitality of Gal and Mr B at the Cambridge Gin Palace on Newmarket Road. Cheers!
\\ \\
Finally, my heartfelt gratitude to my mother, Paula Crowther, and my sister, Anita Crowther, for everything they have done to help me along the way.
\\ \\
\textbf{A note regarding the title of the thesis:} I discovered, after I had made the initial submission of this thesis, that Christian W\"{u}thrich and Nick Huggett have a monograph in the works with a very similar title (being, \textit{Out of Nowhere: The Emergence of Spacetime in Quantum Theories of Gravity}). We have discussed the coincidence, and have agreed that neither title should need to change. I look forward to reading the book!
\\\\

%% file: Intro.tex
\chapter{Introduction: Spacetime and quantum gravity}\label{sect:Intro}

The search for a theory of quantum gravity is the pursuit of a more unified picture of the world. It is a quest to push beyond what is known, one that perhaps leads into the inaccessible. It is a journey guided by principles rather than experiment: principles gleaned from known physics, but which we cannot be sure will carry us as far as we want to travel. One of these principles states that we must be able to return from our journey---if we reach a theory from which we cannot arrive back at the firm ground of established physics, then, whatever we have reached, it is not quantum gravity. 

\paragraph*{}
Moving to a theory of quantum gravity might represent the breakdown of spacetime, in the sense that it is possible that our current conception of spacetime will not feature in the fundamental description of such a theory. If spacetime does not appear fundamentally in quantum gravity, but is to be recovered at some larger-distance (or lower-energy scale), then spacetime is \textit{emergent}. The ``return'' to current physics will represent the process of recovering spacetime. This thesis is concerned with the nature of the breakdown, the process of recovery, and the different conceptions of emergence that it might entail.

\section{Quantum gravity}\label{sub:QG}
Quantum theory, or, more specifically, quantum field theory (QFT), provides an account of all the known fundamental forces of nature---except for one. Gravity stands apart, finding its description in the classical theory of general relativity (GR). Perhaps the most familiar of the fundamental forces, gravity is the dominant force at large distance scales. Its description, provided by GR, is not only incredibly accurate at all known scales, but is conceptually elegant and remarkable in its achievements. By identifying gravity with spacetime geometry, GR transports space and time from the realm of the absolute and unchanging into the realm of motion, affectedness and interaction. 
\paragraph*{}
According to GR, spacetime is not a fixed entity that stands as a background, not a stage for physics to play out upon. Instead, spacetime may be understood as a dynamical entity that influences matter and which itself is influenced by matter. On the other hand, QFT is a theory of fields which are defined on a static, background spacetime. Basically, the theory states that all matter is composed of particles, which are understood as local excitations of quantum fields; the fundamental forces are themselves represented by quantum fields, whose corresponding excitations interact locally with the other particles, depending on their type.\footnote{The idea of locality in QFT is necessary in order to make sense of the axioms of the framework (i.e. unitarity, micro-causality and the Poincar\'{e}-invariant vacuum), and is discussed in \S\ref{sub:QFTrenorm}.}

\paragraph*{}
Quantum gravity is a domain of research that, in some sense, unifies GR and quantum theory. There are many different ways in which this may be interpreted. For example, a QFT that described gravity would be a candidate theory of quantum gravity---this attempt to incorporate gravity into the framework of QFT is known as the ``particle physicists' perspective'' (or the ``high-energy theorists' perspective'') of quantum gravity, since it privileges QFT over the insights of GR (through its use of a background spacetime, for example, as is clarified in \S\ref{sub:BI}). Another candidate would be a theory produced by quantising GR (using standard quantisation procedures, or perhaps some inventive techniques and additional ideas). Quantum geometrodynamics (discussed in \S\ref{subsub:canon}) is an example of this approach, as is loop quantum gravity (\S\ref{sect:LQG}). 
\paragraph*{}
There are more creative approaches, too, and these form the main focus of this thesis. Quantum gravity could be a theory that is neither a QFT nor a quantisation of GR: it may be a theory that is quantum in some sense, but which does not rely on a background spacetime. It may represent a ``small-scale'' (``micro'' or high-energy) theory of spacetime, but without any standard conception of spacetime appearing within it. In such a case, the theory would describe the micro-constituents, or ``atoms'' of spacetime, rather than spacetime itself. If spacetime thus ``breaks down'' at some scale, then familiar quantum theory and QFT breaks down with it. The pressure of unifying gravity and quantum theory may well result in a theory of neither. 

\paragraph*{}
The search for quantum gravity is driven by both conceptual and technical motivations. The main conceptual motivation is the great aspiration of a unified theory. Unification is a traditional ``guiding principle'' in physics, and is often viewed as means of producing successful theories. Familiar examples of this include Maxwell's theory of electromagnetism, which unified light as well as the electric and magnetic forces; the electroweak theory, which unifies the electromagnetic force and the weak force; and even GR, with its identification of inertial mass with gravitational mass, and spacetime with gravity. For those inclined towards unification, the current situation in physics---the dualistic, split picture of the world it presents---is unsettling, and calls us to question the fundamental nature of both GR as well as the framework of quantum theory. 
\paragraph*{}
But not everyone who is interested in the progress of physics is uncomfortable with this dualistic picture of the world, with GR on one hand and QFT on the other: in fact, for many, there is no such divide. Condensed matter theorists such as Philip Anderson see physics not as dualistic in this sense, but \textit{pluralistic}. This view entails recognising that theories in physics form a hierarchy of sorts, ranked according to the distance-scales at which they are useful. The standard model of particle physics, for instance, is useful at very small distance scales compared to those that characterise the systems for which statistical mechanics is an appropriate description. Again, at larger distances, thermodynamics becomes the useful theory, and so on. The picture of the world that results is a layered one: we have a ``tower of theories'' (an idea which is made more precise in \S\ref{sect:EFT}).

\paragraph*{}
As is discussed in \S \ref{sub:CaoSchweb} and \S \ref{sub:effectiveEFT}, this view is often, wrongly, interpreted as standing in opposition to the search for a ``final theory''. In this case, ``final theory'' is construed as a ``theory of everything'', being a unified theory describing all of the known fundamental interactions. Steven Weinberg is well-known for his enthusiasm in support of the search for a final theory. Those who dream of a final theory are those who are uncomfortable with the current lack of a unified framework that incorporates gravity along with the other fundamental forces. The non-dreamers have often been accused of being adversaries to the search for the final theory, but the reality of the position is much more subtle. 
\paragraph*{}
Those who are accepting of the layered picture of the world may be seen as taking a practical, pragmatic position---a position that is open-minded in regards to a final theory. It involves recognising that no theory formulated at small distance scales will be a theory of \textit{everything}.\footnote{This is essentially what distinguishes the pragmatic view I have in mind from the views of a person who accepts the utility of the layered picture, but believes it false and replaceable by a fundamental, high-energy theory of everything. I argue in this thesis that emergence gives us good reasons to deny the redundancy of the layered picture.} Instead, such a theory will describe the world at small distance scales. Moving to larger scales means new degrees of freedom become important, and a new (model or) theory---on a different level of the tower---is required in order to describe them. The theory of the small-scale degrees of freedom, even if it is a complete description of ``everything'', becomes useless. This pragmatic position is the perspective from which this thesis is written; while quantum gravity would no doubt ease the conceptual uncomfortableness of a physics fundamentally divided, it is not expected to make the layered picture obsolete.

\paragraph*{}
The idea of a tower of theories represents an account of \textit{emergence}, in which the theories that are useful at larger length scales are said to emerge from those that are useful at smaller length scales. This relates to the ``physicists' sense of emergence'', introduced shortly, where one theory, $T_N$ (the theory that describes the small-scale degrees of freedom) is said to ``reduce'' to the other, $T_O$ (the theory that describes the large-scale degrees of freedom). $T_O$ is then said to ``emerge'' from $T_N$.\footnote{Note that this is not the only means by which to understand the ``physicists' sense'' of reduction and emergence: for example, there is a (different) sense in which we may speak of a large-scale theory reducing to the small-scale theory, as in thermodynamics ``reducing'' to statistical mechanics (this example is meant as illustrative only, and not to suggest anything about the interpretation nor success of such a reduction).}  The conception of emergence here involves recognising the necessity of theories being framed in terms of the appropriate degrees of freedom for the scale being studied: although QFT, for instance, is supposed to apply at all distance scales, in reality it cannot be used, for example, to describe a game of dice. 
\paragraph*{}
Here, the relevant sense of ``cannot'' needs to be clarified. For philosophers, it is tempting to interpret it as meaning ``cannot \textit{in practice}'', and swiftly categorise this type of emergence as \textit{epistemological emergence}. This category tends to be perceived as less exciting than its contrast of \textit{ontological emergence}, which would take the ``cannot'' to mean ``cannot \textit{in principle}''.\footnote{All of these ideas are explained in \S \ref{sect:Emergence}.} As argued in \S\ref{sect:Emergence} and \S\ref{sub:emergence}, however, the difficulties in grasping and defining these concepts are further complicated by deep issues regarding how our theories and language relate to the world. Rather than jump in and spend lengths of my thesis engaging with these problems, I believe it more sensible to begin by looking at the physics itself, so I have taken this approach here.

\paragraph*{}
It should be noted that although quantum gravity need not represent a ``theory of everything'', such a theory would count as quantum gravity---string theory being the primary example. In spite of this, quantum gravity must, by definition, be a unified theory.\footnote{As stated immediately above, the definition of quantum gravity taken in this thesis is that it is a theory that unifies quantum theory and gravity. And, again, as explained above, there are several ways of understanding the idea of unification and the means of achieving it---a theory that unifies quantum theory and gravity need not be a theory that fits with the framework of QFT, nor one that ``features'' gravity, so long as QFT and GR can be shown to ``emerge'' from it, or be explained by it in some sense.} Thus, quantum gravity can be motivated by the desire for unification, without commitment to a final theory. There is a further motivation, as well, though, and this is the simple desire to advance physics and expand the scope of human knowledge.\footnote{\citet[][p. 1213]{Ashtekar1974} express this sentiment, for example.} This motivation is the wish to understand more of the universe, to push to higher and higher energies beyond what is known, and to explore territory hitherto-unexplored. Also, there is the further tantalising suggestion that quantum gravity promises to profoundly alter our worldview, and this suggestion engenders a strong sense of ``natural curiosity'' which certainly acts as a lure toward investigating quantum gravity \citep[cf.][p. 284]{Rickles2008a}. 
\paragraph*{}
Even without considering quantum gravity specifically, there is a common thought that our current theories are not the final word: this idea comes from some dissatisfaction with QFT as a fundamental theory, due to problems of mathematical coherency and renormalisation, discussed in \S\ref{subsub:axiom}, as well as the feeling that certain patterns and parameter values in the standard model are in need of explanation (or point toward a deeper explanation). This might mean a new theory or framework is required at current energies, to replace QFT and the standard model, or it might mean a new theory is required at higher-energies (this second suggestion, where the higher-energy theory promises a better explanation of current physics, will be appealing for those who feel the pull of reduction).

\paragraph*{}
The physical reasons for seeking a theory of quantum gravity are less clear-cut than the conceptual motivations. This is because there is simply no phenomenon that can be uniquely identified as the result of some combination of general relativity and quantum theory. To make matters worse, as \citet{Butterfield2000} explain, not only is there no data, but there is not even any agreement as to the \textit{sort} of data that would be relevant to quantum gravity. The lack of data is typically cited as due to the extreme inaccessibility of the domain in which quantum gravity is expected to be applicable, as will be indicated shortly (\S\ref{subsub:mot}). Thus, although distant, there are certainly situations for whose descriptions quantum gravity is thought to be necessary; these are situations in which general relativity intersects with quantum theory.

\paragraph*{}
Most characteristically, GR and quantum theory are both necessary in order to describe the case in which a particle of mass $m$ has its Compton wavelength, $l_\text{C} = \hbar/mc$ equal to its Schwarzschild radius, $l_\text{S} = Gm/c^2$, where $G$ is Newton's gravitational constant, $\hbar$ is the reduced Planck's constant, and $c$ is the speed of light. This equality occurs when the mass is the so-called Planck mass $m=m_\text{P}=\sqrt{\hbar c/G}$. A particle's Compton wavelength is a prediction of quantum field theory (\S\ref{sub:QFTrenorm}), which states that localising $m$ to within $l_\text{C}$ uses enough energy to create another (identical) particle of mass $m$. The Schwarzschild radius is a prediction of general relativity; it states that compressing $m$ to $l_\text{S}$ will result in the formation of a black hole. 

\paragraph*{}
Other situations whose full explanations are expected to be provided by quantum gravity include spacetime singularities such as black holes and cosmological singularities---which, as Curiel and Bokulich state, ``are arguably our best windows into the details of quantum gravity'' \citep{Curiel2009}. One of the reasons for this has to do with Hawking radiation from black holes \citep{Hawking1975}, and the idea of Bekenstein entropy \citep{Bekenstein2003, Bousso2002}.\footnote{This is also related to the holographic principle, discussed below in the context of the AdS/CFT duality (\S\ref{subsub:dualities}).} 

\paragraph*{}
Nevertheless, while quantum gravity might be necessary for an explanation (or understanding) of such situations, we might question whether quantum gravity is strictly needed in order to describe them or make predictions\footnote{Cf. \citet{Huggett2001, Wuthrich2005}.}---\citet{Mattingly2009, MattinglyForthcoming}, for instance, argues that a semiclassical theory, or some other approximation to quantum gravity, may be suitable for these purposes.\footnote{This is discussed further in \S\ref{sub:GREFT}, along with some other ``hybrid'' approaches whose aim is to reproduce the results of quantum gravity within accessible energy scales.} While such a theory (or approximation) might fulfil the ``technical aspects'', however, it does not sate the desire for unification or the drive for a fundamental theory---thus, I would submit, the main motivations for quantum gravity are conceptual.

\paragraph*{}
In the absence of unequivocal data, or even consensus as to what sorts of data would be significant, the quest for quantum gravity is a quest guided by principles. These guiding principles provide the definition of quantum gravity, as well as the motivation for the theory; principles may function as restrictions that an acceptable theory must satisfy, or they may be features that are desired of quantum gravity, serving as criteria for theory selection. As should be clear given the variety of different types of theory that would count as quantum gravity, there is no agreed-upon set of principles to adopt. 
\paragraph*{}

Not only are there many different candidates for principles and, hence, many different possible combinations of principles, there are many different ways of interpreting or implementing particular principles. Thus, the range of quantum gravity approaches is a broad one. And yet, according to the \textit{generalised correspondence principle}, which is the principle of core interest in this thesis, each of them must recover GR, and (continuum) spacetime with it, in the regime where GR is known to hold. 

\paragraph*{}
Following Heinz Post, the generalised correspondence principle\footnote{The generalised correspondence principle, of course, takes inspiration from the original ``correspondence principle'' formulated by Niels Bohr, which states (roughly) that quantum mechanics should reproduce the results of classical mechanics for systems involving large orbits and large masses.} can be stated as, ``the requirement that any acceptable new theory $L$ should account for the success of its predecessor $S$ by `degenerating' into that theory under those conditions under which $S$ has been well confirmed by tests.'' \citep[][p. 228]{Post1971}. The principle can also be framed in terms of reduction, and it is perhaps more typical to do so,

\begin{description}\label{def:GCP}
\item[Generalised correspondence principle (GCP)] any acceptable new theory $T_N$ should account for the success of its predecessor $T_O$ by \textit{reducing} to $T_O$ in the domain of applicability of $T_O$ 
\end{description}
The newer theory is broader in scope than the theory it replaces (i.e. it has a larger domain, makes more predictions, or contains additional insights compared to the older theory). The idea of reduction means that the older theory can be ``mapped'' to the newer theory, or shown to be part of it. In other words, the older theory can be deduced from the newer theory, provided we have some appropriate ``correspondence rules'' (also called ``bridge principles'') that enable the definitions of the newer theory to be related to those of the older one. One (perhaps the major) motivation for upholding the GCP is the desire to preserve what is known, and what has been demonstrated by the older theory---the newer theory should increase our knowledge of the world, not render unexplained anything that had been explained.\footnote{In practice, of course (as any historian of science will tell you), this is not always the case.}

\paragraph*{} 
Some oft-cited examples of the GCP in action include the reduction of special relativity to classical mechanics for velocities small compared to the speed of light, and GR reducing to Newtonian gravity in the limit of weak gravitational fields. I will not assess the GCP here, nor argue for it; instead, I will just assume that the GCP is a theory-building principle to uphold in the search for quantum gravity. This thesis is concerned with the emergence of spacetime from quantum gravity, and the GCP provides one way of interpreting the idea of emergence. More specifically, the ``physicist's sense of emergence'' holds that $T_O$ is \textit{emergent} from $T_N$ if $T_N$ and $T_O$ satisfy the GCP.\footnote{This idea is discussed in \S\ref {sub:emerphysics}.} (This is not an exclusive definition---there are other senses of emergence in physics).
\paragraph*{}
Thus, if the GCP is taken as a principle of quantum gravity, then GR must emerge from quantum gravity in the regime where GR is known to hold. Similarly, if quantum gravity is understood as a theory in which there is no background spacetime (e.g. a theory of non-spatiotemporal discrete elements, or ``atoms'' of spacetime), then familiar quantum theory (which does rely on a background spacetime) must also emerge in the domains where it has proven successful as a description of the phenomena.  

\paragraph*{} 
This is not the only sense of emergent spacetime that is explored in this thesis, however, and nor is the ``physicists' sense of emergence'' the only conception of emergence considered. A more philosophically ``weighty'' conception of emergence is developed in \S\ref{sect:Emergence}, and other conceptions of emergence in physics are explored in \S\ref{sect:Univers}, before the specific quantum gravity approaches are considered (in \S\ref{sect:Discrete}, \ref{sect:LQG}). I also discuss models of emergent spacetime that are not models of quantum gravity, and in which spacetime emerges \textit{without} GR (\S\ref{sub:analogue}).  Before introducing these, however, it is helpful to give a basic summary of some of the main conceptual issues in QG and necessary background to the idea of emergent spacetime. 

\paragraph*{}
The rest of this chapter is as follows. I will begin by explaining, briefly, what is meant by spacetime in this thesis (\S\ref{subsub:space}), before outlining the general arguments that are typically taken to motivate the suggestion that spacetime ``breaks down'' at some scale (\S\ref{subsub:mot}). The possibility of quantum gravity being a non-spatiotemporal theory, together with its supposedly extreme experimental inaccessibility leads to concerns regarding its status as a viable physical theory at all (\S\ref{subsub:phys}). Another important issue in understanding the fundamental nature of space and time---or, rather, their breakdown---is the ``problem of time'', introduced in \S\ref{subsub:canon}. 
\paragraph*{}
As stated, the emergence of spacetime is related to the idea of ``recovering'' spacetime from quantum gravity (\S\ref{sub:recovery}); there are two different ``transitions'' that are involved in this, the quantum/classical transition and the micro/macro transition (\S\ref{subsub:trans}). In this thesis I am mostly interested in the micro/macro transition, as I explain in \S\ref{subsub:small-scale}. In regards to the quantum/classical transition, though, it may be that the idea of decoherence plays a role (\S\ref{subsub:deco}). 

\paragraph*{}
Some of the challenges to the idea of emergent spacetime (and gravity), including the Weinberg-Witten theorem (\S\ref{subsub:WW}), are presented in \S\ref{sub:chal}. Also, the idea of emergent spacetime suggested in the AdS/CFT correspondence in string theory, is briefly explored in \S\ref{subsub:dualities}. A theme running throughout this thesis is the relationship between different theories in physics, particularly the cross-fertilisation of high-energy particle physics and condensed matter physics. This is discussed in \S\ref{sub:sand}. Finally, \S\ref{sub:syn} provides a chapter-by-chapter overview of the rest of the thesis.

\section{Spacetime}\label{subsub:space}
As a thesis concerned with the emergence of spacetime, it will be helpful to begin with an explanation of what is meant by spacetime. The best description of spacetime is provided by GR---a theory which famously identifies it with the gravitational field. Spacetime, according to GR, is a dynamical entity that both affects matter and is affected by matter. In this thesis, ``spacetime'' is typically used to denote spacetime as described by GR, i.e. the gravitational field, but I occasionally speak of background spacetime of QFT, or of other theories (when this is done, the context should prevent confusion). 

\paragraph*{}
A model of GR is specified as $\text{M}=\langle \Mani , g, T \rangle$ where $\Mani$ is a four-dimensional manifold of spacetime points, encoding the topology and differentiable structure, $g$ is the Lorentzian metric tensor, encoding the geometry, and $T$ is the energy-momentum tensor. The two tensors satisfy Einstein's field equations,
\begin{equation}\label{eq:Einstein}
G_{\mu\nu}\equiv R_{\mu\nu}[g]-\frac{1}{2}g_{\mu\nu}R[g]+\Lambda g_{\mu\nu} = -8\pi G_N T_{\mu\nu}[\Phi]
\end{equation}
where $G_{\mu\nu}$ is the Einstein tensor describing the curvature of spacetime, $R_{\mu\nu}$ is the Ricci curvature tensor, $-8\pi G_N$ is a coupling constant, proportional to Newton's gravitational constant, and $\Phi$ represents the source(s) of the gravitational field (i.e. matter), whose energy-momenta is described by $T_{\mu\nu}$. 

\paragraph*{}
One approach is to take the manifold as spacetime; it is the set of points on which all the fields---matter and metric---are defined. This view is based in the belief that matter and spacetime are two separate entities: those who take this view point out that we use the concepts of space and time to describe the behaviour of matter, or, in other words, we describe matter as existing in spacetime. On this view, spacetime is seen purely as the container of events. It entails conceiving of the manifold alone as spacetime, because the metric tensor can be interpreted as carrying energy, and thus blurs the line between ``spacetime'' and ``matter'' (and so, taking the metric tensor-plus-manifold as spacetime is undesirable for those who believe there should be a distinction between spacetime and matter). The problem, however, is that the manifold unequipped does not possess any of the properties that, according to GR, spacetime should possess. For example, the light-cone structure is not defined, past and future cannot be distinguished, and no distance relations exist. For this reason, spacetime is better understood as corresponding to the manifold \textit{plus metric}, rather than the manifold alone.\footnote{This idea also more naturally accords with the suggestion that spacetime and matter, as they feature in GR, are interwined. Our only access to spacetime is through the behaviour of matter, which suggests that the geometry of spacetime should be understood as reflecting the relationship between spacetime and matter, rather than the intrinsic nature of spacetime.}

\paragraph*{}
Of course, even this isn't the whole story, because there is a claim to be made that different (diffeomorphically-related) manifold-plus-metric structures correspond to the same spacetime. This is a consequence of the diffeomorphism invariance of GR (which is related to the idea of background independence, discussed in \S\ref{sub:BI}). While the idea of general covariance is familiar as the statement that the laws of the theory are unaffected by a change of coordinates, it can also be understood as an \textit{active} transformation. Instead of relabelling the structures on the manifold with new coordinates, we can imagine that these structures (fields) have actually been dragged along to new positions on the manifold, so that their coordinates are the same as those which they would have possessed had we moved to the new coordinate system. Models related by a diffeomorphism transformation are physically indistinguishable as they agree on all invariant quantities.\footnote{This is the basis of the ``hole argument'', see, \citet{Earman1987, Norton1988}.}

\paragraph*{} 
The fact that a spacetime does not uniquely correspond to a particular field configuration is the ``problem of space''; it has an analogue ``problem of time'' that also owes to the difficulties of interpreting gauge theories \citep{Rickles2006}. This is discussed below, \S\ref{subsub:canon}. Diffeomorphism invariance means that, in order to specify a model of GR, a gauge-invariant \textit{equivalence class} of gauge-variant structures is required, rather than a single tuple of $(\Mani, g, T)$. Nevertheless, when considering the idea of emergent spacetime, several quantum gravity approaches present, or are interested in, only an emergent \textit{metric} structure (e.g. in \S\ref{sub:analogue}). This doesn't completely fly in the face of the lesson of diffeomorphism invariance, however, since, although a given spacetime will not correspond to a particular manifold-plus-metric combination, a given metric may be said to uniquely pick out a spacetime.

\section{Motivations and indications}\label{subsub:mot}
The suggestions, arguments, or ``hints'' that spacetime will break down at some scale (i.e. that there is a fundamental length, or that quantum gravity describes discrete, non-spatiotemporal entities) may be categorised into two different classes: \textit{definitional considerations} and \textit{external considerations}. As will be demonstrated through the examples I present in this sub-section, the definitional considerations are dependent upon taking a particular definition of quantum gravity, i.e. they come about as a result of adopting certain principles (or sets of principles), which, of course, we cannot be sure are warranted. 

\paragraph*{}
On the other hand, external considerations are problems, concerns or observed features of the world that could be neatly explained or solved by the discreteness of spacetime, but aren't necessarily related to quantum gravity. An external consideration, on its own, cannot be treated as evidence of spacetime discreteness, especially without investigating other possible explanations for its appearance. The temptation to ``tally-up'' external considerations in order to make a case for the breakdown of spacetime is potentially dangerous in that it invites us to take the (relatively) easy way out by ascribing to a single, (otherwise) unproven, origin many different concerns with possibly disparate origins: the history of the luminiferous ether should serve to caution against such moves.

\paragraph*{}
The most familiar argument for a fundamental length is the one that also serves as an argument for the scale at which quantum gravity is expected to be important---this is the argument from dimensional analysis for the existence of the Planck length. Dimensional analysis is a commonly used tool in physics, used for order of magnitude estimations and finding appropriate units for various physical quantities. In the context of quantum gravity it involves combining the characteristic constants of GR and quantum theory. Famously, Max Planck demonstrated in 1899 that there is a unique way (apart from numerical factors) to do so, in order to provide fundamental units of length, time and mass. Here, they are designated $l_\text{P}$, $t_\text{P}$ and $m_\text{P}$, respectively\footnote{Quoted from \citet{Kiefer2006}.},

\begin{equation}
l_\text{P}=\sqrt{\frac{\hbar G}{c^3}}\approx 1.62\times 10^{-35}\text{m}
\end{equation}

\begin{equation}
t_\text{P}=\sqrt{\frac{\hbar G}{c^5}}\approx 5.40\times 10^{-44}\text{s}
\end{equation}

\begin{equation}
m_\text{P}=\sqrt{\frac{\hbar c}{G}}\approx 1.22\times 10^{19} \text{GeV}
\end{equation}

As mentioned (\S\ref{sub:QG}), $m_\text{P}$ is the mass at which a particle's Compton wavelength is equal to its Schwarzschild radius. Based on the predictions of GR combined with those of quantum theory, we expect there to be microscopic, rapidly-evaporating black holes at this scale. John Wheeler, in the 1950s, spoke of a ``quantum foam'' at the Planck scale, where quantum fluctuations of spacetime (or fluctuations affecting spacetime) would become significant---geometry at this scale is thought to be ill-defined, or ``fuzzy'' \citep{Wheeler1998}. Thus, we are led to the suggestion that the Planck length is a minimal length, meaning that, on this picture, no distances smaller than the Planck length exist. The arguments based on dimensional analysis are definitional considerations, since they arise from taking as a definition of quantum gravity that it be a theory that combines GR and quantum theory in this way.

\paragraph*{}
Another definitional consideration is the problem of time, which, as discussed below, in \S\ref{subsub:canon}, is a single title used to designate a cluster of problems. These problems stem from the dissimilar ways in which GR and quantum theory treat space and time: attempting to combine the theories leads to conflict and strange results! These results have been taken as suggesting that time does not exist in quantum gravity. More generally, and less controversially, the problem of time does suggest that at least some of the concepts and structures featured in quantum theory and GR will perhaps not be useful in quantum gravity, though it is difficult to determine what these are. 
\paragraph*{}
As discussed below (\S\ref{subsub:canon}), there is a particular manifestation of the problem of time in an approach to quantum gravity that proceeds by quantising the gravitational field (canonical quantum gravity); the equation that is supposed to describe the dynamics of the theory does not feature a time parameter. This context demonstrates how the problem of time may be considered an example of a definitional problem---it stems, in this context, from taking quantum gravity as a theory of a quantised GR.\footnote{To emphasise: this is only one particular aspect of the problem of time. It is explained in a little more detail below (\S\ref{subsub:canon}), along with some other aspects of the problem.}

\paragraph*{}
There are also external reasons for thinking that spacetime might break down; for instance, the fact that many quantum field theories go awry at high-energies (\S\ref{sub:QFTrenorm}) could be avoided if there was a natural high-energy cutoff provided by a fundamental length. This suggestion, historically, arose in the context of studying quantum electrodynamics, and today many of the quantum gravity approaches that describe a fundamental length recognise the utility of doing so in regards to solving the high-energy difficulties of QFT \citep{Hagar2013}. Nevertheless, as argued in \S\ref{subsub:axiom}, the interpretation of QFT is not uncontroversial. Furthermore, although the existence of a fundamental length would help explain the necessity of renormalisation, the necessity of renormalisation does not, on its own, imply the existence of a fundamental length---there are other possible explanations and ways of understanding QFT that do not involve spacetime discreteness.

\paragraph*{}
Another, similar, external consideration is the non-renormalisability of gravity: again, there are ways of ``solving'' this that do not suggest spacetime discreteness (but are not incompatible with spacetime discreteness, \S\ref{sub:asygrav}). GR may be treated in the same way as QFTs are (\S\ref{sub:GREFT}), and the same philosophy---that of \textit{effective} effective field theory, presented in \ref{sub:effectiveEFT}---is applicable. This view is a pragmatic one which entails recognising that there are many different high-energy scenarios that could be responsible for observed low-energy physics, and avoiding making assumptions about which (if any) of these is correct. 
\paragraph*{}
Some additional---though also inconclusive---arguments for discreteness, or a fundamental length scale, are presented in \S\ref{sub:discreteness}. Finally, while foundational and interpretational issues in QFT have been taken as motivating a particular feature of quantum gravity, there is perhaps also---for those vexed by conceptual issues in quantum theory more generally---the suggestion, or hope, that quantum gravity will reveal something about the nature, or origin, of quantum ``weirdness''. 

\section{Quantum gravity as a physical theory}\label{subsub:phys}
The experimental inaccessibility of quantum gravity (in the absence of novel low-energy predictions made by particular approaches) is suggested not only by arguments from dimensional analysis and black hole thermodynamics, but also from the great success of GR at all accessible energies (\S\ref{sub:GREFT}). If scientific theories are supposed to make contact with the empirical realm through predictions and experiment, then this leads to questions regarding the status of quantum gravity as a scientific theory at all: questions such as those that have most publicly been levelled at string theory, for example \citep{Smolin2007, Woit2007}.

\paragraph*{} 
A potentially more serious objection to quantum gravity as a physical theory is raised by \citet{Maudlin2007} and concerns those approaches to quantum gravity that do not feature spacetime. Maudlin claims that, for a theory to be ``physically salient'' it must have spatiotemporal entities as part of its fundamental ontology; otherwise, it is difficult to see how such a theory could ever make contact with the empirical realm. The idea of spatiotemporal entities as being the basic ontology of our scientific theories, ultimately constituting what we observe, comes from Bell's idea of ``local beables'' \citep{Bell1987}. The worry is that, because all observations are observations of local beables---things that exist \textit{somewhere} in space and time---a theory without local beables is unable to account for any observations, and is thus \textit{empirically incoherent}. 

\paragraph*{}
\citet{Huggett2013} argue, and \citet{OritiForthcoming} re-emphasises, however, that a lack of fundamental local beables in a theory does not mean that such a theory is unable to recover local beables in some limit or approximation, and, thus, be testable at that level. On such an account, the local beables would emerge along with spacetime. \citet{Maudlin2007} anticipates such a reply, however, and claims that the empirical contact made by the derived local beables does not establish the ``physical salience'' of the fundamental theory. 

\paragraph*{}
The argument then turns on what is meant by a theory being ``physically salient'', with \citet{Huggett2013} maintaining that we should look to the theory in question itself, the procedure by which the local beables are derived, and its empirical success in order to determine what is physically salient, rather than begging the question against theories that do not feature space and time. I agree with this prescription, as well as the general sentiment that it is based upon, namely that we should first attempt to develop and examine particular approaches toward quantum gravity rather than seek general arguments as to their physical salience---given the diverse range of potential approaches, based on different motivations and guiding principles, any general arguments are themselves unlikely to be salient.

\section{The problem of time}\label{subsub:canon}
The ``problem of time'' in quantum gravity is neither a single problem, nor exclusive to quantum gravity---instead, it is a cluster of problems, at least one of which arises even in classical GR, where it is linked to the interpretation of gauge invariance and is another form of the problem of space (related to the hole argument) described above (\S\ref{subsub:space}). In the context of quantum gravity, though, it stems from the disparate way in which time is treated in GR compared to quantum theory. 

\paragraph*{}
Time, according to quantum theory, is external to the system being studied: it is \textit{fixed} in the sense that it is specified from the outset and is the same in all models of the theory. In GR, however, time forms part of what is being described by the theory. Time in GR is subject to dynamical evolution, and it is not ``given once and for all'' in the sense that it is the same across all models \citep{Butterfield1999}. Reconciling these two treatments of time is not possible, and any attempt at formulating a quantum theory of gravity will thus face the problem of time in some form or another.
\paragraph*{}
The problem of time manifests itself in different guises depending on the different approaches to quantum gravity. The problem is easiest to appreciate in the canonical quantisation program of GR known as \textit{quantum geometrodynamics}, where it arises in several forms, including the one related to the definition of observables and the interpretation of the gauge invariance of GR (which is sometimes called ``the problem of change'').\footnote{For detailed reviews on the problem of time, see \citet{Isham1993, Kuchar1992, Kuchar1999}. For more philosophical introductions to the problem, see \citet{Belot2001, Huggett2013a, Rickles2006}. My discussion here owes to \citet{Butterfield1999, Huggett2013}.} The aim of this program is to produce a straightforward quantisation of GR. To this end, the strategy is to cast GR in Hamiltonian form and then to quantise this Hamiltonian theory using the canonical quantisation procedure. 
\paragraph*{}
Casting GR in Hamiltonian form involves splitting spacetime into space and time, so that the theory describes the evolution in time of the geometry of a three-dimensional spacelike hypersurface, $\Sigma$, i.e. a spacelike ``slice'' of the spacetime manifold. This splitting violates the gauge invariance (general covariance) of GR, with the result that the dynamical equations of the Hamiltonian formulation are not (alone) equivalent to Einstein's field equations. Additional constraint equations are thus imposed on the Hamiltonian system in order to mend the symmetries of GR.

\paragraph*{}
There are two types of constraint required in Hamiltonian GR: the diffeomorphism (or momentum) constraint and the Hamiltonian constraint. These constraints have a geometrical interpretation in terms of motions of $\Sigma$.\footnote{This is true for a given a shift vector $N^a$ and a lapse function $N$.} The diffeomorphism constraint generates (via infinitesimal transformations) diffeomorphisms \textit{on} $\Sigma$: it shifts information tangentially to the slice. The Hamiltonian constraint generates (again, via infinitesimal transformations) symmetries \textit{off} $\Sigma$: it pushes information (in a direction normal to the slice) from the slice to onto one that is infinitesimally close to it. 
\paragraph*{}
The full Hamiltonian of canonical GR (not the constraint) is the sum of the diffeomorphism and Hamiltonian constraints, which together are needed in order to impose full diffeomorphism invariance (general covariance). Taking the Poisson bracket of the full Hamiltonian then gives the ``time evolution'' of the theory, which is just a combination of the two motions, tangential and normal, to $\Sigma$ (because these transformations are symmetries, the motion is unphysical).

\paragraph*{}
The quantisation procedure involves promoting the canonical variables of the classical Hamiltonian theory to quantum operators which satisfy the canonical commutation relations. Because these constraint operators generate the gauge symmetries of the theory, every operator that represents a genuine physical observable must commute with them.\footnote{Part of the definition of a ``physical observable'' is that it be a gauge-invariant quantity.} In other words, the constraints determine the physical Hilbert space of the theory, and only the states which satisfy both constraints are physical states. 
\paragraph*{}
The diffeomorphism constraint is able to be readily interpreted as imposing a ``canonical analogue'' of diffeomorphism invariance, indicating that points in space are not themselves physically meaningful.\footnote{See \citet[][pp. 149--150]{Butterfield1999} for a little more detail on the momentum constraint.} The Hamiltonian constraint, however, is very difficult to make sense of. In quantum geometrodynamics, it is the Wheeler-deWitt equation,
\begin{equation}\label{eq:WdW}
\hat{H}|\psi\rangle = 0
\end{equation}
where $\hat{H}$ is the Hamiltonian operator, and $|\psi\rangle$ are the quantum states.

\paragraph*{}
Since, in a classical theory, the Hamiltonian generates the dynamical evolution of the states, we might expect the Wheeler-deWitt equation to express the dynamical content of quantum geometrodynamics. Indeed, (\ref{eq:WdW}) resembles the familiar time-dependent Schr\"{o}dinger equation that describes the evolution of states in quantum theory---except, of course, the right-hand-side of (\ref{eq:WdW}) has \textit{zero} in the place of the time derivative. This absence of time in the ``dynamical'' equation of the theory is the primary form of the problem of time. 
\paragraph*{}
A second form of the problem of time, which arises even at the classical level of GR, is the \textit{problem of change}. In the context of quantum geometrodynamics, it comes about because, as mentioned above, the constraints generate the gauge symmetries of the theory, and all physical observables (being gauge-invariant quantities) described by the theory must satisfy the constraints. For an observable to satisfy (\ref{eq:WdW}), however, it must be a ``constant of the motion''---conserved through all (gauge) motions and thus unchanging over ``time''. The second form of the problem of time, then, is that all physical observables do not change: the dynamics of the theory is ``frozen''. Any change described by the theory is only a gauge redundancy, an artefact of the mathematical description rather than a reflection of the physics.

\paragraph*{}
It must be emphasised, however, that the Wheeler-deWitt equation is not only hard to make sense of in this respect, but that there are additional interpretative difficulties regarding the quantum state $|\psi\rangle$ (which will be briefly touched on in \S\ref{subsub:deco}). Furthermore, (\ref{eq:WdW}) is difficult to deal with mathematically as well---apparently having no meaningful solutions. Yet, we cannot just dismiss the Wheeler-deWitt equation as meaningless, and nor can we ignore the problem of time. Quantum geometrodynamics represents the most straightforward attempt at formulating a quantum theory of gravity; it uses standard, proven methods to quantise GR. To simply stand back and declare that these methods must be inapplicable in this particular case, given the problems with (\ref{eq:WdW}) would be unjustified. 
\paragraph*{}
If quantum gravity is construed as a theory that combines quantum theory and GR, then quantum geometrodynamics will be of interest because it reveals something about this very combination. (Also, again, it should be noted that at least one form of the problem of time is already part of the interpretation of GR itself, and this takes on new significance when considered in the context of the quantum version of the theory). Thus, we may be tempted to take the problem of time as evidence that time will not appear in quantum gravity.\footnote{There is an additional problem of time in quantum gravity involving the idea of a ``trajectory''. In GR there is no preferred time variable, and time coordinates have no intrinsic meaning (and, since evolving from one slice to another is a symmetry, or a re-labelling, it too has no intrinsic meaning). In classical canonical GR, time is used to parametrise a ``trajectory'' (in configuration space) of the three-geometry $\Sigma$, and this is done in an arbitrary way. The quantum version of this theory has no explicit time parameter, and the trajectories thus ``disappear''. The analogy with the quantum description of a particle is easy to draw (as is done by, for instance, \citet[][p. 171]{Kiefer2000} and \citet[][p. 30]{Rovelli2004}): while a classical description of a particle involves a trajectory, there is no such thing as a trajectory in the quantum description (although, of course, the description of the quantum particle involves a background spacetime, and this is not the case for canonical quantum gravity). }

\paragraph*{}
However, the absence of time in the Wheeler-deWitt equation cannot be interpreted so directly so as to be indicative of the lack of time quantum gravity. This is because quantum gravity needn't be a quantisation of GR.\footnote{\label{foot:QG}For more discussion regarding this point, see \citet{Mattingly2009, Wuthrich2005} and the neat review in \citet{Weinstein2011}.} In this thesis, quantum gravity is taken to be a theory that applies at high-energy and which describes structures ``beyond'' GR. The picture suggested by several of the examples considered herein is that GR is an effective, low-energy approximation to quantum gravity, and according to this conception, quantising the structures of GR will \textit{not} in fact lead us to quantum gravity (\S\ref{sect:EFT}, \ref{sect:GREFT}). According to this view, quantum gravity is not a quantum version of GR---it is a quantum theory that GR approximates in a certain domain, a domain in which the quantum effects of the ``underlying'' theory are able to be neglected. 

\section{The recovery of spacetime}\label{sub:recovery}
As suggested above (\S\ref{sub:QG}), the recovery of spacetime is tied to the recovery of GR (or an approximation to GR, or the metric structure of GR), in accordance with the GCP (\pageref{def:GCP}). While the GCP is a general principle in physics and has often been taken as a criterion of theory acceptance, it acquires unexampled importance in the context of quantum gravity, owing to the lack of empirical evidence available. The motivation for the GCP is the idea that any new theory needs to account for the success of the its predecessor in the domain where the older theory is known to hold. In other words, quantum gravity shouldn't render unexplained anything that GR has already successfully described---the goal is the generation of new knowledge through predictions rather than the un-doing of explication (even if it does demonstrate that however many of the claims that GR makes are only really ``approximate'', in the sense of not being accurate at a ``finer-grained'' level of description).
\paragraph*{}
Moreover, for quantum gravity---being expected to hold at extremely high-energy---the recovery of GR is an important means of ``linking back'' to empirical reality. It is almost a means of making predictions in lieu of actually making (novel) predictions (although, of course, quantum gravity ideally aims at making predictions); thus, the recovery of GR is viewed as a task of great urgency in many of the approaches to quantum gravity.

\paragraph*{}
Conceptually, it is very difficult to imagine how spacetime could be recovered, or emerge, from something non-spatiotemporal. Also, it is difficult technically to demonstrate how a theory of spacetime could be recovered from a quantum gravity theory that describes non-spatiotemporal degrees of freedom. Usual approximations and limiting procedures (including the renormalisation group, which is discussed at length in this thesis), make use of the conception of spacetime. The approach taken in this thesis is to consider the techniques that have been trialled in various quantum gravity approaches toward obtaining a low-energy limit, a spatiotemporal structure or some approximation to GR, and explore these with the hope of making sense of them---or gaining some sliver of conceptual insight into the process, at least. 

\paragraph*{}
While the ``physicists' sense'' of emergence is related to, or even defined by, the GCP, there is the promise that the recovery of spacetime from a theory of non-spatiotemporal degrees of freedom will be interesting to philosophers concerned with emergence: that it will potentially satisfy, or shed light on, the ``philosopher's conception'' of emergence. This suggestion, of exploring the interrelation between physics and philosophy, is taken up in this thesis. Nevertheless, I am concerned primarily with emergence as a relation between theories; when speaking of structures, these are meant as structures of the theories, and degrees of freedom are the degrees of freedom described by the theories. The recovery of spacetime is shorthand for the recovery of a theory of spacetime, or the recovery of an approximation to GR, or the recovery of some spatiotemporal structures, etc.

\section{Two transitions}\label{subsub:trans}
Quantum gravity is not only supposed to be a quantum (in some sense) theory---it is supposed to be a small-scale (high-energy) theory. (Familiar) quantum theory is not itself restricted to certain scales, it is understood as a universal framework in physics.  As is well known, the ``appearance of classicality'', justifying our ability to accurately describe much of our everyday phenomena using classical theories, is to be explained using some additional principle, hypothesis, interpretation or theory. Similarly, as \citet[][p. 168]{Kiefer2000} points out, the smallness of the Planck length should not be used to argue that quantum gravitational effects are small. Although calculations have shown that quantum corrections to GR at low-energies are negligible (\S\ref{sub:GREFT}), an external justification is needed to explain this. 
\paragraph*{}
Thus, the recovery of spacetime from quantum gravity is a two-step process: there is the procedure by which the ``classical appearance'' is recovered, and there is the process of arriving at the large-scale or low-energy limit of the theory, in which we return to known energy scales. In this thesis, the former is known as the quantum/classical transition, while the latter is called the micro/macro transition (terms borrowed from \citet{Hu2009}). 

\paragraph*{}
These two processes (or transitions) are distinct, and may or may not be related to one another. The quantum/classical transition is connected to the \textit{measurement problem} in quantum theory, which is the question of why it is that any measurement on a quantum system finds the system in a definite state even though the system evolves dynamically as a superposition of different states (more detail below, in \S\ref{subsub:deco}). In quantum mechanics, a system's evolution is deterministic and linear according to the Schr\"{o}dinger equation, yet the fact that superpositions are never observed when a measurement is made on a system prompts the postulate of non-linear ``collapse'' of the wavefunction---the measurement problem, then, as framed by Albert (1992), is,
\begin{quotation}
The dynamics and the postulate of collapse are flatly in contradiction with one another [...] The postulate of collapse seems to be right about what happens when we make measurements, and the dynamics seems to be bizarrely \textit{wrong} about what happens when we make measurements, and yet the dynamics seems to be \textit{right} about what happens whenever we \textit{aren't} making measurements. \citep[][p. 72]{Albert1992}
\end{quotation}
\noindent
Thus, one question that is worth asking is how (or if) collapse occurs; there are many different suggestions, related to different interpretations of quantum mechanics.\footnote{For more on the measurement problem, see the introductory-level review provided by \citet{Krips2007} and references therein.} One popular approach is to utilise the idea of decoherence (though decoherence does not, in itself, provide a solution to the measurement problem), which dismisses the postulate of non-linear collapse, and is taken simply as ``standard quantum theory with the environment included''. This is explored shortly, in \S\ref{subsub:deco}. 

\paragraph*{}
Decoherence, as well as the other ways of understanding the non-measurement of superpositions, is supposed to be something that \textit{happens}---an occurrence in time. On the other hand, the processes involved in the micro/macro transition are not supposed to be dynamical---they are theoretical or mathematical procedures used to obtain a low-energy (or large-scale) description of the physics. The micro/macro transition is not something that happens to a system. It may be represented by an approximation procedure, or a  limiting process (the ``continuum limit'') or it may be represented by renormalisation group flow and the other methods of effective field theory, as described in \S \ref{sect:EFT}. 
\paragraph*{}
The idea of treating GR as an effective field theory is focused squarely on this conception of a micro/macro transition: it is a way of understanding how it is that GR can emerge as a low-energy theory from a high-energy theory that describes the ``micro'' constituents of spacetime. Also, it is worth pointing out that there is another sort of transition that may be related to the emergence of spacetime---a phase transition. This \textit{is} a dynamical process (even if instantaneous in some cases). The conceptions of emergence applicable to phase transitions, in particular second-order phase transitions, is explored in \S\ref{sect:Univers}. In the context of quantum gravity and the emergence of spacetime, the phase transition has been termed ``geometrogenesis'', as discussed in \S\ref{sub:pregeo}. 

\paragraph*{}
The importance of considering both the quantum/classical transition as well as the micro/macro transition is recognised by many authors, including \citet[][p. 180]{Kiefer2000} and \citet[][pp. 22--23]{WuthrichForthcoming}, who state that the semiclassical limit of quantum gravity is not sufficient on its own to understand the emergence of the classical behaviour of spacetime. This is because, if the superposition principle holds in quantum gravity, as it is assumed to do, then superposition states of spacetime (or the ``atoms'' of spacetime) will be the generic ones of quantum gravity. \citet{WuthrichForthcoming} uses the work of \citet{Landsman2006}, to argue that a limiting procedure, on its own, will never resolve a quantum superposition. The order in which we consider (or apply) the two transitions may be important, too, depending on the relationship between them (\citet{OritiForthcoming}, for example, argues this point from the perspective of quantum gravity approaches that make use of the idea of geometrogenesis).

\subsection{The small-scale structure of spacetime}\label{subsub:small-scale}
While both the quantum/classical transition and the micro/macro transition are expected to be important in quantum gravity, and the relationship or interplay between the two may be interesting, obviously both need to be studied in some detail and, unfortunately, the spacetime of this thesis is not sufficient. It seems that the two transitions can (at least at this level of generality, i.e. without yet knowing how one may influence the other) be treated independently of one another, and the focus here is on the micro/macro transition. As will be discussed shortly (\S\ref{sub:sand}), many of the quantum gravity approaches considered in this thesis draw inspiration and techniques from condensed matter physics (e.g. the ideas of phase transitions and effective field theories), and these concern the relationship between different energy scales. Similarly, the philosophical conceptions of emergence that I explore are those that apply in these sorts of physical cases.

\paragraph*{}
As stated above, a high-energy theory describing the ``micro-constituents'' of spacetime is perhaps better not even called quantum gravity, given that the conceptions of gravity, or spacetime, will not appear in it. Furthermore, it does not seem as though quantum gravity has to be a theory of a quantised gravitational field.\footnote{See references in Footnote \ref{foot:QG}.} For this reason, \citet{MattinglyForthcoming} prefers the more general term ``micro gravity'' over ``quantum gravity''; the latter which he reserves to refer to only those specific micro gravity approaches that feature a quantised gravitational field. Instead of following this terminology, it should be clear (from the discussion in \S\ref{sub:QG}) that I take ``quantum gravity'' to be the more general term, encompassing all micro gravity theories. The considerations and desires described above as motivations for quantum gravity serve to motivate all types of micro gravity theories.

\subsection{Decoherence}\label{subsub:deco}
The process of \textit{decoherence} provides a way of explaining why it is that we typically describe most of our familiar, macroscopic physical systems using classical, rather than quantum, theories, in spite of the fact that such systems are themselves supposed to be quantum systems. Thus, decoherence is supposed to be a way of explaining how the classical picture \textit{emerges} from quantum theory. If spacetime itself is ultimately a quantum entity, then decoherence might help us understand the emergence of its classical appearance. While I believe that it is likely that decoherence plays an important role in the quantum/classical transition (described above) that is involved in the emergence of spacetime, and yet, I will not spend much time discussing the topic in the rest of this thesis, for reasons that will hopefully be clear by the end of this small subsection.
\paragraph*{}
The central dynamical equation of quantum mechanics is the (time dependent) Schr\"{o}dinger equation,
\begin{equation}\label{eq:Schro}
i\hbar \frac{\partial}{\partial t}|\psi\rangle = \hat{H}|\psi\rangle
\end{equation}
where $\hbar$ is the reduced Planck's constant, $\hat{H}$ is the Hamiltonian operator, and $\psi$ is the wavefunction describing the system. This equation is linear and deterministic: given an initial state of the system, $|\psi\rangle$, we can compute the state at any time $t$. The linearity of (\ref{eq:Schro}) means, however, that if $|\psi_1\rangle$ is a solution and $|\psi_2\rangle$ is a solution, then the superposition state,
\begin{equation}\label{eq:super}
|\psi_3\rangle=\alpha|\psi_1\rangle + \beta|\psi_2\rangle
\end{equation}
is also a solution ($\alpha$ and $\beta$ are coefficients signifying the relative amplitudes, or ``weighting'' of each contribution, the sum of their squares being unity). (\ref{eq:Schro}) tells us that a system described by any initial state, $|\psi\rangle$ will naturally evolve into a superposition. In other words, a superposition is the generic state of a quantum system. Of course, however, superpositions are never directly observed: when we make a measurement on the system, we appear to find it to be in a state associated with a definite value of the property being measured (and superposed states cannot be characterised this way).

\paragraph*{}
The most common way of understanding this phenomenon is via the \textit{Copenhagen interpretation}, which utilises the idea of \textit{wavefunction collapse}.\footnote{Although there may be a variant of the Copenhagen interpretation which makes use of the idea of decoherence rather than wavefunction collapse.} This interpretation states that if we make a measurement on a quantum system in a superposition (\ref{eq:super}), then the superposition is destroyed as the wavefunction $\psi$ collapses into a definite state that depends on what property was measured. A simple example is to imagine a system in a state of superposition where a single particle is localised about two different positions, i.e. the wavefunction is ``peaked'' in two different regions. Performing a measurement in order to determine whether the position of the particle is within a certain region will result in one of the peaks of the wavefunction (the one outside the region in question) collapsing to zero, and the particle being located at a definite position. 

\paragraph*{}
Collapse is instantaneous and irreversible. Interpretations making use of the idea of collapse thus claim a discontinuity: they describe the evolution of a system as smooth and unitary according to (\ref{eq:Schro}), until the act of measurement abruptly collapses the wavefunction. Proponents of decoherence, however, find this disjointed description of a system's evolution unsatisfactory. Rather than an instantaneous, irreversible collapse, the process of decoherence describes how the coherence that characterises pure quantum states becomes heavily suppressed and no longer observable.\footnote{The description of decoherence in this section is based on \citet{Halliwell2005, Joos2003, Zurek1991}.} The process is supposed to describe a strictly unitary interaction, and thus be only \textit{practically} irreversible \citep{Schlosshauer2007}.\footnote{Other authors go on to tie its irreversibility to an increase in entropy and the second law of thermodynamics \citep[e.g.][]{Kiefer2000, Zeh2007, Zurek1991}.} It calls on us to recognise that the vast majority of macro systems are not isolated, but are continually interacting with other entities surrounding them; for instance, a very simple macroscopic system comprising a single dust particle will have molecules (from the air) bouncing off it\footnote{This system is considered in the calculation by \citet{Joos1985}: one of the original papers on decoherence.}, and it will have photons (from sunlight) bouncing off it. Such external entities not being taken as part of the system are called the \textit{environment}. Decoherence says that interactions with the environment are responsible for suppressing the quantum nature of a system. 

\paragraph*{}
In quantum theory, when we need to express our ignorance of the state of the system before measurement, a \textit{density matrix} is the appropriate tool. It describes the probability distribution for the alternative outcomes of a measurement on the system. For a system characterised by $\psi(x)$, the density matrix is,
\begin{equation}\label{eq:dens}
\rho(x,y) = \psi(x)\psi^*(y)
\end{equation}
This density matrix corresponds to a \textit{pure state}, also known as a coherent state:  it corresponds to a system that can only be described by quantum mechanics. On the other hand, a system that involves both quantum and statistical mechanics is called a \textit{mixed state}. The density matrix for a mixed state where we don't know exactly which quantum state the system is in, but we know the probability $p_n$ that the system is in quantum state $\psi_n$ is given by \citep[][p. 98]{Halliwell2005},
\begin{equation}
\rho(x, y)=\sum_n p_n\psi_n(x)\psi_n^*(y)
\end{equation}
For a superposition state, as in (\ref{eq:super}), the density matrix is (with $\alpha=\beta=\frac{1}{\sqrt 2}$, for normalisation),
\begin{equation}\label{eq:sdens}
\rho^c(x,y)=\frac{1}{2}[\psi_1(x)\psi_1^*(y)+\psi_2(x)\psi_2^*(y)+\psi_1(x)\psi_2^*(y)+\psi_2(x)\psi_1^*(y)]
\end{equation}
Where the superscript $c$ stands for ``coherent''. The last two terms represent the interference effects, and it is these that prevent us from saying that the system is either in state $|\psi_1\rangle$ or $|\psi_2\rangle$. In order to describe classical probabilities, these off-diagonal terms (of the density matrix) need to be cancelled to produce the \textit{reduced density matrix},
\begin{equation}\label{eq:rdens}
\rho^r(x,y)=\frac{1}{2}[\psi_1(x)\psi_1^*(y)+\psi_2(x)\psi_2^*(y)]
\end{equation}

\paragraph*{}
The reduced density matrix $\rho^r$ expresses classical ignorance (equating to the statement that the system is either in state $|\psi_1\rangle$ or $|\psi_2\rangle$). Decoherence is the process by which the pure, or coherent, state described by $\rho^c$ becomes the mixed state $\rho^r$. A large system comprising many degrees of freedom is more strongly coupled to its environment than a small system with fewer degrees of freedom, and so decoherence occurs much faster for larger systems. Since macroscopic objects (being systems with a large number of degrees of freedom) are difficult to isolate from their environments, decoherence tells us that the interference effects rapidly ``leak'' out as the system goes from being in a pure state to a mixed state with its environment. In other words, the coherence is ``distributed'' over a large number of degrees of freedom characterising the system-plus-environment, making it unobservable (heavily suppressed) at the level of the system itself. The local suppression of interference is the reason why quantum effects are typically not observed for macroscopic objects.

\paragraph*{}
If we believe that spacetime itself is a quantum entity, then---once we have an account of the micro-degrees of freedom---the idea of decoherence may be thought to play a role in explaining the absence of (detected) quantum effects; the quantum properties of spacetime may be suppressed via decoherence. As explained above (\S\ref{subsub:canon}), quantising spacetime leads to the Wheeler de-Witt equation (\ref{eq:WdW}). Like the Schr\"{o}dinger equation (\ref{eq:Schro}), the Wheeler-deWitt equation is linear and so admits solutions that are superpositions. The aim of decoherence is to explain why these superpositions are not observed in our universe \citep{Padmanabhan1989}. 

\paragraph*{}
Here, we need to introduce an important distinction between \textit{quantum gravity}, which describes spacetime as a quantum system, and \textit{quantum cosmology}, which describes the entire universe as a quantum system. Motivation for quantum cosmology also comes from decoherence and the universality of quantum theory: for instance, \citet[][p. 167]{Kiefer2000} states that if a system is coupled to its environment, and this environment is itself coupled to an environment, then it seems that the only closed system is the entire universe. Understanding decoherence in the context of the entire universe seems immediately problematic for exactly this reason, however, as the universe (by definition) does not exist in an environment---there can be no interactions with external degrees of freedom. Nevertheless, philosophers interested in quantum cosmology have argued that there are ways to define decoherence in this context, and that the idea of decoherence offers a better explanation of the non-appearance of universal superpositions than its alternatives.\footnote{See, e.g. \citet{Kiefer2000, Ridderbos1999, WuthrichForthcoming}.}

\paragraph*{}
While research on decoherence in quantum cosmology often intersects with (or conflates quantum cosmology with) quantum gravity \citep[see, e.g.][]{Craig2010, Halliwell1989, Kiefer2000, Seidewitz2007}, it should be clear that in this thesis I am only concerned with quantum gravity. \citet[][p. 267]{Rickles2008a} demonstrates the logical independence of quantum cosmology and quantum gravity by pointing out that the fact that our universe contains gravity is \textit{contingent}. There could be another universe in which gravity is not present, and, although it would be a very different type of universe, we could still talk about quantum cosmology in that universe, even though gravity does not exist.\footnote{Also, quantum gravity is usually taken to include a kind of vacuum state that would, at least in some sense, correspond to a world without gravity. To emphasise: quantum gravity and quantum cosmology, while distinct fields of research, certainly (and very importantly) inform one another.}

\paragraph*{}
Another difference between research in quantum cosmology compared to quantum gravity is that researchers in quantum cosmology who are interested in decoherence will typically consider the very early universe, when decoherence is thought to have occurred. By contrast, in quantum gravity we are interested in the micro-structure of spacetime as it is at present. Yet, if decoherence of spacetime occurred soon after the birth of the universe, it might seem odd that decoherence could still be of interest in quantum gravity. This concern is misplaced, however, as it overlooks one of the assets of decoherence compared to collapse interpretations---decoherence doesn't ``turn'' a quantum system into a classical one, instead the coherence ``dissipates'' over a large number of degrees of freedom. The interference effects might still be relevant at extremely small distance scales---say, the Planck scale. Also, we might expect that if we were to appropriately isolate some quantum gravitational system (or micro-spacetime system) that the quantum properties (i.e. interference effects) would be manifest. Decoherence would then explain why it is that such interference effects aren't (relevantly) present in non-isolated quantum-gravitational systems.\footnote{It is worth pointing out, too, that the relationship between decoherence and quantum gravity is not limited to the use of decoherence in attempts to recover the appearance of spacetime. Quantum gravity may help reveal new insights about decoherence: it might demonstrate that certain interpretations of quantum theory are better than others \citep[as in e.g.][]{Ridderbos1999, Singh2009}, and perhaps gravity itself is involved in the decoherence of quantum systems that aren't exclusively quantum gravitational \citet{Gambini2004, Kok2003}.}

\paragraph*{}
To summarise: I believe that in order to understand the emergence of spacetime from quantum gravity we require both a quantum/classical transition as well as a micro/macro transition. Although it is possible that the two transitions be somehow intertwined, they are certainly conceptually distinct and can therefore be discussed (more or less) independently of one another. It may be that decoherence is involved in the quantum/classical transition---yet it seems plausible that just as we can explore the relationships between known theories in physics (even quantum theories) and their low-energy limits without needing to invoke the idea of decoherence (or any other interpretation of the quantum/classical transition), so too we can explore the relationship between (a theory of) spacetime and (a theory of) its micro-structure.\footnote{\citet{Anastopoulos2008, Anastopoulos2007} make a similar point, though perhaps do so by downplaying the significance of the role of decoherence.}

\paragraph*{}
It seems preferable to investigate the emergence of spacetime in quantum gravity independently of decoherence. One reason for this is that if quantum gravity is understood as a theory of the micro-degrees of freedom of spacetime---degrees of freedom that are not spatiotemporal---then an attempt to recover a theory of spacetime from quantum gravity using the idea of decoherence seems difficult to implement. This is because decoherence is a dynamical process: it occurs \textit{in time}. Rather than engage with this difficulty, we can make a start in studying the micro/macro transition unclouded by the additional complexity of the quantum/classical transition. Furthermore, it seems that it is only once we are equipped with a proper solution to the measurement problem that we will be able to understand the emergence of classicality (in any context), and solving the measurement problem is clearly beyond the scope of this thesis.\footnote{This thesis is not entirely neglectful of ideas related to the quantum/classical transition, however---there are brief discussions in some of the chapters where such ideas are of particular interest or importance, including \S\ref{sect:Discrete} and \S\ref{sect:LQG}.}

\section{Emergent gravity}\label{sub:chal}
Speaking of the recovery of GR as a low-energy limit of quantum gravity means more than just the emergence of spacetime. As \citet{Carlip2012} highlights, GR is not exhausted by the structures we identify as spacetime; the principles of GR, insofar as they are necessary for the metric formulation of GR, serve as constraints on an effective theory of gravity. \citet{Carlip2012} is interested in outlining some general principles that are necessary for a model of emergent gravity---these are the key principles of GR, together with some additional requirements that make natural an emergent metric structure---and expounding the restrictions on the model that then follow from these. He does this in order to demonstrate how these conditions create obstacles that any model of emergent gravity must overcome if it is to reproduce the predictions of GR in the appropriate domain. 

\paragraph*{}
To make his argument, Carlip starts with the physical basis for treating GR as a metric theory. This comes as a result of combining several principles, including local Lorentz invariance, which implies the existence of a field of light cones and thus establishes a causal structure, topology, and a conformal structure: an equivalence class of metrics (that differ only by local rescalings) for which the paths of light rays are null geodesics. Next, the equivalence principle (understood as the universality of free fall) determines a set of preferred paths in spacetime (being the trajectories of freely falling objects), which gives us an equivalence class of affine connections (those for which the preferred paths are geodesics). 
\paragraph*{}
The compatibility between these principles leads to the fixing of a Weyl structure, an equivalence class of conformal metrics and affine connections in such a way that it would lead to an effect (the ``second clock effect'') that is not in fact observed in our world. Thus, a further condition is imposed to eliminate the appearance of this effect (there are several options to choose from). Together, these conditions provide the kinematic setting of GR---implying that motion in a gravitational field can be described as geodesic motion in a Lorentzian spacetime, with a metric providing a full description of the field.

\paragraph*{}
As Carlip explains, there are several different ways of then obtaining the dynamics of GR. One route is to recognise that the absence of any non-dynamical background structures in GR implies general covariance, in the sense that gravity be described by diffeomorphism invariant expressions involving only the metric and other dynamical fields. Then, in order to eliminate the possibility of additional dynamical fields mediating between matter and the metric, we impose the condition that any non-metric degrees of freedom be decoupled from gravitational dynamics---meaning that the gravitational effective action should depend on the metric alone. Finally, the methods of effective field theory can then be used to formulate the action. At low-energies, the effective action includes all interactions allowed by the symmetries of GR and will match the Einstein-Hilbert action. These are the steps required in order to construct an effective (emergent) theory of gravity.

\paragraph*{}
The restrictions that \citet{Carlip2012} presents as consequences of these conditions are rather severe; he concludes that the key principles of GR---local Lorentz invariance, the equivalence principle, diffeomorphism invariance and background independence---are ``not easy to mock up''. Rather, because all these features are intertwined, he argues, it seems a successful theory of emergent gravity might require some more fundamental guiding principle---one which we are yet to uncover. One of the restrictions, or ``challenges'' for emergent gravity that Carlip presents is the Weinberg-Witten theorem, discussed below (\S\ref{subsub:WW}); the others are not dealt with generally here, but are important in discussing the discrete approaches to quantum gravity (\S\ref{sect:Discrete}).

\subsection{Weinberg-Witten theorem}\label{subsub:WW}
One constraint that is worth briefly mentioning is the Weinberg-Witten theorem, since it is often taken as presenting an insurmountable obstacle for the formulation of emergent gravity \citep{Weinberg1980}. If gravity is to be treated as a QFT, then its associated particle, the \textit{graviton}, is massless and of spin-2. Consider $|p>$ and $|p'>$, being one-particle, spin-2, massless states labelled by their 4-momenta, and having the same Lorentz-invariant helicity $\pm 2$. If $T^{\mu\nu}$ is a Lorentz covariant, conserved current, and (hence) the matrix elements $<p'|T^{\mu\nu}|p>$ Lorentz covariant, then the Weinberg-Witten theorem states,
\begin{equation}\label{eq:WW}
\lim_{p' \to p} <p'|T^{\mu\nu}|p>=0
\end{equation}
The case of importance, of course, is where $T^{\mu\nu}$ is the stress-energy tensor, and in this case (\ref{eq:WW}) states that the graviton cannot carry observable energy or momentum.\footnote{This presentation is based on \citet{Barcelo2011, Carlip2012, Jenkins2009}.} Thus, naively, this theorem seems to rule out any theory, including GR, in which the gravitational field carries energy.
\paragraph*{}
Obviously, GR gets around the Weinberg-Witten theorem, and it does so thanks to a subtle interaction between gauge invariance and Lorentz invariance that means the matrix elements $<p'|T^{\mu\nu}|p>$ are non-covariant.\footnote{For details, see \citet{Carlip2012}.} There are other means of avoiding getting caught by the Weinberg-Witten theorem; in particular, a theory will get around the theorem if it:
\begin{enumerate}
\item[(a)] lacks a stress-energy operator, or
\item[(b)] has non-relativistic gravitons, or
\item[(c)] has emergent (effective) gravitons, with emergent gauge invariance, propagating in an effective spacetime distinct from the background spacetime.
\end{enumerate}
The emergent (effective) gravitons must exist in an effective spacetime, since a spin-2 gauge invariance in the background spacetime would prevent the gravitational energy from being locally observable. This would be a problem given that an array of separated mass scales is a requirement for formulating an effective field theory (as will be discussed in \S\ref{subsub:EFTformalism}). First-quantised string theory falls under category (a), since $T^{\mu\nu}$ is unable to be defined.\footnote{There is no consistent, off-shell definition of the string action $S$ in the background spacetime with metric $g_{ab}$, so the object $T^{ab}=\frac{1}{\sqrt{-g}}\frac{\delta S}{\delta g_{ab}}$ is undefined (Jenkins, 2009, p. 4).} The AdS/CFT correspondence (discussed below, \S\ref{subsub:dualities}) falls under category (c).
\paragraph*{}
Many of the approaches to quantum gravity that are inspired by condensed matter physics feature emergent gravitons and fall under category (b). In the case of the general condensed matter approaches discussed in \S\ref{sect:GREFT}, there are no emergent gravitons, but only an effective geometry (a curved Lorentzian metric), and so the Weinberg-Witten theorem has nothing to say. However, for the condensed matter approaches in  \S\ref{sub:analogue} that do consider gravitons (i.e. quantum fluctuations of the effective geometry), namely Volovik's superfluid models, the escape route is via (c). Still, those models in which the dynamics is implemented via the inclusion of quantum effects (along the lines of Sakarov's ``induced gravity'') are explicitly excluded from the Weinberg-Witten theorem, which ``clearly does not apply to theories in which the gravitational field is a basic degree of freedom but the Einstein action is induced by quantum effects'' (Weinberg \& Witten, 1980, p. 61). This is also the case in several of the discrete approaches considered in \S\ref{sect:Discrete}.

\subsection{AdS/CFT duality}\label{subsub:dualities}
A \textit{duality} is an (exact or approximate) physical equivalence, which may hold between two different theories (featuring very different structures), or between two different regions of the parameter space of a single theory, e.g. a gauge freedom. The existence of dualities holding between theories with different pictures of spacetime has led to the suggestion that spacetime may not be fundamental (for instance, this claim is made by \citet{Seiberg2007}). 
\paragraph*{}
Perhaps the most significant example of such a duality is the \textit{AdS/CFT duality}, also known as the \textit{Maldacena conjecture}. This duality comes from superstring theory, which describes the fundamental constituents of matter as extended one-dimensional objects (strings) propagating on a ten-dimensional background spacetime. The strings can be open (i.e. lines), in which case they correspond to gauge particles, or closed (with ends joined to form a loop), in which case they correspond to gravitons.\footnote{The \textit{graviton} is the hypothetical particle that mediates the force of gravity, when gravity is treated in the framework of quantum field theory.} The AdS/CFT duality is an exact physical equivalence between a theory which features gravity and a Yang-Mills theory in which gravity does not feature among the fundamental degrees of freedom; hence, the AdS/CFT duality is also called a gauge/gravity duality.\footnote{The description of the AdS/CFT in this section is based on \citet{Rickles2013, Horowitz2009}. See also \citet{Klebanov1998}.}

\paragraph*{}
The AdS/CFT correspondence is a concrete example of the \textit{holographic principle}, which is a more general idea that has developed from concerns and suggestions in a number of different areas, including black hole physics as well as string theory and other approaches to quantum gravity.\footnote{\citet{Bousso2002} provides a review. For more discussion on the holographic principle as a principle of quantum gravity, see, e.g. \citet{Bigatti2000, SierokaForthcoming}.} Its enunciation in regards to string theory came from \citet{Susskind1995}, owing to earlier ideas from \citet{Bekenstein1973}, \citet{tHooft1993} and \cite{Thorn1992}. The holographic principle states that, for a gravitational theory defined over some region of spacetime, called the \textit{bulk}, a complete description of the physics can be provided by a theory defined on the \textit{boundary} that contains the bulk spacetime. The boundary will be of lower dimension than the bulk region of spacetime, for instance a two-dimensional surface bounding a three-dimensional spacetime region.
\paragraph*{}
The holographic principle is embodied in the AdS/CFT duality, since the gravitational theory involved describes closed strings propagating on a spacetime, while the physically equivalent gauge theory is defined on the boundary of this spacetime. This equivalence was first presented as a conjecture by \citet{Maldacena1998}. More technically, it states that the physically observable properties of a particular string theory in anti-de Sitter space (AdS) are equivalent to those of a particular conformal field theory (CFT) defined on the (conformal) boundary. The claim that dual theories are physically equivalent, means there exists a bijection between states, operators, and correlation functions of the two theories.\footnote{Note: this is a restricted definition of what it means to have dual theories. Different, or more general, definitions may be used in other cases.} Symmetries should also be preserved across the dual theories\footnote{Though the theories can have different gauge redundancies, since the gravitational theory is diffeomorphism invariant while the gauge theory is not.}, and \citet[][footnote 10]{Rickles2013} points out that this fact can aid in identifying genuine duality claims.

\paragraph*{}
Even though the relationship between the two theories is symmetric, the duality is generally interpreted as implying that the gauge theory on the boundary is fundamental, while the higher-dimensional bulk spacetime is \textit{emergent} from it. On this interpretation, one spacetime emerges from another spacetime (of lower dimension), and so it does not mean that spacetime itself is not fundamental, only that a particular type of spacetime is not fundamental. Although the dual theories may (depending on your definition of theoretical equivalence) be thought of as \textit{different theories}, a compelling case can be made for considering the relationship between them as akin to a gauge transformation, or a change of variables in a single theory---we can imagine that both theories describe the same physical situation using different concepts \citep{Horowitz2009}. 

\paragraph*{}
If this interpretation is adopted, then the claim of emergence must be re-evaluated in turn. \citet{Horowitz2009} suggest that we can still understand the gravitational theory as emergent, on the grounds that it is only approximately understood, whereas the gauge theory is exactly understood, and in that sense provides a better explanation of the physics.\footnote{The idea that emergent ``higher level'' theories are approximations to more ``fine grained'' or fundamental descriptions will be explored (and debunked) in \S\ref{sect:EFT}.} 

\paragraph*{}
\citet{Rickles2013} argues that this position stands at odds with the suggested interpretation of the relationship between the dual theories (being akin to a change of coordinates); and I agree also with his claim that this interpretation is the preferable one given the physical equivalence of the theories in question. In defending the interpretation that the dual theories represent the same physics, \citet[][p. 8]{Rickles2013} appeals not only to the existence of the duality mapping (by which one theory is mapped to the other), but also to the fact that the theories share the same symmetries. 
\paragraph*{}
These shared symmetries represent a \textit{deeper structure} which underlies the two representations, and which will only be fully revealed by yet another theory---one that would be more suited to being called ``fundamental'' by those  inclined to use the term. If we accept this interpretation, then, it seems that we might again have a basis for the claim that the AdS/CFT duality implies emergent spacetime---in fact, we might say that it implies two different emergent spacetimes (as approximations, or representations, of the more fundamental physics).

\section{The world in a grain of sand}\label{sub:sand}
A theme running throughout this thesis is what Nambu, in his 2008 Nobel Prize speech, referred to as the ``cross-fertilisation'' of high-energy physics and condensed matter physics \citep{Nambu2008}. The cross-fertilisation considered in this thesis spreads even further, though, as we move into the realm of quantum gravity, and draw inspiration (as well as techniques) from not only particle physics and condensed matter physics, but statistical physics and thermodynamics as well. This thesis represents, in its approach, an attempt at cross-fertilisation between physics and philosophy, exploring what we can learn of the notion of emergence by considering its use in both domains, and hopefully, in doing so, enriching the philosophy of emergence as well as moving toward understanding the emergence of spacetime in quantum gravity.

\paragraph*{}
As described in \S\ref{sub:andersonwein}, the idea of emergence in physics that I am interested in gained the attention of philosophers following Anderson's ``More is Different'' \citeyearpar{Anderson1972}, which argued that there are different ``levels'' in science, defined by the energy scales at which different theories or descriptions are applicable; and one level is no less fundamental, or important, than another simply by virtue of its applying at higher-energy. According to \citet{Anderson1972}, a small-scale description of a physical system is not usefully applied at large-scales, because new degrees of freedom emerge at large-scales---describing the emergent phenomena requires new laws and new concepts not featured in the higher-energy theory and not easily obtained from the higher-energy theory. 
\paragraph*{}
Anderson's views are well-supported by the examples considered in this thesis. We find that low-energy theories are not only importantly novel compared to the high-energy description (having different equations of motion and describing very different degrees of freedom), but also robust under changes in the high-energy physics---they depend, for the most part, only minimally on the small-scale happenings at the ``level below'' (in length scale). 

\paragraph*{}
Yet, in spite of this, we find that there is much that is ``carried over'' between levels. The techniques of the renormalisation group and the formalism of effective field theory were developed in quantum field theory, then shown to usefully apply in condensed matter physics. Nowadays, the renormalisation group is used in fluid mechanics, nanotechnology and cosmology (and, in this thesis, the ideas of the renormalisation group and effective field theory are explored in the context of quantum gravity). The universe at large, it seems, shares some features (even if only structural features) in common with the world at the smallest scales. 
\paragraph*{}
What is particularly interesting about the renormalisation group is that it explains its own success across these different levels with their very different laws and concepts. As described in (\S\ref{sect:EFT}, \ref{sect:Univers}), the renormalisation group reveals that only minimal aspects of the high-energy physics need to be accounted for at low-energies, and that the means of ``taking into account'' these interactions disguises their origins in terms of the robust low-energy degrees of freedom.

\paragraph*{}
The approaches to quantum gravity discussed in this thesis not only utilise the ideas of the renormalisation group and effective field theory, but also draw inspiration, techniques and principles from statistical physics, thermodynamics and hydrodynamics. Again, the aspects that are drawn out and borrowed owe their power to the way they ``pick out'' certain features of the high-energy physics and demonstrate how these translate, very generally, to low-energy phenomena: they do so in a way that leaves the details of the high-energy system \textit{underdetermined} by the low-energy physics. In other words, the high-energy system contains far more information than is required in order to explain the \textit{emergent} features of the low-energy physics.\footnote{I apologise for the vagueness of these remarks, but I promise they will be explained and (hopefully) made concrete in the body of this thesis!} 

\paragraph*{} 
There are tantalisingly strong analogies between quantum field theory and condensed matter physics, and exploring how these might extend into quantum gravity is interesting as well as exciting. \citet{Volovik2003} demonstrates how we might understand the standard model of particle physics as emergent, at low-energies, from superfluid helium, and, as discussed in \S\ref{sect:GREFT}, other condensed matter approaches can produce the effective curved spacetime (metric) of GR. Yet, in attempting to make sense of the success of these analogies, in order to potentially learn something of the unknown high-energy realm, it is frustrating to understand the reason for their success as also being a tremendous hinderance to our gaining insight into the details of this realm. 

\section{Synopsis}\label{sub:syn}
\textit{Chapter \ref{sect:Emergence}: Emergence}
\\The relationship between spacetime (or GR) and quantum gravity has often been called ``emergence'', yet the term is notoriously ill-defined. In light of this, the second chapter of this thesis is dedicated to exploring the meaning of the term in philosophy and in physics. Rather than canvassing the many different uses of ``emergence'' in philosophy, the focus is on understanding emergence as it is used to describe an inter-theoretic relation in the philosophy of science and the philosophy of physics. Accounts of emergence in philosophy typically appeal to the ideas of reduction and derivation; after explaining this, I outline the difficulties with such accounts and begin to give an indication of why I do not find these ideas useful (at least not for the current project). 

\paragraph*{}
Following this, I introduce the ``physicists' debate'', between Steven Weinberg and Philip Anderson (and others) which was responsible for sparking much of the recent curiosity in the topic of emergence in the philosophy of science. While I do not engage with the debate directly, it is of interest because the physical examples that were appealed to in making the claims of emergence are essentially the same as those that I am concerned with. Having already attempted to discourage thinking about emergence in terms of reduction, I present some indication of how we might understand emergence without appeal to reduction. This alternative sense of emergence is based in the \textit{novelty} and \textit{autonomy} of the emergent theory given the theory it emerges from. After explaining this, I briefly look at how the term ``emergence'' is used in physics; finally, I consider the idea of fundamentality (i.e. what it means for a theory to be fundamental) in regards to the new conception of emergence being advocated.

\paragraph*{}
\textit{Chapter \ref{sect:EFT}: Effective field theory}
\\If we take quantum gravity to represent a micro-theory of spacetime, and we are to uphold the GCP, then GR is to be understood as an \textit{effective} theory, meaning it is supposed to be valid only at length-scales that are large compared to the characteristic length-scale of quantum gravity (perhaps the Planck length). The framework of \textit{effective field theory} is a means of formalising this idea for a certain class of theories under certain assumptions. An effective field theory (EFT) may be said to be \textit{emergent} from the micro-theory that ``underlies'' it. The idea of emergence in EFT is non-standard and has been controversial in the philosophy of physics literature, so the purpose of this chapter is to explore the idea of emergence in EFT independently of quantum gravity. 
\paragraph*{}
The chapter comprises two parts: in the first, I provide a basic introduction to EFT and its development, which stemmed from  ``the problem of renormalisation'' in QFT and the discovery of the renormalisation group (RG). The second part of the chapter deals with the philosophy of EFT, including the idea of emergence. Because much of the philosophy of EFT has centred around some controversial claims made in the presentation by \citet{Cao1993}, I begin by examining these. I argue that the controversy stems from a confusion between EFT as it applies \textit{in principle} and EFT as it actually applies \textit{in practice}. 
\paragraph*{}
I then go on to propound a philosophy of EFT based in how the formalism applies in practice, emphasising that we are not justified in asserting that many of the ``in principle'' claims hold true in physics---especially when it comes to inaccessibly high-energy scales. I explain how this view has consequences for our understanding of QFT. Finally, I present an account of emergence in EFT compatible with this philosophy. Owing to a subtlety in the necessity of EFT, I find that a conception of emergence defined by reduction is uninteresting (or inapplicable) in this case. The appropriate account is one based simply on the novelty and autonomy of an EFT compared to its high-energy theory.

\paragraph*{}
\textit{Chapter \ref{sect:Univers}: Universality, higher-organising principles and emergence}
\\ Inspired by two recent accounts of emergence in physics \citep{Batterman2011, Morrison2012} associated with phase transitions and critical phenomena (in which the RG plays a crucial role), I turn to explore the conception of emergence applicable in these physical examples. This is significant given that some of the approaches to quantum gravity imply, or claim, that spacetime geometry emerges following a phase transition. There are bases for emergence, I find, in the ideas of \textit{universality} and \textit{higher-organising principles}. 
\paragraph*{}
Universality in critical phenomena is the fact that large classes of different systems exhibit the same behaviour when undergoing a second-order phase transition. This idea is essentially related to that of \textit{multiple realisability}, which has itself featured in philosophical discussions of emergence, so the chapter begins with a brief discussion of the relationship between these ideas. Following this, the idea of a higher-organising principle is introduced, with symmetry-breaking in phase transitions presented as an example.

\paragraph*{}
The case of critical phenomena is then outlined, and I argue that we can associate a conception of emergence both with the idea of symmetry-breaking as a higher-organising principle and with the idea of universality associated with fixed points in the RG flow (i.e. they each furnish an account of emergence). Following this, I consider the conceptions of emergence of \citet{Batterman2011} and \citet{Morrison2012}. Although each account has its own nuances---with Batterman emphasising the role of limiting relations and mathematical singularities, and Morrison stressing the importance of symmetry breaking in addition to the RG-based explanation---I argue that the interesting aspects of emergence are actually provided simply by the idea of universality. 

\paragraph*{}
The idea of universality is tied to several other examples of emergence in physics, including EFT and hydrodynamics, and so I explore the relationship between these as well. Such investigation is important given the suggestions (explained later in \S\ref{sub:asygrav}) that gravity is asymptotically safe (meaning it is represented by a fixed point in the RG flow and has an associated notion of universality), as well as other claims that GR is analogous to hydrodynamics (these are examined in \S\ref{sect:GREFT}, \ref{sect:Discrete}).

\paragraph*{}
\textit{Chapter \ref{sect:GREFT}: Spacetime as described by EFT}
\\ This chapter examines the possibility of treating GR as an EFT, and, drawing from the ideas presented in the previous three chapters, \S\ref{sect:Emergence}--\ref{sect:Univers}, explores what we might learn of emergent spacetime through the framework of EFT. The idea of treating GR as an EFT is natural not only from the acceptance of the GCP (as argued above), but because of the desire for unification that the search for quantum gravity represents. This desire for unification leads us to attempt to incorporate gravity into the framework of QFT and treat it as we do other fields. There are two perspectives from which we can approach GR as an EFT---``top-down'', where we start with a high-energy theory and attempt to recover GR as a low-energy EFT, and ``bottom-up'', where we start with GR as the low-energy EFT and seek to discover the micro-theory---and I look at examples of each in this chapter.
\paragraph*{}
Firstly, I consider examples of ``analogue models of GR'' which present the metric structure of GR (described by an EFT of quasi-particles) emergent from a condensed matter system. These provide concrete examples of spacetime emergent as the low-energy collective excitations of very different micro-degrees of freedom. I explain how (and to what extent) these models illustrate the conception of emergence (as novelty and autonomy) in EFT outlined in the previous chapters. Interestingly, these models provide us with emergent spacetime, rather than emergent GR. I also argue that, in accordance with the philosophy of EFT propounded in \S\ref{sect:EFT}, we should be wary of drawing too much from the analogy between condensed matter physics and QFT. 
\paragraph*{}
Secondly, I look at examples of the bottom-up approach to GR as an EFT. I again argue that, due to the conception of emergence suggested by EFT, we are restricted in how much we can draw from these theories. I finish the chapter by outlining the asymptotic safety scenario, which is an important conjecture that comes from treating GR in the same way we treat other QFTs, and relates not only to the idea of a fixed point, but has inspired many different ``discrete'' or ``background-independent'' approaches to quantum gravity, some of which will be discussed in the next chapter.

\paragraph*{}
\textit{Chapter \ref{sect:Discrete}: Discrete approaches to quantum gravity}. 
\\In this chapter I consider several examples of ``discrete'' approaches to quantum gravity---including causal set theory, causal dynamical triangulations, quantum graphity and quantum causal histories. These are approaches that describe discrete basic elements that, in some sense, constitute spacetime at high energy. In many of the approaches, spacetime is conceived of as an effective, low-energy manifestation of very different high-energy degrees of freedom. Some of the theories thus draw inspiration from the techniques used in condensed matter physics.
\paragraph*{}
After briefly outlining each approach, I examine the means by which each attempts to recover spacetime (and/or GR) as well as the potential bases for emergence that they present. I find that the conceptions of emergence that these approaches suggest are similar to those already considered in this thesis. Interestingly, many of them also provide evidence of a phase transition, and, by analogy with the conceptions of emergence explored in the previous chapter \S\ref{sect:Univers}, may provide examples of diachronic novelty as well as autonomy. 

\paragraph*{}
\textit{Chapter \ref{sect:LQG}: Loop quantum gravity}
\\ Loop quantum gravity (LQG) is one of the most well-established quantum gravity programs (along with string theory). Proponents of LQG hold that the most important lesson of GR is that the gravitational field is diffeomorphism invariant, and so LQG seeks to preserve diffeomorphism invariance at the high-energy level of quantum gravity. Like the discrete approaches, LQG describes the small-scale structure of spacetime as being discrete. However, the proponents of LQG claim that, rather than being a \textit{postulate} of the theory (as it is in the discrete approaches considered in the previous chapter), fundamental discreteness is a \textit{prediction} of the theory. It is not clear that this is indeed the case, though, because the discrete operators described by LQG are not physical observables as they stand. 
\paragraph*{}
This chapter is concerned with the conception of spacetime described by LQG: I explore the micro-structure of space, as well as that of spacetime, suggested by the theory. I then consider the semiclassical limit and the attempts to recover spacetime in LQG, before discussing the potential bases for emergence in the theory.

%% file: Emerg.tex
\chapter{Emergence and reduction}\label{sect:Emergence}
\section{Introduction}
My aim in this project is to look at the relation of spacetime to the high-energy theory we believe underlies it. This relation has often been termed ``emergence'', but there are a variety of different things going on in the different examples of approaches to quantum gravity. In this chapter I look at the philosophical literature on emergence in order to see how the term applies and decide whether it is a potentially useful one to appeal to in making sense of the emergence-claims in the physical examples. I find that, although the word ``emergence'' has many different uses, it is preferable that I do not, in fact, begin by framing my thesis in terms of it.  
\paragraph*{}
Instead, as I explain, the better option is that I explore the different relations in the physical examples on their own merits rather than as candidates for a relation of emergence according to any particular philosophical conception. Of course, this does not mean that I reject any possible correspondence or comparison of these relations with other philosophers' conceptions of emergence; rather, given the many different conceptions of emergence available, remaining neutral as to the applicability of the term should facilitate a clear and open discussion.
\paragraph*{}
If I were to begin with an aim of interpreting the physical examples in terms of emergence, I fear that this would place me in a tight spot. On the one hand, I would need to capture enough of what philosophers take to be important in defining a conception of emergence for the concept to still be understandable as emergence, but, on the other hand, I would not want to be so tied to a prior conception of emergence that we'd be able to learn nothing significant from the physical examples (other than to what extent they may be said to embody the conception of emergence we begin with). The difficulty would be compounded by the fact that it is very hard to provide a general definition of emergence (as we shall see); indeed, the most sensible philosophical stance is to admit that there is no single ``best'' definition of emergence that applies across the board, but rather many different conceptions, each with its own advantages and domain of applicability. 
\paragraph*{}
Even though ``emergence'' is so widely-used as to elude a precise general definition, it is typically associated with the ideas of supervenience and reduction---and I find that these ideas are not straightforwardly applicable in the context of the physical examples I look at. Also, at the heart of emergence-discussions in philosophy lies a distinction between ontological- and epistemological-emergence, and this forces us to focus on the ideas of deduction and derivability, which, I argue, is an unhelpful (or at least uninteresting) focus to take when considering the physical examples.
\paragraph*{}
The conclusions I arrive at in this chapter are not completely alien. \citet{Butterfield1999}, after a thorough exploration of three different candidate relations for emergence (reduction understood as definitional extension, and supervenience), find a ``heterogeneous picture of emergence'', and suggest that it is best to bear in mind the variety of different ways in which theories may be related (with particular emphasis on limits and approximations), rather than seek a general definition of the term. More recently, \citet{Butterfield2011a, Butterfield2011b, ButterfieldForthcoming} argue, using several physical examples, that emergence is independent of both reduction and supervenience. Finally, \citet[][p. 637]{Silberstein2012}, in reviewing three large recent edited collections of articles on the topic, states that different cases require different conceptions of emergence, and that it is absurd for philosophers to try and argue otherwise.
\paragraph*{}
Butterfield (2011a, p. 924), also endorses pluralism: admitting that he is ``not an essentialist when it comes to how to use `emergence'". By focusing primarily on the physical examples themselves and exploring the relations of emergence they suggest, rather than beginning with a precise definition of the term, I am taking the ``science-first approach''. In doing so, I am following Silberstein (2012, p. 638), who promotes this approach as the antidote to the current trend in the literature on emergence, which, he says, consists in ``cross-talk between a bewildering variety of analytic metaphysicians and philosophers of science, but not necessarily those practicing the science-first method''.
\paragraph*{}
This chapter proceeds as follows: I begin by attempting to give some indication of what philosophers mean by ``emergence'', and, in doing so, reveal emergence-talk as a vast and thorny thicket. Next, I outline the difficulties with typical accounts of emergence, being those that are somehow linked to reduction and/or derivation, and explain that the tendency to think of emergence as a failure of reduction (or derivation) is related to the desire to classify cases of emergence as either ontological or epistemological. I argue that we are better to consider the science first, rather than immediately getting tangled up in questions regarding how our theories relate to the world.
\paragraph*{}
I then introduce the ``physicists' debate'' that sparked much of the recent interest in the topic of emergence, at least in philosophy of science. I am not so interested in examining the debate itself, but the physical examples that were appealed to in making the claims of emergence are very similar to (or even representative of) those that I am concerned with; as is shown in later chapters, these physical cases exemplify some interesting aspects of emergence and other inter-theory relations. Most importantly, the lessons that can be drawn from them are also applicable to modern approaches to spacetime and quantum gravity. I fear that much of what is interesting about these cases (i.e. those that are involved in the ``physicists' debate'', not quantum gravity) has been neglected precisely because it does not match-up to philosophers' prior conceptions of what counts as emergence. 
\paragraph*{}
Following this, I attempt to give some characterisation of the features of the positive relations or aspects of emergence that I feel are more useful than the treatment of emergence as failure of reduction in some sense, and then I explain how the term ``emergence'' is typically used by physicists (or, at least, how I remember my physics lecturers used the term). Finally, I outline the implications of this discussion for the conception of fundamentality, i.e. what it means for a theory to be \textit{fundamental}.

\section{Emergence}\label{sub:Eemergence}
In philosophy, the topic of emergence is currently a very popular one; just some evidence for its ``academic trendiness'' is the recent publication of three large edited collections on emergence \citep{Howhy2008, Bedau2008b, Corradini2010} as reviewed by Silberstein (2012). The vast literature on emergence is almost matched in size, though, by the wildly diverse range of uses (and definitions) of the term itself.  As Silberstein (p. 627), with some apparent vexation, notes: philosophers tend to bristle upon hearing the word `emergence', feeling it ``too multifaceted, vague or ambitious to be coherent''. 
\paragraph*{}
Working from the articles in these edited volumes---which range in subject from the emergence of classical physics from quantum physics, emergence associated with singular limits and phase transitions, emergence of life from chemistry, emergence of embodied cognition, emergence of group cognition, emergence of consciousness, to the emergence of souls---Silberstein (2012) develops a taxonomy of emergence comprising a total of seven different claims. 
\paragraph*{}
The most basic expression of emergence, ``$X$ is emergent with respect to $Y$'', is the idea, very crudely, that $Y$ is some presumably more fundamental property (phenomenon, system, theory, etc.), upon which $X$ depends in some sense and from which $X$ has autonomy in some sense; according to Silberstein, the emergent $X$ is typically understood as being \textit{in some sense} less fundamental than its base $Y$, and \textit{in some sense} not reducible to its base $Y$ 
\paragraph*{}
The most general definition of emergence may thus be taken as comprising two claims. \citet[][p. 375]{Bedau1997} puts them thus (although the labels ``Dependence'' and ``Autonomy" are my additions),
\begin{description}
\item[Dependence (or Linkage)] Emergent phenomena are somehow dependent on, constituted by, generated by, underlying processes. (A less-fraught term is \textit{Linkage}, where we might say that the emergent phenomena is, in some appropriate sense, linked to the underlying system or processes).\footnote{My definition of Dependence is provided on p. \ref{def: mine}, and is the claim that the emergent theory is related to the theory it emerges from via the RG and EFT techniques.}
\item[Autonomy] Emergent phenomena are somehow autonomous from underlying processes.
\end{description}
Typically, the first of these claims, \textit{Dependence}, is relatively uncontroversial, and it is the second claim that serves to distinguish different conceptions of emergence. In the vast majority of cases, \textit{Autonomy} involves ascribing one or more of the following features to the emergent phenomenon (property or theory): irreducibility, unpredictability, causal independence, or unexplainability given its base. In other words, the claim that some phenomenon is emergent is usually understood as the claim that the phenomenon is \textit{in some sense} not reducible to (i.e. deducible from) its base. 

\paragraph*{}
This leads us to the core distinction in the emergence literature. \textit{Ontological emergence} is the thought that this failure of reduction (in whatever sense is meant) is a failure \textit{in principle}: that there are genuinely emergent phenomena (properties, systems, theories, etc.). It is emergence in a strong sense. This stands in contrast to \textit{epistemological emergence}, which is a failure of reduction \textit{in practice}, meaning that the apparent emergence is (somehow) really only an artefact of our computational limitations. Epistemologically emergent phenomena are not genuinely emergent, but, for whatever reason, it is very difficult for us to explain, predict or derive them on the basis of their underlying system(s); as \citet[][p. 186]{Silberstein1999} put it, ``Epistemologically emergent properties are novel only at a level of description.'' This leads to epistemologically emergent properties being termed \textit{predictive} or \textit{explanatory} emergent properties.
\paragraph*{}
There are two general classes that fall under the category of explanatory emergence in Silberstein's 2012 taxonomy, one of the definitions that comes under the first of these classes is Bedau's ``weak emergence'' \citep[see, e.g.][]{Bedau1997, Bedau2002, Bedau2008}. In developing this conception of emergence, Bedau is interested in complexity science, which deals with systems that are extremely sensitive to their initial conditions. Given the micro-details of such a system (including the micro-dynamics), together with the initial conditions plus all other external conditions, the macro-description of the system can be derived but only by \textit{simulation}. This involves inputting a continual stream of successive boundary conditions into the equations governing the micro-dynamics. 

\paragraph*{}
These boundary conditions, Bedau (1997, p. 379) emphasises, are extensively contingent, and derivations that depend on simulations are ``awash with accidental information''. Such derivations are too detailed and unstructured to impart any sort of understanding of the relation between the micro- and macro-levels, and may in fact obscure simpler macro-level explanations of the physics, but, nevertheless, the macro-level description is able to be derived from the micro-level description plus external conditions. This is equated with ``explanatory incompressibility'' (Bedau, 2008).
\paragraph*{}
The second class of explanatory emergence is associated with the representational resources needed to understand some phenomena.\footnote{As we shall see, \ref{sect:EFT} explanatory emergence is one conception of emergence that may be said to apply to effective field theory: for instance, we would, presumably, want to say that the effective theory describing the hadronic states is explanatorily emergent from the ``underlying'' quantum chromodynamics---it is derivable in principle (we believe), but nevertheless necessary for imparting any understanding of the low-energy phenomena.} As Silberstein puts it: 
\begin{quote}
Certain wholes (systems) exhibit features, patterns, behaviors or regularities that cannot be fully represented and understood using the theoretical and representational resources adequate for describing and understanding the features and regularities of their parts and reducible relations. Even when the properties of the whole are metaphysically or otherwise determined by the properties of the proper parts of the whole, we might not be able to model the properties of the whole in terms of the vocabulary that we use to model the properties of the parts. (Silberstein, 2012, p. 633).
\end{quote} 

\paragraph*{}
I've presented these two conceptions of explanatory emergence because they demonstrate, clearly, just how complicated and nuanced the focus on derivability can be (and typically is). The difficulty in articulating a conception of Autonomy usually becomes the difficulty in articulating the conception of reduction that is not being exemplified. This is not a helpful shift, because, as Silberstein (2012, p. 627) states, the use of the term ``reduction'' is as equally ubiquitous and heterogeneous as ``emergence''. Rather than stressing the relation of emergence as a failure of reduction (in any sense) I want to focus on the other, positive, aspects of the idea. 

\paragraph*{}
The following three sub-sections are intended to motivate the shift from thinking in terms of reduction to an emphasis on a positive conception of emergence as dependence plus novelty and autonomy: i.e. the emergent, macro-level theory is novel and autonomous from the high-energy theory that is related to it via the relevant physics (the RG and EFT techniques, which will be introduced later). In accordance with my acceptance of there being multiple different conceptions of emergence, however, the following are very general problems and are not supposed to be conclusive arguments. I would not mind if the ideas of ``novelty and autonomy", isolated from concerns regarding reduction, were dismissed \textit{as an account of emergence} by any philosopher whose view of emergence is so ineluctably bound up with reduction---however, I would be disappointed if these ideas were to fail to be recognised as useful (in the philosophy of physics) on account of not according with such a view of emergence. If it is not to be thought of as emergence, may it be seen as an \textit{alternative} to emergence that is applicable in certain physical cases. The positive account of ``novelty and autonomy'' arises very naturally from the physical examples considered in this thesis, when, by contrast, the attempts to apply an account based on reduction and derivability yield results that are complex, hazy and forced.

\subsection{Distinguishing \textit{in principle} from \textit{in practice}}\label{subsub: in practice}
One reason we might be tempted to shift (or, at least, be open to shifting) from talking about a definition of emergence based on reduction to talking about a definition based in other notions is that emergence is wounded by the great cut that runs through it, dividing it into ontological and epistemological cases. The main problem with distinguishing emergent phenomena as either ontologically or epistemologically emergent is that, in many interesting cases, it is unclear whether our failure to derive, explain or predict them given their base is a failure \textit{in principle} or merely \textit{in practice}. Indeed, it is unclear even how the distinction is supposed to be decided in most cases when we are talking about successful scientific theories. Presumably it is not simply that we are unable to make the deduction from micro-theory to macro-physics unaided as humans, but then we seem to shift the question into one about computational limitations. 

\paragraph*{}
If we were to feed the relevant micro-theory---plus boundary conditions---into a supercomputer having the maximum possible, but nevertheless finite, computational power, and the computer was unable to provide an intelligible description of the macro-physics, presumably we would not want to count this as evidence for the macro-physics being ontologically emergent. A quick example of this is an attempt to derive molecular structure from the Schr\"{o}dinger equation. Although we take this as possible \textit{in principle}, it would require infinite computational power: if the amount of computer memory necessary to represent the quantum wavefunction of one particle is $N$, then the memory required to represent the wavefunction of $k$ particles is $N^k$ \citep{Laughlin2000}. The idea of the derivation being possible in any meaningful sense threatens to evaporate.

\paragraph*{}
In spite of this difficulty, however, it is of course entirely plausible that one could find a definition or a some other means of distinguishing ontological from epistemological cases of emergent scientific theories. Yet, the inclination and cost of doing so should be queried. We are not doing metaphysics or philosophy of mind, and do not need to carry over certain concepts unless they are useful. A focus on the question of whether an account of emergence is ontological or epistemological is potentially distracting, and, in spending our time trying to make sense of the categories given the physical theories, we risk overlooking or ignoring more interesting and tangible relations; we end up going metaphysics-first rather than science-first in our methodology. Therefore, as stated earlier, the accounts of emergence presented in this thesis are supposed to apply to theories (or models), and the question of whether they are to be understood as epistemological or ontological is not properly addressed.

\subsection{A varied landscape where less is different}\label{subsub:varied}
\citet{Butterfield2011a, Butterfield2011b, ButterfieldForthcoming}, using an assortment of physical examples, argue that reduction and emergence (and supervenience) are independent of one another: that is, we can have emergence with reduction, as well as emergence without reduction (and we can have emergence with supervenience, as well as without it). Emergence is defined simply as novel and robust behaviour relative to some comparison class, and reduction is defined as deduction aided by appropriate definitions or bridge principles (i.e. as \textit{definitional extension}). 
\paragraph*{}
Rather than considering Butterfield's (2011b) physical examples here, it will simply do to state that they all involve taking limits, so that the system exhibiting emergent behaviour is a limit of a sequence of systems as some parameter, $N$, goes to infinity, $N \rightarrow\infty$ (or some other value, $x$, $N \rightarrow x$), and the behaviour is said to be emergent with respect to that of the non-limit system. The strategy that Butterfield employs is to perform the deduction of the novel behaviour after the limit has been taken. This does not need imply that the $N =\infty$ limit is physically real, however, because the next step that Butterfield takes is to point out that there is, in all cases, sufficiently novel and robust behaviour that appears before we reach $N=\infty$.
\paragraph*{}
Because the argument in Butterfield (2011b), that emergence is compatible with reduction, is made by appeal to physical examples, and all these examples involve limits, it is tempting to say yes, certainly the examples are well-framed and correct, but the conclusion does not extend beyond these select examples. In other words, a critic might argue that emergence \textit{is} incompatible with reduction \textit{except} in certain physical cases that happen to all involve limits. While I believe that Butterfield would be disheartened by this conclusion, of course it does not affect his take-home message that emergence is not in all cases incompatible with reduction.\footnote{\citet{Butterfield2011a, Butterfield2011b} takes emergence to be novel and robust behaviour, and states in the abstract to (2011a) that the two papers together ``rebut two widespread philosophical doctrines about emergence. The first, and main, doctrine is that emergence is incompatible with reduction. The second is that emergence is supervenience; or more exactly, supervenience without reduction.''} 
\paragraph*{}
Butterfield (2011b, p. 1068), as we have seen, does not believe there is a single best definition of emergence, and is happy to admit numerous different ones: he does not believe that emergence always involves taking a limit, nor that taking a limit always results in emergence. However, it seems perhaps the physical examples he appeals to in order to make his argument do show that in cases where emergence is compatible with reduction, a limit is necessary. The physical examples I present in this thesis will hopefully refute this conclusion. Thus, the message I believe we should take from Butterfield's papers is the one he promotes: that there is no necessary connection between the relations of emergence and reduction.

\subsection{The problem of defining ``derivation''}\label{subsub:derivation}
The problem of defining ``reduction'' is one I return to shortly (\S\ref{sub:reduction}), however, there is a similar problem in defining ``derivation''. As we shall see in considering effective field theory (\S\ref{sect:EFT}), it is tempting to relate emergence to a failure of derivation \textit{in practice}, and even to attempt to strengthen this account by stressing the derivational independence of the theories in question. \citet{Bain2013a, Bain2013b}, for instance, argues that the use of approximations and heuristic reasoning in arriving at the macro-theory from the micro-theory, together with the fact that specification of the equations of motion (plus boundary conditions) for the micro-theory will fail to specify solutions to the equations of motion for the macro-theory, mean that there is no sense in which we can arrive at the macro-theory by means of a \textit{derivation} from the micro-theory. 
\paragraph*{}
Clearly this conception of emergence then just depends on what we would wish to count as a derivation. Typically in physics, derivations \textit{do} involve some use of approximation and/or additional assumptions (even if it it just those involved in definining the ``correspondence principles'' that are required in addition to the theory), so it perhaps becomes a matter of degree. However, the question of how strongly a derivation in physics must rely on approximation or additional assumptions before it ceases to count as a derivation and instead is to be classed as something else, is one that I feel is irrelevant to the question of emergence, or, at least irrelevant to the important or interesting aspects of the relation. Again, appealing to \S \ref{subsub: in practice}, it may be difficult to tell whether, in any particular case, the use of approximation is necessary in principle or merely in practice, and, again, we might question the point of even attempting to distinguish between the two scenarios.

\section{The Anderson/Weinberg debate}\label{sub:andersonwein}
The tradition of discussing emergence in physics that I am interested in here began with Philip Warren Anderson's classic 1972 article \textit{More is Different}, which was written expressly to defend the intrinsic value of condensed matter physics, against Steven Weinberg's claim (which encapsulates an attitude very common even today) that high-energy (i.e. particle) physics is somehow ``more fundamental'' than other areas of science. This ``physicists' debate'' was provoked by the issue of funding: particularly, the issue of funding the proposed (but never built) Superconducting Super Collider, upon which scientists from various disciplines were called to testify.\footnote{For more on the physicists' debate, see \citet{Cat1998, Schweber1993}.} It was in this context that Weinberg famously stated,
\begin{quotation}
In all branches of science we try to discover generalizations about nature, and having discovered them we always ask why they are true. I don't mean why we believe that they are true, but why they \textit{are} true. Why is nature that way? When we answer this question the answer is always found partly in contingencies, that is partly in just the nature of the problem we pose, but partly in other generalizations. And so there is a sense of direction in science, that some generalizations are ``explained" by others[\dots]
\paragraph*{}
There are arrows of scientific explanation which thread through the space of all scientific generalizations. Having discovered many of these arrows, we can now look at the pattern that has emerged, and we notice a remarkable thing: perhaps the greatest scientific discovery of all. These arrows seem to converge on a common source! Start anywhere in science and, like an unpleasant child, keep asking ``Why?" You will eventually get down to the level of the very small.
\paragraph*{}
[\dots]All I have intended to argue here is that when the various scientists present their credentials for public support, credentials like practical values, spinoff, and so on, there is one special credential of elementary particle physics that should be taken into account and treated with respect, and that is that it deals with nature on a level closer to the source of the arrows of explanation than other areas of physics. \citep[][p. 434]{Weinberg1987}
\end{quotation}
In response, Anderson defended the view that the laws and principles he studied as a condensed matter physicist were emergent: entirely different from, yet of no lower status, than those studied in particle physics.
\begin{quotation}
The reductionist hypothesis may still be a topic of controversy among philosophers, but among the great majority of active scientists I think it is accepted without question. The workings of our minds and bodies, and of all the animate or inanimate matter of which have any detailed knowledge, are assumed to be controlled by the same set of fundamental laws, which except under certain conditions we feel we know pretty well.
\paragraph*{}
[\dots]The main fallacy in this kind of reasoning is that the reductionist hypothesis does not by any means imply a ``constructionist'' one: The ability to reduce everything to simple fundamental laws does not imply the ability to start from those laws and reconstruct the universe. In fact, the more the elementary particle physicists tell us about the nature of the fundamental laws, the less relevance they seem to have to the very real problems of the rest of science, much less to those of society.
\paragraph*{}
The constructionist hypothesis breaks down when confronted with the twin difficulties of scale and complexity. The behaviour of large and complex aggregates of elementary particles, it turns out, is not to be understood in terms of a simple extrapolation of the properties of a few particles. Instead, at each level of complexity entirely new properties appear, and the understanding of the new behaviors requires research which I think is as fundamental in its nature as any other. That is, it seems to me that one may array the sciences roughly linearly in a hierarchy, according to the idea: The elementary entities of science X obey the laws of science Y.
\paragraph*{}
[\dots]But this hierarchy does not imply that science X is ``just applied Y''. At each stage entirely new laws, concepts, and generalizations are necessary, requiring inspiration and creativity to just as great a degree as in the previous one. Psychology is not applied biology, nor is biology applied chemistry. (Anderson, 1972, p. 393)
\end{quotation}
Many philosophers have attempted to flesh out Anderson's views into a clear philosophical position, but most, after a degree of struggle, conclude, with some bewilderment, that Anderson has made a simple mistake of confusing epistemological emergence with ontological emergence \citep{Mainwood2006}. Nevertheless, I'm not interested in pinning down exactly what Anderson had in mind, nor am I interested in arguing over the best interpretation of his text. Rather, I am interested in looking at the \textit{physical examples} that inspired Anderson and his colleagues---whom, following Mainwood (2006), I will refer to as New Emergentists\footnote{Other condensed matter theorists who presented views similar to Anderson's include Robert Laughlin, David Pines and Piers Coleman. See, e.g. \citet{Laughlin2005, Coleman2003, Laughlin2000}. \citet{Mainwood2006} refers to these, and their followers, as the ``New Emergentists''.}---to speak of emergence, and the physical mechanisms that underlie these. The reason I am interested in these examples is because they demonstrate how theories emerge from one another at different energy scales---and, if we conceive of quantum gravity as a small-scale theory of spacetime (i.e. a theory that is supposed to replace GR at high-energies), then this idea of emergence is important.

\paragraph*{}
I have mentioned the Anderson/Weinberg debate and the tradition of the New Emergentists not just because references to their claims pervade the philosophical literature on emergence in physics, but because the physical examples and mechanisms the New Emergentists were inspired by are exactly those that continue, today, to inspire physicists to speak of emergence. The emergence-claims have flowed along the direction of Weinberg's arrows, however: no longer are they confined to the level of condensed matter physics, but, as we shall see, they appear even in the domain of high-energy physics.

\section{Emergence in physics}\label{sub:emerphysics}
Usually in physics the idea of one theory, $T_1$, being emergent from another, $T_2$, is taken to mean that $T_1$ approximates (i.e. approximately reproduces the results\footnote{As Butterfield and Isham (2001) note, ``results'' here can include theoretical predictions as well as larger structures such as derivations and explanations.} of) $T_2$ within a certain limited domain of $T_2$'s applicability. Physicists' terminology runs backwards to the philosopher's, as physicists would say that in this case the ``more fundamental'' theory $T_2$ \textit{reduces to} the emergent theory $T_1$ within the domain where the latter is applicable. An example is Newtonian mechanics emergent from special relativity: the latter ``reduces'' to the former in the classical limit, $(v/c)\rightarrow 0$ or (where $v$ is the velocity of the system and $c$ the speed of light), which is just to say that the emergent theory, Newtonian mechanics, can be derived from the more fundamental theory, special relativity, within the domain (i.e. for particular values of pertinent quantities) where the former is known to hold.\footnote{This relates to the GCP (\ref{def:GCP}), which I am taking as a principle to uphold in the search for quantum gravity.} In philosopher's jargon: Newtonian mechanics is reducible to special relativity. 

\begin{figure}[htbp]
\centering
\includegraphics[height=10cm]{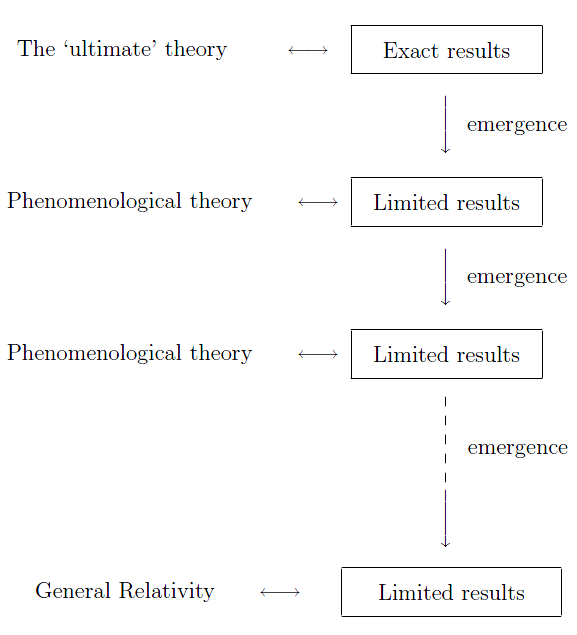}
\caption{Tower of emergent theories. (Adapted from Butterfield and Isham, 2001, p. 79)}
\label{fig:tower}
\end{figure}

\paragraph*{}
The physicists' conception of emergence in such cases is thus very different to that of the philosophers, being more akin to reduction rather than to a failure of reduction. A nice illustration of this conception is provided by Fig.\ref{fig:tower}, adapted from \citet[][p. 79]{Butterfield2000},
This figure represents a ``tower of theories'', where each emerges from the one above it (we might think of the theories toward the top of the tower---i.e. closer to the ``ultimate'' theory---as being applicable at higher energies than those at the bottom---i.e. the phenomenological theories), essentially the same as the hierarchy described by Anderson (\S\ref{sub:andersonwein}) and his fellow New Emergentists Laughlin and Pines (2000, p. 30), who state ``Rather than a Theory of Everything we appear to face a hierarchy of Theories of Things, each emerging from its parent and evolving into its children as the energy scale is lowered''. 
\paragraph*{}
These ideas are made more precise in considering effective field theory (\S\ref{sect:EFT}), where the idea of a tower of theories is very natural, yet much-debated. Although Fig.\ref{fig:tower} makes reference to ``the ultimate theory'', we needn't be committed to its existence in order to comprehend or utilise this conception of emergence. Also, although Fig.\ref{fig:tower} shows only a single tower, we expect that, in general, there will be many different towers branching off from any given theory (Butterfield and Isham, 2001, p. 79). 
\paragraph*{}
\citet{Hu2005, Hu2009} makes an interesting suggestion that we might interpret as representing an orthogonal perspective on the tower. He distinguishes between two types of theories: those that describe the basic constituents of matter and their interactions, and those which describe the collective behaviour (dynamics) of these constituents. Hu (2009, p. 5) suggests that these two different types of theories represent almost orthogonal perspectives, so, if we regard the chain of:
\begin{itemize}
\item quantum gravity
\item grand unified theory (GUT)
\item chromodynamics
\item quantum electrodynamics
\item \ldots
\end{itemize}
as a vertical progression depicting the hierarchy of basic constituents, there is also a horizontal progression depicting the collective states of matter: 
\paragraph*{}
stochastic -- statistical -- kinetic -- thermodynamic/hydrodynamic
\\ \\
A particular theory in this horizontal chain will yield similar results when applied at different levels on the vertical progression---for example, the macroscopic behaviour of electron plasma is, in many respects, similar to that of the quark-gluon plasma. In the long-wavelength regime, the behaviour of both types of matter can be adequately described by hydrodynamics, even though their underlying micro-theories are different. The idea of treating low-energy hydrodynamical descriptions as emergent is discussed in \S\ref{sect:Univers}, and later in discussing spacetime in \S\ref{sect:GREFT}.
\paragraph*{}
Now, taking the science-first approach does not entail deferring to the scientists, especially when it comes to philosophical issues. However, since the aim of my project is to elucidate the relations between particular theories---relations that purportedly represent emergence---I will not dismiss as uninteresting those cases that exemplify the physicists' conception of emergence rather than the philosophers'. Unfortunately, it is almost as difficult to articulate the physicists' conception of emergence as it is the philosophers'; Butterfield and Isham (1999) attempt to make it precise in terms of \textit{reduction}, and then in terms of \textit{supervenience} and find that neither is able to do the job. Here I will consider only their exploration of reduction.

\section{Reduction}\label{sub:reduction}
Butterfield and Isham's (1999) exploration of reduction as a candidate for emergence is made from the perspective of the physicists' sense of emergence. Nevertheless, if we take the philosophers' conception of emergence as a failure of reduction, then this discussion is of relevance, as it demonstrates the difficulty in articulating an appropriate definition of reduction. In this sense, the contents of the present sub-section may be taken as presenting a further difficulty for those approaches to emergence that attempt to tie the concept to reduction, and so another reason for my suggesting that we abandon the tradition of linking emergence and reduction.
\paragraph*{}
The intuitive idea of reduction that \citet{Butterfield1999} work with is that one theory $T_1$ is reduced to another $T_2$ if $T_1$ is shown to be a part of $T_2$. Reduction is taken as deduction of one theory from another, typically with the reducing theory being augmented with appropriate definitions or bridge-principles linking the two theory's vocabularies, i.e. as \textit{definitional extension}.\footnote{This is also how Butterfield (2011a, 2011b) articulates his idea of essentially-Nagelian reduction.} This basic idea of reduction, as deduction aided by definitions or bridge-principles, is essentially the traditional account of reduction proposed by Ernest \citet{Nagel1961}\footnote{Nagel's account of reduction is not uncontroversial, as Butterfield points out. Various objections have been raised against it, but it has also been defended by many philosophers, including: \citet{Endicott1998, Marras2002, Klein2009, Needham2010, Diz2010}.}  In order to use this idea, however, we must understand theories via the \textit{syntactic} conception, that is, we must treat the postulates of a theory as sets of sentences closed under deduction. This contrasts with the \textit{semantic} conception, according to which theories are classes of models satisfying the axioms. So, already, by taking this path we are forced to justify closing the door not only on conceptions of emergence as a relation between \textit{things other than theories}, but also on a very popular rival view of theories as sets of models.

\paragraph*{}
Butterfield (2011a, pp. 926--927) goes to some length to justify his choice of working with the syntactic conception, including arguing that it is capable of describing perfectly well the phenomena in scientific theorising that advocates of the semantic view tout as the merits of models, but, nevertheless, also emphasising that he does need, for his own purposes in making a crucial point about supervenience, to later switch to the semantic conception and use the idea of a class of models that is not the set of models of a given (syntactic) theory. Mainwood (2006, p. 32), however, points out optimistically that we may assume that nothing hangs on the choice of taking the syntactic conception rather than the semantic one, other than that the former provides a neat definition of reduction. Without going into the depths of details---it thus seems, simply, that we could perhaps avoid all this trouble if we were to avoid talking about reduction. 
\paragraph*{}
Next, having taken the syntactic conception of theories in order to get our nice definition of reduction, we define definitional extension; $T_1$ is a definitional extension of $T_2$ iff one can add to $T_2$ a set $D$ of definitions, one for each of $T_1$'s non-logical symbols, in such a way that $T_1$ becomes a sub -theory of the augmented theory $T_2\bigcup D$. In other words, in the augmented theory we can prove every theorem of $T_1$ The issues then regard the construction of the definitions: these are chosen with a view to securing the theorems of $T_1$, and can require a great deal of creativity and skill. 

\paragraph*{}
Also, there is no requirement that the definitions be brief: ``A definition or deduction might be a million pages long, and never formulated by us slow-witted humans" (Butterfield, 2011a, p. 932). Finally, if we take a definition, for a predicate, to be a statement of co-extension (and for a singular term, co-reference), then this scheme does not obviously extend the domain of quantification (i.e. there are objects in $T_1$ distinct from those of $T_2$), and so Butterfield (p. 932) describes three tactics, which I will not go into, for dealing with this problem.
\paragraph*{}
Definitional extension, while being a nice candidate for reduction because it is both intuitive and precise, is, unfortunately, sometimes too strong and sometimes too weak to do the job. Both of these points, which are apparently widely-recognised, are made by Butterfield and Isham (1999, p. 118). In regards to definitional extension sometimes being too weak for reduction, the problem has to do with the fact that, even though $T_1$ and $T_2$ are strictly co-extensive in their predicates, we might think that there are some aspects of $T_1$ that ``outstrip'' $T_2$, for instance that $T_1$ might have aspects to do with explanation, or modelling, or heuristics, that aren't captured by $T_2$, and definitional extension is inadequate to capture these. 
\paragraph*{}
Of course, however, we could then perhaps choose to label these additional aspects ``novelty'', which case we might say that their presence indicates a failure of reduction, rather than that definitional extension is insufficient for reduction. Such a tactic might be taken by an advocate of explanatory emergence, described earlier. More typically, however, the move is to add supplementary conditions to definitional extension in order to bolster it as a candidate for reduction, Nagel himself, for instance, included some additional clauses regarding explanation, i.e. that $T_2$ should explain $T_1$, where explanation is understood in deductive-nomological terms. In some cases, authors choose to add clauses that prohibit other candidates for novelty, so that it becomes controversial which supplementary clauses are correct (see Butterfield, 2011a, p. 930).
\paragraph*{}
In regards to definitional extension sometimes being too strong for reduction, the criticism is that, on taking reduction as deduction aided by appropriate bridge principles, there are intuitive cases of $T_2$ reducing $T_1$ in spite of there being considerable conceptual and explanatory disparities between the two theories. Butterfield and Isham (1999, p. 120) quote Feyerabend's \citeyearpar{Feyerabend1981} example of Newtonian gravity theory reducing Galileo's law of free fall: although the reduction goes through, the Newtonian theory is inconsistent with Galileo's law because it says (contrary to Galileo's law) that the acceleration of a body increases as it falls towards the earth. There are other standard examples, namely, special relativity and Newtonian mechanics, or statistical mechanics and thermodynamics: as Butterfield and Isham point out, while some authors cite these examples as paradigmatic of reduction, others cite them as examples of replacement or incommensurability. The moral is that reduction often requires approximation: in many cases, reduction involves $T_2$ including some sort of analogue, $(T_1)^\ast$, of $T_1$, where $(T_1)^\ast$ is required to be close enough to $T_1$ that we are happy to say that $T_2$ reduces $T_1$.\footnote{Following their discussion of reduction, Butterfield and Isham (1999) turn to the relation of \textit{supervenience} as a candidate for emergence, which some metaphysicians have found promising, given its apparent ability to sidestep controversial issues such as property-identity and explanation that, as we have seen, have to be addressed in order to provide an analysis of reduction. They find that, although supervenience promises the advantages of being weaker than reduction (it allows for definitions that are infinitely long as well as finitely long) though also quite precise, these advantages are often illusory; firstly, it is not clear whether supervenience is, indeed, weaker than definitional extension, and, secondly, it is sometimes too weak for emergence. Butterfield (2011a) finds other difficulties with the notion, and argues that it is unrelated to emergence. I accept these results, and work with them/build upon them.} 

\paragraph*{}
Butterfield and Isham (1999, p. 125) conclude that rather than seeking a definition of emergence framed in terms of reduction (or supervenience), we should instead bear in mind the variety of ways in which one theory may be emergent from another, particularly focusing on the notions of limits and approximations. I believe that this is correct. I have presented Butterfield and Isham's exploration of reduction simply to illustrate that even a logical cut-and-dried formulation of reduction can be quite thorny and quickly lead us into difficulties. I do not mean to say that it is impossible to formulate a workable conception of reduction upon which to base a conception of emergence, but simply that it appears not an easy task: if we can avoid having to take this route in defining emergence, it seems preferable to do so. It is not at all clear what there is to be gained from it.

\section{Emergence as dependence plus novelty and autonomy}\label{sub:novelty}
So far I have presented a case for disassociating emergence and reduction, and have suggested that we instead focus on the positive aspects of emergence. However, insofar as so many accounts of emergence are associated with a failure of reduction, and insofar as emergence is characterised by the ontological/epistemological divide, perhaps the positive aspects of the relation that I am to focus on, taken separately from these other issues, are better to not be termed ``emergence''. Hence, in this section at least, I just call them what they are: novelty and autonomy. It is important to note, though, that when I speak of ``novelty and autonomy'', I mean this as shorthand for ``The emergent, macro-level theory is novel and autonomous compared to the high-energy theory that is related to it via the relevant physics'' (here, the ``relevant physics'' are the RG and EFT techniques, which are introduced in the next chapter). In other words: please bear in mind that there is also a notion of \textit{dependence} involved, which relates the theories to one another.
\paragraph*{}
For future reference, here is an explicit statement of the definition of emergence that is used in this thesis.
\begin{description}
\item[Dependence] \label{def: mine}The low-energy theory is related to the high-energy theory via the physics of the RG and EFT techniques (this relation may or may not be classed as a derivation, see Chapter \ref{sect:EFT}). Alternatively, if one has an attachment to the concept of supervenience, we might understand Dependence as involving supervenience: the system described by the low-energy (macro-) theory \textit{supervenes} on that of the high-energy (micro-) theory, where supervenience is understood as the claim that there cannot be two objects that are alike in all high-energy respects (i.e. two systems that are the same as described by a particular, appropriate, high-energy theory), but differ in respect to their low-energy physics.\footnote{The most basic characterisation of supervenience states that a set of properties $A$ supervenes upon another set $B$ just in case no two things can differ with respect to $A$-properties without also differing with respect to their $B$-properties (see, e.g. \citet{McLaughlin2011}.)} (This use of supervenience requires a little more explanation, given Butterfield's (2011a) demonstration that we can have emergence without supervenience, so I comment on it again shortly).
\item[Independence] The low-energy physics is \textit{novel} and \textit{autonomous} with respect to the high-energy description.
\end{description}
I take emergence to be a relation between physical theories. When I speak of ``systems'', I mean just the systems (putatively) described by the (models of the) theories. I leave aside the question of how the theories are related to the systems they describe; discussions of scientific realism are beyond the scope of this thesis.\footnote{These questions are, of course, non-trivial! And my stance described here is not unproblematic, as will be particularly evident in Chapter \ref{sect:Univers}, where I discuss the thermodynamic limit. }

\paragraph*{}
A note regarding my admittance of the concept of supervenience as potentially a part of my definition of emergence: although \citet{Butterfield2011a} demonstrates that we can have emergence (with or) without supervenience, I believe that supervenience---as I have stated it above---holds in all of the cases I consider in this thesis. As I have put the claim above, it states that two systems that are the same according to a particular high-energy theory (whatever the appropriate one may be for the  system and energy under consideration), can be described as having the same physics (as one another) by the appropriate low-energy theory. Butterfield (2011a, \S 5.2.2) has two examples of emergence without supervenience. Briefly: the first example involves recognising the work of philosophers such as \citet{Silberstein1999}, who present entangled quantum states as cases of emergence without supervenience (i.e. the entangled states are emergent and do not supervene on the states of the entangled particles individually). This involves a different notion of supervenience (mereological supervenience) than the one I've admitted here, and emergence is taken as a failure of supervenience. Certainly, however, accounts of quantum gravity which describe entanglement may present emergence and a failure of supervenience in the sense argued for by \citet{Silberstein1999}, and it is unfortunate that I have not been able to explore this in the thesis.
\paragraph*{}
The second case that Butterfield (2011a) presents as an example of  emergence without supervenience is counterfactual, and involves the possibility of ``configurational forces'': fundamental forces that come into play only when the number of bodies (or particles, or degrees of freedom) exceeds some number, or when the bodies etc. are in certain states. Although we know of no such forces (science does not describe any such forces), and Butterfield (2011a) acknowledges that physics does not require any configurational forces, it is possible that configurational forces are required to explain some chemical or biological phenomenon or phenomena. If this were the case, then, Butterfield (2011a) argues, we would again have emergence without supervenience, because the emergent chemical or biological facts would not supervene on the micro-theory (quantum mechanics of the particles), instead also requiring the facts about the configurational forces.
\paragraph*{}
The existence of configurational forces would represent a problem for basing the notion of Dependence, part of my definition of emergence, on supervenience. This is because such forces would mean that we could have two systems which were the same according to the particular high-energy theory appropriate for describing them, yet which differed according to the low-energy physics---if the low-energy physics depended in some way on the configurational forces, and these differed for the two systems. Configurational forces would not feature in the high-energy theory, but would need, presumably, to be described according to some new theory, applicable at some ``intermediate'' energy scale, between the domains of the high-energy and low-energy theories being considered. As it stands, however, we have no reason for thinking that there are such forces at work in the physical examples I consider in this thesis, or, indeed, in physics at all. Thus, because my definition of emergence is only meant to apply to the physical examples I consider in this thesis, the counterfactual example of Butterfield (2011a) presents no problem here.\footnote{Butterfield (2011a) also emphasises one major reason for reservation regarding the use of the concept of supervenience. This is the necessity of being very careful in actually applying it: a supervenience claim needs to define precisely what properties or predicates are in the subvening set (i.e. the set $A$).} 

\paragraph*{}
\textit{Novelty} is taken as robust behaviour exhibited by the macro-system (appropriately described by the emergent theory) but not present in the micro-system (described by the micro theory).\footnote{\label{foot:micro} I use the term ``macro'' just to contrast with ``micro'' here: I mean these only as relative terms (i.e. not to imply that the emergent phenomena must always be confined to the macro-realm), and will often say ``upper-level'' or ``low-energy'' to denote the same thing, being the phenomena that emerge from the ``lower-level'', ``underlying'' or ``high-energy'' system.} The emergent theory is novel compared to the theory it emerges from if it is formally distinct from the latter, describing different physics and different degrees of freedom. A similar conception is propounded by Butterfield (2011a, p. 921), who supposes novelty to be something like ``not definable from the comparison class'', and maybe ``showing features (perhaps striking features) absent from the comparison class'', and robustness he supposes to be something like ``the same for various choices of, or assumptions about, the comparison class''. I agree with Butterfield's intuitive presentation of novelty and robustness. The idea of novelty is further clarified by considering the physical examples, as in \S\ref{sect:Univers}, \ref{sect:GREFT}, and \ref{sect:Discrete}.

\paragraph*{}
The idea of novelty, free of any implications concerning reduction, might strike many philosophers as uninteresting, or ``too weak'' to represent a conception of emergence. Morrison (2012, p. 148), for instance, wants to distinguish emergent phenomena from ``resultant'' ones. Emergence carries connotations of ``the whole being greater than the sum of its parts"---the suggestion being that, again, an emergent structure must be one that is somehow irreducible to its components or underlying description. As emphasised throughout this chapter, I do not feel the need to restrict our conception of emergence in physics by tying it to ideas carried over from metaphysics or the philosophy of mind---the idea of emergence as a failure of reduction is not helpful in all cases. As Butterfield demonstrates, and this thesis hopefully does too, we can have novelty representing emergence with or without reduction.

\paragraph*{}
In regards to \textit{autonomy}, we may say that a particular level of the ``tower'' is autonomous from the one above it (i.e. the higher-energy theory underlying it) if impervious to changes in the high-energy system. Usually there is not \textit{absolute} autonomy, but rather \textit{quasi-}autonomy, meaning that the level is independent of \textit{much} or \textit{most} of the high-energy physics. This idea is made more precise in the discussion of the physical examples, but it may be characterised as the high-energy theory being \textit{severely underdetermined} by the low-energy physics. 
\paragraph*{}
Like novelty, autonomy is a feature that, divorced from issues of reduction and deduction, philosophers would be reluctant to class as emergence. Consider \citet[][p. 413]{Morrison2012}, who says that this sort of autonomy is a necessary but not sufficient condition for emergence, because ``the fact that we need not appeal to micro phenomena to explain macro processes is a common feature of physical explanation across many systems and levels'' (I return to discuss these views in \S\ref{sect:Univers}.). In the physical examples considered in this thesis, the idea of autonomy is certainly a pervasive one, but I do not think it should be overlooked simply for this reason; if anything, the pervasiveness should inspire us to examine the relation more closely and ask why it is so widely exemplified in nature. In particular, autonomy is very important, and its explanation certainly not trivial, in the literature on EFT, as explored in (\S\ref{sect:EFT}).

\section{Fundamentality}\label{sub:fundamentality}
As indicated by even the initial, very crude statement of emergence provided by Silberstein---that an emergent phenomenon is in some sense less fundamental than its base, and in some sense not reducible to its base (\S\ref{sub:Eemergence})---discussions of emergence typically involve some reference to the idea of fundamentality, and the discussion in this chapter has some implications for our understanding of the notion. Anderson's (1972) suggestion, as we have seen, \S\ref{sub:andersonwein} is to treat the basic laws that govern each of the levels in the tower as each being as fundamental as any other.\footnote{Although I must point out that Anderson bases this view on our inability to derive the laws of any one level from those of the one beneath it; and it is left to Mainwood (2006) to explicate the relevant sense of ``ability''.} Similarly, \citet{Cao2003}, argues that the definition of a fundamental theory as being one from which all other theories can be derived, has lost its meaning thanks to the ideas of effective field theory and the renormalisation group (\S\ref{sect:EFT}).\footnote{Instead, \citet[][p. 28]{Cao2003} proposes a new definition, of a fundamental theory being one which can be derived from no other theory. This definition enables QFT and GR to be classed as fundamental theories. I am not convinced that we are justified in classifying QFT (understood as the Standard Model) and GR as fundamental theories, however---certainly they are \textit{currently} unable to be derived from anything, but, as outlined in \S\ref{sect:Intro}, we have reasons for believing there may be quantum gravity, in which case, if the GCP applies, then GR will cease to be fundamental. Similarly, there are reasons (discussed in \S\ref{sub:effectiveEFT}) for not viewing QFT as fundamental---not just in the sense that there may be a theory ``underlying'' it, from which it may be derived, but perhaps that the framework itself is flawed or incomplete (as adherents of Axiomatic QFT argue).} 
\paragraph*{}
I am sympathetic to these views, which favour the condensed matter theorists' perspectives on the world, rather than the particle physicist's ones. Even though I think the idea of the renormalisation group is enough to define a meaningful sense of ``direction'' in the way Weinberg conceives, the levels are novel and autonomous enough that seeking an explanation of one in terms of the one above it (in energy) seems an exercise in futility: such an explanation imparts no understanding of the important physics that characterises the level of interest (in other words, there is emergence, even if it is just ``explanatory emergence''). 

\paragraph*{}
It is tempting, because of the ``direction'' imparted by the RG (pointing in the opposite direction to ``emergence'' in Fig.\ref{fig:tower}), to follow Weinberg and take ``more fundamental'' to simply mean ``higher-energy''. This is not a useful view to have, though, since it is suggestive of there being a single ``source of the arrows'', a ``final theory'' (to again use Weinberg's terms), and taking this ultimate theory to be just the one that is valid at the highest-possible energy scales. Recognising that such a theory would simply be a theory that is valid at the highest-possible energy scales (it is an open question whether the RG-based arguments for ``direction'' would be applicable at such scales, see \S\ref{sub:effectiveEFT}) rather than one that \textit{explains} all low-energy physics, would, I think, make the attribution of fundamentality uninteresting. 
\paragraph*{}
Of course, equating ``more fundamental'' with ``higher-energy'' does not necessarily carry the implication that there is a ``most fundamental'' level, but neither does it make the idea of fundamentality particularly interesting. Because I am not sure what we should take ``fundamental'' to mean, I prefer to just avoid using the term. Instead, I will endeavour to simply distinguish between levels based on their relative energy-scales, using the phrasing micro- and macro-, with the disclaimer made above (Footnote \ref{foot:micro}) regarding their use simply as contrastive labels.\footnote{If ``fundamental'' is used to describe any theory herein, it may be understood as simply meaning ``higher-energy'', or as referring to a particular theory from which a given macro-theory is supposed to emerge.}

\section{Conclusion}
As explained in the previous chapter, the purpose of this thesis is to explore the relationship between quantum gravity and GR (or, perhaps, the structures described by GR). In beginning this exploration, I am taking the GCP (defined on p. \pageref{def:GCP}) as a principle to be upheld by quantum gravity. This means that quantum gravity and GR are supposed to be related by the ``physicists' sense of emergence''---in other words, it is taken as a requirement of quantum gravity that GR is able to be \textit{derived} from it. In other words, according to the GCP, reduction is supposed to hold \textit{in some sense}.  Yet, the idea of emergence in philosophy is typically taken as a failure of reduction \textit{in some sense}. 
\paragraph*{}
Rather than attempting to make sense of these different senses, I leave aside the conception of emergence as a failure of reduction, and join Butterfield in accepting the diversity of emergence and other relations. Emergence can hold with or without reduction, and we have no good reason to restrict our attention to either case! Indeed, I have argued that we have motive for keeping an open mind in our investigation of the physical examples. So, instead of focusing on articulating the relevant failure of reduction, I look at the ways in which GR (and the structures it describes) might be said to be novel and autonomous from quantum gravity (while still being related to quantum gravity, through the notion of Dependence defined on p. \ref{def: mine}). The idea of treating novelty and autonomy as bases for a conception of emergence is natural given the physical examples that are considered in this thesis---not just the approaches to quantum gravity, but other examples of EFTs. 

\paragraph*{}
This chapter gave a very brief, very general overview of the philosophical understanding of emergence, on the advice that we proceed via the science-first approach rather than dragging excess weight in the form of previous metaphysical debates regarding ontological and epistemological emergence. Several suggestions of the potential difficulties with applying an account of emergence in terms of reduction were made, intending only to motivate the shift from holding a restrictive prior understanding of emergence to adopting one that is more flexible. This idea of emergence as independent of reduction is very similar in spirit to that of \citet{Butterfield2011a, Butterfield2011b}, but the particular account of novelty and autonomy presented here is to be made more precise, and extended using different examples, than in these papers.

%% file: Eff.tex
\chapter{Effective field theory}\label{sect:EFT}

\section{Introduction}\label{sub:EFTintro}
As explained in the first chapter of this thesis, if we take quantum gravity to be a micro-theory of spacetime, then GR must be understood as an \textit{effective} theory, meaning it is thought to be valid only at large length-scales compared to the characteristic length-scale of quantum gravity (perhaps the Planck length). The framework of \textit{effective field theory} is a means of formalising this idea for a particular class of theories (field theories, including quantum field theories and field theories in condensed matter physics), under particular assumptions (including a clear separation of scales); though the term ``EFT'' is also used to refer to a given EFT that satisfies this framework. An EFT is typically contrasted against the higher-energy theory ``underlying'' it (often termed the ``(more) fundamental'' theory), and, when discussing emergence, will be said to be emergent from this higher-energy theory. The idea of emergence in EFT is interesting and nuanced enough itself that it is worth exploring in detail before considering how it may apply in quantum gravity, and this is the purpose of the present chapter.

\paragraph*{}
The framework of EFT, and its associated philosophy, arose as a solution to ``the problem of renormalisation'' in QFT, and so this chapter begins with an introduction to QFT and renormalisation (\S\ref{sub:QFTrenorm}), before describing the idea of the renormalisation group (\S\ref{sub:RG new})---the development of which was central to EFT (\S\ref{sub:EFT}). A simple pedagogical example illustrating the RG using the nearest-neighbour Ising model is presented in \S\ref{sub:Blockspin}. There are two ways of constructing theories in EFT, and these are outlined in \S\ref{sub:tower}. Some comments regarding the idea of a ``cutoff'' energy scale in our theories are made in \S\ref{sub:reconcutoff}, before the philosophy of EFT is explored in the remaining part of the chapter. 

\paragraph*{}
Most of the philosophers who have considered EFT have responded to the presentation by \citet{Cao1993}, which contains some controversial claims regarding emergence and fundamentality. I begin by examining these in \S\ref{sub:CaoSchweb}, arguing that the controversy and debates they've generated stem from a confusion between EFT as it applies \textit{in principle} and EFT as it actually applies \textit{in practice}. I then go on to propound a philosophy of EFT based in how the formalism applies in practice, arguing that we are not justified in asserting that many of the ``in principle'' claims hold true in physics---especially when it comes to inaccessibly high-energy scales. This view has consequences for our understanding of QFT (\S\ref{sub:effectiveEFT}). Finally, I examine the conception of emergence in EFT, and present the appropriate account given the lengthy discussion that led up to this section: an account based on the \textit{novelty} and \textit{autonomy} of an EFT compared to its high-energy theory.

\section{Quantum field theory and renormalisation}\label{sub:QFTrenorm}
In the early days of its development, QFT was plagued, infamously, by divergences. These divergences stem from the heart of the formulation of QFT: its combination of quantum mechanics and special relativity. The vacuum of QFT is not a void, but rather a plenum---it is envisaged as the lowest-energy (ground) state of a collection of quantum fields. A particle is understood as a localised excitation of its corresponding field. This localisation, however, means that the uncertainty in the position of the particle is zero.\footnote{My statements here are very brief and heavily simplified, intended only to provide a heuristic introduction to understanding (one standard interpretation of) QFT. Things are more complicated, of course. For instance, a Fock vacuum state is a type of QFT vacuum state that need not be interpreted as the ground state of a collection of quantum fields; there (arguably) isn't a one-to-one correspondence between particles and fields; and particle localisation is a difficult concept.} According to the position-momentum uncertainty principle,
\begin{equation}\label{eq:uncert}
\Delta x \Delta p \geq (\hbar /2)
\end{equation}
(where $\Delta x$ is the uncertainty in position, $\Delta p$ the uncertainty in momentum and $\hbar$ is the reduced Planck's constant) this means that that the momentum is completely indeterminate. Taken together with the energy-momentum relation of special relativity, 
\begin{equation}
E^2=m^2c^4+p^2c^c
\end{equation}
the arbitrarily high value of momenta implies both that the vacuum of QFT contains an arbitrarily large amount of energy, and that this energy is available for the creation of particles.\footnote{To see this another way, consider that a particle of mass $m$ in QFT is unable to be localised to within one half the distance defined by its reduced Compton wavelength, $\hbar/mc$. In order to probe a distance smaller than this, an energy greater than $mc^2$ is required (by the uncertainty relation, using the relativistic momentum), and a new particle of mass $m$, strictly identical to the original, is thereby created. Hence, QFT, by treating particles as local (pointlike) objects, holds that an arbitrary amount of energy is available for the creation of particles, and thus implies the existence of an infinite number of multi-particle states.} In this way, the quantum fields, being local operators, imply the existence of an infinite number of multi-particle states. A perturbative calculation\footnote{Since realistic quantum systems are typically very difficult to find exact solutions for, perturbation theory is utilised as an approximation. This involves taking a simple system for which exact solutions are known, and adding a ``perturbing'' Hamilitonian which represents a weak disturbance of the system. The various physical quantities associated with the perturbed system can then be expressed as ``corrections'' to those in the simple system, and can be calculated using approximate methods including expansion in terms of an asymptotic power series. So long as the expansion parameter is very small (i.e. the corrections to the physical quantities in the ``simple'' system are very small compared to the quantities themselves), the results will seem to converge to the exact values when summed to higher-order.} 
of any particular physical process involves a summation over all possible intermediate states (and this is done at all orders of perturbation theory)\footnote{In practice, often only the first few terms are taken and the higher-orders, it is hoped, decay rapidly.} because QFT implies that there are an infinite number of such states, and the terms are not sufficiently suppressed, perturbative calculations within the theory lead to divergent integrals.\footnote{Really, the situation is more subtle than my statements here suggest. The infinities that arise in QFT are due to an interplay of four different factors: the integration over all the momentum-energy states, the local nature of the dynamics (i.e. point sources and interactions), the type of interactions, and the dimension of the momentum space. (Thanks to Sebastian Rivat for helpful discussion on this point).} 

\paragraph*{}
In order to give meaning to the theory, divergent integrals must be tamed by removing all those states with energy larger than some finite but arbitrarily large cutoff value, $\Lambda_0$---this may be done, for instance, by integrating only up to $\Lambda_0$ rather than taking infinity as the upper limit of the integrand. The cutoff is also called the \textit{regulator}: its introduction being known as the \textit{regularisation} of the theory. Being introduced in this manner, by hand and for the purpose of regularising divergent integrals, the cutoff may be viewed purely as a necessary mathematical imposition upon the theory.\footnote{This view was the general one when the process of renormalisation was first introduced to QFT in the late 1940s. Later, however, following the development of the renormalisation group, QFT and the cutoff with it, enjoyed a re-conceptualisation, in which QFTs be understood as effective field theories – the divergences being a consequence of the realisation that our theories are not ``fundamental'' or ``complete'' in the sense that they are not valid to arbitrarily high energies. The cutoff then gained physical significance as a ``parametrisation of ignorance'', signifying the breakdown of our theory \citep{Zee2010}.}  This is a natural view of the cutoff given its role in the process of renormalisation, which I outline now, although later we shall see that in considering the renormalisation group and effective field theory, the cutoff plays a physically significant role.
\paragraph*{}
Even once a QFT has been regularised, it is still not useful as it stands, because it is not formulated in terms of observable quantities. In order that it be physically relevant, the QFT must be reformulated in terms of physical parameters.\footnote{This description of renormalisation borrows from \citet{Lepage1989}, \citet{Delamotte2004} and Zee (2010).}  Once this has been done, the cutoff energy is mathematically taken to infinity in order to recover a continuum theory (i.e. a theory without any highest energy, or, equivalently, shortest distance). The physically measurable quantities in the theory should not depend on the value of the cutoff, and so are kept constant as this limit is taken. This process is called \textit{renormalisation}. The cutoff aids in the process of reparametrising our QFT in terms of physical quantities, but at the end of calculations, $\Lambda_0$ is removed from the theory.
\paragraph*{}
In a little more detail, consider the example of quantum electrodynamics (QED), which is defined by its Lagrangian density, $L_0$, and the regulator $\Lambda_0$. The Lagrangian density contains parameters known as \textit{coupling constants} which are values that determine the strength of particular (particle) interactions. The coupling constants in $L_0$ are ``bare'' coupling constants, $e_0$ and $m_0$, which, owing to the divergence of the theory, are not physically measurable. In renormalising QED, these bare coupling constants are replaced by ``renormalised" or ``physical" coupling constants, $e_R$ and $m_R$, 
\begin{equation} \label{eq:e_0 e_R}
e_0\rightarrow e_R = e_0+\delta_e
\end{equation}
\begin{equation}\label{eq:m_0 m_R}
m_0 \rightarrow m_R = m_0 + \delta_m
\end{equation}
Where $\delta e$ and $\delta m$ are counter-terms which serve to cancel the divergences that appear as we carry through the renormalisation procedure in higher-orders of perturbation theory. Once the renormalisation procedure has been carried through successfully to all orders the cutoff is removed by taking the limit $\Lambda _0 \rightarrow \infty$. The renormalised coupling constants $e_R$ and $m_R$ are identified as, respectively, the charge and mass of the electron. The values of these constants are not calculable from theory and must be determined by comparing the calculations of two QED processes with the results of experiment at a given energy. Once these parameters are fixed, the theory is well-defined and accurate up to errors of order $1/(\Lambda_0)^2$. 
\paragraph*{}
It is remarkable that the renormalisation of QED only requires the reparametrisation of the two bare coupling constants that feature in the unrenormalised theory. For many---in fact, most---other QFTs, it turns out that the renormalisation procedure requires more and more constants to be added at each order of perturbation theory in order for the divergences to be absorbed and the theory to be capable of yielding meaningful predictions. If the theory requires that we add an infinite number of parameters to the theory, it is said to be \textit{non-renormalisable}.  If, on the other hand, it is sufficient to add a finite number of parameters, the theory is \textit{renormalisable}.
\paragraph*{}
Of course, the divergences in the theory do not magically disappear; they have been absorbed by the renormalised coupling constants. The divergences are encoded in the relationship (\ref{eq:e_0 e_R}--\ref{eq:m_0 m_R}) between the renormalised constants and the bare parameters: the bare coupling constants diverge as $\Lambda_0$ is taken to infinity. The counter-terms $\delta_e$ and $\delta_m$ balance the infinite values of the bare parameters and are, essentially, responsible for the finite values of the physical coupling constants. Although not technically problematic, this procedure of subtracting infinities from infinities in order to produce finite results is deeply unsatisfying conceptually.\footnote{The first dedicated book on the philosophy of QFT is Teller's \citeyearpar{Teller1995} \textit{An Interpretive Introduction to Quantum Field Theory}. See also the critical review by \citet{Huggett1996}, and Teller's \citeyearpar{Teller1998} reply.} 
\paragraph*{}
The interpretation, basically, is that the observed charge and mass of the electron result from the presence of the unavoidable fluctuations (i.e. multi-particle states) in the quantum fields dealt with by the theory---the ``bare'' charge and mass may be infinite, but these are not physical quantities (i.e. they are not measurable, and, following the ``new conceptualisation'' that comes with effective field theory, as we shall see below, are thought to stem from a misinterpretation regarding the validity of our theories at arbitrarily high energies). 
\paragraph*{}
Feynman\footnote{See, for example, \citet{Feynman2006}.} was vocal in expressing his belief that renormalisation theory is physically suspect, as was Dirac, who called it an ``illogical'' process,
\begin{quotation}
This [process of infinite renormalisation] is quite nonsense physically, and I have always been opposed to it. It is just a rule of thumb that gives results. In spite of its successes, one should be prepared to abandon it completely and look on all the successes that have been obtained by using the usual forms of quantum electrodynamics with the infinities removed by artificial processes as just accidents when they give the right answers, in the same way as the successes of the Bohr theory are considered merely as accidents when they turn out to be correct. \citep[][p. 55]{Dirac1983}
\end{quotation}
The great success of renormalisation in salvaging QFT and producing accurate predictions, however, is indisputable. Furthermore, the procedure was thought helpful in providing a means of selecting, out of a large number of potential QFTs, correct physical theories. Weinberg in his 1979 Nobel lecture, admits,
\begin{quotation}
I thought that renormalizability might be the key criterion, which also in a more general context would impose a precise kind of simplicity on our theories and help us pick out the one true physical theory out of the infinite variety of conceivable quantum field theories. [\dots] I would say this a bit differently today, but I am more convinced than ever that the use of renormalizability as a constraint on our theories of the observed interactions is a good strategy. \citep[][p. 547]{Weinberg1979}
\end{quotation}
Thus, renormalisability was implemented as a criterion for theory selection. On this view, non-renormalisable theories, given that they would require an infinite number of experiments (and infinite energy) to fix the values of all their parameters, were disregarded on the assumption of being not predictive.

\section{Renormalisation group and the re-conceptualisation of QFT}\label{sub:RG new}
The new conceptualisation of renormalisation which led to the acceptance of non-renormalisable theories (by actually making the term ``non-renormalisable theory'' redundant\footnote{\label{foot:renorm}Nowadays, thanks to the re-conceptualisation of QFTs as EFTs, we no longer distinguish between renormalisable and non-renormalisable \textit{theories}; rather, the distinction is made in regards to the terms that appear in the theories. If a coupling constant has dimension $[mass]^d$, with $\hbar=c=1$, then the integral for a process of order $N$ will behave at high-energy like $\int p^{A-Nd}dp$ where $A$ depends on the details of the process, but not $N$. Hence, the (naively) non-renormalisable terms are those whose couplings have negative mass dimension, i.e. $d<0$.}) came with the development of RG, and is closely tied to the interpretation of the cutoff.  To see the role that the cutoff plays in defining the theory, take QED defined by $\Lagr_0$ and $\Lambda_0$ and consider what happens when a new cutoff, $\Lambda \ll \Lambda_0$, is introduced, thereby removing from the theory all states having energy (or momenta) higher than $\Lambda$ (Lepage, 1989, p. 4). For reasons that will soon be elucidated, we can only consider the theory at energies much lower than $\Lambda$, so it is in this range that we explore how the original theory must be modified in order to compensate for the removal of those states with energy greater than $\Lambda$. The new theory must, after all, reproduce the predictions of the old theory within this low-energy regime. 

\paragraph*{}
It turns out that the effects of the neglected high-energy ($\Lambda > \Lambda_0$) states can be accounted for by the addition of a new interaction term $\delta \Lagr$ to $L_0$, so that the new theory is defined by Lagrangian density $\Lagr= \Lagr_0+ \delta \Lagr $  and the cutoff $\Lambda$. Adding the new interaction term amounts to replacing the bare parameters $e_0$ and $m_0$ in the original theory with new coupling constants, $e_\Lambda$ and $m_\Lambda$ where,
\begin{eqnarray}
e_\Lambda = e_0+e_0 c_0 \left(\frac{\Lambda}{\Lambda_0}\right)  \\
m_\Lambda=m_0+m_0\tilde{c}_0 \left( \frac{\Lambda}{\Lambda_0}\right)  
\end{eqnarray}
(and $c_0$ and $\tilde{c}_0$ are dimensionless parameters that depend only on the ratio $\Lambda/\Lambda_0$). The new theory gives the same results as the original theory up to corrections of order $1/\Lambda^2$. Thus, a change in the cutoff can be compensated for by changing the bare couplings in the Lagrangian density so as to leave the low-energy physics unaffected (Lepage 1989, p. 8). 
\paragraph*{}
In practice, what we have called the original cutoff  $\Lambda_0$, in QED is a large energy scale, while what we have called the new cutoff, $\Lambda$, represents the energy at which we are interested in working. The coupling constants $e_\Lambda$ and $m_\Lambda$ vary as the cutoff $\Lambda$ varied, and are said to ``run'' or ``flow'' as more or less of the state space is included in the theory. The equations which govern this flow are called the renormalisation group (RG) equations. In this case we have,
\begin{eqnarray}
\Lambda(\frac{de_\Lambda}{d\Lambda}) = \beta(e_\Lambda)\\
\Lambda(\frac{dm_\Lambda}{d_\Lambda})= m_\Lambda \gamma_m (e_\Lambda)
\end{eqnarray}
More generally the RG equations take the form of Beta functions,
\begin{equation}
\beta(C)=\Lambda\frac{\partial C}{\partial \Lambda}=\frac{\partial C}{\partial \ln C}
\end{equation}
Where $\Lambda$ is the energy scale under consideration, and $C$ the coupling constant. For a set of coupling constants, $\{C_k\}$, we have,
\begin{equation}\label{eq:betafn}
 \{C^\prime_k\} = \beta (\{C_k\})
\end{equation}
Where the Beta function is said to induce the flow of the set of coupling constants, $\{C_k\}$.
\paragraph*{}
Perhaps the most striking feature of renormalisation group flow is its very existence; it tells us that physical parameters which were previously (i.e. before the development of RG) taken to be ``fundamental'' constants in nature (the mass and charge of the electron, in this case) are not unchanging, but instead depend on the energy at which they are measured. The interpretation of this is based in the idea of the QFT vacuum as a plenum: probing higher energies means that more energy is available for the creation of particles. Virtual particle--antiparticle pairs, of energy $E$ are created from the vacuum, and can exist for time $t$ as allowed by the uncertainty relation (\ref{eq:uncert}), before annihilating one another. 
\paragraph*{}
The virtual electron--positron pairs become polarised in the vicinity of an electron, i.e. the virtual particles of opposite charge to the particle are attracted to it, while particles of like charge are repelled. This has the effect of partially cancelling out, or \textit{screening}, the electron's charge, meaning that we measure an effective charge that is lessened by the presence of the charged virtual particle--antiparticle pairs. Getting closer to the electron, we see the effects of fewer virtual particles-antiparticles, and the effective charge increases. This is expressed by the beta function for QED being \textit{positive}.\footnote{\label{foot:asy}In quantum chromodynamics (QCD), the effect of virtual quark--antiquark pairs is to screen the colour charge of the quark. The force-carrying particles in QCD, gluons, also carry colour charge (in a different way to the quarks) and an anti-colour magnetic moment. The net effect of polarisation of virtual gluons is not screening, but \textit{antiscreening}, which means that the effective colour charge of a quark is weaker the closer we are to it. In other words, the beta function for QCD is \textit{negative}. Furthermore, the coupling constant for QCD decreases logarithmically, which leads to a phenomenon called \textit{asymptotic freedom}: given the weakened interactions of the quarks and gluons at high-energies, perturbation theory becomes an increasingly better approximation for QCD. This is explained below, \S \ref{subsub:fixed points}.}
\paragraph*{}
Although we may not know what particles exist at high energy, the renormalisation group accounts for their effects upon the mass and charge of the electron at relatively low energy. This is true more generally: the RG tells us that the coupling constants which feature in any QFT are artefacts of interactions, and thus change with the magnitude of these interactions. As we change energy, the contributions of the various interactions responsible for the values of the coupling constants in our theory are accounted for by the RG equations, so long as we are working at energies well below the cutoff, $\Lambda$. 
\paragraph*{}
As we move to energies approaching the cutoff, however, the RG equations start to break down as the theory begins to behave as though it contains non-renormalisable terms. This is taken as a hint that unknown physics is coming into play; the effects of some high-energy interactions are no longer able to be incorporated by a reparametrisation of the currently employed coupling constants: a new QFT must be constructed whose Lagrangian contains extra terms in order to explicitly account for the new high-energy interactions. This theory will then, in turn, be applicable only at energies well below some new cutoff, of energy greater than that of the original theory.
\paragraph*{}
The cutoff, rather than being a merely formalistic device, is, in RG, a physically significant feature of QFT. It represents the edge of our theory's domain of validity. It is a marker by which to judge low- from high-energy physics: a road-sign that states, ``beyond this point lies physics unknown''.  The new philosophy of QFT, based on the development of the RG, thus holds that our theories are \textit{effective}---they are not expected to be applicable at arbitrarily high-energies, but are valid only at a particular scale (i.e. within a particular range of energies).

\subsection{Block-spin RG}\label{sub:Blockspin}
\begin{figure}
  \begin{center}
    \subfigure[\textit{Block spin formation:} Ising model with lattice spacing of 1 (small round dots). The spins are grouped into blocks of 4 and replaced by block spins (large square dots), the block spins having a lattice spacing of 2.]{\label{fig:bspin1}\includegraphics[]{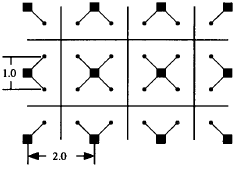}}\\
    \subfigure[\textit{Rescaling:} the lengths are shrunk in order to recover the original lattice spacing of 1.]{\label{fig:bspin2}\includegraphics[]{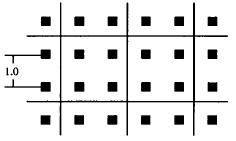}} \\
    \subfigure[\textit{Renormalisation:} interaction parameters are adjusted in order to preserve the large-scale properties of the original lattice.]{\label{fig:bspin3}\includegraphics[]{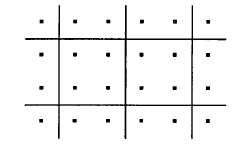}}
  \end{center}
  \caption{Kadanoff block-spin approach to the 2-dimensional nearest-neighbour Ising model \citep[Figures adapted from][p. 1043]{Batterman2011}}.
  \label{fig:blockspin}
\end{figure}

The simplest pedagogical example of RG is provided by the block-spin approach devised by Leo Kadanoff in 1966 \citep{Kadanoff1966}. The central idea of the Kadanoff approach is that of a scale transformation: we map the original system onto another at a larger scale, and then compensate for this by adjusting the original parameters so that all large-scale properties are preserved. Cases where we can in fact do this (i.e. cases where we can compensate for the scale transformation by simply altering the values of the coupling constants in the Lagrangian describing the system) are called renormalisable.
\paragraph*{}
Consider the physical picture of the two-dimensional nearest-neighbour Ising model: a perfect square array of atoms (spins) in a two-dimensional solid at a particular temperature, $T$, where the atoms interact only with their nearest neighbours, with the strength of this interaction given by a particular coupling, $C$, as in Fig. \ref{fig:bspin1}, below. Kadanoff suggested that we combine the effects of several spins into a block, so that each block has a single magnetic moment (the block spin) and interacts with its nearest neighbour. Kadanoff then showed that the block variables can also be modelled as an Ising system---i.e. they interact according to the same laws as the original system---but with a different (effective) value for temperature, $T^\prime$ and a different valued coupling, $C^\prime$.\footnote{Although the Kadanoff approach is pedagogically useful and yields much insight, it is worth pointing out that it is not entirely accurate in its assumptions. In particular, it works only as a crude approximation for the Ising model. As \citet[][p. 157]{Mainwood2006} points out, if we want to faithfully reproduce the physics of the original lattice we would need to take into account not just nearest-neighbour interactions, but also the next-nearest-neighbour, and introduce four-spin terms (since the nearest-neighbour interactions in the original lattice would also have to be accommodated). Successive iterations would lead to ever more complex terms, and the theory would quickly become unmanageable.} 

\paragraph*{}
In a real-space approach to the Ising model, this process can be illustrated schematically in three-steps. Firstly, we course grain the original lattice according to some rule (e.g. summation, partial integration, decimation, etc.), so that a new lattice is produced with fewer spins, Fig.\ref{fig:bspin1}. We then (spatially) rescale so that the new lattice can be compared to the original, Fig.\ref{fig:bspin2}, and renormalise the interaction parameter(s) to reproduce the large-scale properties, Fig.\ref{fig:bspin3}.
Because physical systems normally contain a great number of atoms, the process is iterated many times, meaning the observation scale increases and the number of atoms decreases. Iteration induces an RG flow of the coupling constant(s) that appear in the Lagrangian describing the system, as in (\ref{eq:betafn}). 

\subsection{Fixed points}\label{subsub:fixed points}
It often happens that, as we move to larger observation-distance scales (i.e. decreasing the value of $\Lambda$ in the RG equations), the set of physical coupling constants, $C^\prime$, flow to a \textit{fixed point}, so that further increase of $\Lambda$ produces no change in the value of $C^\prime$. When the fixed point is reached this way, by decreasing the energy scale, it is known as an infrared (IR) fixed point. This will be important when discussing the conception of emergence based on critical phenomena, \S\ref{subsect:criticalphen}. 
\paragraph*{}
On the other hand, an ultraviolet (UV) fixed point is one that is approached as the value of $\Lambda$ is increased. Theories which possess a UV fixed point and have a finite number of couplings attracted to it are called \textit{asymptotically safe}, since they are well-defined at arbitrarily small distance scales. This is true of many non-Abelian gauge theories, including QCD, which has the Yang-Mills coupling and the quark masses asymptotically approach the zero in the UV. Note that QCD is special because the UV fixed point that its couplings flow to is zero, and so the theory is called \textit{asymptotically free}. This is because the fixed point value zero indicates no interaction at all, and so corresponds to a \textit{free} field theory. Because of its asymptotic safety, QCD is not typically considered to be an effective field theory---it is well-behaved with no divergences at any high energy scale (although this is not the case at low energies, as will be discussed in the next section).

\section{Effective field theory}\label{sub:EFT}
Treating our QFTs as effective field theories (EFTs) relieves concerns regarding the appearance of divergences and the necessity of renormalisation; it is an acknowledgement of the limitations of our theories. The fact that a given QFT goes awry at high-energy is not conceptually problematic if we recognise that, of necessity, our theory does not fully capture the underlying high-energy physics, but only accounts for its effects at \textit{comparatively low-energy}. Recall that we fix the parameters in our theory based on input from low-energy experiments. We cannot conveniently ``screen off'' the influence of high-energy interactions when conducting experiments---the short-distance physics is, ultimately, \textit{somehow} responsible for the low-energy physics we observe. 
\paragraph*{}
Rather, the results of our low-energy experiments already include the effects of physics that underlies them, and we thus build-in the influence of physics unknown (with respect to the theory we are working on) when constructing our (renormalised) QFT. For instance, in QCD the mass of the proton depends on the mass of the top quark, as does the electromagnetic coupling constant at low-energy. And yet, in spite of this dependence, the mass of the top quark is totally unnecessary for studying, for example, the hydrogen atom---a system which has the electromagnetic coupling, the proton mass and the electron mass as parameters of the theory \citep{Manohar1997}. The values of these parameters are determined through low-energy experiments, and, as such, already include the low-energy influence of the top quark mass. 
\paragraph*{}
In QCD, it is not just a great convenience to be able to study low-energy systems without knowledge of the high-energy particles---it is a practical necessity.  As we flow to low-energy in QCD, the parameters tend toward infinity and so give us no indication of the EFT we are interested in.\footnote{This increase of the QCD coupling constant (corresponding to colour force) as we move to lower energy is known as ``infrared slavery'' or quark \textit{confinement}, and is the flip-side to asymptotic freedom mentioned in Footnote~\ref{foot:asy}, where the coupling constant decreases rapidly at high-energy.  See, for instance, \citet{Zee2010}.} Identifying the correct low-energy degrees of freedom is a very difficult task, and experimental input is required. In the case of QCD, the low-energy degrees of freedom are pions and nucleons, rather than quarks and gluons. 
\paragraph*{}
There are many other examples where this is the case, especially within condensed matter physics. The high-energy (being the Coulomb energy scale $1\sim 10eV$) description of a condensed matter system is quantum mechanics of electrons and nuclei, but this is utterly unhelpful for describing low-energy phenomena such as superfluidity and superconductivity, where instead we require an EFT of quasi-particles (phonons) which arise as collective excitations and are simply not present at short-distance \citep{Zhang2004}. Thus, although QFT understood as EFT is closely tied to---and thus limited by---current experimental capabilities, this should not be viewed as a defect of the theory, but rather a side-effect of its great utility. 
\paragraph*{}
An EFT may be approached in either a \textit{bottom-up} or \textit{top-down} manner, with the terms bottom-up and top-down referring to relative positions and motion along the energy scale. The \textit{top-down} approach is taken when the high-energy theory (of cutoff $\Lambda_2$) is known, but a low-energy EFT (of cutoff $\Lambda_1< \Lambda_2$) is required (for reasons that will be explained in \S\ref{sub:tower t-d}). As stated above, the trick lies in identifying the appropriate degrees of freedom for the low-energy system being studied, and writing the theory in terms of these \citep{Manohar1997, Pich1998}. Interaction terms corresponding to heavy particles, i.e. those of mass greater than $\Lambda_1$, are integrated out and replaced by a finite series of local (non-renormalisable) interactions among the light particles.\footnote{There is an alternative method of constructing EFTs in which regularisation does not involve integrating out the heavy particles, but instead employs a mass-independent subtraction scheme. This program, known as \textit{continuum} effective field theory, will be discussed in \S\ref{subsub:continuum}}. The non-renormalisability of the resulting EFT is not problematic, however, as we shall see in a moment.
\paragraph*{}
The \textit{bottom-up} approach is taken either in the case where there is no theory amenable to the system being studied, or there is a low-energy theory, but its high-energy counterpart is required. In the first of these cases, an effective Lagrangian is constructed based on symmetry constraints and the interactions assumed to be relevant at the energy scale under consideration.\footnote{For examples see \citet{Kaplan1995, Manohar1997, Zee2010}.} In the case where the low-energy theory (with cutoff $\Lambda_1$) is known, and a higher-energy theory (with cutoff $\Lambda_2>\Lambda_1$) is sought, the renormalisation group equations can be used to adjust the parameters already included in the Lagrangian density, but this procedure is complicated by the fact that as we approach $\Lambda_1$, the theory behaves as though it contains non-renormalisable terms. This is taken as an indication that new physics is to appear between $\Lambda_1$ and $\Lambda_2$. Additional terms must be included in the theory to take into account new interactions, not just those arising from new particles of mass greater than $\Lambda_1$, but different interactions of particles of mass less than $\Lambda_1$ that become relevant at higher energy. In the majority of cases, experimental input is required in order to fix these additional parameters. 
\paragraph*{}
Notice that the $\Lambda_2$ theory is constructed so as to save the phenomena covered by the lower-energy theory, and thus will be empirically equivalent to the original theory at energies less than $\Lambda_1$, yet it will have different numerical values for its coupling constants, representing particle masses and charges \citep{Hartmann2001}. There are other sources of underdetermination in QFT, as will be discussed in \S\ref{subsub:underdet}.  For now, I consider the construction of an EFT in a little more detail, to show how the terms in our theory change as we flow to different energy scales.

\subsection{Basic formalism}\label{subsub:EFTformalism}
Formally, the basic idea behind EFT is that a physical process with some characteristic energy $E$ can be described in terms of an expansion in $E/\Lambda_i$ , where $\Lambda_i$ are various physical scales involved in the system which are larger than $E$ \citep{Pich1998}. The effective Lagrangian, $\Lagr$, is written as a sum of operators, $\mathcal{O}$, which are allowed by symmetry constraints of the theory,
\begin{equation}\label{eq:EFTsum}
\Lagr_{eff}=\sum_ic_i\mathcal{O}_i
\end{equation}
where $c_i$ are coupling constants of magnitude dependent on the heavy energy scale, the dimension of spacetime, and the dimension\footnote{The dimension of an operator characterises its scaling properties.}, $d_i$, of the $i$th operator, $\mathcal{O}_i$, so that:  $c_i \sim 1/\Lambda^{d_i-4}$ (the $4$ in the exponent represents the dimension of spacetime, here we are working in four-dimensional relativistic field theory). This can be more clearly written by defining dimensionless coupling constants: $\lambda_i =  c_i\Lambda^{d_i-4}$, so that the order of the $i$th term in (\ref{eq:EFTsum}) is $\lambda_i(E/\Lambda)^{d_i-4}$.
\paragraph*{}
The operators fall into three categories, based on their dimension.\footnote{My discussion here borrows from \citet{Pich1998, Polchinski1993}} Operators with $d_i < 4$ have positive-dimensional coefficients and are called \textit{relevant} because they are relevant at low energies, but get smaller as $E\rightarrow \Lambda$. There are only a few different operators which fall into this category in four-dimensional relativistic field theory:  $d = 0$ being the unit operator, $d = 2$ being boson mass terms, and $d = 3$, being either fermion mass terms or 3-scalar interactions. Terms featuring these operators are renormalisable. \textit{Irrelevant} operators have $d_i > 4 $ and thus negative-dimensional coefficients. Terms featuring irrelevant operators are non-renormalisable. At low-energies these terms are suppressed by powers of $(E/\Lambda)$, but become important as $E\rightarrow\Lambda$. Finally, operators with $d_i = 4$ and dimensionless couplings are called \textit{marginal}, and are constant at high- and low-energies. Marginal terms, however, will almost always have their dimension altered once quantum corrections are taken into account and hence become either relevant or irrelevant (although it is not clear whether this is true in quantum gravity).
\paragraph*{}
Typically, as indicated by dimensional analysis, the expansion (\ref{eq:EFTsum}) will contain only a finite number of marginal and relevant operators. It can, however, contain arbitrarily many irrelevant operators. The EFT approach is premised on there being a large gap between the energy being studied, $E$, and the energy scale at which new physics appears, with the latter characterised by a heavy mass, $M$. In the regime where $E\ll M$, the irrelevant operators (non-renormalisable terms) are suppressed by powers of $(E/M)$ in the expansion. Conceiving of QFTs as EFTs thus means accepting that---although all our theories contain both non-renormalisable and renormalisable terms---provided we are working in situations where $E\ll M$, our EFT will behave for all practical purposes like a renormalisable QFT. We are not prevented from making predictions. 

\subsection{Appelquist-Carazzone decoupling theorem}\label{subsub:Appelquist}
As should be clear, the EFT formalism relies on there being a range of heavy mass scales that are sufficiently far apart from one another that we could call them ``neatly separated''. The basic approach just described (\S\ref{subsub:EFTformalism}) finds formal legitimacy in the decoupling theorem of \citet{Appelquist1975}. If we have a renormalisable theory describing coupled fields of different masses, the theorem shows that at energies small compared to the heavy masses, the heavy masses \textit{effectively decouple}.  This means that the effects of the heavy fields can always be included in the low-energy theory through the process of renormalisation (i.e. a simple change of the parameters of the theory), provided that: 1. we begin with a complete renormalisable theory, 2. the mass scales are neatly separated, and 3. we are working at energies small compared to the heavy masses. 

\section{Tower of theories}\label{sub:tower}
The two directions from which to approach EFT each have their own philosophy as well as methodology. These will be briefly considered in the following two sub-sections. The relationship between quantum gravity and GR may be explored from both directions, as will be demonstrated in \S\ref{sect:GREFT}. As well as these two directions, there are also two different approaches to EFT---so far, I have only discussed one, being a \textit{mass-dependent} approach, but there is also a \textit{mass-independent} approach, which does not rely on a cutoff energy in renormalising the theory. Although this approach should produce EFTs that are empirically equivalent to those of the mass-dependent scheme, they are conceptually different; this approach is introduced in \S\ref{subsub:continuum}, and some of its implications are discussed at times throughout \S\ref{sub:CaoSchweb} and \S\ref{sub:effectiveEFT}.

\subsection{Bottom-up}\label{sub:tower b-u}
As long as we are working at energies well below the cutoff, $M$, our EFT will be approximately renormalisable, and we can act as though the heavy particle $M$ does not exist. Non-renormalisable interactions due to the heavy particle are negligible, being suppressed by powers of $E/M$. As we move to energies approaching $M$, however, the non-renormalisable terms grow and the heavy particle can no longer be ignored. At some stage, as the theory starts to show evidence of breakdown, a new theory must be constructed that explicitly includes the heavy particle. This new theory will, in turn, be valid at energies that are small compared to some heavier mass serving as the cutoff, and will be well-behaved in this regime, until we again move to higher energies where non-renormalisable terms become non-negligible and we need to construct a new theory. In this way, we can imagine the world described by a tower of EFTs, with the mass of each particle potentially representing a boundary between theories. This situation, in which each heavy mass $M$ corresponds to a new cutoff scale, is known as the ``extreme view'' of EFT (Georgi 1989, p. 455).\footnote{While it is Georgi himself who terms this the ``extreme view'' (see also Georgi, 1993, p. 212),  it is not a view that he finds at all problematic, for reasons that will become clear in \S\ref{sub:effectiveEFT}. Other authors, however, in following a possible misrepresentation of this view by Cao and Schweber (1993) have taken it as controversial. This will be discussed shortly \S\ref{sub:CaoSchweb}.}

\paragraph*{}  
Notice that this tower is constructed \textit{bottom-up}---we don't have the high-energy theory, but we inch our way up, adding more and more terms to the theories we have in order to account for the effects of unknown interactions. This method of constructing a tower of EFTs is borne by practical necessity; it is a pragmatic response to the current experimental situation. In the absence of a high-energy theory, we have to construct our theories as descriptions of the low-energy realm that is accessible to us. We then extend these theories as higher-energy physics becomes important. This view of tower construction has become the standard conception of EFT; usually when authors refer to a tower of EFTs, it is the bottom-up tower that they have in mind. If we had a high-energy theory beyond the standard model, the expectation is that the less-desirable bottom-up tower of EFTs could be replaced. For example, Hartmann (2001, p. 296),  states that instead of accepting a complicated bottom-up tower, ``theorists will search for a more fundamental theory which will reduce the contingency that goes along with the tower construction strategy.'' 

\paragraph*{}
As we have seen, the construction of the high-energy theory from a low-energy EFT is a complicated procedure. The low-energy theory must be modified so as to take into account any new interactions between the already-featured particles that become relevant at higher energy, and this may involve modifying the original parameters of the theory as well as adding new ones. The new theory must also include the masses, charges and other coupling constants of any new particles that are able to be created at the higher energy. While QFT supplies a tool-box that enables us to construct the form of the new terms in the Lagrangian density, the values of the coupling constants must be determined by experiment. Thus, the tool-box of QFT can do little to guide us through the murky waters beyond the boundary of experimentally-accessible energies. This problem is compounded by EFT, whose nature means that there is very little that ties the low-energy effective theory to the high-energy theory that underlies it (this is related to the idea of emergence, to be discussed in \S\ref{sub:emergence}). 

\paragraph*{}
Condensed matter physics furnishes two well-known examples which illustrate typically how little a low-energy prediction depends on the high-energy theory. Both examples are related to the BCS theory of superconductivity \citep[][p. 8]{Burgess2004}. The BCS theory is an approximation which ignores most of the mutual interactions of the electrons and focuses only on a particular interaction due to phonon exchange. In spite of this, it works surprisingly well in many situations, with its predictions often in agreement with experiment to within several percent. The success of the BCS theory can be explained by EFT---it can be shown that only the specific interactions used by the BCS theory are relevant at low energies. All the rest of the interactions---important in the high-energy theory---are suppressed at low energies by powers of a small energy ratio (Burgess, 2004, p. 9).  

\paragraph*{}
The second example is the success of the BCS theory as applied to the Josephson Effect, used to determine the value of the fine structure constant in precision tests of QED. In this case, the agreement with experiment is in terms of parts per million! Remarkably, this success is not due to any of the details of the BCS model except for one important symmetry-breaking pattern it predicts. As Burgess explains, the Josephson prediction follows on general grounds from the low-energy limit of this broken gauge symmetry---any theory with the same low-energy symmetry-breaking pattern shares the same predictions. Considering these cases, it becomes clear how difficult it is, given only the low-energy theory, to determine the high-energy theory.  The relevant details of the high-energy theory at low energy severely underdetermine the high-energy theory. 
\paragraph*{}
Importantly, this is also true of the standard model of particle physics. As Gross (2004, pp. 60-61) explains, the standard model is the inevitable low-energy manifestation of \textit{any} high-energy theory, so long as the high-energy theory is local and contains the gauge symmetry we observe. If we begin with the assumption that we have a local quantum field theory defined just below some very large characteristic energy scale $\Lambda$ then we need to assume nothing else of whatever theory describes the world at or beyond this scale. The most general QFT just below $\Lambda$ has an infinite number of arbitrary parameters describing all possible fields and interactions. However, at low energies (of order $E$) the EFT will contain only a finite number of couplings, with the RG determining how these parameters flow with the energy. The EFT we obtain will be equivalent to the standard model, up to corrections of order $E/\Lambda$, which are negligible at the low $E$ for which we are working. The standard model thus tells us very little of the high-energy theory beyond---a theory which need not even be a QFT.
\paragraph*{}
The bottom-up tower of EFTs has been the subject of much controversy. As will be discussed in \S\ref{subsub:finaltheory}, the tower of EFTs has mistakenly been interpreted as a threat to the possibility of there being an ultimate fundamental theory of physics. As most (if not all) authors on the topic recognise, however, the success of the EFT approach does not actually imply anything about the existence of a final theory. As will be argued, the philosophy of EFT that I advocate in this thesis means that we should remain agnostic about physics at high energies, apart from certain broad characteristics that the low-energy physics demands.

\subsection{Top-down}\label{sub:tower t-d}
Although the bottom-up tower of theories is taken as the standard understanding of EFT (in philosophy), there is an alternative way of constructing a tower, via the top-down method. While a tower of EFTs constructed bottom up might provide indication, or ``hints'', of the high-energy physics, as we shall see in considering GR as an EFT (\S\ref{sect:GREFT}), the top-down aims to capture the relevant physics at low energies. Also, while a bottom-up tower is typically perceived as bulky and inelegant, owing to the need to add more and more parameters to account for the effects of the unknown (high-energy) physics (\S\ref{sub:EFT}), the top-down approach produces effective theories that are simple, elegant and heuristically useful.\footnote{\citet{Shankar1999} and Zhang (2004) provide examples of top-down effective field theories that reveal simplicity and beauty in the low-energy physics compared to the high-energy systems.} Even when we are in possession of a high-energy theory, it is \textit{appropriate} and \textit{important}, when studying a system at low energy, to use a theory formulated at low energy.\footnote{\label{foot:G}``The two key words here are appropriate and important. The word important is key because the physical processes that are relevant differ from one place in parameter space to another. The word appropriate is key because there is no single description of physics that is useful everywhere in parameter space." (Georgi, 1993, pp. 211--212)}

\paragraph*{}
There are two main reasons why it is appropriate and important to use top-down EFT. Firstly: typically, the high-energy theory is too complicated to apply at low-energy.  Consider the example (from Pich, 1998) of the electromagnetic interaction as the basis of the laws of chemistry: using QED of quarks and leptons to describe chemical bonds among atoms is both unnecessary as well as extremely difficult. Instead, a simplified description in terms of non-relativistic electrons orbiting a nuclear Coulomb potential provides a better understanding of the relevant physics; for a first approximation, all that is required is the electron mass and the fine-structure constant, and only the proton mass is needed to calculate the leading corrections. Yet, even this simplified picture quickly becomes too cumbersome to apply to condensed matter or biological systems. 
\paragraph*{}
Secondly, because the high-energy theories fail to provide understanding of the relevant phenomena, they do not allow us to study the low-energy system in question.  EFTs produce local, intuitive understanding and avoid the problem, as Hartmann (2001, p. 294) puts it, of ``being too far away from the phenomena". EFTs are the appropriate way of doing physics at low energy, because they are formulated in terms of the low-energy degrees of freedom. 

\paragraph*{}
Effective theories are sometimes spoken of as approximations to higher energy theories, but it is important to remember that they are typically neither less accurate nor less apt than their higher energy counterparts. Although a higher-energy theory presents a more fine-grained description of the ontology, this is not only unnecessary, but also unhelpful when we are studying low energy systems. The fact that the low-energy theory describes degrees of freedom that simply do not exist at high-energy is an important part of idea of emergence in EFT that I advocate, (\S\ref{sub:emergence}).

\subsection{Wilsonian versus Continuum EFTs}\label{subsub:continuum}
The EFT formalism that I have presented thus far is known as the \textit{Wilsonian} EFT approach: these are EFTs that employ a mass-dependent renormalisation scheme.\footnote{Named after Kenneth Wilson, who played an integral role in developing the conceptual and mathematical apparatus of RG. Wilson was awarded the 1982 Nobel Prize for this work.} As we have seen, in order to construct a top-down EFT in this approach, the heavy fields are demarcated with respect to the cutoff energy $\Lambda$, and are then integrated out of the high-energy Lagrangian. There is an alternative way to construct EFTs, known as continuum EFT, which employs a mass-independent renormalisation scheme. I base this brief outline on \citet{Georgi1993}. In constructing a top-down EFT in continuum EFT, we start with a theory defined at some large energy scale, $\mu$, with Lagrangian density of the form,
\begin{equation}\label{eq:continuum}
\Lagr_H(\chi,\phi)+\Lagr(\phi)
\end{equation}
Where $\Lagr(\phi)$ describes the light fields, and $\Lagr_H(\chi, \phi)$ describes everything else: $\chi$ being the set of heavy fields of mass $M$. 
\paragraph{}
We then use the RG to scale the theory to lower energy. When $\mu$ goes below the mass, $M$, of the heavy fields, we move to an EFT that does not include these fields. The parameters of the theory will change, and new, non-renormalisable interactions may be introduced. The new Lagrangian has the form,
\begin{equation}
\Lagr(\phi)+\delta\Lagr(\phi)
\end{equation}
Where $\delta\Lagr(\phi)$ is the ``matching correction'' that encodes all the changes (Georgi, 1993, p. 228).  These changes to the theory are calculated by ``matching'' the physics just above $M$ with the physics just below $M$, so that the physics does not change across the boundary $\mu = M$. We can calculate the matching correction by expanding $\delta\Lagr(\phi)$ as a complete set of local operators, in the same way as is done in Wilsonian EFTs (\ref{eq:EFTsum}), so that,
\begin{equation}\label{eq:contEFTsum}
\Lagr_{eff}=\Lagr (\phi)+\sum_i \delta \Lagr^i(\phi)
\end{equation}
We can now use dimensional analysis to determine how the terms in (\ref{eq:contEFTsum}) scale as we change energy, just as in Wilsonian EFTs.
\paragraph*{}
This approach is known as continuum EFT because it does not rely on a cutoff energy in renormalising the theory, and, as such, there is no associated ``discretisation''.  It is important to notice that the decoupling theorem (\S\ref{subsub:Appelquist}) does not apply in continuum EFT. While Wilsonian EFT is concerned with how a theory changes as we integrate out the heavy fields and move to larger distance scales, continuum EFT is concerned with how we need to modify a theory, using a mass-independent renormalisation scheme, in order to get the physics right at lower energies. Thus, the idea is to put in ``by hand" as much of the scale dependence as we can \citep[][p. 215]{Georgi1993}. The decoupling of the high-energy theory from the low-energy theory is put in by hand: it is ensured by the matching process. 

\section{Reconceptualising the cutoff}\label{sub:reconcutoff}
As stated earlier, the new conceptualisation of QFT also entails a new understanding of the cutoff in our theories as a physically meaningful parameter. I have emphasised, and will continue to emphasise, that the best way to understand this claim is simply: the cutoff represents the upper edge of the domain of validity of the theory. The physical mechanism responsible for the breakdown of the theory at energies approaching the cutoff scale is a question external to the theory itself. In all cases, remember that at distances far from the cutoff, the high-energy physics and the actual mechanism responsible for the existence of the cutoff are irrelevant, except insofar as the value of the cutoff determines the values of the parameters of the theory. Although in the preceding discussion, the cutoff has been interpreted as a massive particle, we cannot expect this to always be the case, especially when we are talking about theories beyond the Standard Model.
\paragraph*{} 
This situation stands in contrast to the one in condensed matter physics, where the physical mechanism behind the cutoff is clear. The cutoff corresponds to the range of interaction of the constituents of the model, for example the lattice spacing for spins, or the average molecular distance for fluids. An example of a low-energy description of condensed matter phenomena is an EFT of phonons in a crystal; in this case the EFT breaks down as we approach energies comparable to the atomic lattice spacing---eventually the phonons have short enough wavelength to detect the discrete nature of the underlying system, and the continuum description in terms of fields is unworkable. This will be important in the discussing analogue models of spacetime in condensed matter physics (\S\ref{sub:analogue}). In QFT the nature of the cutoff is less clear.
\paragraph*{}
Recall that in ``traditional'' renormalised QFT, imposing the cutoff is akin to placing the theory on a lattice: the underlying spacetime no longer continuous, but breaking down at some short length scale. At the end of calculations, the cutoff was restored to infinity and a continuum theory was recovered. In RG, of course, the cutoff remains in the theory. This is again equivalent to placing the theory on a lattice: Poincar\'{e} covariance, translation and rotation invariance are violated. This is unproblematic so long as we remember the philosophy of EFT---our theories are not expected to hold to arbitrarily high-energy. Whatever is going on beyond our theory, whatever is responsible for the expected failure of the theory and the apparent violation of Poincar\'{e} covariance at high-energy is unknown and irrelevant to the low-energy physics. 

\paragraph*{}
As \citet[][p. 123]{Wallace2011} explains, the cutoff could actually be imposed by a Poincar\'{e} covariant theory (string theory is Poincar\'{e} covariant, for example); but even if the underlying theory is not Poincar\'{e} covariant, this needn't be problematic---we can manage fine with effective (phenomenological) Poincar\'{e} invariance in our EFTs. The apparent violation of Poincar\'{e} covariance, however, is probably just an artefact of the way the theory has been constructed. \citet{Bain2013}, for instance, argues that \textit{continuum} EFT---although it generates theories that are empirically equivalent to those of Wilsonian EFT---does not violate Poincar\'{e} covariance because it does not implement a mass-dependent renormalisation scheme.

\section{Philosophy of EFT: Cao and Schweber (1993)}\label{sub:CaoSchweb}
The most influential philosophical commentary on EFT is one of the earliest. Cao and Schweber (1993) controversially present the EFT program as supporting ``a pluralism in theoretical ontology, antifoundationalism in epistemology and an antireductionism in methodology" (p. 69). The arguments for this statement stem from the authors' claim that the bottom-up approach, being of practical necessity given that we do not have an ultimate high-energy theory, leads to ``an endless tower of theories, in which each theory is a particular response to a particular experimental situation, and none can ultimately be regarded as the fundamental theory" \citep[][p. 66]{Cao1993}. The claims of \textit{ontological pluralism}, \textit{epistemological antifoundationalism} and \textit{methodological antireductionism} are all interconnected, and have each been attacked on several fronts by subsequent authors on the topic. 

\paragraph*{}
I consider each of these claims and the criticisms of them separately (although their interconnectedness means that the order in which I do so is a bit muddled). Overall I find that Cao and Schweber's (1993) presentation of EFT, together with other authors' responses to it, has resulted in a thorny characterisation of the approach, apt to be misunderstood. The central problem is a mix-up between EFT as it applies in principle and EFT as it is currently used in physics. I argue that it is only the latter application of EFT that is important. Once we recognise this, the confusion surrounding Cao and Schweber's claims is dissolved and we can see clearly which is justified on this view and which is not.

\subsection{Epistemological antifoundationalism}\label{subsub:epistanti}
The current lack of a high-energy theory beyond the standard model necessitates a bottom-up approach to high-energy physics. We have to construct a picture of the world based on the energies accessible to us, and gradually extend our theory as higher energy interactions become relevant (as described in \S\ref{sub:tower b-u}). Cao and Schweber argue that this leads to an endless tower of theories, and that none of these theories can be regarded as \textit{the} fundamental theory. 
\paragraph*{}
The claim of ``epistemological antifoundationalism'' that Cao and Schweber make is based in three other ideas that these authors present: the idea that EFT gives us an endless tower of theories, the idea that the physics described each of these theories is to a large extent independent of the physics at other energies, and the idea that EFT does not allow us to get from one such theory to another by way of derivation. The first of these ideas, that there is no final theory, will be examined shortly, \S\ref{subsub:finaltheory}, where I find that EFT is best understood as neutral with respect to the existence of  final theory. The second and third of these ideas, which represent Cao and Schweber's conception of emergence, I explore in \S\ref{subsub:quasiaut} and \S\ref{subsub:rel layers}, respectively, finding that they are essentially correct. After propounding my own view of emergence in EFT, in \S\ref{sub:emergence}, I state briefly its relation to these ideas---although they are certainly related to emergence in EFT, we do not need to make the additional claim of epistemological antifoundationalism (mostly because, as explained in \S\ref{sub:fundamentality}, I have nothing to say about what it means for a theory to be ``fundamental'').

\subsection{Final theory}\label{subsub:finaltheory}
The success of the EFT program does not imply anything about the possible existence of a final theory (taking ``final theory'' to be a theory valid at the ``highest energy-scale'' and as providing an ultimate base for reductionism). This fact has been emphasised by a number of authors responding to Cao and Schweber, namely: 
\citet{Castellani2002, Hartmann2001, Huggett1995, Robinson1992}. This is an \textit{external criticism} of Cao and Schweber, however, since, as will be explained shortly, the \textit{bottom-up} EFT program, in conjunction with the EFT philosophy, \textit{does} imply that there can be no final theory.\footnote{To be explicitly clear: I argue that, although the bottom-up EFT program does imply that there can be no final theory, we should take a pragmatic view of EFT where the formalism itself is viewed as effective. We should recognise that EFT is not really a ``program'' that requires die-hard subscription and fights in opposition to the search for a final theory, but rather a pragmatic necessity, and wonderful aid to progress in physics. The success of EFT does not mean there is no final theory (as most other authors have recognised) and viewing EFT as effective means remaining open to all possibilities (a view that other philosophers have not considered, but one that is hinted at by some physicists in their papers on EFT).} As such, EFT has been set up as standing in opposition to a final theory, and is seen to be in competition with the approaches to physics that aim at a final theory. The focus in the philosophy of EFT has become a debate between whether there is a never-ending tower of EFTs or whether there is a final theory. This is an unfortunate and unnecessary way to understand EFT. Bottom-up EFT is the only secure foothold we currently have on high-energy physics: it allows progress inch-by-inch, bound by experimental capabilities. It is not aimed at producing a final theory, but, as I will argue, neither should it be understood as in competition with those approaches that do aim at a final theory.
\paragraph*{}
The reason why the authors responding to Cao and Schweber make the claim that we cannot exclude the possibility of one day finding a final theory must be because they believe we are not confined to the bottom-up EFT program.  Although Cao and Schweber may be criticised for ignoring the possibility of there one day being a final theory, they are not incorrect in characterising the bottom-up EFT approach as leading to an endless tower of theories. The endlessness of the bottom-up tower of theories, Cao and Schweber (1993, p. 66) state, is entailed by the local operator formulation of QFT: recall that locality implies that an arbitrarily large amount of energy is available for the creation of particles (\S\ref{sub:QFTrenorm}). Of course, this is a basic assumption of QFT and one of the foundations of the theory. 
\paragraph*{}
If we stick with QFT, then, the bottom-up tower should be endless. Further, if we accept the philosophy of EFT, then we accept our inability to ever claim discovery of a final theory. Cao and Schweber's point is that the tower will continue to grow with the limits of our experimental capabilities. Within this picture, we know that any theory we arrive at must be considered \textit{effective}: apt to be replaced as higher energies are probed.\footnote{Consider \citet[][p. 186]{Huggett1995}, ``So we have the following picture; at low energy we have a renormalisable EFT, but as we increase the energy it starts to behave as if it has non-renormalisable terms, which grow in importance until they must be replaced by introducing a new renormalisable QFT particle. But the end of this story leaves us in an identical position to the beginning; again we have a renormalisable theory. It must be a possibility that as the energy increases non-renormalisable interactions will appear, and our theory with the heavy particle will turn out to be an EFT of another theory with an even heavier particle. And mightn't that turn out to be an EFT?''} 
In order to talk about the possibility of a final theory we have to step outside the bottom-up EFT approach.  
\paragraph*{}
EFT is generally recognised as being a pragmatic necessity---without experimental testing we cannot make headway in the quest for a final theory. Philosophers understand it as the way to progress in physics; however, many are reluctant to accept the accompanying picture of the world---viewing it, in some sense, as a cop-out. \citet[][p. 40]{Redhead1999} captures the sentiment adequately when he says, that ``To subscribe to the new EFT programme is to give up on this endeavour [of searching for an ultimate underlying order] and retreat to a position that is admittedly more cautious and pragmatic and closer to experimental practice, but is somehow less intellectually exciting". Huggett and Weingard (1995), in spite of not wanting to commit to either the ``assumption of plurality" or the ``assumption of unity", admit EFT as the way to continue to expand the scope of physics. They think that this is unproblematic, so long as we opt for a ``mixed strategy", wherein we also continue to work on those approaches that are aimed at a final theory. 
\paragraph*{}
Rather than diverting the discussion onto methodology (although we will get there in the next sub-section), we should pause to appreciate the real problem here. We recognise that if we are to progress in physics we need to use the EFT approach. We recognise the theories produced by this approach as being the most accurate physical theories ever. We realise that taking it seriously means giving up on the dream of a final theory, but at the same time we believe that the dream is still viable---although the EFT approach may exclude it, we cannot say that nature does not. How can this be? Are we just in denial, clinging desperately to the dream while glossing over the success and necessity of EFT? 

\paragraph*{}
Instead, I think we must realise that EFT is \textit{not} actually in opposition to a final theory. This idea, which I have hinted at throughout this section, and argue for more thoroughly in \S\ref{sub:effectiveEFT} means recognising EFT itself as \textit{effective}: we can accept EFT as providing our best current physics, while remaining open to the possibility of it being revised.\footnote{I am hesitant to just take EFT (with its associated philosophy of a never-ending tower) as ``wrong''. Given the fact that these theories work so well, and that we do not know what the future holds, it would be presumptuous to make any such assertion (assuming that the way we define ``wrong'' in this case is simply by using it to describe an overthrown theory once its successor has been implemented). On the other hand, by labelling EFT ``effective'', I do not mean to imply that it is ``true'', either---I would very much like to avoid getting into any debate as to what ``true'' means.} Indeed, a common thought seems to be that QFT will cease to be applicable at some energy and new physics will take over.\footnote{As I will discuss in \S\ref{sub:effectiveEFT}.}
\paragraph*{}
Although EFT provides an incredibly useful and accurate description of the world at the energies accessible to us, it is a mistake to expect the formalism to hold to arbitrarily high energy. Just as EFT counsels us to remain agnostic about the physics beyond any particular theory, we should, in turn, remain agnostic in regards to the status of EFT at energies far beyond those currently accessible. In this way, we can understand EFT as truly neutral with respect to the existence of a final theory. 

\subsection{Methodological antireductionism}\label{subsub:meth anti}
Cao and Schweber (1993) claim that the development of EFT marks a shift in the methodology of high-energy physics, from being driven by the search for an ultimate renormalisable theory, to one that accepts theories as locally-valid, contingent and phenomenological descriptions of nature. Both the authors making this claim, as well as those reacting to it, further entrench the view of EFT antithetical to the search for a final theory.\footnote{One exception is Hartmann (2001) who, from the outset, presents EFTs, theories and models as different things, and argues that each plays a different, essential role in scientific practice.}  In this section I want to show that it is wrong to treat EFT as somehow in competition with other approaches to high-energy physics. Huggett and Weingard (1995) present the most forceful explicit response to methodological antireductionism, so I will focus on discussing their view here, but the comments I make are supposed to hold more generally, against any position that views EFT as being opposed to other possible ways of doing high-energy physics.
\paragraph*{}
Huggett and Weingard (1995) describe three different ways of understanding EFT, and three corresponding ways that ``physics should proceed''. The first view they present is the idea that the world is described by a unified theory. On this view, EFTs are \textit{unphysical} and physics should proceed by searching for a single renormalisable theory.\footnote{It is left unclear exactly what ``unphysical'' means, but, apparently Huggett and Weingard take it to refer to any theory that is not valid at the energy scale of the ``final theory'', given their labelling of QED as unphysical because it is not asymptotically safe. ``Today, careful searches have convinced most that it is not asymptotically safe at any fixed point. In this sense then QED is unphysical - its applicability is restricted to a certain energy range. This is probably surprising news, but despite being the most predictively accurate theory ever, QED probably cannot be true.'' (p. 184)} The second possibility that Huggett and Weingard present is that EFTs are approximations to the complete ``true physics'', and physics should proceed by searching for new theories which accommodate EFTs but ``overcome the problems which afflict them" (1995, p. 185). The third possibility is what Huggett and Weingard label ``the EFT view"---the view that ``nature is described by a genuinely never-ending tower of theories, and that the competing possibilities of unification and new physics be abandoned" (p. 187).\footnote{Actually, Huggett and Weingard (1995, p. 187) present two different variants of this view. The regular strength version, which they attribute to Cao and Schweber, is the one they focus on, and is the one I present here. The mild strength version they attribute to \citet{Georgi1989}, and is simply the caution that although there may be a final theory, it might be very far away from the energies currently accessible to us.}  
\paragraph*{}
Now, it seems obvious that we can only evaluate the question of ``how physics should proceed'' if we have some fixed goal in mind. EFT, however, has a different goal than the other two ``alternatives''---it does not aim at a final theory---so setting it up alongside these other views is an immediate mistake. Huggett and Weingard beg the question against EFT, too, in reserving the status of physicality for theories whose domains of applicability are not restricted to certain energy scales. This idea that only theories that apply at all energy scales can be physical means that, from the outset, EFT is not on the same footing as the other ``alternatives''. The other problem that Huggett and Weingard seem to believe that EFT suffers from is complexity, i.e. the need to add more terms to our theories as we move bottom-up.\footnote{Of course, although this is typically how a bottom-up tower is constructed, it is not always the case that higher-energy QFTs will be more complicated than their low-energy counterparts. For example, the different mesons were reconceptualised following the development of QCD, as being the states of ``more fundamental'' entities.} Viewing this as a problem also begs the question against EFTs: the need to add more terms is undesirable because we are lead, often, to more and more complicated theories rather than in the direction of a single, simple theory.\footnote{Ultimately, this three-way rivalry presented by Huggett and Weingard boils down to a clash between two prejudices: unity and plurality (where unification is associated with simplicity, while plurality is seen as undesirable complexity).  If unification and pluralism are taken as alternative paths toward the end-goal of physics, then the question of what this goal is is left open. This is a debate to be played out elsewhere, for although these issues may be relevant for EFT, EFT does not bear on these issues.}
\paragraph*{}
I maintain that the best way to understand EFT is to not view it as an ``alternative" to the search for a final theory. It cannot be judged against the same set of criteria as these programs. EFT can be thought of as doing something different to these programs, and need not be seen as a hindrance to them. One way to understand the different role played by EFT, and not see it as in competition with those approaches which aim at a final theory, is to treat it as being in a category of its own. 
\paragraph*{}
Hartmann (2001) for instance, treats EFT as being on a different level to that of theories, and argues that EFT and theories each play an important, but different, role in scientific practice. While theories have a wide scope of applicability and provide a coherent, unified account of a large range of phenomena, Hartmann says, they often fail to provide a real understanding of the physics we are interested in, because they are ``too far away'' from the phenomena. EFTs, on the other hand, are heuristically useful and provide a local, intuitive account of a given phenomenon. As well as failing to provide a local understanding of the relevant physics, a final theory could not be applied to a low-energy system without carrying out a low-energy expansion, which is where top-down EFT would play an important role. Furthermore, Hartmann (p. 296) points out, we will probably never arrive at a final theory without the ``supporting scaffolding" provided by various models and EFTs. 
\paragraph*{}
Bottom-up EFT may be understood as a practical means of reproducing the low-energy predictions of the high-energy theory (where by this I mean any theory that applies beyond the standard model). EFT may also be understood as the only way of doing high-energy physics that is firmly grounded in experiment. EFT allows physics to proceed with the frontier of accessible energies, while those approaches which aim at a final theory do not follow this path. EFT seems to make progress in one area while the other approaches are focused on another. 
\paragraph*{}
This is essentially the view underlying the ``mixed strategy" of Huggett and Weingard (1995). It is presented as the ``way physics should proceed" according to the ``mild strength version"” of the EFT view, which is simply the caution that although there may be a final theory, it might be very far away from the energies currently accessible to us. Hartmann (2001) also advocates a mixed strategy, wherein we continue to build up our EFT tower, because it can be systematically tested, but agrees with Redhead's \citeyearpar[][p. 49]{Redhead1999} argument that we do not give up on the idea of a final theory because it acts as a ``powerful aesthetic ingredient'' in a regulative ideal motivating the progress of science.
\paragraph*{} 
While a mixed strategy in regards to methodology might be advocated, the philosophy underlying it needs subtle clarification in order to be properly neutral.  EFT should not be seen as of purely instrumental value, as a means of keeping the predictions flowing while the search for a final theory continues elsewhere. There is, after all, the possibility that physics never finds a final theory. Although EFT does not fulfil the same desiderata as a final theory, there is the possibility that it fill its role. There is the possibility that EFT will be the final word in physics.

\subsection{Ontological pluralism}\label{subsub:ont plu}
Cao and Schweber's (1993) idea of ontological pluralism is based in the observed success of the EFT approach in presenting a picture of the physical world as organised into ``quasi-autonomous domains" (p. 72).  According to this picture, each domain has its own theoretical ontology, and this ontology is largely independent of those of the other layers. Accepting this pluralistic picture of the world, they say, is a consequence of taking seriously the philosophy of EFT and the decoupling theorem. It entails, they say, an acceptance of emergence, and the belief that the reductionist program is an illusion (p. 71). Although Cao and Schweber recognise that the RG equations enable the decoupling theorem to describe the ``causal connections'' between the layers, they reject the idea of these causal connections being of direct relevance to the practice of science, given that these connections alone do not allow us to arrive at the ontology at other layers. The authors emphasise the necessity of empirical input in inferring ``the complexity and the novelty that emerge at the lower energy scales from the simplicity at the higher energy scales" (p. 72). 
\paragraph*{}
Cao and Schweber's presentation of emergence in EFT has been the source of much confusion and controversy. Thus, in this section I will introduce this concept of emergence and give some indication of the confusion that surrounds it. In later sections (\S\ref{sub:emergence} and \S\ref{sect:Univers}) I will explore other notions of emergence in EFT and develop a new conception of emergence---or, rather, a more flexible cousin of emergence, as \textit{novelty and autonomy}---that is similar, in some respects, to Cao and Schweber's account of emergence, but which explains and dispels the confusion associated with their presentation. 

\paragraph*{}
Recognising that EFT entails ``quasi-autonomous domains'' is one step towards an argument for ontological pluralism. The second step, which is the question of whether or not we can accept quasi-autonomous domains without accepting ontological pluralism, is one that is unrelated to EFT and will not be discussed. Similarly, the results of my discussion of epistemological antifoundationalism (\S\ref{subsub:meth anti}) will have implications for the idea of ontological pluralism, if epistemological antifoundationalism is taken as support for the latter, but these will not be explored here. 

\subsection{Emergence: quasi-autonomous domains}\label{subsub:quasiaut}
Cao and Schweber's conception of emergence in EFT is composed of two parts: the idea that the world is layered into quasi-autonomous domains, and the idea that we cannot get from one layer to another working solely from within the framework of EFT. Both of these ideas have, I feel, been unfairly dismissed by subsequent authors, and Cao and Schweber's notion of emergence has yet to be properly examined. In this sub-section I will focus on making sense of Cao and Schweber's idea of quasi-autonomous domains, and in the next sub-section I discuss how the links between these domains can be understood as representing emergence. In both cases I find that the cloud of confusion that has been stirred up in the wake of other authors' criticisms can be dispelled by understanding EFT as effective, and focusing on how it is actually used in physics, rather than how it applies in principle.

\paragraph*{}
As stated above, Cao and Schweber take their notion of emergence to rest essentially on the observation that EFT describes the world as layered into \textit{quasi-autonomous domains}. These domains are remarkably stable: largely, but not absolutely, independent of the physics at higher energies. Quasi-autonomous domains correspond, essentially, to the levels of the tower of EFTs. Each domain is described by its own theory (or theories), and each of these theories describe very different degrees of freedom. Recall the lesson of RG that is at the heart of EFT, the low energy physics depends on the short distance theory only through the relevant and marginal couplings (and possibly through some leading irrelevant couplings if one measures small enough effects \citep{Polchinski1993}). The minimalism of the influence of the high-energy theory on the low-energy theory adjacent to it in the tower, together with the distinct degrees of freedom described by each, means that, given the low-energy theory, the high-energy physics is severely underdetermined, and so there is some basis for treating EFT as describing novel and robust behaviour (this is returned to in \S\ref{sub:emergence} and later in \S\ref{sect:Univers}). 
\paragraph*{}
Cao and Schweber (1993, pp. 64--65) are careful to emphasise the \textit{quasi}-autonomous nature of these domains: high-energy effects do manifest themselves in relevant and marginal terms, but thanks to the decoupling theorem, these effects are suppressed at low energy. Thus, Cao and Schweber explicitly appeal to the validity of the decoupling theorem in arguing for their picture of quasi-autonomous domains. Recall, however, that for the decoupling theorem to hold, we must start with a renormalisable (high-energy) theory and the mass scales of the system must neatly separate (\S\ref{subsub:Appelquist}). 

\paragraph*{}
Hartmann (2001, p. 298) argues that Cao and Schweber, having shunned the possibility of an underlying renormalisable theory in their insistence that the tower of EFTs is endless, are unable to appeal to the decoupling theorem, and, as such, must forfeit their claim to a picture of the world as layered into quasi-autonomous domains. It seems that the inconsistency in Cao and Schweber's view highlighted by Hartmann's criticism is a result of their switching between speaking of bottom-up EFT and speaking of top-down EFT.  As stated in \S\ref{subsub:finaltheory}, the view of the world as described by a never-ending tower of theories is an implication of constructing EFT bottom-up. On the other hand, discussion of the decoupling theorem is based on \textit{top-down} EFT. 

\paragraph*{}
\citet{Bain2013} responds Hartmann's criticism of Cao and Schweber, arguing that it only goes through if Cao and Schweber are referring to Wilsonian EFT, as opposed to continuum EFT. In continuum EFT, recall (\S\ref{subsub:continuum}), the decoupling theorem does not apply: quasi-autonomous domains are ensured by the matching conditions, which are put in by hand. I believe that Bain is mistaken with this comment, however, since, although in continuum EFT we do not need the decoupling theorem to hold in order to have quasi-autonomous domains (and so we do not need to require there be a high-energy renormalisable theory in order that the decoupling theorem hold), we still require that there be a renormalisable theory at high-energy in order to use continuum EFT (or that we begin at sufficiently low-energy that the non-renormalisable terms have been washed out, just as in Wilsonian EFT). This is clear from the presentation in \citet[][p. 213, 228-229]{Georgi1993}.

\paragraph*{}
\citet{Bain2013} states that Cao and Schweber's (1993, p. 64) description of EFT construction sounds like it refers to \textit{continuum} EFT and even quotes them as saying,``The EFT can be obtained by deleting all heavy fields from the complete renormalizable theory and suitably redefining the coupling constants, masses, and the scale of the Green's functions, using the renormalization group equations.'' So, Bain has perhaps missed the point of Hartmann's criticism, being that Cao and Schweber are illegitimately relying on there being a complete renormalisable theory, not that they are illegitimately relying on the decoupling theorem.

\paragraph*{}
Perhaps, though, one might think it reasonable to argue in the reverse: that commitment to the required posit of a high-energy renormalisable theory, is warranted because it explains the observed stability of our current theories. Such commitment, however, places a restriction on physics at energies beyond those to which we have access, and, as I have already begun to argue, we should understand EFT as cautioning us to avoid such commitments. In this case, the commitment to a high-energy renormalisable theory leads to a contradiction, thanks to the clash between the top-down and bottom-up approaches; we are committed to a high-energy renormalisable theory because it explains the success of our bottom-up theories, and yet, if we take these theories seriously, we are led to a picture in which there is no high-energy renormalisable theory. 

\paragraph*{}
The solution is, of course, to treat EFT as effective and not take its success as implying that we are stuck in an endless tower of theories. But, conversely, neither should we take its success as implying that there is a high-energy renormalisable theory. We do not need to posit the existence of a high-energy renormalisable theory in order to account for the observed stability of our low-energy physics. There can be decoupling even if the high-energy theory is non-renormalisable, provided that we are working at low enough energy that the non-renormalisable interactions are too small to be relevant.\footnote{An example of this is GR treated as an EFT, as discussed in \S\ref{sub:GREFT}.}
\paragraph*{}
Actually, I believe that so long as Cao and Schweber restrict their talk of quasi-autonomous domains so as to refer to only currently known physics, then they do not need lay any bets concerning unknown physics---such talk is justified regardless of whether or not there turns out to be a high-energy renormalisable theory beyond the Standard Model, and regardless of whether we use the continuum or the Wilsonian approach to construct our EFTs. We do not need to appeal to unknown high-energy physics in order to recognise that EFT, as it is used at the energy scales we are familiar with, includes decoupling as an essential feature. QFT is the basis of some of the most accurate and successful physical theories ever, and, with these theories being understood as EFTs, we see QFT describe the world as layered into quasi-autonomous domains. 

\paragraph*{}
While \citet{Hartmann2001} argues that Cao and Schweber are wrong to appeal to the validity of the decoupling theorem, given that we do not know whether or not there is a complete renormalisable theory beyond the Standard Model (and, for the top-down EFT formalism to hold, such a theory is required), I believe that Cao and Schweber needn't have cited the decoupling theorem. Cao and Schweber only appeal to the decoupling theorem in order to justify the observed success of the EFT story, which paints a picture of the world as layered into quasi-autonomous domains. But surely the success of this picture in describing the world is justification enough to accept it---indeed, this is surely the more important sense of justification in science---than the validity of some mathematical theorem. Although we may not have decoupling \textit{in principle} in all cases, we do have \textit{effective} decoupling (i.e. decoupling \textit{in practice}). Rather than talking about how EFT applies in principle, we should be talking about \textit{how it actually applies in physics}.

\subsection{Emergence: relations between the layers}\label{subsub:rel layers}
The idea of quasi-autonomous domains in Cao and Schweber's account is not on its own responsible for the notion of emergence these authors present. Emergence is based in the observed fact that we are unable to move from one such domain to one beneath it while working purely from within the formalism. Typically (especially in informal EFTs), an external source of guidance is required in order to identify the relevant low-energy degrees of freedom, and this usually takes the form of experimental data. \citet{Bain2013} expounds this view of emergence in EFT, which emphasises the idea that each layer has its own \textit{distinct theory}, and argues that these theories are \textit{derivationally independent} of one another, i.e., specification of the equations of motion (plus boundary conditions) of a theory will fail to specify solutions to the equations of motion of its EFT. 

\paragraph*{}
Reformulating the steps involved in the construction of an EFT in the form of a derivation, Bain says, is, in general, difficult, if not impossible. This is due to the fact that the construction of an EFT (of either the Wilsonian or continuum type) typically involves approximations and heuristic reasoning. I discuss Bain's account of emergence later (\S\ref{sub:emergence}). Here, I focus on Cao and Schweber's conception of emergence, and characterise it, simply, as the idea that EFT presents us with a picture of the world as layered into quasi-autonomous domains, and within this picture we cannot get from any given layer to one at lower energy without invoking external inspiration.

\paragraph*{}
Cao and Schweber's idea of emergence is essentially antireductionist: accepting emergence, they say, entails rejecting the reductionist program as illusory.  Notice that this account of emergence is very much based in how EFT is actually used, rather than in theory. As such, it is missed by authors who focus only on the formalism. \citet[][p. 265]{Castellani2002}, for example, argues that EFT does not imply antireductionism because the EFT schema allows ``definite connections'' between layers. She quotes Georgi (1989, p. 455), who states that if we had a complete renormalisable theory at infinitely short distances, we could work our way down to the physics at any lower energy in a ``totally systematic way''.\footnote{I assume that Georgi makes this claim simply to demonstrate the power of the continuum EFT formalism, since, presumably, of course, a complete (renormalisable) theory valid at any large energy scale should do the trick.} Thus, Castellani concludes that empirical input must only be required because we do not have the complete renormalisable theory at infinitely short distances: a reconstruction of the low energy physics is possible \textit{in principle}. 

\paragraph*{}
This \textit{in principle} idea is, however, so far away from reality and current possibility so as to be utterly uninteresting. In order to accept it, we must accept that there is a complete renormalisable theory valid at infinitely high energy, when, in fact, we have no reason to even suppose that such a theory could exist. Furthermore, as I will argue in the next section, Castellani's conclusion is mistaken: even if we did have such a theory, it's not clear that we would be able to use it to arrive at the physics at any lower energy without appealing to external resources. Georgi's quote that Castellani appeals to is based in the formalism of EFT, while EFT as applied in physics warns us against assuming that this formalism holds to infinitely high energy.

\section{EFT as effective}\label{sub:effectiveEFT}
To expect our theories to hold at energies far beyond that at which they were formulated is to miss the greatest insight of EFT. This is not a strange claim, in fact, it is the way Georgi understands EFT. He writes, for instance, that,
\begin{quotation}
In this picture [i.e. effective field theory], the presence of infinities in quantum field theory is neither a disaster, nor an asset. It is simply a reminder of a practical limitation---we do not know what happens at distances much smaller than those we can look at directly.\paragraph*{}
Whatever happens at short distances, it doesn't affect what we actually\textit{ do }to study the theory at distances we can probe. We have purged ourselves of the hubris of assuming that we understand infinitely short distances. This is the great beauty of effective field theory language. (Georgi 1989, p. 456).
\end{quotation}
And again, in a well-quoted passage from a later paper,
\begin{quotation}
It is possible, I suppose, that at some very large energy scale, all the nonrenormalizable interactions disappear, and the theory is simply renormalizable in the old sense. This seems unlikely, given the difficulty with gravity. It is possible that the rules change dramatically, as in string theory. It may even be possible that there is no end, simply more and more scales as one goes to higher and higher energy. 

Who knows? 

Who cares? 

In addition to being a great convenience, effective field theory allows us to ask all the really scientific questions that we want to ask without committing ourselves to a picture of what happens at arbitrarily high energy. \citep[][p. 215]{Georgi1993}.
\end{quotation}
Although Georgi warns us against speculating about physics at distances below the Planck length, he still states that if we had a theory at infinitely short length scales we could use it to systematically derive the physics at any larger length scale (recall from \S \ref{subsub:rel layers}). This is the main difference between my view and Georgi's: I believe we should refrain from making any such assertion. Georgi, presumably, believes it because it is justified by the formalism of continuum EFT.  I have already emphasised the distinction between EFT as it applies in principle and EFT as it is actually used in physics---there are cases where the ``in principle'' formalism is borne out in the world, but there are also cases where it is not. 

\paragraph*{}
Really, we should recognise this as distinguishing between different types of EFTs.\footnote{Thanks to Jonathan Bain for suggesting this to me.} Firstly,\label{informal} there are EFTs that are constructed from theories where the formalism can be systematically applied---we might call these \textit{formal EFTs}. In these cases, the low-energy degrees of freedom can be readily identified, and constructing an EFT is relatively straightforward. Secondly, there are those cases where we lack a formal way of identifying the appropriate low-energy degrees of freedom directly from the high-energy theory, and have to do a lot of work ``by hand'', utilising other methods and data---call these \textit{informal EFTs}. This distinction is discussed throughout this section (especially \S\ref{subsub:efEFT high}).

\paragraph*{}
Thus, because of the existence of informal EFTs, I've argued against taking conclusions derived from the formalism as necessarily being applicable to EFT as it is used. Now I make a further claim: we should not infer from the success of the formalism as it currently applies in physics that it will hold any more exactly at higher energies. In other words, if the required assumptions for EFT to hold at infinitely high energy were fulfilled (i.e. if we were to have a complete renormalisable theory with distinct mass scales), this is no guarantee that we could use it to systematically derive the physics of any lower energy. We should recognise EFT as itself applying \textit{effectively}. 

\paragraph*{}
In this section, I will firstly clarify my claim above, that we should remain agnostic regarding high-energy physics. This will tie in, later, with my discussion of emergence (\S\ref{sub:emergence}). The idea of effective EFT, however, does not simply mean high-energy agnosticism, but also cautions against strong commitments even at energy scales we are familiar with. I examine this claim shortly (\S\ref{subsub:effQFT}).

\subsection{Effective EFT and high-energy physics}\label{subsub:efEFT high}
As emphasised, the difficulty in constructing EFTs is that, in many cases (those which I've called ``informal EFTs''), the high-energy theory does not provide any indication of the low-energy physics. Condensed matter physics furnishes numerous examples, (and see, for instance, Shankar, 1999), but the most commonly cited instance is QCD.  As we flow to low-energy in QCD, the parameters tend toward infinity, which makes calculations in this regime extremely difficult. The further difficulty is that, although we want a low-energy theory formulated in terms of hadrons, we cannot readily derive the hadronic interactions from the QCD Lagrangian--- the high-energy theory provides no indication of the relevant low-energy degrees of freedom. Fortunately, because we know the symmetries of QCD we can write down an effective Lagrangian in terms of the hadronic states, and parametrise the unknown dynamics in terms of a few couplings. The resulting EFT is known as chiral perturbation theory (for more information, see Pich, 1995).
\paragraph*{}
Chiral perturbation theory is an example of an informal EFT (according to the definition on p. \pageref{informal}). It represents a case where we cannot straightforwardly derive the low-energy physics given only the high-energy theory, and thus defies the formalism of EFT as it applies in principle (according to which, if we have a complete, approximately renormalisable high-energy theory, then we are able to systematically derive the physics at any lower energy). Of course, although we are unable to analytically or perturbatively solve QCD in the strong-coupling (low-energy) regime, we still believe that the theory contains all the information needed in order to make low-energy predictions. Researchers are slowly pushing forward in the derivation of the hadron masses, mostly using numerical methods or the lattice QCD approximation.\footnote{See, e.g. \citet{Durr2008}.} The claim is not that it is impossible to derive the low-energy physics given only the high-energy theory, but just that the systematic formalism of EFT does not always hold as strictly as other authors claim that it does. Instead, we should recognise that there are many different types of EFTs---and the differences between them are important in understanding EFT in general. 

\paragraph*{}
Consider again the claim that, if we had a complete renormalisable theory at infinitely short distances, we could work our way down to the physics at any lower energy in a ``totally systematic way''. It is a claim based on the formalism of EFT as it applies in principle, but I've argued that conclusions derived from the formalism do not necessarily represent EFT as it is actually used. Rather, we have examples of ``informal'' as well as ``formal'' EFTs, and in constructing informal EFTs, we require assistance in identifying the relevant degrees of freedom. The situation at higher energy will not be different purely by virtue of being higher energy. Instead, presumably, it is the hypothetical \textit{completeness} of a theory at infinitely high-energy that is expected to bolster the formalism enough that we no longer require external input. However, it is unclear why it is thought that the need for empirical input is due only, or even in part, to the ``incompleteness'' of our current theories. In fact, it is unclear even what is meant by referring to our theories as incomplete---after all, our theories do provide a complete description of the relevant physics for a particular system at a particular energy scale. Any unknown high-energy interactions, if they affect the physics, have their effects accounted for. 

\paragraph*{}
Recall the example of QCD: although it is understood as a complete theory with no divergences at any high energies, we are unable to derive the low-energy physics. Although this might be an issue peculiar to QCD, its existence nevertheless demonstrates that completeness and high-energy validity of a theory are no guarantee of being able to construct an EFT in the absence of external assistance. I can see no justification for the belief that the deficiencies in EFT as it is used (or, instead, the existence of informal EFTs) are due to the non-fulfilment of the required assumptions of the formalism.
\paragraph*{}
Remaining agnostic in regards to the high-energy physics does not mean simply that we do not assume that the requirements for the EFT formalism (i.e. a high-energy complete renormalisable theory and a range of well-separated mass scales) will hold at high-energy; it means that even if these requirements\footnote{Although I have argued that the assumed requirement of a high-energy complete renormalisable theory is not a necessary one (\S\ref{subsub:quasiaut}).}  are fulfilled, this is no guarantee of being able to systematically derive the low-energy physics given the high-energy theory alone. 

\subsection{Effective QFT}\label{subsub:effQFT}
There are conceptual issues in QFT that give us reason to suspect that it will not be the final word; these are reasons for treating the entire framework of QFT---as opposed to individual theories---as effective. On its own, this claim is perhaps not so controversial, after all, there is the general sentiment among physicists, expressed, for example, by Georgi (1989) and \citet{Gross1999}, that QFT has run its course. It is viewed as a mature, well-understood subject, and many authors believe it unable to provide answers to the questions we anticipate at high-energy---questions such as those regarding the unification of forces, the origin of the lepton-quark families, and the explanation of the parameters of the standard model (Gross, 1999, p. 62). Also, although the neglect of spacetime curvature in QFT is not thought to be problematic given the tiny length-scales involved, the conceptual issue of relying on a framework developed without reference to GR, our best theory of spacetime, seems a pressing one.
\paragraph*{}
This general sentiment, that QFT is, in some sense, provisional in the absence of ``new physics" at high-energy, is closely tied with the new understanding of renormalisation, wherein the conceptual difficulties with our theories are accepted as stemming from the unknown physics beyond. Wallace, for example, 
\begin{quotation}
This, in essence, is how modern particle physics deals with the renormalization problem: it is taken to presage an ultimate failure of quantum field theory at some short length scale, and once the bare existence of that failure is appreciated, the whole of renormalization theory becomes unproblematic, and indeed predictively powerful in its own right. (Wallace, 2011, p. 119)
\end{quotation}
In contrast to Gross (1999) and Wallace (2011), I do not think it appropriate to expect the formalism to break down, but simply that we should be open to the possibility of its failure---in other words, that we should not expect QFT to remain applicable at high-energy. There is the possibility that we continue to use bottom-up EFT to push forward and perhaps never find a final theory, in the sense of being one framed in terms of ``new physics'' that answers the questions we want to ask at arbitrarily high-energy.
\paragraph*{}
 Weinberg \citep[e.g.][]{Weinberg2009}, for example, argues that new physics may not be needed, and that QFT (the Standard Model, and GR understood as an EFT) may indeed be the fundamental theory. In this case, problems related to renormalisability are solved by appeal to the idea that gravity is asymptotically safe: its couplings flow to a fixed point at some high-energy (discussed in \S\ref{sub:asygrav}). If, on the other hand, we do find a final theory outside of EFT, we will still need to use EFT to gain an understanding of the low-energy physics---recalling \citet{Georgi1989}, and the discussion in \S\ref{sub:tower t-d}: EFTs are the appropriate and important means of describing nature (this idea is expanded upon as a basis for emergence, in \S\ref{sub:emergence}) . This will likely lead to new insights regarding current QFTs.
\paragraph*{}
The claim that QFT is effective, understood as the statement that we should remain open to the possibility of the framework breaking down at high-energy, should not be controversial---although the lack of acknowledgement of the view by philosophers might be thought to suggest otherwise (recall from \S\ref{sub:CaoSchweb} the difficulties that can be clarified or avoided by recognising QFT as effective). The view does, however, have some implications that are less familiar and perhaps in conflict with conventional wisdom. For one, the idea that the framework of QFT itself, not just individual theories, should not be expected to hold to arbitrarily high-energy means that even theories that are not generally considered to be EFTs should, in fact, be considered effective. QCD, for example, being a complete theory with no sign of breakdown at any high-energy is not usually thought of as an EFT, but on this position we are supposed to remain agnostic in regards to its applicability at energies far beyond those with which we are now familiar.
\paragraph*{}
There is another way in which to understand the idea of QFT as effective, however. This (more controversial) view means remaining open to the revision (or replacement) of our theories even at the energy scales where they currently apply. Some reasons for this relate to the underdetermination of our theories, discussed in \S\ref{subsub:underdet}, and the (perhaps hypothetical) possibility of an alternative QFT framework, which does not accept that the problems of renormalisation are best explained as being due to external high-energy physics, but rather maintains that they are indicative of a deficiency of the formalism of QFT. This potential alternative framework is axiomatic quantum field theory, discussed next \S\ref{subsub:axiom}.\footnote{Another reason for taking this view is, interestingly, blocked by commitment to the GCP. This is the possibility of finding a high-energy theory (perhaps, but not necessarily, a final theory) that urges us radically re-evaluate our understanding of the world. If such a theory is discovered, and the GCP is satisfied, then there would be a sense in which our current QFTs were recovered as effective theories in the regimes where they hold---they would not necessarily need to be revised at all.}

\paragraph*{}
I argue that taking QFT as effective should mean taking both these positions, that is: (1) high-energy agnosticism and, (2) being not so committed to our current QFTs that we believe them necessarily permanent. It may seem a strange view, because if QFT is, in the future, replaced with something else in the regime where it currently holds, then, it seems likely that QFT would be abandoned, and the question of what happens at high energies would be completely transformed. It seems likely that, in such a case, an agnostic position in regards to the high-energy applicability of QFT would no longer be appropriate. In other words, accepting (2) may weaken our commitment to (1). Although the position might seem strange, however, it is neither ridiculous nor inconsistent---after all, there is the possibility that the framework of QFT breaks down at some scale, and there is the possibility that our QFTs are replaced at current scales\footnote{The likelihood of QFT being replaced by something else in the regime where it holds does seem minute. This is because it's doubtful that any new theories could be as successful as our current ones, as well as being as appropriate and important (recalling the words of \citet{Georgi1993}, cited in Footnote \ref{foot:G} in this chapter). Still, however, it is a possibility.}, even if there is not the possibility that \textit{both} scenarios occur.

\subsection{Axiomatic quantum field theory}\label{subsub:axiom}
Standard, or conventional QFT (previously, and henceforth, QFT) is notorious for being mathematically ill-defined, due, mainly, to its reliance upon the informal procedures of renormalisation \citep[see, e.g.][]{Fraser2009, Wallace2006}. The program of axiomatic quantum field theory---of which the main representative is algebraic quantum field theory (AQFT)---is the attempt to reformulate QFT in a mathematically rigorous way by defining a set of physically necessary and mathematically precise axioms, and constructing QFTs which satisfy them. The problems regarding the infinities that originally plagued QFT and led to the institution of renormalisation do not arise in AQFT because the axioms demand that the quantum fields cannot be defined at points. They can be defined on arbitrarily small spacetime regions, however, meaning that any QFT found to satisfy the axioms will be well-defined to arbitrarily high-energy.

\paragraph*{}
The ill-definedness of QFT means that foundational and philosophical projects intended to interpret it face peculiar difficulties---for instance, it is not clear how to understand the ontology of the theory. While \citet{Wallace2006} has attempted to demonstrate that the lack of mathematical rigour of QFT is not dire enough to preclude foundational work, \citet{Fraser2009, Fraser2011} maintains that any interpretation of QFT should be based in the well-defined axiomatic variant. \citet{Wallace2011}, of course, disagrees; the main argument being based on the fact that, currently, \textit{there does not exist any physically realistic interacting AQFT in four dimensions}. Thus far, the only realistic four-dimensional models that have been found in accordance with the axioms are free-field theories.

\paragraph*{}
Because QFT (as in the standard model) is a very well confirmed theory, and AQFT doesn't even present us with a realistic theory, Wallace (2011) argues we should take conventional QFT seriously. As Fraser (2011), points out, however, in all cases where AQFT models have been constructed, they produce the same predictions as QFT. If the AQFT program is successful in finding four-dimensional interacting field theories to rival those of current QFT, we would expect these theories to be theoretically (conceptually) distinct from those of the rival program, but empirically indistinguishable \textit{as applied at familiar energy scales}. 
\paragraph*{}
The situation would be different for high-energies, though, for while QFTs are unable to be appealed to at energies beyond the cutoff (and, indeed, at energies approaching the cutoff), AQFTs remain finite, yielding well-defined predictions. It is for this reason that Wallace (2011) does not think the enterprise of AQFT is necessary, or, indeed, useful. Wallace (p. 120) argues that we have good reasons for thinking that spacetime breaks down at some high-energy scale, so we have no reason to trust, or even be interested in, what AQFT would say about energies beyond the cutoff of our current theories. The necessity of renormalising (most of) our QFTs is one such reason for expecting spacetime to break down, and is in consilience with suggestions from quantum gravity that spacetime is discrete. 
\paragraph*{}
It must be emphasised that the nature of the cutoff in QFT (recall \S\ref{sub:reconcutoff}, \ref{sub:effectiveEFT}) does \textit{not} give us reason to suppose that spacetime breaks down at some scale. Firstly, the QFT framework \textit{itself} doesn't require a cutoff, only particular theories do. Also, importantly, as Wallace admits, there are several different physical scenarios that could explain the necessity of renormalisation (i.e. why many of our theories fail at high-energy): this could be another field theory, a non-field theory (i.e. new physics), a real lattice structure, a discretisation of spacetime, or some other as-yet-unimagined solution beyond the particular QFT in question. Wallace argues, however, that various approaches to quantum gravity make the suggestion that spacetime is discrete, so this, presumably, provides us with evidence in favour of spacetime breaking down at high-energy. This is not a valid move, however. As we shall see in \S\ref{sub:discreteness}, evidence for discreteness is not conclusive, and, indeed, in many approaches, it is taken simply as a postulate of the theory (although, of course, the success of the theory can then be seen as evidence in support of its postulates).
\paragraph*{}
Another fact, not considered by Wallace, that has been taken to mean that spacetime breaks down at high-energy is the non-renormalisability of gravity; this is interpreted as the dominance of large quantum fluctuations as we approach the Planck energy scale. I will have more to say on this in Chapter \ref{sect:GREFT} and Chapter \ref{sect:Discrete}, but, ultimately, again, the failure of the perturbative approach to quantum gravity needn't be taken as suggestive of spacetime breaking down. Rather, it might be, as Weinberg (1972) suggested, that there is a UV fixed point for gravity, similar to that of QCD (except that it is assumed to be non-Gaussian, i.e. non-zero), in which case we have a field theory that is able to be dealt with using known QFT techniques at high-energy. Understanding QFT as effective means remaining agnostic as regards to high-energy physics---we cannot say certainly that the formalism will cease to be valid, but neither should we expect it to hold at arbitrarily high-energy. The fact that QFT is not mathematically well-defined provides a motivation for treating QFT as effective, not just in regards to high-energy physics, but also at the energy scales we are familiar with. Also, we should take this view of our current theories if we are to remain open to the possibility of AQFT being successful.

\subsection{Underdetermination}\label{subsub:underdet}
There are several streams of underdetermination in QFT: one between QFT and AQFT, between QFT ``without cutoffs" (i.e. the traditional renormalisation method, in which the cutoff was taken to infinity at the end of calculations) and QFT ``with cutoffs" (i.e. the newer means of constructing QFTs, in which the cutoff is retained), and between Wilsonian EFT and continuum EFT.\footnote{\label{foot:gap}Fraser (2011, p. 127) attempts to explain the propensity of QFT for underdetermination by arguing that in the context of QFT, there is a particularly distinct gap between the theoretical and empirical levels. The theoretical content of the theory comprises principles that are used in deriving S-matrix (scattering matrix) elements, and the S-matrix elements are then used in predicting results of scattering experiments. The empirical content of the theory is taken as the S-matrix elements, and the results of the scattering experiments represent the empirical evidence in support of this empirical content. The theoretical principles of QFT are, according to Fraser (p. 127), ``background apparatus", remote from the derived S-matrix elements. On this picture, the gap between the theoretical principles and empirical content means there is room for underdetermination of the theoretical content by empirical evidence.} 

\paragraph*{}
The debate between Fraser and Wallace outlined briefly in the preceding section is centred on AQFT versus QFT ``with cutoffs" (which, for the remainder of this section I will label CQFT, following Wallace, 2011; the \textit{C} standing for \textit{cutoff} or \textit{conventional}). CQFT and AQFT distinct theories in the sense that they have distinct theoretical principles, and, as such, make different ontological claims. For instance, CQFT supports a particle (quanta) interpretation, whereas AQFT does not. The S-matrix elements that can be derived from the two formalisms, however, are the same, and so they both reproduce the same empirical results.\footnote{The empirical results here refer to the cases in which theories from the two variants of QFT are applied at the same energy scale.} 

\paragraph*{}
Wallace (2011, p. 120) flatly rejects this underdetermination claim on the grounds that there are no physically realistic 4-dimensional interacting theories of AQFT. Given that we live in a world that has interactions and 4 (or more) spacetime dimensions, Wallace states that AQFT makes no empirical predictions whatsoever.\footnote{``Even being charitable: the only empirical predictions of AQFT are general results (the spin-statistics theorem, the CPT theorem, etc.) which are also derivable (by the usual standards of used in theoretical physics) in CQFT (perhaps only as extremely good approximations, depending on whether the world is Poincar\'{e}-covariant at the fundamental level)." Wallace (2011, p. 120)} Because of this, he says, we only have the mere \textit{possibility} of underdetermination (p. 121). Wallace's rejection of AQFT as being unable to make predictions given that the only existing theories are unrealistic, however, is perhaps a bit quick: Newton's laws, for instance, involve claims that are strictly false of our world, but, of course, this does not mean they are unable to make predictions.

\paragraph*{}  
Nevertheless, if we follow the debate, the question is whether or not we have reason to believe that there will be physically realistic models of AQFT to rival those of QFT. Wallace argues that we do not. To do this, he appeals to the theories' different ontological interpretations as well as the unrivalled empirical success of CQFT. The no-miracles style argument claims that because CQFT is incredibly successful we should tentatively accept its central claims as (approximately) true. Given that one of CQFT's central claims is that ``field degrees of freedom are frozen out at sufficiently short length scales", we should tentatively accept this claim as (approximately) true (p. 120). 
\paragraph*{}
Wallace concludes that, because AQFT does not tell us that the field degrees of freedom are frozen out at any short length scale, its basic structure is wrong, and we should thus not expect it to produce physically realistic theories. By ``frozen out", Wallace (p. 118) means that there is \textit{some} physical reason, external to the particular CQFT in question, that the theory fails at high-energy.\footnote{However, as stated, Wallace is careful to clarify that the CQFT framework \textit{itself} doesn't require a cutoff, only that certain theories within this framework do (i.e. QCD with its asymptotic freedom does not require a cutoff, and so this no-miracles argument is not supposed to apply to it).} Fraser's (2011, p. 133) response is to point out that empirical success is not supportive of any particular ontological claims in this case, because, she argues, there is a``gap'' between theory and empirical prediction in the context of QFT.\footnote{See Footnote \ref{foot:gap}.} This gap is manifest by the implications of RG methods: the high-energy physics is severely underdetermined by low-energy experiments.\footnote{Here, we are speaking of CQFT; it is not clear that this is the case in AQFT.} 

\paragraph*{}
In spite of the indisputable empirical success of current QFTs, I agree that the no-miracles argument does not pull weight in this case. This is not necessarily because of the ``gap'' that Fraser believes exists, but simply because of the very nature of what is being claimed: our theories go haywire at energies approaching their cutoff, and so cease to be reliable at some point before this. Surely the fact that these theories cease to be reliable at some point is not a reliable basis for any ontological claim regarding the nature of the degrees of freedom at energies well beyond the domain covered by the theories in question. As we have seen, Wallace attempts to bolster his argument by claiming we have external evidence that accords with the ``freezing out" of the high-energy degrees of freedom in CQFT, this being the suggestion from quantum gravity that spacetime is discrete. As I have stated, this move is illegitimate---the evidence from various quantum gravity proposals is nowhere near conclusive, nor is it based in experimental results. 

\paragraph*{}
Furthermore, as discussed in later chapters (\S\ref{sub:asygrav}; \S\ref{sect:Discrete}), the non-renormalisability of gravity itself is not standardly taken as evidence for the reality of the cutoff (as a smallest length scale): indeed, most of the proponents of the discrete approaches to QG believe the non-renormalisability of gravity may simply be indication of the misapplication of perturbative techniques. Given that this is the case, it seems strange to suggest that the non-renormalisability of other particular QFTs is evidence for a smallest length scale.\footnote{To be clear: this is not Wallace's argument, which is simply that the ``freezing out'' of high-energy degrees of freedom in particular QFTs is compatible with the existence of a smallest length scale. I have pushed the point just to highlight the fact that the existence of a smallest length scale is not obviously established (or, at least not so by the arguments that Wallace appeals to).}

\paragraph*{}
In the end, the debate between CQFT and AQFT comes down to a disagreement over the admissibility of external considerations in constructing (or trusting) our theories. Fraser, representative of the AQFT-camp, believes that taking QFT seriously means finding the best formulation of it on its own merits---that is, as a combination of quantum theory and special relativity, not as a theory intended to treat quantum gravity. On this view, it is unacceptable to blame the difficulties with QFT on external (high-energy) physics. Wallace, on the other hand, as a supporter of CQFT, argues that we cannot ignore the reality of quantum gravity at high-energy. Supporting CQFT means adhering to the EFT philosophy: accepting the difficulties with our theories and holding that they are due to the influence of unknown physics. 

\paragraph*{}
While CQFT is content with its theories only being applicable in a limited domain, AQFT strives for theories that are not restricted in this sense. In either case, it seems we do best to take a pragmatic attitude toward CQFT. On the one hand, if we are to believe in the possibility of success of AQFT, then we should understand CQFT as effective even within its current domain of validity. This might mean taking an instrumentalist view, in which we trust the RG methods, given that they illuminate the empirical structure, but perhaps are cautious of the theoretical principles and ontological commitments of CQFT. If, on the other hand, the problems with CQFT are, in fact, due to external physics, then, on this view, too, we should be open to the possibility that the theory be revised in the light of future discovery.
\paragraph*{}
More generally, the underdetermination (perhaps permitted by a ``gap'' between the theoretical and the empirical levels in QFT, as Fraser suggests) provides reason, in itself, to take a pragmatic attitude toward QFT. The underdetermination between QFT with cutoffs and QFT without cutoffs is another case in point: the empirical predictions afforded by the derived S-matrix elements are the same in each theory (again, where there are representative theories of each, applicable at the same energy scale), but the theoretical formalism is different, and supports different ontologies. QFT with cutoffs supports a quanta interpretation, for example, whereas (due to the lack of firm mathematical foundation) it is not clear that QFT without cutoffs does \citep{Fraser2009}.  

\paragraph*{}
Also, while QFT without cutoffs is Poincar\'{e} covariant, QFT with cutoffs is not. Fraser points out that taking QFT with cutoffs seriously means taking seriously the claim that spacetime is discrete.\footnote{Whether this is a logical necessity, however, is not obvious: it might be that some field provides a natural cutoff and regularises the others, without demanding that spacetime itself be discrete, for example.} However, even this is underdetermined; as we have seen, there is an alternative method of constructing EFTs: continuum EFT, and this uses a mass-independent renormalisation scheme. In continuum EFT, the scaling variable $\mu$ (i.e. the renormalisation scale that appears in the RG equations) plays the role of demarcating the low- from high-energy physics of the theory, but the scaling variable is not responsible for regulating the divergent integrals in the theory \citep{Bain2013}. Thus, taking QFT with cutoffs seriously means first specifying which variant is being referred to: the continuum method does not support an ontology on which spacetime is discrete. Yet, as stated, Wilsonian and continuum EFT produce theories that are empirically equivalent. This seems to undermine the motivation for taking one or the other version seriously.

\subsection{Taking EFT effectively}
The debate between Fraser and Wallace is interesting because it is framed as the question of how best to take particle physics seriously. As discussed above, proponents of CQFT maintain that taking particle physics seriously means accepting our current theories as they are, given that they are so successful. This means accepting, however, that our theories are only approximations, in the sense that they are effective at large distance-scales, where the high-energy details have been washed out. If we are to follow Wallace and Gross in expecting new physics to be necessary, and thus believe that the framework of QFT will break down at some scale, then it seems we should consider QFT as a sort of intermediary. 
\paragraph*{}
QFT might be a stepping-stone to the high-energy theory, or it may be provisional in the sense that the high-energy theory will reveal a more appropriate low-energy description. Proponents of AQFT maintain that the basic structure of our current theory is incorrect, and if we are to take particle physics seriously, we should be looking for a theory that does not presage its own failure. Concerns of new physics at high-energy is, on this view, external to QFT, and we need not take it into account.
\paragraph*{}
CQFT, understood as EFT, means constructing theories that are valid at a particular energy scale, building in the low-energy effects of the high-energy physics. It means accepting that there is unknown physics beyond. AQFT means ignoring (for the purposes of constructing QFT) unknown physics; we needn't take particle physics to be an approximation of something else. In this context, it is an interesting question, I think, to ask how many of our reasons for thinking there must be unknown physics beyond QFT, are based in the difficulties with QFT itself (for instance, the requirement of renormalisation, the desire to know what is responsible for the values of the parameters in our theory, etc.). It would seem that, if AQFT is, in fact, a viable alternative, then many of these reasons may be undermined.\footnote{The other reasons for postulating unknown physics at high-energy come, of course, from the desire for a quantum theory of gravity. As discussed in \ref{sect:Intro}, the strongest motivation for this is the idea of unification.} This is yet another reason for treating EFT itself as effective, and remaining agnostic in regards to unknown physics, rather than expecting QFT to break down.
\paragraph*{}
Also, of course, this view means remaining open to alternatives. Taking EFT \textit{effectively} means not taking particle physics so seriously that we are locked into it. Already, this isn't a strange view; recall from \S\ref{subsub:finaltheory} that other authors, in evaluating the implications of QFT (understood as EFT), actually take a viewpoint outside of QFT (understood as EFT). Hartmann (2001, p. 298) also asserts that, ``We are not trapped in the language game of one theory", and Robinson (1992) argues that we are able to build up ontological commitments in the absence of a theory. Although we should not expect it to fail, our current picture of the world, as described by this particular tower of EFTs, is apt to be replaced. Should this picture be replaced, however, the idea of EFT would still be necessary; in the case of top-down EFT, it would be necessary because of its ability to capture the novelty and (quasi-)autonomy of the low-energy physics, and, in the case of bottom-up EFT, it may also be necessary for probing energies beyond the domain covered by whatever the successor theory happens to be.

\section{Conclusion: Emergence}\label{sub:emergence}
Articulating the conception of emergence in EFT has proven difficult because it represents a case in which emergence (if we are to call it ``emergence'') decouples from reduction. As indicated in the discussion above and in \S\ref{sect:Emergence}, there has been a preoccupation with the ideas of derivability and deduction in EFT as related to emergence, and a consequence of this has been to mostly overlook the genuine and interesting relations in EFT. In addition to the general difficulties in formulating the distinction between reduction (and derivability) \textit{in principle} and \textit{in practice}, outlined in \S\ref{sect:Emergence}, there is the fact that the literature on EFT is ambiguous between EFT as the formalism applies \textit{in principle} and EFT \textit{in practice}. 

\paragraph*{}
As I have argued, the \textit{in principle} considerations are irrelevant, and it is best to focus on EFT as it is actually used in physics. Philosophers risk talking past one another, for instance Castellani (2002, p. 265) rejects the claim that EFT vindicates Anderson's view of emergence, as Cao and Schweber claim, because she believes that \textit{in principle} EFT allows for a reconstruction of low-energy physics from the complete high-energy theory. Cao and Schweber (and perhaps Anderson, as well), however, are best understood as speaking of EFT as it is actually used in physics (and, as argued in \S\ref{subsub:rel layers}, Castellani's ``in principle'' claim is misleading, perhaps even mistaken, given that a complete renormalisable high-energy theory is no guarantee of being able to arrive at the low-energy theory---QCD being a case in point). 

\paragraph*{}
\citet{Bain2013, Bain2013a}, on the other hand, is concerned with EFT as applied, and argues for an account of emergence based on the \textit{derivational independence} of the low-energy theory from its associated high-energy theory. The basis for Bain's claim of derivational independence is that the high-energy theory and the low-energy theory are distinct: represented by different Lagrangian densities describing very different dynamical variables, meaning that specification of the equations of motion for the high-energy theory (together with pertinent boundary conditions) will fail to specify solutions to the equations of motion of the EFT. 
The derivational independence of the EFT, Bain (2013a) argues, is ensured by the fact that it is obtained by eliminating degrees of freedom from the higher-energy theory. Bain (2013) also emphasises the role of approximations in building an EFT and the (often non-trivial) task of first identifying the relevant degrees of freedom: ``This suggests that, in general, it will be difficult, if not impossible, to reformulate the steps involved in the construction of an EFT (of either the Wilsonian or continuum types) in the form of a derivation." 

\paragraph*{}
While I believe that Bain is right to consider EFT as it is actually used, and that the distinctness of the EFT from its high-energy theory is indeed significant in understanding emergence in EFT, I think the focus on derivability is uninteresting. Although the point of focusing on emergence as a failure of derivability is to have the account align with the typical philosophical understanding of emergence, such a concession is not necessary, for reasons argued in \S\ref{sect:Emergence}---we can have emergence with or without reduction, so long as we take emergence as novelty and autonomy. As is demonstrated shortly, a strong conception of novelty and autonomy is granted by the very features of EFT that Bain cites to support his argument of derivational independence. Furthermore, there is a subtlety in the necessity of EFT that threatens to make trivial any claim of emergence that is based on a failure of reduction or derivation. 

\paragraph*{}
Consider the example of QCD again: it is believed that QCD does, in principle, contain all the information required in order to make low-energy predictions (given that we use it as a basis for approximations that do, however stubbornly, yield low-energy predictions). As stated in \S\ref{subsub:efEFT high} researchers are slowly pushing forward in the \textit{ab initio} derivation of the hadron masses, mostly using numerical methods. In other words, although often extremely difficult in practice, it is possible in principle to obtain low-energy results from the high-energy theory. In the typical situation where we cannot in practice obtain results directly from the high-energy theory (presumably because of our own fallibility), EFT is just one of several different techniques for doing so. In the case of QCD, the other methods include the lattice approximation \citep{Bazavov2010, Callaway1983, Callaway1982, Durr2008}, QCD sum rules \citep{Shifman1998} and the Nambu-Jona-Lasinino model \citep{Nambu1961, Nambu1961a}.

\paragraph*{}
For this reason we should be wary of speaking of EFT as though it is strictly necessary for obtaining low-energy predictions (and, consequently, the interest in debating whether or not an EFT is able to be strictly derived from its high-energy theory seems greatly diminished). EFT, however, is necessary in a more subtle sense: its theories are formulated in terms of the appropriate degrees of freedom for the energy being studied, and are necessary for an understanding of the low-energy physics. Because the low-energy degrees of freedom do not exist at higher energy, EFT presents a clear picture of relevant physics, a picture that the high-energy theory is unable to provide. Thus, we can distinguish between an EFT's role in producing quantitative predictions in the low-energy regime---a role which, in principle, could be fulfilled by the high-energy theory (or some other method of approximating the high-energy theory)---and its role in facilitating an understanding of the low-energy physics, by providing a low-energy theory formulated in terms of the appropriate degrees of freedom---a role which could not be fulfilled by the high-energy theory. 

\paragraph*{}
The idea of emergence in EFT that I advocate is thus based on the characteristics of EFT that are responsible for its great utility, and the utility and significance of an EFT comes about because it is novel and (quasi-)autonomous compared to the high-energy physics. The idea of \textit{novelty} in EFT is related to the fact that an EFT is formally distinct from its high-energy theory (the theory that it emerges from, and that it is linked to via the RG and EFT techniques): the degrees of freedom of the effective Lagrangian density are associated with states that are formally distinct from those associated with the degrees of freedom of the original (high-energy) Lagrangian density (as Bain (2013) points out). The low-energy degrees of freedom simply do not exist at high-energy. This is clear in the examples of analogue spacetime from quantum superfluids (discussed in \S\ref{sub:analogue}), which present EFTs of phonons (quasiparticles) that are nothing but collective excitations of the underlying superfluid, and cease to exist at energies approaching the characteristic scale of the atoms in the superfluid. The idea of novelty just described is the reason why an EFT may be thought of as derivationally independent from its high-energy theory, and helps form part of the conception of emergence in EFT presented by \citet{Bain2013, Bain2013a} and \citet{Zhang2004}. 
 
\paragraph*{}
The idea of \textit{autonomy} (or, rather, quasi-autonomy) in EFT comes from the fact that the low-energy theory is largely independent of the details of the high-energy theory. There is extra information contained in the high-energy theory, far over-and-above that required in order to describe the low-energy behaviour of significance. As the examples of analogue spacetime demonstrate, the low-energy physics actually depends on very little of the high-energy physics: some particular interactions and symmetries are important, but the details of the high-energy theory are not. This is the hallmark of the EFT program. Thanks to the action of the RG, the high-energy effects that filter down to low-energy are typically able to be realised by any number of different systems, and, as such, the high-energy theory is severely underdetermined by the low-energy physics. The relation between the ideas of underdetermination and autonomy is explored further in the next chapter (\ref{sect:Univers}).

\paragraph*{}
It is these two ideas of novelty and autonomy---the significant distinction between the theories, and the subtle way in which the high-energy theory affects the low-energy theory---that together are responsible for the difficulty in constructing EFTs. They are the reasons why, in many cases, the EFT cannot be constructed given \textit{only} a description of the high-energy physics. 
This sense of (in practice) ``underivability'' or ``irreducibility'' is compatible with Cao and Schweber's (1993) account.\footnote{I believe this suggestion also accords with Anderson's views (\S\ref{sub:andersonwein}).} As explained in \S\ref{subsub:ont plu}, Cao and Schweber's idea of emergence is best understood as describing the way EFTs are actually constructed in physics, rather than the \textit{in principle} implications of the formalistic procedure. Recall that, although Cao and Schweber were happy to maintain that there are ``causal connections'' between the levels, they rejected the idea that these causal connections allow us to get from one level to another. In other words, the fact that the high-energy physics influences the low-energy physics does not mean we are able to construct the low-energy theory given the high-energy theory alone.

\paragraph*{}
A final point on the nature of emergence in EFT: owing to the ``subtlety'' regarding the necessity of EFT described above, it seems appropriate that it fall under \textit{epistemological emergence}. Silberstein and McGeever (1999, p. 185) state that epistemological emergence means that the ``higher-level'' description is ineliminable (emergent) in its nature. Once we have identified a case of epistemological emergence, Silberstein and McGeever prompt us to then ask \textit{why} the higher-level description is ineliminable: we should also ask, they say, how exactly the higher-level description manages to have explanatory or predictive value, if, in fact, ontological emergence is not in play. I think the answers to these questions should now be clear in the case of EFT.

%% file: Univ.tex
\chapter{Universality, higher-organising principles and emergence}\label{sect:Univers}

\section{Introduction}
In this chapter I explore some conceptions of emergence associated with universality and higher-organising principles. The main focus is the idea of universality in critical phenomena, being the fact that large classes of different systems exhibit the same behaviour in undergoing a second-order phase transition. In approaching this topic, I am inspired by the recent ``science-first'' work of \citet{Morrison2012} and \citet{Batterman2011}, who each advance their own conceptions of emergence in critical phenomena. \citet{Batterman2002, Batterman2011} advocates an idea of emergence based in the idea of fixed points in the RG flow, emphasising the role of limiting relations between the theories (i.e. between the emergent theory and the one it emerges from) and the associated mathematical singularities, which he views as a crucial feature of emergence. Morrison (2012) is also interested in emergent phenomena associated with fixed points in the RG flow, but argues that Batterman downplays the importance of symmetry breaking as a dynamical mechanism for producing the stable behaviour characteristic of universality. 

\paragraph*{}
I argue that the interesting aspects of emergence are actually provided by the simple idea of \textit{universality}, rather than the nuances of either Morrison's or Batterman's accounts. The idea of universality is essentially related to multiple-realisability and the idea of underdetermination. It is tied to several different purported examples of emergence in physics, which I will summarise at the end of the chapter (\S\ref{sub:Uconcl}). Exploring these examples is of particular interest given the suggestions that gravity is asymptotically safe, meaning it is represented by a fixed point in the RG flow and has an associated notion of universality (see \S\ref{sub:asygrav}). Similarly, there are claims that general relativity is analogous to hydrodynamics: the near-universal long-wavelength limit of a wide range of different potential ``micro''-theories (\S\ref{subsub:AEmergence}, \S\ref{subsub:ecauset}).

\paragraph*{}
This chapter begins with a brief introduction to the idea of multiple-realisability and its relation to universality, \S\ref{subsect:multiplereal}. Following this (\S\ref{sub:HOP}) is a basic explication of Laughlin and Pines' (2000) idea that emergent properties are ``transcendent'' and associated with ``higher organising principles''. The relation between universality and emergence is explored in \S\ref{sub:UniversEmergence}. I then (\S\ref{subsect:criticalphen}) present the physical example of second-order phase transitions and critical phenomena, where the conception of emergence may be tied to  either the idea of symmetry-breaking as a higher-organising principle or the idea of universality, associated with fixed points in the RG flow. As described in (\S\ref{subsect:Morrison}), Morrison (2012) presents an account of emergence that emphasises the importance of both the RG-fixed point story and the symmetry-breaking story. While I agree that both features provide an important basis for understanding emergence in physics, Morrison's (2012) definition of emergence suffers several problems that result from its being an ontological characterisation that emphasises the failure of reduction. I explain these problems in \S\ref{subsect:Morrison} and \S\ref{Unired}.
\paragraph*{}
Batterman (2011) maintains that mathematical singularities are necessary in explaining emergence in the RG-based account\footnote{``The singularities that appear when considering the relationships between micro-theories at some high energy/short distance scale and those theories at smaller energies/larger lengths are absolutely necessary for an understanding of emergent phenomena in physics.'' (Batterman, 2011, p. 1049)}, but I argue that he does not prove that emergence is necessarily associated with singularities, given the alternative explanation based in symmetry-breaking. Rather, I find it more appropriate to say that, although a description of critical phenomena might necessarily involve singularities and divergences, an \textit{explanation} of universality need not (\S\ref{subsect:Batterman}). After considering Batterman's argument I thus look more closely at the idea of associating emergence with singular limits, \S\ref{subsect:singularities}. I find that my conclusion here is in accordance with that of \citet{Butterfield2011b}, who claims that a singular limit is neither necessary nor sufficient for emergence.
\paragraph*{}
In \S\ref{subsect:universalEFT} I consider how the idea of emergence associated with universality is related to the idea of emergence associated with EFT (presented in \S\ref{sub:emergence}). I argue that both of these conceptions of emergence have their idea of autonomy based in the underdetermination of the micro-physics given the macro-physics. The difference between the two cases pertains to the idea of novelty: in the case of critical phenomena, the low-energy physics is novel compared to the description of the system before the phase transition, whereas in the more general case of EFT, the low-energy physics is novel compared to the high-energy description of the system at a given point in time. 
\paragraph*{}
In \S\ref{Unired} I consider a difficulty that results from taking a definition of emergence as a failure of reduction (as Batterman (2011) and Morrison (2012) do) when considering cases of universality. I argue that this difficulty is another reason for preferring my more general account of emergence, whose definition is not based on a failure of reduction. In \S\ref{subsub:hydro} I consider the idea of emergence in hydrodynamics; this represents a theory which counts as emergence on the more general definition, but not on Batterman's or Morrison's definitions.

\section{Multiple-realisability}\label{subsect:multiplereal}
The thesis of multiple-realisability is most familiar as described in the philosophy of mind by Hilary \citet{Putnam1967}, where it is the acknowledgement that a particular mental state can be ``realised by'' (associated with) many distinct physical states; for instance, the mental state of ``being in pain'' can occur in different mammals, where it is associated with different brain states, and it can plausibly occur in animals that are not mammals---for instance, molluscs (e.g. octopuses)---where it would be associated with neural states that are different, again, from those ``realising'' brain states in mammals.

\paragraph*{}
Jerry \citet{Fodor1974} uses an argument from multiple-realisability to make the claim that a higher-level science (e.g. psychology) is unable to be reduced to a lower-level science (e.g. neurology). Reductionism in this case is taken as (at least in part), the assertion that every singular occurrence that a higher-level science can explain can also be explained by a lower-level science, and, that every law in a higher-level science can be explained by laws in a lower-level science.The idea of multiple-realisability, more generally, then, can be framed (as in \citet{Fodor1997}) as the idea that any given \textit{macro} -state can be realised by any one of a multiplicity of different \textit{micro} -states.

\paragraph*{}
Multiple-realisability is thus essentially similar to the idea presented when considering EFT, of the micro-physics being underdetermined by the macro-physics, and also accords with the idea of autonomy developed in (\S\ref{sub:emergence}). The association of multiple-realisability with emergence has some history in the metaphysics and the philosophy of science literature, where Jaegwon \citet{Kim1992}, in forming part of his response to the arguments against reductionism made by Putnam and Fodor, presents the \textit{Causal Inheritance Principle} designed to ensure that there are no emergent causal properties. The Causal Inheritance Principle states that, if higher-order property $H$ is realised in a system at time $t$ in virtue of physical realisation base, $P$, the causal powers of \textit{this instance} of $H$ are identical with the causal powers of $P$. 
\paragraph*{}
These arguments, which have been of great influence in the philosophy of mind, concern mainly the ideas of causality, scientific kinds, and reductionism in science.\footnote{See \citet{Bickle2008} for a concise presentation of the main arguments and replies.} It won't be useful to look at them in any detail here, except to point out that the fact that Sober's \citeyearpar{Sober1999} arguments sucessfully demonstrate that multiple-realisability is compatible with reductionism is further evidence for a suggestion made in earlier chapters of this thesis (\S\ref{sect:Emergence}, \S\ref{sect:EFT})---that the link between reductionism and emergence is not (in all cases) an interesting one.

\paragraph*{}
The interesting question, instead, concerns \textit{why} there is multiple-realisability. Fodor (1997, p. 161) as well as \citet[][p. 196]{Silberstein1999} recognise this question, with the latter authors suggesting that one answer ``would be the existence of ontologically emergent properties (relational holism, fusion, etc.) that constrain or supersede the intrinsic properties of the parts. Otherwise [the multiple-realisability] must be in principle explainable in terms of the intrinsic properties of the parts.'' The physical examples they present as possible illustrations of these properties are chaotic (non-linear dynamical) systems.\footnote{More specifically, the structures that Silberstein and McGeever identify as representing the ontologically emergent properties are ``strange attractors", though I won't discuss these.} Silberstein and McGeever argue that the causal story underlying the dynamics of the emergent structures (i.e. those structures that Silberstein and McGeever identify as representing the emergent properties),  may be the non-linear nature of the relations between these structures. 

\paragraph*{}
Silberstein and McGeever quote \citet[][p. 20]{Scott1995}, ``In a non-linear situation the effect from the sum of two causes is not equal to the sum of individual effects. The whole is not equal to the sum of its parts''. Another example in physics where this is the case is GR: the non-linearity of the Einstein equations means that the combined gravitational effects of two bodies are not equal to the sum of the effects of each of these two bodies separately. Silberstein and McGeever's (1999) conception of ontological emergence is the failure of part-whole reductionism, and they argue that because the behaviour of non-linear systems is not explained in terms of the intrinsic properties of their components, these systems represent genuine cases of ontological emergence. 
\paragraph*{}
These dynamical, ontologically emergent properties are supposed to explain the phenomenon of multiple-realisability because they are robust under changes in the states of the micro-system (Silberstein \& McGeever, 1999, pp. 195-196). In other words, the emergent features are \textit{autonomous} from the details of  the high-energy physics. Although I agree with linking the idea of emergence to multiple-realisability and the idea of autonomy associated with it, unfortunately I will not explore the explanation of emergence in terms of non-linearity here.\footnote{Also, as should be clear, I am not interested in ``ontological emergence'', nor the failure of part-whole reductionism.} Nevertheless, I believe the ``ontologically emergent properties'' that Silberstein and McGeever (1999) present may be essentially similar to what \citet{Laughlin2000} call ``higher organising principles''---a concept I discuss shortly (\S\ref{sub:HOP}).

\paragraph*{}
\citet{Batterman2000} attempts to answer the question (of how multiple-realisability is possible) by suggesting we relate multiple-realisability to \textit{universality} in physics. The definition of universality that is usually provided by philosophers (or alluded to, in Batterman's case) is simply that ``universal behaviour'' is identical behaviour exhibited by many different physical systems---in other words,  universality is just multiply-realised behaviour in physics. However, as \citet[][pp. 187--188]{Mainwood2006} points out, this definition is rather too vague to be satisfactory, as it covers cases of multiple-realisability that physicists do not, in practice, refer to as universality: the fact that the period of a pendulum is independent of its micro-structure is not typically counted as universality, for example.
\paragraph*{}
Instead, the common approach in philosophy is to follow the physics textbooks and to just use the term ``universality'' to refer to\label{def:univ} those properties that an RG analysis is (apparently) amenable to elucidating (Mainwood, 2006). I follow this convention, as I believe it has the advantage of explaining the required sense of direction: the micro-physics influences (in \textit{some sense}) the macro-physics, but not vice-versa. Thus, the appeal to the RG in defining universality serves to ground the link between the micro- and the macro-levels necessary in order to claim that the former emerges from the latter.\footnote{In the example of the pendulum, for instance, there is no such link between the micro-structure of the pendulum and the period of the pendulum. This is true of other cases of purported multiple realisability where the multiply-realised property is simply a very \textit{general} one, such as ``having a shape'' or ``being coloured'', etc.} 

\paragraph*{}
Batterman takes the textbook example of universality in physics (critical phenomena, which is explained below, in \S \ref{subsect:criticalphen}), outlines its explanation, then distils the general explanatory strategy that underlies this particular explanation. This general explanatory strategy is then presented as an answer to the question of how it is possible that the properties of the special sciences can be heterogeneously multiply realised. The explanatory story that Batterman presents is grounded in the RG. As Batterman (2000, p. 129) argues, the RG provides a method for extracting just those features of a system that are stable macroscopically under perturbation of their micro-details. The general explanation of universality (multiple-realisability) he suggests thus consists in developing reasons for ignoring any detailed microphysical information about any actual system or systems being investigated.\footnote{Mainwood (2006, pp. 118-206)  is a detailed critique of Batterman (2000), in which Mainwood claims that Batterman's presentation conflates the Kadanoff approach with the Wilsonian approach to the RG---and argues that it is the latter that adequately explains universality. Mainwood also maintains that Batterman's argument does not represent a truly original explanation of multiple-realisability---as is Batterman's claim---compared with the suggestions already present in the metaphysics literature.}

\paragraph*{}
While Batterman thus maintains that multiple-realisability is explained by the RG, Morrison (2012) builds upon this account, emphasising the role of symmetry-breaking in explaining the emergent phenomena.

\section{Higher organising principles}\label{sub:HOP}
Anderson's fellow condensed-matter theorists Laughlin and Pines (2000) refer to emergent domains (levels) of physics as ``protectorates"---stable states of matter that are effectively independent of their micro-- (or high-energy--) details. According to Laughlin and Pines, the generic low-energy properties of such a state ``are determined by a higher organising principle and nothing else" (p. 29). The idea of a higher organising principle is not clearly defined, but illustrated only by examples; however, it seems we can take a higher organising principle (HOP) as being one that \textit{provides an explanation for the observed low-energy physics in terms of phenomena, features or mechanisms that are essentially independent of the details of the high-energy system}.  

\paragraph*{}
A major piece of evidence for the existence of protectorates is the fact that, often, we are able to predict exact results using approximate models. One case of this, the BCS theory of superconductivity, has already been introduced (\S\ref{sub:tower b-u}).\footnote{And is returned to in \S\ref{subsub:symmetrybreaking}.} Laughlin and Pines list several other physical theories that yield exact results, including the Josephson quantum (effect), the predictions of which are exact because they are determined by the principle of continuous symmetry breaking; another is the quantum Hall effect, which, they say, is exact because of localisation. Laughlin and Pines (2000, p. 28) make the dramatic assertion that ``Neither of these things can be deduced from microscopics, and both are transcendent, in that they would continue to be true and to lead to exact results even if the Theory of Everything were changed".\footnote{Presumably, the changes in the ``Theory of Everything'' would be alterations in details that preserved the relevant symmetries and other features responsible for the HOPs.}

\paragraph*{}
Just as the idea of universality provides an account of autonomy where the high-energy theory is underdetermined by the low-energy physics, so too does the idea of a HOP. Different systems (with different high-energy details) can have their low-energy behaviour governed by the same HOP, where the HOP, rather than the high-energy theory, provides the \textit{explanation} of the low-energy physics. It is interesting, then, to ask whether the RG, as it features in EFT, is an example of a HOP. The answer to this question comes after more thoroughly comparing the different accounts of emergence (\S\ref{subsub:SBHOP}).

\section{Universality and emergence}\label{sub:UniversEmergence}
In exploring the idea of emergence associated with universality, I begin by comparing the accounts of emergence presented by Batterman (2011) and Morrison (2012) with the conception applicable to EFT (as developed in \S\ref{sub:emergence}). Initially, it might appear that one obvious similarity between these conceptions of emergence is the revelation of the idea of deduction (or derivability) from the micro-theory as uninteresting. Morrison (2012, p. 162) for instance, believes that the derivability \textit{in principle} of the macro-theory from the micro-theory is irrelevant to describing emergence in the context of universal phenomena, stating that ``If we suppose that micro properties could determine macro properties in cases of emergence, then we have no explanation of how universal phenomena are even possible''. This is because, of course, the large-scale universal behaviour emerges from systems that are very different from one another at small scales. Therefore, if we focus on deriving the predictions of critical phenomena from the micro-theory describing any one of these systems, we will miss the crucial aspect of this account of emergence, namely, the fact that such phenomena appear in a variety of different systems under different conditions. 

\paragraph*{}
Batterman (2011, p. 1049) makes this point too,
\begin{quote}
If the goal is to answer the ``how is it possible'' question about universal behaviour, it seems highly unlikely that a scheme that depends upon a mathematical derivation from the finite, ``true'' theory of everything can fulfil that role. After all, such a derivational story must, of necessity start form the detailed microstructural constitution of the individual molecules in a \textit{particular} fluid. But \textit{why should that individual derivation} have any bearing on a completely different individual derivation for a different fluid with a potentially radically different microstructural constitution?
\end{quote}
Although my discussion of EFT involved arguing that the idea of a derivation \textit{in principle} is irrelevant, Morrison and Batterman do not find the idea of derivation irrelevant, but instead base their accounts of emergence on its failure. Reduction to any particular micro-theory fails to capture the \textit{universality}. Hence, the positive conception of emergence here, I think, is based in the universality: any explanation of the universality will be a story that transcends the micro-physics, and will thus provide an account of emergence. Importantly, the universal phenomena cannot be derived from any given micro-system (for the very simple reason that such a derivation does not account for the universality), but it can be derived from what Laughlin and Pines call higher-organising principles. 

\paragraph*{}
In framing her conception of emergence, Morrison (2012, p. 143) distinguishes between epistemological and ontological independence, arguing that, while the former is necessary for emergence, it is the latter that is sufficient for emergence. Morrison's definitions of the terms are:
\begin{description}\label{def:Mor}
\item[Morrison's \textit{Epistemological independence}] We \textit{need not} appeal to micro-details in order to explain or predict the emergent phenomena
\item[Morrison's \textit{Ontological independence}] We \textit{cannot} appeal to micro-details in order to explain or predict the emergent phenomena
\end{description}
Morrison (2012, pp. 143, 161) thinks epistemological independence is uninteresting because ``it is a common feature of physical explanation across many systems and levels''. I believe, however, that the fact that it is widespread in the physical sciences should not diminish the marvel of, in many cases, being able to describe the world at any particular scale without requiring knowledge of the levels ``underlying'' it. Indeed, as we have seen, the development of EFT goes a long way toward explaining how this is possible, rather than simply enabling us to take the idea of ``levels'' for granted. 

\paragraph*{}
Morrison believes the distinction between ontological and epistemological emergence is important to distinguish emergent properties from those that are ``merely resultant'' (while Batterman (2002, 2011) takes the appearance of singularities in the mathematical formalism to distinguish the emergent from the merely resultant\footnote{The relation between singularities and universality will be explicated in the next sub-section.})---but, I think what does the work in distinguishing emergent from resultant properties in this case is just the universality. The fact that many different types of systems exhibit the same behaviour demonstrates that the behaviour is not simply resultant.\footnote{The conception of emergence suggested here seems to recall Silberstein and McGeever's (1999) notion of ontological emergence as a failure of part-whole reductionism---which would explain Morrison's choice of terminology in defining her preferred conception of emergence.} To focus on the resultant is to make the same mistake as focusing on the derivation from one micro-system---it is to miss the universality (or, rather, the autonomy associated with universality) that characterises the emergence. Accordingly, I will speak of this conception of emergence as explicitly identified with multiple-realisability (universality or HOPs), even though Morrison (2012) and Batterman (2011) both employ extra tenets. 

\section{Fixed points and critical phenomena}\label{subsect:criticalphen}
\begin{figure}[h]
\centering
\includegraphics[height=8cm]{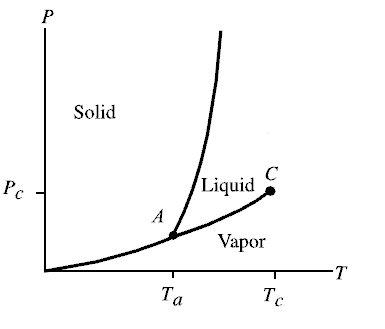}
\caption{Temperature-pressure phase diagram for a fluid. (Adapted from Batterman, 2011, p. 1035)}
\label{fig:criticalpoint}
\end{figure}

\begin{figure}[htbp]
\centering
\includegraphics{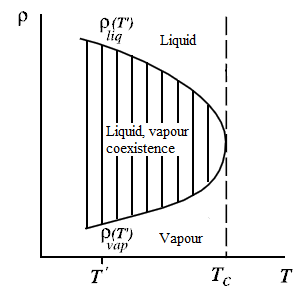}
\caption{Coexistence curve for a fluid. (Adapted from Batterman, 2011, p. 1036)}
\label{fig:coexistence}
\end{figure}

Recall from \S \ref{subsub:fixed points} the idea of a fixed point in the RG flow: for many condensed matter systems, an IR fixed point corresponds to a second-order phase transition. While first-order phase transitions are familiar as qualitative changes in the state of a system (for instance ice melting to liquid water, liquid water vaporising to steam), second-order phase transitions represent conditions under which there is no real distinction between two states of the system (for instance, between the liquid and vapour phases of water). Consider the phase diagram,  Fig.~\ref{fig:criticalpoint}, below.
Above a particular value of temperature, called the critical temperature $T_c$, and a particular pressure $P_c$, the distinction between liquid and vapour no longer makes sense. The point marked by these coordinates is known as the \textit{critical point}, and represents the second order phase transition.\footnote{In first-order phase transitions there is a discontinuity in the value of the order parameter (about to be defined) as it changes with temperature. In second-order phase transitions there is no such discontinuity.}
\paragraph*{}
At the critical point there is no distinguished length scale. As is typical of systems at a UV fixed point, the system does not change as we view it at smaller length scales; the system is \textit{scale invariant}. In the case of water, for example, this means that the system resembles a fractal structure, exhibiting \textit{critical opalescence}: within the bubbles of steam float liquid water droplets, and these droplets appear full of bubbles of steam, which, in turn, appear full of little droplets of liquid, and so on... until we reach the scale of atoms. 
\paragraph*{}
Critical behaviour is characterised by an \textit{order parameter}, which for fluids is the difference in densities between the different coexisting phases. An order parameter is a quantity which is introduced in order to distinguish between two states of a system: its magnitude being zero in one phase and non-zero in the other. Generally, a non-zero order parameter corresponds to a state of broken symmetry (the zero value representing the symmetric state of the system), however, order parameters can be introduced in phase transitions not involving symmetry-breaking,  such as in this example (i.e. the liquid-to-gas transition).  Along the line A-C in Figure~\ref{fig:criticalpoint} the order parameter,$\Psi $, is the difference between the liquid and vapour densities,
\begin{equation}
\Psi = |\rho_{liq} - \rho_{vap}|
\end{equation}
If the system is to pass from its vapour to its liquid phase while at any temperature below $  T_c $, it must go through a state in which both liquid and vapour are simultaneously present. The order parameter will thus be non-zero for temperatures below $ T_c $. Above $ T_c $, the order parameter is zero. Figure~\ref{fig:coexistence}, below, is a plot of density versus pressure known as a coexistence curve.
\paragraph*{}
\paragraph*{}
The region within the curve indicates the values of temperature and pressure at which liquid and vapour coexist. The vertical ``tie lines” show that at a given temperature, $ T' $, the liquid density is $ \rho(T')_{liq} $ and the vapour density is $ \rho(T')_{vap} $. At $ T_c $ the difference between these two values vanishes, and so $ \Psi = 0 $. 
Remarkably, many distinct fluids, of differing microstructures, have the same shape coexistence curve near their respective critical temperatures. The reduced temperature, \textit{t}, is defined as,
\begin{equation}
t = \left| \frac{T-T_c}{T_c}\right| 
\end{equation}
The reduced temperature shows how far from criticality a particular system is, and allows us to compare the critical behaviour of systems with different values of $ T_c $. Now, for every fluid, the order parameter, $ \Psi $, vanishes as some power, $ \beta $, of \textit{t}, as is indicated by the fact that they have the same shape coexistence curve, so that,
\begin{equation} 
\Psi = \left|\rho_{liq}-\rho_{vap}\right| \propto t^\beta
\end{equation}
$\beta$ is called a \textit{critical exponent}: the fact that different systems exhibit the same behaviour as they approach $T_c$  (i.e. universality) is the statement that these systems all share the same value of $\beta$. The explanation for universality comes from the RG, but in order to see why, we need first understand the idea and behaviour of $ \xi $, known as the \textit{correlation length} of the system. 
\paragraph*{}
The correlation length is the average distance over which one microscopic variable is correlated with another. If all other parameters are fixed, the correlation length is a function of temperature. In the Ising model (\S\ref{sub:Blockspin}), for example, neighbouring spins on the lattice interact in a way that tends to align them with each other. At low temperatures these correlations will result in regions (or blocks) of parallel spins, which accounts for the ferromagnetic nature of the system. The correlation length is a measure of the size of the correlated regions. Thus, at low temperatures, in this case, the correlation length is large. As the temperature is increased, however, thermal energy tends to randomise the direction of the spins, overpowering the tendency of the spin-spin interactions to align with each other. At high temperatures, then, there is no net magnetisation and the correlation length is small.
\paragraph*{} 
More generally, a small correlation length (i.e. the correlation function rapidly decreases as we move from one point to another) means that faraway points of the system are relatively uncorrelated: the system is dominated by its microscopic details and short-range forces. If the correlation length is large (correlation function decreases slowly), the system displays order on the macroscopic level, since faraway points have a large degree of influence on one another.
\paragraph*{}
Importantly, as the system approaches the critical temperature, the correlation length diverges:
$\xi(T)\rightarrow\infty$ as $T\rightarrow T_c$.
This means that near $T_c$ far distant points become correlated and long-wavelength fluctuations (i.e. those much larger than the lattice spacing) dominate. The range of influence of any one variable is extremely large, and a great number of degrees of freedom are coupled together. The RG is thus employed in order to make calculations tractable: knowing that the long-wavelength fluctuations are the ones of importance, we can integrate out the short-wavelength modes. As in the block-spin model of \S\ref{sub:Blockspin}, we reduce the number of parameters by increasing the observation length (spatial rescaling) and renormalising so that the large-scale behaviour is preserved. This amounts to reducing the correlation length by some (spatial rescaling) factor, \textit{b}, so that,
\begin{equation}\label{eq:tauxiinf}
\tau(\xi)=\frac{\xi}{b}
\end{equation}
\paragraph*{}
As the RG transformation is iterated, and the number of degrees of freedom reduced, we obtain a sequence of Hamiltonians that have the same lattice spacing but with smaller and smaller correlation lengths. This sequence of Hamiltonians can usefully be pictured as defining a trajectory in a space coordinatised by the system's parameters. These parameters include the temperature, the external fields, and the coupling constants (which correspond to the various interaction strengths between spins). Because the coarse-graining process of the RG introduces new kinds of couplings, this space will have to be of high-(typically infinite-) dimensionality (being the space of all possible Hamiltonians). 
\paragraph*{} 
Consider a given system characterised by a given set of parameters undergoing a phase transition. As we vary the temperature (approaching $T_c$), the point representing the system under consideration moves about in the space of all Hamiltonians: a flow \textquotedblleft generated" by the RG transformation $\tau$.\footnote{It should be emphasised that the temperature is not singled out  as a special way of parametrising the system in the Hamiltonian formalism---it is just one parameter among several, all on equal footing. The presentation here is slightly misleading in this regard, but it is an intuitive way of arriving at the concept of a fixed point.} The path it takes may be called the ``physical line" (following \citet[][p. 12]{Pfeuty1977}). We can draw surfaces, $S_{\xi}$ in the parameter space corresponding to constant values of the correlation length; because the correlation length changes as we change the temperature, the physical line intersects several of these surfaces, as shown in Fig.~\ref{fig:physicalline} below.
\begin{figure}[htbp]
\centering
\includegraphics[height=8cm]{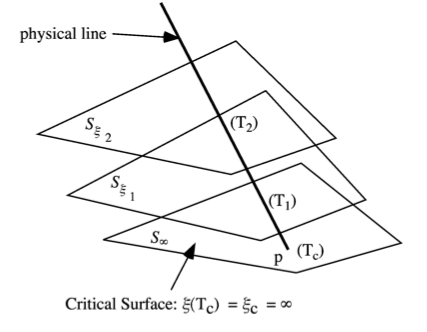}
\caption{The \textquotedblleft physical line" of a system, parametrised by temperature, intersecting several surfaces of constant correlation length in parameter space. Where $\xi(T_1)=\xi_1$, $\xi(T_2)=\xi_2$, with $\xi_1 > \xi_2$, and $\xi(T_c)=\infty$}
\label{fig:physicalline}
\end{figure}

\paragraph*{}
The surface $S_{\infty}$  is the \textit{critical surface}: it corresponds to Hamiltonians (defined by their various parameters) having infinite correlation length. The critical temperature $T_c$ is the point at which the system under consideration (with its given parameter values) intersects the critical surface, denoted as point $p$ in Fig.~\ref{fig:physicalline}. 
Every point on the physical line is mapped, under the RG, to another point in the space of Hamiltonians. Notice that any point on the critical surface (for instance, in this case, $p$), will necessarily be confined to that surface, since, according to (\ref{eq:tauxiinf}), we have,
\begin{equation}
\tau(\xi_c)=\frac{\xi_c}{b} = \frac{\infty}{b} = \infty
\end{equation}
The correlation length remains infinite as further applications of the RG transformation are performed---in other words, at the critical point the system is invariant under the RG. The critical point, $\infty$, is a fixed point, indicating scale invariance. As Batterman (2011) is careful to point out, the fixed point should be understood as a property of the RG transformation itself, rather than a property of the system under consideration. The relation between fixed points in the RG and critical phenomena was discovered by Wilson, building on the insights of Kadanoff. Michael Fisher, a colleague of Wilson's, also contributed greatly to our understanding in this area.\footnote{See, for instance, \citet{Wilson1971, WilsonK1974}.} 
\paragraph*{}
The fixed points are defined by the equation,
\begin{equation}
\tau(H^*)=H
\end{equation}
Where $H^*$ is the Hamiltonian at the fixed point.  Recall that the Hamiltonian can be coordinatised by its set of coupling constants, $K$; the RG transformation $\tau$ relates the renormalised couplings, $K'$ to the original couplings $K$ via the \textit{recursion relations},
\begin{equation}
K'=\tau K
\end{equation}
So that the fixed points correspond to couplings that are left unchanged by the renormalisation transformation:
\begin{equation}
K^*=\tau K^*
\end{equation}

\paragraph*{}
Much of the physics near critical points can be understood by the transformations of the couplings. Performing a linearisation of the couplings in the neighbourhood of the fixed point (on the critical surface) shows how the RG transformation acts upon points that differ only slightly from the fixed point itself. This analysis enables us to calculate the values of the critical exponents (recall that the critical exponents describe the behaviour of a system near the critical point) because it determines how lengths are scaled, locally, in different directions surrounding the fixed point. The linearisation reveals that each fixed point has an associated \textit{basin of attraction}, being all those points in the parameter space (the space of all possible Hamiltonians), which, under the RG transformation, $\tau$, eventually flow into that fixed point. Now, \textit{universality} is explained: systems with the same set of critical exponents lie in the basin of attraction of the same fixed point. These systems are each described by a different set of couplings (i.e. they are represented by different points in the parameter space), but have the same scaling function.
\paragraph*{}
The couplings of systems that belong to the same universality class are \textit{irrelevant}, since they correspond to differences in these systems that ultimately have no impact on the behaviour of the systems at the critical point. The linearisation procedure allows us to classify the couplings of a system as irrelevant if they decrease under the RG transformation and the system flows toward the fixed point. Conversely, the couplings are \textit{relevant} if they increase under the RG transformation and the system flows away from the fixed point, and \textit{marginal} if their scaling behaviour is unable to be determined by the analysis.\footnote{See \citet{ButterfieldForthcoming}.} It is worth noting that the coincidence of the critical exponents across different types of systems was inexplicable prior to the development of the RG.

\section{Batterman's account of emergence}\label{subsect:Batterman}
There are two facets to Batterman's conception of emergence, both based on mathematical singularities and divergences, but which can, I think, be neatly separated. Accordingly, I will discuss the first aspect of his conception of emergence, which is tied to universality, here, and discuss the second, which is based in a failure of limiting relations between theories, in \S\ref{subsect:singularities}. As we have seen \S\ref{subsect:multiplereal}, Batterman's (2000, 2002) strategy for explaining universality is to develop ``principled physical reasons'' for ignoring the detailed microphysical information about any actual system or systems being investigated, and, in the physical cases being considered, these reasons are based in the RG.
\paragraph*{}
Batterman (2011) argues that the universality is explained as above, i.e. that the systems in the same universality class have the same set of critical exponents, and, by the RG are shown to lie in the basin of attraction of the same fixed point. The fixed point is shown to be independent of the choice of initial Hamiltonian; the high-energy details of the system are not important. The differences between systems in the same universality class correspond to irrelevant parameters, those that are washed out as the scaling process is iterated, and are unable to be appealed to in understanding critical phenomena. 
\paragraph*{}
Batterman emphasises the necessity of the role played by the mathematical divergences in this explanatory story, in particular the divergence of the correlation length, which, he points out, is associated with the \textit{thermodynamic limit}. This is the limit as the number of particles in a system approaches infinity, and is an essential idealisation invoked in accounting for phase transitions; although, of course, the phase transition occurs in a finite system, the statistical mechanics of a finite number of particles cannot exhibit the non-analytic behaviour necessary to represent the qualitatively distinct phenomena that occur when the system undergoes a phase transition \citep[][pp. 238--239]{Kadanoff2000}. 

\paragraph*{}
The thermodynamic limit is associated with the divergence of the correlation length as the system approaches the critical temperature, $\xi(T)\rightarrow\infty$ as $T\rightarrow T_c$. Batterman (2011) argues that this divergence is an essential aspect of the explanation of universality, since, without it, there would be no way for us to compare the different systems (in the same universality class) with one another: unless we are dealing with infinite systems, there will always be some characteristic length scale (atomic spacings, for instance) that will differ from system to system. The divergence of the correlation length removes the characteristic length scales that otherwise serve to distinguish the systems, and allows for the comparison of the different fluids at criticality by allowing for scaling or self-similar solutions (Batterman, 2011, p. 1037). Since this account of emergence is, essentially, an explanation of universality, we would thus have no emergence without the infinities that facilitate the comparison of different theories. 
\paragraph*{}
Because the RG story provides such a neat account of the emergent phenomena, Batterman dismisses Laughlin and Pines' (2000) suggestion that symmetry breaking is capable of providing, on its own, an explanation of universality. He states that ``It seems hardly satisfactory to appeal to symmetry breaking as an organizing principle independent of microdetails when we have such a profoundly successful story about why the microdetails in fact are largely independent or irrelevant'' (2011, p. 1038).\footnote{This statement clarifies the ambiguity in a later reiteration of Batterman's argument, ``Laughlin and Pines are correct to hold that from-first-principle derivations the existence of protectorates are not to be had. They are wrong to claim that the existence of such protectorates depend \textit{only} upon higher organizing principles like spontaneous symmetry breaking. One can locate the necessary explanatory features in the mathematical singularities that appear in the thermodynamic limit.'' (Batterman, 2011, pp. 1047--1048)} But, as will be argued, \S\ref{subsect:Morrison}, Laughlin and Pines are correct in characterising symmetry breaking as a HOP, which may be understood as providing, on its own, an explanation of the ``emergent'' low-energy phenomena. Symmetry breaking provides an account of the multiple-realisability, demonstrating that the micro-details of a system are unimportant; and this account is one that can stand independently of the RG-based account. Systems that embody the same symmetry breaking pattern exhibit the same low-energy behaviour. 
\paragraph*{}
Batterman (2011) presumably rejects symmetry-breaking as an account of emergence, too, because his conception of emergence is defined as involving thermodynamic singularities. As will be explained in \S\ref{subsect:universalEFT}, Batterman uses the presence of a thermodynamic singularity to distinguish ``emergent'' from ``resultant'' physical phenomena; however, the case of critical phenomena is interesting because (as I argue) the universality can be explained either by the RG-based account or the symmetry-breaking account.  Even if Batterman's argument is correct, and the RG-based account of emergence necessarily involves reference to divergences, it does not---given the alternative explanation provided by the symmetry-breaking account---demonstrate that an explanation of emergence (i.e. universality) necessarily involves mathematical divergences. Rather, I think it better to suppose that, although the divergences feature essentially in an account of critical phenomena, they do not play an essential role in \textit{the explanation of universality}. Batterman's mistake is, ultimately, just a failure to recognise that we are better to admit a variety of different conceptions of emergence rather than claim there is a single (best) one that applies in all cases.

\section{Morrison's account of emergence}\label{subsect:Morrison}
To demonstrate her conception of emergence, Morrison uses the example of superconductors. Many of the features of superconductors (e.g. heat capacity and critical temperature) are dependent on the type of metals they are composed of. There are emergent universal features, however, that are common to all superconductors: infinite conductivity (currents that can circulate for years without perceptible decay, due to extremely low electrical resistance), flux quantization (the magnetic field is quantised in units of $ h/2e $) and the Meissner effect (the expulsion of a magnetic field from a superconductor). These emergent features are said to be \textit{exact}, meaning that they can be predicted with extraordinary accuracy. 

\paragraph*{}
As mentioned in \S\ref{sub:tower b-u}, the prediction of the fine structure constant in precision tests of QED using the Josephson effect is one example of such a result. Recall that the BCS model used to obtain this result is actually an approximation, and the reason for its success is because it embodies a particular symmetry-breaking pattern. This is the same in the case of predicting exact results in superconductors---as \citet[][p. 43]{Weinberg1986} states, the high-precision predictions about superconductors actually follow not from the details of the models themselves, but more generally from the fact that these models exhibit a spontaneous breakdown of a symmetry. Because they are exact results, Morrison (2012, p. 151) says, they must follow from general principles rather than the details of the approximations used to model them. This is in accordance with Laughlin and Pine's (2000) claim that these properties of superconductors represent emergent phenomena (\S\ref{sub:HOP}).
\paragraph*{}
While the BCS model provides a story about the micro-details of how superconductivity occurs (and provide the basis for approximate quantitative calculations, including the critical temperature\footnote{This is the temperature at and above which superconducting properties are no longer present.}), when it comes to deriving the universal characteristics of superconductivity, these micro-details are not essential in themselves. What is important is simply the breaking of the symmetry (electromagnetic gauge invariance) of the system. In this respect, the micro-story is significant only insofar as it serves to fill in the structure of the symmetry-breaking pattern. As was the case in \S\ref{sub:tower b-u}, the high-energy details themselves are irrelevant, being severely underdetermined by the low-energy physics. 

\paragraph*{}
According to the BCS model, while the system is in its symmetric state, above the critical temperature $T_c$, the electrons behave as free particles, repelled from one another because of their like charge. At $T_c$ the electrons, on this model, form Cooper pairs---two electrons become correlated with one another, interacting via phonon exchange. When this occurs, the electromagnetic gauge invariance is broken and the system becomes superconducting, with its electromagnetic properties dominated by the Cooper pairs. The order parameter in this case is related to the macroscopic ground state wave function of the Cooper pairs, $<\varphi>$. In the non-superconducting phase, $<\varphi>=0$ since there are no Cooper pairs, while in the superconducting phase $<\varphi>\neq 0$ since the Cooper pairs have formed. The nonzero order parameter indicates that the symmetry has been broken and a phase transition has occurred. The coherence length $\xi$ this case is the mean separation of electrons at which pair correlation becomes effective.\footnote{The value of which is between 100 and 1000 nanometres, being three or four orders of magnitude larger than the lattice spacing.} 

\paragraph*{}
As Weinberg (1986) explains, and Morrison (2012) re-emphasises, we do not need the story about the formation and behaviour of the Cooper pairs in order to derive the most important exact consequences of superconductivity. To do this, we need only the assumption of broken electromagnetic gauge invariance, and nothing depends on the specific mechanism by which the breakdown occurs. A basic outline of this assumption, together with those quantities required for the derivation of the characteristic properties of superconductors, are now presented. Following this, I show briefly how these can be used to understand the phenomenon of infinite conductivity (for the corresponding derivations of the Meissner effect and flux quantisation, see Weinberg, 1986).
\paragraph*{}
The electromagnetic gauge group, $ U(1) $ is defined as the group of multiplication of fields $\psi(x)$ of charge $q$ with phases $\Lambda$,
\begin{equation}
\psi(x)\rightarrow exp(i\Lambda q/\hbar)\psi(x)
\end{equation}
We assume that all charges $q$ are integer multiples of the electron charge $-e$, so that the phases $\Lambda$ and $\Lambda + 2\pi\hbar/e$ are identical.\footnote{This discussion is based on Weinberg (1986).}  This group $U(1)$ is spontaneously broken to $Z_2$, the subgroup consisting of $U(1)$ transformations with $\Lambda = 0$ and $\Lambda = \pi\hbar/e$.
\paragraph*{}
Now, any system described by a Lagrangian with symmetry group $G$, when in a phase in which $G$ is spontaneously broken to a subgroup $H\subset G$, will possess a set of Nambu-Goldstone excitations. These excitations are described by fields that transform under $G$ like the coordinates of the coset space $(G/H)$.\footnote{This means that the Nambu-Goldstone fields transform under $G$ like the coordinates used to label the elements of $G$ itself, but with two field values identified if one can be transformed into the other by multiplication with an element of the unbroken subgroup $H$.}  In this case, we will have a single Nambu-Goldstone excitation, described by a field $\varphi(x)$ that transforms under $G = U(1)$ like the phase $\Lambda$. Because the $U(1)$ group has the multiplication rule $g(\Lambda_1)g(\Lambda_2) = g(\Lambda_1 + \Lambda_2)$, under a gauge transformation with $\Lambda$, the field $\varphi(x)$ will undergo the transformations,
\begin{equation}
\varphi(x)\rightarrow \varphi(x)+\Lambda
\end{equation}
Since $\varphi(x)$ parameterises $U(1)/Z_2$ rather than $U(1)$ itself, we may take $\varphi(x)$ and $\varphi(x)+\pi\hbar/e$ to be equivalent values,
\begin{equation}
\varphi(x)\equiv\varphi(x)+\pi\hbar/e
\end{equation}
The characteristic property of a system with a broken symmetry is that the quantity $\varphi(x)$ behaves as a propagating field; the second variational derivative of the Lagrangian with respect to $\varphi(x)$ has non-vanishing expectation value.
\subparagraph*{}
When we turn on the electromagnetic fields \textbf{B} and \textbf{E}, their interaction with the superconductor is governed by the principle of local gauge invariance, under which the field $\varphi(x)$ transforms as before under $U(1)$, but now with space-time-dependent phase
\begin{equation}
\varphi(x)\rightarrow \varphi(x)+\Lambda(x)
\end{equation}
The potentials transform as,
\begin{equation}
\textbf{A}(x)\rightarrow\textbf{ A}(x)+\bigtriangledown\Lambda (x)
\end{equation}
\begin{equation}
A^0(x)\rightarrow A^0(x)-\partial\Lambda(x)/\partial t
\end{equation}
And all other field operators are gauge-invariant.\\
The Lagrangian for the superconductor plus electromagnetic field may be written as
\begin{equation}
L = \frac{1}{2}\int d^3x(\textbf{E}^2-\textbf{B}^2)+L_m\left[\bigtriangledown\varphi - \textbf{A}, \dot \varphi + \textbf{A}^0, \tilde \psi\right]
\end{equation}
Where $L_m$ is the matter Lagrangian, an unknown functional of the gauge-invariant combinations of $\partial_{\mu} \varphi$ and $A_{\mu}$ as well as the unspecified gauge-invariants $\tilde \psi$ representing the other excitations of the system. From $L_m$ we can obtain the electric current and charge density as,
\begin{equation}
\textbf{J}(x)=\frac{\delta L_m}{\delta \textbf{A}(x)}
\end{equation}
\begin{equation}\label{eq:epsilson}
\epsilon (x) =\frac{\delta L_m}{\delta A^0(x)}= - \frac{\delta L_m}{\delta \dot\varphi (x)}
\end{equation}
We assume, finally, that in the absence of external electromagnetic fields, the superconductor has a stable equilibrium configuration with vanishing fields,
\begin{equation}
\bigtriangledown \varphi - \textbf{A} = \dot \varphi +\textbf{A}^0 = 0
\end{equation}
This assumption is equivalent to the requirement that the leading terms in $L_m$ are at least of second order in $\bigtriangledown\varphi - \textbf{A}$ and $\dot \varphi+\textbf{A}^0$ (i.e. there are no first-order terms of these quantities). Furthermore, the assumption that electromagnetic gauge invariance is spontaneously broken is equivalent to the statement that the coefficients of the terms in $L_m$ of second order in $\bigtriangledown\varphi-\textbf{A}$  and $\dot \varphi+\textbf{A}^0$ have non-vanishing expectation values, so that $\varphi$ behaves like an ordinary physical excitation.
\paragraph*{}
Now, simply from the general assumptions and quantities defined above, we can derive all the important characteristics of superconductors. On this picture, $\varphi(x)$ is not taken to be related to the wavefunction of the Cooper pairs, as it is in the BCS model, but, rather, it is a Nambu-Goldstone field that accompanies the breakdown of electromagnetic gauge invariance. As Morrison (2012, p. 155) states, we do not need a micro-story of electron pairing: Planck's constant simply does not appear in the differential equation that govern $\varphi$.
\paragraph*{}
As an example of how the characteristics of superconductors follow simply from these assumptions and quantities, consider the idea of infinite conductivity. According to (\ref{eq:epsilson}), the charge density is the dynamical variable canonically conjugate to $\varphi$. In the Hamiltonian formalism, the matter Hamiltonian $H_m$ can thus be taken as a functional of $\varphi(x)$ and $\epsilon(x)$ rather than of $\varphi(x)$ and its first derivative with respect to time. The time-dependence of $\varphi$ is given by,
\begin{equation}
\dot \varphi(x) = \frac{\delta H_m}{\delta -\epsilon(x)}
\end{equation}
The change in the energy density per change in the charge density at any point may be regarded as the ``voltage'', $V(x)$ at that point,
\begin{equation}\label{eq:V(x)}
V(x)\equiv\frac{\delta H_m}{\delta\epsilon(x)}
\end{equation}
Hence, the time-dependence of the Nambu-Goldstone field at any point is simply the voltage,
\begin{equation}
\dot \varphi(x) = - V(x)
\end{equation}
One immediate consequence is that a length of superconducting wire carrying a steady current, with time-independent fields, must have zero voltage difference between its ends. If the voltage difference were not zero, then, according to (\ref{eq:V(x)}), the gradient $\bigtriangledown\varphi(x)$  would have to be time-dependent, and this would lead to time-dependent fields. A zero voltage difference for finite current is just what is meant by infinite conductivity.
\paragraph*{}
As has now been shown, infinite conductivity depends only on the spontaneous breakdown of electromagnetic gauge invariance. Although the micro-story of the BCS approximation provides a detailed mechanism about how this breakdown occurs, namely, by the formation of Cooper pairs, its details are not necessary for understanding infinite conductivity or the other important properties of superconductors---these being consequences of the broken symmetry. Importantly, this means that we could change the irrelevant details of the micro-story and still have superconductivity, with its characteristic properties, emerge. For instance, as Weinberg (1986, p. 49) points out, infinite conductivity would, presumably, occur even if the charged particles whose pairing resulted in the breakdown of the symmetry were bosons rather than fermions.\footnote{For this reason, Weinberg (1986, p. 49) states that it is misleading for textbooks to relate the infinite conductivity of superconductors to the existence of an energy gap separating a Fermi sea of paired electrons from their excited unpaired states. Furthermore, there are also examples of superconductors without an energy gap. The one respect in which Fermi statistics play a necessary role in superconductors is that the Fermi surface enhances the long-range effects of the phonons exchanged in the electron pairing. But, again, this fact is only important in showing how the Cooper pairs are formed (and symmetry broken), rather than in deriving the general consequences of the broken symmetry.} 
\paragraph*{}
Now, according to Morrison (2012, p. 156), the idea of emergence associated with spontaneous symmetry breaking encompasses and clarifies both the ontological and epistemological aspects of her account (\ref{def:Mor}). This is because, she says, even though superconductors are constituted by their microscopic properties, their defining features are immune to changes in those properties, for instance, replacing fermions with bosons. Although Morrison doesn't spell out more fully the relation between her conceptions of epistemological and ontological independence as it applies in this example, I believe the picture is clear enough for us to do so. In the case of epistemological independence, then, the example of superconductors demonstrates that we \textit{need not }appeal to the micro-properties, because (presumably) even though we have a micro-story (the BCS model) that is capable of providing an account of superconductivity and its associated (emergent) phenomena, we can also derive this phenomena by a different method (via the assumption of broken electromagnetic gauge invariance). 
\paragraph*{}
The case of ontological independence, however, means saying that we \textit{cannot} appeal to the micro-properties. 
On the face of it, this might seem strange or even incorrect, given that the RG account just demonstrates that some micro-properties are relevant to the macro-behaviour, while most are often irrelevant: it does not say that all the micro-details must be disregarded.
However, the idea of emergence here is associated with \textit{universality}. Recall from \S\ref{sub:UniversEmergence}, that any micro-description, purely by virtue of being a particular micro-description, will fail to capture the multiple-realisability, or universality, of the emergent phenomena. Recall that Laughlin and Pines (2000, p. 28) present symmetry breaking as an example of a HOP, and the phenomena that depend on it are ``transcendent''---insensitive to, and unable to be deduced from, micro-details. As Morrison (2012, p. 149) states, too, the notion of symmetry breaking is not itself linked to any specific theoretical framework, rather, it functions as a structural constraint on many different kinds of systems in high-energy physics, condensed matter physics and cosmology.

\section{Comparing the accounts}\label{subsect:universalEFT}
I've argued in this chapter that universality and HOPs provide a basis for the autonomy that characterises emergence---the fact that vastly different micro-systems can exhibit the same multiply-realised macro-behaviour demonstrates that we cannot account for the behaviour by appeal to the micro-dynamics (the ``dependence'' aspect of emergence is provided by the fact that the macro-behaviour is related to the micro-dynamics through the RG and EFT techniques). In this section I compare the conception of emergence associated with universality to the ideas offered in EFT. I also discuss the differences between the conception of emergence based on the RG account with the conception of emergence associated with the symmetry-breaking account. Following this, I present some problems with Morrison's and Batterman's definitions of emergence as being a failure of reduction, and argue that the more general, positive conception that I've developed (throughout this thesis) is preferable. Finally, I consider the example of hydrodynamics, which represents emergence on the more general account, but not for Batterman or Morrison.
\subsection{Universality and EFT}
Although the conception of emergence associated with the idea of universality presented by Morrison (2012) and Batterman (2011) pertains only to EFTs, it is not supposed to hold generally of EFTs. The claim is of limited scope, meant only to apply to descriptions of critical phenomena. Morrison (2012, p. 142), for instance, prefaces her account by stating that, on it, condensed matter physics is not emergent from particle physics in all cases, ``but rather that there are certain kinds of phenomena, both in condensed matter physics and indeed high-energy physics itself, whose explanation can only be understood from within the emergentist picture''. 
\paragraph*{}
\citet{Bain2013} argues that the conception of emergence presented by \citet{Batterman2002} is not able to be directly extended to describe the emergence of an EFT from its high-energy theory. This is in spite of there being a compelling analogy between Batterman's (2002, p. 123) example of thermodynamics emergent from statistical mechanics and the example of a QFT ``with cutoffs'' treated as emergent from its (renormalisable) continuum version, i.e. the QFT ``without cutoffs''.\footnote{This analogy is clearly laid out in Bain (2013, p. 29).} The latter example, of course, is not representative of an EFT as emergent from an underlying high-energy theory, and, indeed, would register neither on Bain's account (since the QFT with cutoffs and the QFT without cutoffs are formally identical), nor, indeed, on Batterman's (2011) account of emergence (since it does not involve the appearance of qualitatively distinct behaviours).

\paragraph*{}
Nevertheless, I believe there is plausible basis for comparison of accounts of emergence in critical phenomena and the account I presented above for EFT. Morrison's (2012, p. 161) description of emergence, for instance, sounds very familiar,
\begin{quote}
Emergence is characterized by the fact that we cannot appeal to microstructures in explaining or predicting these phenomena despite their microphysical base. RG methods reveal the nature of this ontological independence by demonstrating the features of universality and how successive transformations give you a Hamiltonian for an ensemble that contains very different couplings from those that governed the initial ensemble. 
\end{quote}

\paragraph*{}
Recall that the idea of novelty in the account of emergence in EFT is related to the distinctness of the high- and low-energy theories, and the fact that they describe very different degrees of freedom. Another key feature of the EFT account is the autonomy of the EFT, due to the tenuous link between a high-energy theory and the emergent low-energy physics. The high-energy theory is severely underdetermined by the low-energy EFT. Universality (as in critical phenomena) seems to present us with concrete examples of such underdetermination: as systems with different micro-details give rise to the same phenomena at criticality. On both the universality and EFT cases, the high-energy details correspond to irrelevant parameters and are washed out as we flow with the RG to lower-energies, and the consequent underdetermination of the high-energy physics given the low-energy theory leads to a conception of autonomy. While Batterman and Morrison each rely on the nuances of their own conceptions of emergence to articulate the relevant sense of autonomy that distinguishes the emergent from the merely resultant, I have argued that the work is done by the universality itself.

\paragraph*{}
In both the examples of critical phenomena, as well as the case of EFT more generally, the high- and low-energy theories are distinct and (quasi-)autonomous, but, one might argue, as Morrison (2012) and Batterman (2011) do, that the examples of critical phenomena represent a stronger sense of \textit{novelty} as ``qualitatively distinct behaviour'' than EFT (or hydrodynamics, discussed shortly, \S\ref{subsub:hydro}). This stronger sense of novelty, though, I argue, comes about from comparing the relevant low-energy theory not with the high-energy theory, as in EFT, but with the high-energy theory describing the system \textit{before} the phase transition. Thus, we may distinguish between a \textit{synchronic} conception of novelty, where the low-energy physics is novel compared to the high-energy system at a given time, and a \textit{diachronic} conception of novelty, where the low-energy physics is novel compared to the high-energy physics following some dynamical change in the state of the system.

\subsection{Symmetry breaking and higher-organising principles}\label{subsub:SBHOP}
The conception of emergence furnished by the symmetry-breaking account of critical phenomena is based in diachronic novelty, as well as autonomy that stems from the fact that symmetry-breaking serves as a HOP in the example of superconductors presented above. The fact that we can have exact results---results that can be shown to be described simply by the symmetry-breaking pattern itself, rather than the high-energy theory---demonstrates that the behaviour is emergent with respect to the high-energy description of the system in question. 
\paragraph*{}
One reason for considering symmetry breaking as a HOP is that it is not linked to any specific theoretical framework; rather, as Morrison (2012)  states, it functions as a ``structural constraint'' on many different kinds of systems in high-energy physics, condensed matter physics and cosmology. Mainwood (2006, p. 114) also makes this point, stating that, 
\begin{quotation}
Principles such as symmetry-breaking crop up across very different areas of physics: in the Standard Model of particle physics, in many-body quantum mechanics and in large-scale classical and celestial mechanics. Such principles that apply across physics of very different scales, give a sense of unity to physics, which implies no prejudice in favour of microphysics.
\end{quotation}
Symmetry-breaking thus furnishes two different, and independent, conceptions of emergence: as explaining multiple-realisability (or universality, which it has in common with the RG fixed-point account) and as functioning as a higher-organising principle.
\paragraph*{}
Symmetry-breaking is a dynamical physical occurrence; it is a process that occurs in time. On the other hand, the RG equations simply provide a description of the system at different length scales: the changes that the RG describes are not dynamical. Although the RG is often said to ``induce a flow" in the parameters of a system, this way of speaking is misleading, as the RG is not something that can properly be said to act.  The parameters that ``flow'' are those whose values are seen to differ in the space of all Hamiltonians as we change (by hand) the energy at which we view the system. Put crudely, the RG is a means of describing a system, while symmetry-breaking is a mechanism that happens within a system. This difference shows that the RG-based account and the symmetry-breaking account are two different forms of explanation.\footnote{Bain (2013b) suggests that the RG-based account represents a ``unifying explanation'', while the symmetry breaking account is a ``causal/mechanical explanation''.} Recognising that the RG is better understood as a description (with, at best, a ``pseudo-dynamics'' describing the zooming in and out of the energy scale, $\Lambda$), we find an answer to the question posed in the introduction to this chapter, regarding whether or not the RG might be considered a higher-organising principle. The RG describes how the high-energy physics influences the low-energy physics: it is not that the low-energy physics is dependent on the RG.
\paragraph*{}
It is perhaps tempting to say that this is true of symmetry-breaking as well, though: that symmetry-breaking shows how the low-energy physics depends on some \textit{particular aspects} of the high-energy physics (i.e. those that are responsible for ``filling in'' or realising the symmetry-pattern). The difference is, however, that so long as we focus on the multiple-realisability\footnote{This applies to cases where the universality class may be defined as the set of systems that share the same low-energy symmetry breaking pattern.}, we could say that the low-energy behaviour does depend on the symmetry-breaking \textit{and nothing else}, whereas we would, presumably, not say that the low-energy EFT depends on the RG and nothing else. Also, the RG does not lead to exact results, as symmetry-breaking does. This is evidenced by the example of the BCS theory (\S\ref{sub:tower b-u}): the RG explains why it yields a very good approximation of the low-energy physics, but it is only the symmetry-breaking pattern it predicts that is responsible for the exact results predicted in the Josephson effect. For these reasons I do not think it is appropriate to describe the RG as a higher-organising principle.

\subsection{Universality and reduction-based definitions of emergence}\label{Unired}
As explained in \S\ref{sub:UniversEmergence}, Batterman (2011) and Morrison (2012) define emergence as a failure of reduction, and argue that this is exemplified in cases of universality because any single reduction of the macro-description to a micro-description will not explain the universality (the fact that there are many other systems, with different micro-descriptions are amenable to the same macro-description). However, there is a major problem with focusing on the reduction's failure to explain the universality: that it means we can never classify the behaviour of any single system as emergent. On these accounts, we need to have several examples of other physical systems with the same low-energy physics (but different high-energy physics) in order to say that there is emergence. It would be preferable, surely, to address a case of emergence without the need to consider additional, external systems. This is one of the benefits that my account offers. The positive conception of emergence as being simply the novelty and autonomy of the low-energy physics compared to the high-energy physics (plus the two levels being related by the RG and EFT techniques) means we can consider individual cases of emergence. This definition of emergence is more general than those upheld by Batterman and Morrison, as it allows us to recognise emergence even in ``unrealised'' cases of universal behaviour (i.e. cases where the emergent physics depends so minimally on the micro-theory that the latter could be any number of things, even when we do not have examples of these other high-energy systems). 

\paragraph*{}
There are particular difficulties with Morrison's (2012) account that result from its dependence upon universality and the issue mentioned above, that we cannot consider individual cases of emergence. It would seem, on such an account, that if we only had a single system (whose low-energy behaviour is described by an EFT), then we could say just that it exhibits ``epistemological independence'' (using Morrison's definition, above) and not emergence. Although its low-energy properties depend only tenuously on the high-energy physics (and in such a way that we need not know the high-energy physics), Morrison would claim that these properties are still ``resultant'' rather than emergent. Yet, as soon as another system, with different micro-composition but the same low-energy behaviour was discovered, the systems would both exhibit ``ontological independence'' (using Morrison's definition) and demonstrate emergence. Suddenly, the micro-story becomes impotent: explanatory one minute, useless the next. Furthermore, it is especially problematic for Morrison's account of emergence, which claims to be an ontological account, that emergence depends on our state of knowledge.

\paragraph*{}
Consider that if we had only a single example of a superconductor, composed of a particular type of metal, then we would not call its superconductivity an emergent phenomenon on Morrison's (2012) account. The properties of the superconductor are derived using the BCS model, which tells an explanatory micro-story about the formation of Cooper pairs at the critical temperature. We might recognise that the derivation of the superconductivity does not rely on the details of the micro-story but rather on the broken gauge symmetry alone, but this would not change our classification that the behaviour is resultant: after all, it is the micro-physics that actually exhibits (``fills in'' or ``realises'') the symmetry, and the fact that it does so is the reason we are able to recognise that the system exhibits the relevant pattern. Morrison (2012, pp. 162-163) argues that the symmetry breaking cannot be classified as fundamental physics (i.e. part of the micro-theory) and so the fact that superconductivity can be derived from it does not count relevantly as reduction. I do not want to argue about what counts as derivation or reduction, nor what should be classed as fundamental physics, but, in the absence of other examples of superconductors we cannot appeal to universality in order to recognise the superconductivity as emergent. Again, however, as soon as a different superconductor is discovered, the superconductivity becomes emergent on this account.

\paragraph*{}
We can avoid these problems, of not being able to deal with individual cases of emergence, and having emergence depend on our knowledge of the existence of other systems, by adopting the more general, positive account of emergence I've expounded in this paper. Because it applies to all EFTs, rather than just those which describe more than one low-energy system, the positive conception of emergence admits cases of ``potential'' or ``unrealised'' universality. It means that our ascription of emergence will not change once we discover that there are other systems with the same low-energy behaviour. This conception of emergence also fits more naturally with the physics by encompassing EFT generally (as applied), rather than a select group of EFTs (those which we happen to know apply to multiple systems). As stated above, Morrison (2012) rejects EFT and its ``epistemological independence'' because it is common. I hope to have shown that this dismissal is hasty and disloyal to the physics---yes, the fact that we typically need not appeal to micro-details in order to explain macro-behaviour in science is certainly a pervasive one, but its great frequency and usefulness should make it more interesting, not less. It is a testimony to the power of the RG and EFT methods used in physics, and deserves attention; yet it is overlooked thanks to our fixation on reduction, derivation and ontology.

\paragraph*{}
Regardless of this, however, if Morrison (2012) wishes to retain a more exclusive definition of emergence, yet avoid the ``individual case'' problems, then less burden should be put on the idea of universality in her account. For example, Morrison could appeal to the fact that the characteristic properties of superconductors are exact results, and thus tie emergence to the ``transcendent'' HOPs of \citet{Laughlin2000}. Not all instances of EFT produce exact results, but the examples involving symmetry breaking discussed above do. Alternatively, Morrison could appeal to the fact that emergence associated with critical phenomena and symmetry breaking phase transitions is a diachronic sense of emergence, whereas in other cases of EFT the low-energy behaviour is novel and autonomous compared to the high-energy system at a given time. This is an important point that distinguishes Morrison's (2012) account, which emphasises the role of symmetry breaking in addition to the RG, from Batterman's (2012) account, which favours the RG fixed point story and dismisses the significance of symmetry breaking. This is because, as explained above, symmetry breaking is a dynamical process, while the RG is better understood as a description.

\paragraph*{}
Either of these two strategies would lessen the dependence of Morrison's (2012) account on the idea of universality, and, if developed, could preserve Morrison's exclusive, ontological account of emergence while guarding against the ``individual case'' problems. Taking either of these strategies would exclude those cases of EFT that are not cases of universality, but also exclude some cases of universality that don't involve symmetry breaking---for instance, hydrodynamics (explained in the next sub-section).

\paragraph*{}
The other difficulties with Morrison's (2012) conception of emergence are related to the fact that it is supposed to represent an ontological account of emergence. Presumably, Morrison wants to take the emergent properties as real, but her explanation of how they are arrived at employs both the RG fixed-point story as well as symmetry breaking. The RG fixed-point story involves an obviously unphysical infinite idealisation (the thermodynamic limit), which we should not be naively committed to.\footnote{Apparently, there may also difficulties with naive realism when it comes to trusting our theories of spontaneous gauge symmetry breaking, see \citet{Elitzur1975} and the recent response \citet{Friedrich2013}.} So, expounding an ontological account of emergence requires some indication of how we are to commit to our theories---why are we supposed to be committed to some parts of our theories and not others (even though they are apparently parts upon which the ontologically trustworthy parts rely)? The positive conception of emergence that I've presented in this paper, of course, avoids any such questions and holds that the development and criticism of realist positions is to be undertaken separately to the project of understanding emergence in physics.

\subsection{Hydrodynamics}\label{subsub:hydro}
Recall from \S\ref{sub:emerphysics} the ``horizontal'' chain introduced by Hu (2009), describing the collective states of matter, namely, the stochastic -- statistical -- kinetic -- thermodynamic/hydrodynamic. The idea is that systems with different micro-constituents will behave similarly when viewed at low-energy where long-wavelength fluctuations dominate and it is only the collective (macro-) behaviour that is of interest. As discussed in the next chapter, on treating spacetime in EFT (\S\ref{sect:GREFT}), several physicists have explored the suggestion that general relativity is the hydrodynamic limit of some higher-energy theory, and have taken it to be a case of emergence.\footnote{For instance, \citet{Hu2005, Hu2009, Barcelo2001, Barcelo2001E}.} 
\paragraph*{}
Rather than trying to relate these ideas to reduction, however, I believe the conception of emergence suggested here is best understood in terms of novelty and autonomy (recalling the discussion earlier in \S\ref{sub:novelty}). This is because treating the relationship between thermodynamics and statistical mechanics as reduction is not straightforward nor uncontroversial \citep[see, e.g.][]{Yi2003}. Yet, in spite of this, many philosophers and physicists consider the relationship between thermodynamics (viewed as the macro-theory) and statistical mechanics (viewed as the micro-theory, although, as stated, it is one that describes the collective behaviour of the constituent particles, and thus already abstracts away from the details that otherwise characterise these constituents) to be the paradigm example of inter-theory reduction. It is perhaps because of this that philosophers are loathe to recognise thermodynamics/hydrodynamics as describing emergent behaviour: it is ``resultant'' rather than ``genuinely novel''. \footnote{But cf. e.g. Butterfield.}

\paragraph*{}
What is interesting is that thermodynamics/hydrodynamics does present us with something akin to universality, i.e. it is low-energy behaviour that is multiply-realisable across many different micro-systems, and its description is provided by the RG. Thus, as with the examples above, we may tie the (quasi-)autonomy of hydrodynamics to the underdetermination of the high-energy physics given the low-energy physics. Other philosophers will, at this stage, appeal to the nuances of their own particular conceptions of emergence in order to provide an explanation for their disinterest in hydrodynamics. 
\paragraph*{}
Batterman (2011, p. 1048), for instance, considers non-critical thermodynamic systems that exhibit universal behaviour from the point of view of statistical mechanics, taking the example of the Gaussian probability distribution. This distribution, which may be described as having ``critical exponent'' of $ \frac{1}{2} $, characterises a number of quantities in a large variety of systems away from their critical points. According to Batterman, this behaviour should not be understood as emergent, however, because it does not feature a \textit{qualitative change} in the state of the system, the way that a phase transition does. More importantly, on Batterman's account, such phenomena are not admissible as emergent because their explanatory story does not contain reference to any thermodynamic singularity. 
\paragraph*{}
Morrison's (2012) conception of emergence, however, is perhaps not strong enough to distinguish between thermodynamics (which, presumably, she would want to class as ``resultant'' and exclude from her account) and critical phenomena. This is because, as I have argued, the work done by her conception of emergence, being what she terms ``ontological independence'' (``we cannot appeal to micro-details in order to explain or predict the emergent phenomena''), is done simply by universality. Thermodynamics demonstrates universality just as critical phenomena does. 
\paragraph*{}
One might object that the relevant difference is that in the case of thermodynamics we do have a derivation that captures the universality: the reduction to statistical mechanics might be thought to explain the universality of the thermodynamic quantities. This will not do, however, because it doesn't explain why a large variety of micro-systems of differing constituents are able to be described by statistical mechanics; statistical mechanics is a framework for theories and already abstracts away from micro-details. In other words, the reduction of thermodynamics to statistical mechanics (if we believe it goes through) does not represent an ``appeal to micro-details'' and so is unrelated to Morrison's conception of emergence; more work is needed. 
\paragraph*{}
Batterman's appeal to the \textit{qualitatively distinct phenomena} that appear when systems undergo a phase transition, as being genuine novelty is an interesting suggestion for distinguishing critical phenomena from thermodynamics. The appropriate comparison class, however, is what I think is important here: critical phenomena is qualitatively distinct from the behaviour the system exhibits when it is not at the values of temperature and pressure that characterise the critical point. In other words, in looking at critical phenomena, we are comparing the system before the (second-order) phase transition, i.e. the system at non-critical values of temperature and pressure, with the system when it undergoes the (second-order) phase transition, i.e. the system at the critical point. In the case of thermodynamics, the system is kept under the same conditions throughout, but its behaviour at different length scales is examined. For this reason, we can say that the appropriate conception of novelty when considering emergence in critical phenomena is diachronic, while in hydrodynamics (as in EFT), it is synchronic novelty. 

\section{Singularities, limiting relations and emergence}\label{subsect:singularities}
Batterman expounds an additional account of emergence not associated with universality (though it is one that he does not consider separate from the conception of emergence as universality), based on a failure of derivation. This conception of emergence, unlike the one presented earlier (\S\ref{subsect:Batterman})  \textit{does} rely on singularities for its explanation. Batterman (2005, 2002) argues that if we take emergence as a failure of the philosopher's sense of reduction (i.e. the Nagelian sense of reduction where the laws of the reduced theory can be derived from those of the reducing theory), then, we have emergence of thermodynamics when it is applied to cases of critical phenomena. 
\paragraph*{}
The philosopher's sense of reduction, Batterman (2005, 2002) argues, is related to the physicist's sense of reduction, which is that the ``finer-grained'' or ``newer'' theory (e.g. special relativity) reduces to the ``coarser'' or ``older'' theory (e.g. Newtonian mechanics) as some appropriate limit is taken (e.g. as the limit as the speed of light is taken to infinity): the philosopher's sense of reduction will hold, he says, only if the physicist's one does. The appropriate limiting relation that facilitates the reduction of thermodynamics to statistical mechanics is the thermodynamic limit, $N, V\rightarrow\infty$ (while holding $N/V$ constant).
\paragraph*{}
Now, given that the correlation length diverges when a system is at a critical point, the limiting relation between statistical mechanics and thermodynamics becomes singular at this point. Because of this, Batterman (2005, p. 227, 2002, p. 124) argues, the physicist's sense of reduction fails---there cannot be any sort of  ``derivational'' connections between thermodynamic laws and statistical mechanics for systems at criticality---and so the philosopher's sense of reduction will fail. Batterman (2002, p. 125) then ties his conception of emergence to this failure of reduction, stating that ``The novelty of emergent phenomena---their relative unpredictability and inexplicability---is a direct consequence of the failure of reduction.'' And so, he argues, emergent phenomena (as opposed to ``merely resultant'' phenomena), are dependent on the existence of singularities. Batterman is not the only author to argue that emergence is due to singularities; \citet{Rueger2000, Rueger2001}, also argues that novel properties that appear in the lower-energy (macro-) theory cannot be explained in terms of the higher-energy (micro-) theory when a singular limit is present, and ties this to emergence. 
\paragraph*{}
\citet{Wayne2012} argues against this view, that a singularity is sufficient for emergence, by using a physical example from Rueger (2001, 2000) of the van der Pol nonlinear oscillator.\footnote{The one-dimensional van der Pol oscillator was originally investigated as a model of the human heart \citep{Pol1928}, but has subsequently been used to describe some oscillatory vacuum tube and electronic circuits (Wayne, 2012).} This is an example of a system that exhibits different phenomena at different energy levels, with some of the ``upper-level'' (macro-) properties being qualitatively different from those at the ``base'' (micro-) level, and in which there is a singular limit relation between the levels. Wayne (2012) admits the novelty of the macro-phenomena, but demonstrates that the presence of the singular limit does not preclude there being a full explanation of the macro-phenomena entirely in micro-level terms. In doing so, he undercuts Batterman and Rueger's claims that macro-theories separated from their respective micro-theories by a singular limit are emergent, by arguing that the provision of a full explanation of the macro-phenomena in micro-level terms means that the novel macro-phenomena cannot be classed as emergent.
\paragraph*{}
Butterfield (2011b, p. 1088) also presents his own example of singular limit without emergence (related to fractals) in order to demonstrate that a singular limit is insufficient for emergence. As we have seen, \citet{Butterfield2011a, Butterfield2011b} takes emergence simply as novel and robust behaviour, and so would disagree with one of the core assumptions of Batterman's (2002) argument, namely, that the novelty that characterises emergence is a direct consequence of a failure of reduction (Butterfield also maintains that emergence is compatible with reduction). Of course, it should not be surprising that I also reject this tenet of Batterman's argument, and believe that Wayne and Butterfield are successful in their demonstrations that emergence does not necessarily result from a singular limit. 
\paragraph*{}
At first glance, there is a potential conflict between Wayne's and Butterfield's accounts, however, since Butterfield takes emergence as novelty, whereas Wayne believes we can have novel phenomena without there being emergence, provided we can explain it entirely in micro-level terms. The cases that Butterfield (2011a,b) examines, however, are all systems where a limit (not in all cases singular) is involved, and he shows that we cannot describe the behaviour in the limit in terms of the non-limit system (although Butterfield argues also that we can have a weaker, phenomenological, form of emergence \textit{before} the limit). Butterfield's examples, by taking emergence to occur in systems that involve a limit, and having the comparison class as the systems \textit{without} the limit having been taken, can be said to describe ``genuine'' novelty; this is particularly evident with the example in Butterfield and Bouatta (2011), of phase transitions.

\section{Conclusion}\label{sub:Uconcl}
In summary, I have examined four different examples of emergence in physics, each of which present us with slightly different bases for novelty and autonomy. This is shown in the Table \ref{table:emergence}, below.
\begin{table}[!h]
\centering
    \begin{tabular}{l|c c c c}

        ~                           & EFT   & RG fixed-points & SB & Hydrodynamics \\ \hline
        Autonomy                    & Quasi & Yes           & Yes               & Yes         \\ 
        Underdetermination          & Yes   & Yes             & Yes               & Yes           \\ 
        Universality                & Sometimes    & Yes             & Yes               & Yes           \\ 
                HOP & No    & No              & Yes               & No            \\
        Novelty           & S   & S/D       & D               & S            \\ 
    \end{tabular}
    \caption{Comparing four examples of emergence. SB~=~Symmetry breaking; HOP~=~Higher-organising principle; S = Synchonic novelty; D = Diachronic novelty.}
    \label{table:emergence}
\end{table}
\paragraph*{}
A few comments are required. In EFT, the quasi-autonomy of the micro- and macro-levels is due to the fact that the EFT breaks down as the energy at which the system is probed approaches the cutoff, indicating the effects of high-energy physics. The full autonomy of the micro- and macro-levels yielded by RG-fixed points and hydrodynamics is tied to the \textit{universality} that these frameworks describe. In the case of symmetry-breaking, the full autonomy of the micro- and macro-levels is due to the fact that it acts as a \textit{higher-organising principle} that transcends the micro-details (i.e. the low-energy physics depends on the symmetry-breaking pattern only): this is evidenced by the exact results that symmetry-breaking predicts.\footnote{The statement that ``Yes'' symmetry-breaking represents universality is made because the phenomena that symmetry-breaking describes is multiply-realised at the micro-level and is able to be described using the RG. Thus, the phenomena that symmetry-breaking describes accords with the definition of universality (p. \pageref{def:univ}), even though I have presented symmetry-breaking as distinct from the RG fixed-point as a basis for emergence.} In all examples we have \textit{underdetermination} of the high-energy physics given the low-energy physics, and this is associated with the \textit{autonomy} (quasi- or otherwise) of the micro- and macro-levels. 

\paragraph*{}
In regards to \textit{novelty}, I believe that \textit{in each of the cases} we have a plausible conception of novelty as robust behaviour exhibited by the macro-physics that is not exhibited by the micro-physics (in the case of thermodynamics/hydrodynamics, the micro-physics is taken to be the theory describing the details of the constituent particles, not statistical mechanics). This idea of novelty comes from the fact that the micro- and macro-theories, in all cases, are formally distinct, describing very different degrees of freedom, and are (relatively) autonomous. However, in some cases, for instance those involving critical phenomena, where a dynamical change occurs within the system, we might say we have \textit{diachronic novelty}---where the comparison class is no longer the micro-theory, but the system \textit{before} the dynamical change occurs. In the cases of EFT and hydrodynamics, the comparison is between the micro- and macro-levels at a given time, so the novelty is said to be \textit{synchronic}.
\paragraph*{}
For the RG fixed-point account, the novelty is listed as diachronic/synchronic because it is not always the case that an RG fixed-point corresponds to a critical point (infra-red fixed points often correspond to second-order phase transitions, whereas ultra-violet fixed points typically do not). In the cases where the RG fixed point is a critical point, there is basis for synchronic novelty, and in the other cases, there is diachronic novelty. Finally, there is  ``sometimes'' listed for universality in EFT because the underdetermination of the high-energy theory is not always physically manifest, as it is in the other cases (i.e. we do not always posses concrete examples of different micro-systems yielding the same EFT).

%% file: analogue.tex
\chapter{Spacetime as described by EFT}\label{sect:GREFT}

\section{Introduction}
This chapter examines the idea of treating GR as an EFT, and, drawing from the ideas presented in the previous three chapters, \S\ref{sect:Emergence}--\ref{sect:Univers}, explores what we might learn of emergent spacetime through the framework of EFT. In this chapter I consider examples of both top-down and bottom-up EFT; the former case is represented by analogue models of (and for) gravity, which describe spacetime (an effective curved geometry, to be more precise) as emergent from a condensed-matter system at high-energy. Although they have only recently attracted philosophical interest (thanks to \citet{Bain2008}), analogue models of gravity based on EFTs have a long history dating from the earliest days of general relativity and are successful in replicating much general relativistic phenomena and QFT in curved space (see \citet{Barcelo2011} for a review).\footnote{\label{foot:modern}The earliest instance the authors identify is Gordon's 1923 use of an effective gravitational metric field to mimic a dielectric medium. A later example is Unruh's ``Experimental black hole radiation'' \citeyearpar{Unruh1981}, which used an analogue model based on fluid flow to explore Hawking radiation from actual GR black holes. \citet[][p. 42]{Barcelo2011} present this example as being the start of what they call the ``modern era'' of analogue models.} 

\paragraph*{}
The bottom-up approach to treating gravity as an EFT, on the other hand, starts with GR at low-energy and aims to calculate the quantum corrections to the theory from the unknown high-energy physics. The only real obstacle to treating GR as we do other QFTs has been the non-renormalisability of gravity, however, the new conceptualisation of QFT as EFT resolves this difficulty. As should be clear from the discussion of EFT in \S\ref{sub:reconcutoff}, the non-renormalisability of the gravitational couplings is not a problem in the low-energy regime experimentally accessible to us.
\paragraph*{}
The success of these approaches suggests that the strong analogy between condensed matter physics and QFT naturally carries over to GR and cosmological phenomena. The analogy between condensed matter physics and QFT has usefully been employed in the past, and described as the ``cross-fertilisation'' of these disciplines \citep{Nambu2008}.  A well-known example of this is the idea of spontaneous symmetry breaking, as is the RG flow. It is tempting to move from the fact that both condensed matter physics and QFT can be described using the same theoretical apparatus to the conclusion that they share some other, more profound, physical similarities. Our desire to unify GR with the rest of fundamental physics, to treat gravity on par with the way we treat other fields, naturally leads us to attempt to incorporate it into the framework of QFT also. We are thus led toward an analogy between condensed matter physics and spacetime, one that is attractive given its potential to allow us insight into the universe at large by studying the universe at small (a sentiment expressed, for example, by \citet{Hu1988, Volovik2003, Zhang2004}).
\paragraph*{}
I argue that this analogy, however, owes its strength to the physical underpinning of EFT and the power of the RG, which, in turn, place strict limitations on how much we are entitled to draw from it. As should be clear from the previous chapters, the different senses of emergence in EFT, related to the ideas of (quasi-) autonomy and underdetermination, mean that there is very little that ties the low-energy EFT to the high-energy theory that underlies it: most of the details of the micro- theory are not relevant, apart from some particular symmetries and interactions. 
\paragraph*{}
Hence, drawing too strong an analogy between condensed matter systems and spacetime, by carrying over superfluous details, is liable to be dangerous or misleading---we are entitled to commitment to only some theoretical structures, and should remain agnostic about the rest. This accords with my previous argument, that we should understand EFT as effective---as being a pragmatic, heuristic way of speaking about the world. EFT allows us to make predictions at familiar scales without making assumptions about what happens at other scales.
\paragraph*{}
As stated, it is tempting to take the analogy between condensed matter physics and QFT as support for strong physical claims. The most obvious of these is the assertion, already encountered, that spacetime breaks down at some scale. This is a mistake. Although analogue models make concrete the analogy between condensed matter physics and spacetime, they fail to motivate the claim that spacetime breaks down---in fact, these models do not even support the claim that GR is an EFT, for reasons that will become clear in \S\ref{subsub:analogue?}. 
\paragraph*{}
Also, as already argued in \S\ref{sub:effectiveEFT} (and will be re-enforced by arguments in \S\ref{sect:Discrete}), an appeal to the analogy without reliance upon the analogue models will not help either. EFT counsels us to remain agnostic about the details of the high-energy physics. Our QFTs, understood in the framework of EFT, are not expected to hold to arbitrarily high-energies: they cease to be valid at some point, when the effects of unknown, high-energy physics become important. Although the breakdown of QFT in this sense may be the result of the breakdown of spacetime, it does not, on its own, motivate the claim that spacetime breaks down.
\paragraph*{}
The recovery of GR in the domain where we know GR to be applicable is one of the only generally agreed-upon criteria of acceptability for a good quantum gravity proposal. In spite of this, the recovery of GR is no indication of a theory's truth---if GR is an EFT, the low-energy degrees of freedom will probably be able to be realised by any of a number of different systems.  For this reason, trying to find a good candidate theory by working top-down toward known physics is possibly misguided. As other authors have recognised, the bottom-up approach is the better one, for pragmatic reasons, i.e. it is systematic and intended simply to produce testable results within experimentally accessible energy ranges \citep{Georgi1993, Hartmann2001}. The bottom-up approach from GR, described in \S\ref{sub:GREFT}, treats GR in the same way we treat other QFTs, attempting to quantify the higher-order corrections that result from neglected high-energy physics. The aim of the bottom-up approach is not to find an elegant high-energy theory underlying GR, but rather just to reproduce the predictions such a theory would make at low energy.
\paragraph*{}
The structure of this chapter is as follows. In \S\ref{sub:analogue}, I consider analogue models of spacetime as an example of the top-down approach to EFT. These models illustrate the conception of emergence in EFT outlined in the previous chapters. Interestingly, these models provide us with emergent spacetime, rather than emergent GR. I also argue that we should be wary of drawing too much from the analogy between condensed matter physics and QFT. In \S\ref{sub:GREFT} I consider two different examples of the bottom-up approach to GR as an EFT. I again argue that, due to the conception of emergence suggested by EFT, we are restricted in how much we can draw from these theories. Finally, in \S\ref{sub:asygrav} I outline the asymptotic safety scenario, which is an important conjecture that comes from treating GR in the same way we treat other QFTs. The suggestion, made by \citet{Weinberg1979, Weinberg2009}, is that the couplings for gravity approach a fixed point at high-energy, in a similar way to QCD. 

\section{Top-down: Analogue models of (and for) gravity}\label{sub:analogue}
Modern\footnote{See Footnote \ref{foot:modern}.} analogue models of spacetime begin with a quantum fluid (such as a Bose-Einstein condensate) and use an EFT to describe the behaviour of the quasiparticles (phonons) that emerge as low-energy collective excitations when this system is probed with a small amount of energy. The simple conceptual picture is to imagine the quasiparticles floating on top of the underlying condensate (i.e. the quasiparticles possess additional degrees of freedom to the particles in the condensate). The quasiparticles are subject to an effective curved-space metric, meaning they behave as though they ``exist in'' curved spacetime, oblivious to the underlying (flat) surface of the condensate. As energy is increased, however, the quasiparticles eventually have short enough wavelength to detect the discrete particles of the condensate, and the EFT ceases to be valid.
\paragraph*{}
Bain (2008) presents a simple example of relativistic spacetime emergent from a Bose-Einstein condensate (BEC) of particle density $\rho$ and coherent phase $\theta$. In constructing the analogue model, these variables are linearly expanded about their ground state values, $\rho = \rho_0 + \delta\rho, \theta = \theta_0 + \delta\theta$, where $\delta\rho$ and $\delta\theta$ represent fluctuations in density and phase above the ground state. These variables are then substituted into the Lagrangian describing the BEC, and the high-energy fluctuations are identified and ``integrated out'' so that only the low-energy interactions are included in the theory. The result is, schematically, a sum of two terms:
\begin{equation}\label{eq:Leff}
\Lagr = \Lagr_0[\rho,\theta] + \Lagr_{eff}[\delta\theta]
\end{equation}
where $\Lagr_0$ is the Lagrangian describing the ground state of the BEC and $\Lagr_{eff}$ is the effective Lagrangian describing the low-energy fluctuations above the ground state. $\Lagr_{eff}$ is formally identical to the Lagrangian that describes a massless scalar field in (3+1)-dimension spacetime, and the curved effective metric depends on the velocity, $v_i$ of the underlying superfluid.
\paragraph*{}
As Bain (2013) points out, given the substantial difference between $\Lagr_0$ and $\Lagr_{eff}$---the former being non-relativistic, the latter relativistic---we can treat the original Lagrangian and the effective Lagrangian as describing two different theories. The analogue models show us that emergent Lorentz invariance is incredibly easy to obtain from a variety of different systems; the high-energy theory is severely underdetermined. \citet{Barcelo2001E} have demonstrated that an effective curved spacetime is a generic feature of the linearisation process used in constructing the analogue models. All that is needed is a Lagrangian, $\Lagr(\partial_\varphi,\varphi)$, depending on a single scalar field, $\varphi(t,\textbf{x})$, and its first derivatives.

\subsection{Gravity in superfluid superfluid $\tensor*[^{3}]{\mathrm{He}}{}-A$}
Another interesting analogue model is Volovik's \citeyearpar{Volovik2003, Volovik2001} example in which gravity as well as the standard model of particle physics are emergent from superfluid helium 3-A.\footnote{Since this thesis is concerned with emergent spacetime, I will not Volovik's model's replication of the standard model in any detail.} Being fermions, the $\tensor*[^{3}]{\mathrm{He}}{}$ atoms must form pairs in order to condense as a BEC. These bosonic pairs are similar to the Cooper pairs of electrons described by the BCS model of superconductivity (\S\ref{sub:tower b-u}, \ref{subsub:symmetrybreaking}), except that the $\tensor*[^{3}]{\mathrm{He}}{}$ Cooper pairs have additional spin and orbital angular momentum degrees of freedom, and this allows for a number of distinct superfluid phases.\footnote{In particular, the \textit{A-}phase of $\tensor*[^{3}]{\mathrm{He}}{}$ is characterised by pairs of $\tensor*[^{3}]{\mathrm{He}}{}$ atoms spinning about anti-parallel axes that are perpendicular to the plane of their orbit. See Volovik (2003) or the short review in Bain (2008).}
\paragraph*{}
The non-superfluid $\tensor*[^{3}]{\mathrm{He}}{}$ liquid (and, at higher temperatures, gas) phase possesses all the symmetries possible of ordinary condensed matter systems: translational invariance, global $U(1)$ group, and two global $SO(3)$ symmetries, of spin and orbital angular momentum. \citet[][p. 3]{Volovik2003} calls this $U(1)\times SO(3) \times SO(3)$ the analogue of the ``Grand Unification'' group (although, of course, the actual Grand Unification group in particle physics is supposed to be much larger). Decreasing the temperature, to the critical value, $T_c$ (around 1mK), results in the $\tensor*[^{3}]{\mathrm{He}}{}$ becoming superfluid. At this point, the analogue ``Grand Unification'' symmetry breaks, and the only symmetry the system possesses is translational invariance (being a liquid). Decreasing the temperature even further, however (approaching 0K), the $\tensor*[^{3}]{\mathrm{He}}{}$ acquires new symmetries, including an analogue of Lorentz invariance, local gauge invariance, and elements of general covariance.
\paragraph*{}
Volovik (2003) explains that the appearance of these symmetries at low-energy owes to the \textit{universality class} of the Fermi liquid, $\tensor*[^{3}]{\mathrm{He}}{}$. At low-energy, \textit{any} condensed matter system in this universality class will describe chiral (left- and right-handed) fermions as quasiparticles and gauge bosons as collective modes. The universality class is determined by the topology of the quasiparticle energy spectrum in momentum space, where the quasiparticle energy spectrum is obtained by diagonalising the Hamiltonian that describes the $\tensor*[^{3}]{\mathrm{He}}{}$ Cooper pairs. This Hamiltonian takes the schematic form,
\begin{equation}\label{eq:He3A}
H_{\tensor*[^{3}]{\mathrm{He-A}}{}}=\chi^\dagger\mathcal{H}\chi , \mathcal{H}=\sigma^bg_b(\textbf{p}) , b=1,2,3
\end{equation}
where $\chi$ and $\chi^\dagger$ are non-relativistic 2-spinors that encode creation and annihilation operators for $\tensor*[^{3}]{\mathrm{He}}{}$ atoms, $\sigma^b$ are Pauli matrices, and $g_b$ are three-functions of momentum that encode the kinetic energy and interaction potential for $\tensor*[^{3}]{\mathrm{He-A}}{}$ Cooper pairs. (\ref{eq:He3A}) is essentially the standard BCS Hamiltonian, but modified to account for the extra degrees of freedom of the $\tensor*[^{3}]{\mathrm{He}}{}$ Cooper pairs.\footnote{For details see Volovik (2003, pp. 82, 96). This summary follows Bain (2008, pp. 309--311).} 

\paragraph*{}
The energy spectrum in momentum space vanishes at two points, known as \textit{Fermi points}, which may be represented as, $p^{(a)}_i, i=1,2,3, a=1,2$. The Fermi points arise via a symmetry-breaking process, and are stable features of the system in the sense that small perturbations will not remove them. Because the Fermi points define topologically-stable singularities in the one-particle Feynman propagator, $\mathcal{G}=(ip_0-\mathcal{H})^{-1}$, their existence is protected by the topology. The quasiparticle energy spectrum is given by the poles in the propagator,
\begin{equation}\label{eq:spectrum}
g^{\mu \nu}(p_\mu-p^{(a)}_\mu)(p_\nu-p^{(a)}_\nu)=0
\end{equation}
where $g^{\mu \nu}=\eta^{bc}e^\mu_b e^\nu_c$ and $\eta^{bc}=\mathrm{diag}(-1,1,1,1)$.
\paragraph*{}
While the existence of the Fermi points is insensitive to small perturbations of the system, however, their positions in the energy spectrum can change as a result of such perturbations. The positions of the Fermi points are given by the values of $p^{(a)}_\mu$. For a bosonic quasiparticle (a collective mode of the ``fermionic vacuum'' represented by the underlying $\tensor*[^{3}]{\mathrm{He}}{}$ system), the motion that shifts the position of the Fermi point corresponds to the gauge field \textbf{A}. The small perturbation can also change the slope of the curve of the energy spectrum in momentum space, and this forms the metric tensor field, $g^{\mu\nu}$ (Volovik, 2003, p. 100). The Lagrangian density corresponding to the energy spectrum (\ref{eq:spectrum}) can be written as,
\begin{equation}\label{eq:Lquasi}
\Lagr'_{\tensor*[^{3}]{\mathrm{He-A}}{}}=\bar{\varPsi}\gamma^\mu(\partial_\mu-q^{(a)}A_\mu)\varPsi
\end{equation}
Where$\gamma^\mu=g^{\mu\nu}(\sigma_\nu \otimes \sigma_3)$ are Dirac $\gamma$-matrices, the $\varPsi$'s are relativistic Dirac 4-spinors (constructed from the pairs of 2-spinors in (\ref{eq:He3A})) and $q^{(a)}A_\mu=p^{(a)}_\mu$. This Lagrangian describes massless Dirac fermions interacting with a 4-vector potential $A_\mu$ in a curved Lorentzian spacetime with metric $g_{\mu\nu}$.
\paragraph*{}
The topology of the energy spectrum (\ref{eq:spectrum}) in momentum space determines a \textit{universality class} that essentially characterises the type of EFT that describes the system at low-energy. As Volovik (pp. 99--100) explains, systems with elementary Fermi points (those with topological charge $N_3=+1$ or $N_3=-1$) have the remarkable property that Lorentz invariance always emerges at low-energy, even if the system itself is non-relativistic. Thus, in the vicinity of the Fermi point, the massless quasiparticles are always subject to $g_{\mu\nu}$. While the micro-details of the underlying system---for instance, the superfluid velocity and density---play a role in specifying the energy spectrum, these details are lost in the hydrodynamic limit where the EFT completely describes the low-energy physics. The EFT is characterised by the universality class, which itself depends only on symmetry and topology (Volovik, 2003, p. 5).
\paragraph*{}
Because of the universality of the low-energy theory, we are unable to reconstruct the micro-structure of the underlying condensed matter system from the low-energy collective modes (for example, we cannot reconstruct the atomic structure of a crystal from its low-energy acoustic waves because all crystals have similar acoustic waves describe by the same equations of the same EFT). Quantising the low-energy collective modes produces phonons, not atoms; in other words, the QFT produced by quantising the classical effective fields is still an EFT, and does not provide information on the high-energy theory, except for its symmetry class. What is important for $\Lagr'$, describing the effective dynamics, as well as the low-energy properties and degrees of freedom, is not the details of the micro-physics, but only the symmetry and topology of the condensed matter system (Volovik, 2003, pp. 6-7). This is essentially the example of superfluidity presented earlier (\S\ref{subsub:symmetrybreaking}).
\paragraph*{}
In order to produce the Einstein-Hilbert Lagrangian of GR, Volovik follows an approach similar to that of Sakharov's \citeyearpar{Sakharov1967} ``induced gravity'' proposal, in which the Lagrangian density (\ref{eq:Lquasi}) is expanded in small fluctuations in the effective metric $g_{\mu\nu}$ about the ground state and then the high-energy terms are integrated out. Unfortunately, in the case of the $\tensor*[^{3}]{\mathrm{He-A}}{}$ effective metric, the result contains higher-order terms dependent on the superfluid velocity, $v_i$, and these dominate the Einstein-Hilbert term. This is a consequence of the fact that the Fermi points arise from a spontaneously broken symmetry.
\paragraph*{}
In order to reproduce the Einstein-Hilbert action, the effects of the broken symmetry must somehow be suppressed (Volovik, 2003, pp. 8, 113). Because the superfluid velocity is inversely proportional to mass, Volovik (pp. 130-132) considers the limit in which the mass of the $\tensor*[^{3}]{\mathrm{He-A}}{}$ atoms goes to infinity, and thus $v_i\rightarrow 0$. In such a system the terms dependent on $v_i$ are suppressed and the Einstein-Hilbert action recovered. Unfortunately, such a system does not represent a superfluid. Volovik thus states that the physical vacuum cannot be completely modelled by a superfluid---a conclusion reasserted by Bain (2008).

\subsection{The quantisation of gravity}\label{subsub:qmisguided}
The major implication of spacetime emergent through EFT is that any attempt to construct a quantum theory of gravity by quantising some aspect of general relativity is misguided \citep{Barcelo2001E, Hu2009, Visser2008, Volovik2003}. If spacetime is emergent in this way, quantising it will not help us identify the fundamental (i.e. high-energy) degrees of freedom---by analogy, we would arrive at a theory of phonons rather than a description of the underlying atoms of the condensate. The typical sentiment is expressed by Visser,
\begin{quotation}
There is a possibility that spacetime itself is ultimately an emergent phenomenon, a near-universal ``low-energy long-distance approximation'', similar to the way in which fluid mechanics is the near-universal low-energy long-distance approximation to quantum molecular dynamics. If so, then direct attempts to quantize spacetime are misguided---at least as far as fundamental physics is concerned. In particular, this implies that we may have totally mis-identified the fundamental degrees of freedom that need to be quantized, and even the fundamental nature of the spacetime arena in which the physics takes place. (Visser, 2008, p. 1)
\end{quotation}
If programs involving the quantisation of the metric tensor produce theories of particles analogous to phonons, then it is unsurprising that they should break down at high-energy. Their breakdown can motivate the search for a high-energy theory beyond GR, but we cannot say that the degrees of freedom of the high-energy theory would themselves need to be quantised in order to produce a theory of quantum gravity.

\subsection{Analogue models of gravity?}\label{subsub:analogue?}
Notice that the analogue models of gravity do not actually contain ``gravity''. These models produce an effective Lorentzian curved spacetime geometry, but not the Einstein field equations. For this reason, \citet[][p. 799]{Barcelo2001E} state that we have analogue models \textit{of} general relativity rather than \textit{for} general relativity. The conceptual picture we arrive at is an unusual one: we are used to obtaining the metric as a solution to the Einstein equations, which, in turn, are supposed to describe the dynamics of GR. Instead, in this picture we obtain the metric field as part of an EFT from an underlying condensate which itself defines an approximately flat (non-relativistic) spatiotemporal structure.
\paragraph*{}
Although it sounds strange, it is perhaps possible that we could effectively model a background independent theory (within a certain low-energy range) using a background dependent one (just as, for example, we are able to effectively model a discrete system at low-energy as a continuous one). What is important here is just that the emergent, low-energy physics is able to be treated independently of a background spacetime \textit{in a particular low-energy regime}---meaning that, at low-energies, the theory appears background independent, but may be revealed as background dependent at higher-energy scales. There are some attempts along these lines, for instance Barcelo, Visser and Liberati's (2001) demonstration that something suggestive of an effective dynamics can be produced by the inclusion of one-loop quantum effects, along the lines of Sakharov's (1967) ``induced gravity'' proposal, and recent work by \citet{Sindoni2009, Sindoni2011}.
\paragraph*{}
Bain (2008), however, argues that the analogue models fail to replicate not just the dynamical, but also the kinematical aspects of GR. The analogue models have Lorentzian spacetime emerge from a prior spacetime structure and depend, in some way, on the properties of the background structure (for instance, the velocity dependence of $\Lagr_{eff}$ in (\ref{eq:Leff})). Bain (p. 308) thus claims that insofar as general solutions to the Einstein equations are background independent, they will not be modelled effectively by analogue models that are background dependent. Bain does not explain specifically what idea of background independence he has in mind, but it certainly is not obvious that the dependence of the effective theory on some aspect of the underlying system would preclude us from effectively modelling general solutions to the Einstein field equations.\footnote{The idea of background independence is discussed further in the next chapter, \S\ref{sub:BI}.}
\paragraph*{}
Further, Bain (2008, p. 308) claims that, to the extent that the Einstein equations are diffeomorphism invariant, they will not be modelled effectively by an analogue spacetime, insofar as the EFT of the latter is not diffeomorphism invariant. The background geometry of the condensed matter system provides a privileged coordinate frame, so it is natural to suspect that diffeomorphism invariance is not preserved. However, just as we could potentially model a background independent theory effectively, so too we could potentially model a diffeomorphism invariant theory effectively. 
\paragraph*{}
\citet[][p. 105]{Barcelo2011} point out that active diffeomorphism invariance is maintained for a low-energy observer ``within'' the system, (i.e. an observer who can only perform low-energy experiments involving the propagation of the relativistic collective fields). Invariance under active diffeomorphisms is equivalent to the claim that there is no prior geometry, or that the prior geometry is undetectable. In this case the prior structure is undetectable to an internal observer, and so, in this sense, diffeomorphism invariance is effectively maintained at low-energy scales (even though, at high-energies, the theory is revealed as not diffeomorphism invariant).
\paragraph*{}
One difficulty with interpreting the analogue models, however, is that if we are to accept that they give us emergent spacetime, we must identify spacetime with the Lorentzian metric structure. If we have some other conception of spacetime, for instance, an equivalence class of diffeomorphism invariant four-geometries, the analogue models fail to give us emergent spacetime.

\subsection{Emergence}\label{subsub:AEmergence}
The sense in which the EFT describing spacetime in the analogue models is autonomous from the micro-theory of the condensed matter system relates both to the conception of autonomy relevant to EFT more generally (\S\ref{sub:emergence}), as well as the conception associated with hydrodynamics (\S\ref{subsub:hydro}). As is typical of EFT, the low-energy theory depends on very little of the high-energy theory, and so the high-energy theory is underdetermined by the low-energy physics. Furthermore, however, in models such as Volovik's $\tensor*[^{3}]{\mathrm{He-A}}{}$, it is only the symmetry and topology of the high-energy system that is important in determining the low-energy physics. These determine a universality class, and any system within this universality class will exemplify the same low-energy physics. For those models where this is the case, we can say that the EFT describing spacetime does not depend on the high-energy theory at all.
\paragraph*{}
The strength of the condensed matter approaches to quantum gravity is that they are able to demonstrate the limitations of any quantum gravity theory that conceives of GR as an EFT. As Volovik (2003, p. 7) states, because we are familiar with the condensed matter structure at many different scales (including the inter-atomic spacing, which is taken to be analogous to the Planck length in quantum gravity), the condensed matter approaches to quantum gravity may help indicate which quantities in quantum gravity are able to be calculated within EFT, and which quantities depend essentially on the details of the trans-Planckian physics.
\paragraph*{}
Conceiving of these approaches heuristically, while remaining conscious of their limitations, accords with the philosophy of EFT more generally, as described in \S\ref{sub:effectiveEFT}. In particular, we should be cautious in speaking of specific properties of BECs as though they are necessarily required for emergent spacetime. We are thus warned against following \citet{Hu2005}, for example, who is willing to bite the bullet and accept any ``radical conclusions'' that result from pushing the analogy between condensed matter physics and spacetime.\footnote{These include considering the Planck temperature (1032 K) as ``low temperature'', given that BECs only exist at very low temperatures and spacetime is supposed to exist at this temperature in the early universe, for example.} Instead, when we refer to the underlying ``condensate'' we should do so in a symbolic sense, taking it to refer to whatever (unknown) entity it is that possesses the relevant symmetries and mathematical structure.

\section{Bottom-up: GR as an EFT}\label{sub:GREFT}
The bottom-up approach to GR as an EFT is a highly pragmatic exercise. Instead of worrying about what happens at high-energies, people engaged in this program attempt to quantify the effects of the unknown physics upon GR at experimentally accessible energies. In this case, the EFT framework is embraced in the spirit described in \S\ref{sub:effectiveEFT}---it is not in competition with those approaches to quantum gravity that seek a final theory or new physics, indeed, it could assist such searches by providing quantitative predictions that any other quantum gravity approaches would be expected to reproduce. On the other hand, however, it may well be that we continue with the EFT approach, and that new physics is not found (again, as stated in \S\ref{subsub:meth anti}). This latter suggestion gains support from the observation, yielded by treating GR as an EFT and calculating the quantum corrections, that GR shows no signs of breakdown as far as we can see: the quantum corrections are small, and there is no (urgent) problem of quantum gravity in this respect.
\paragraph*{}
Similarly, Weinberg (2009) has suggested that there may not be new physics beyond the standard model and GR; it is possible that the appropriate high-energy degrees of freedom just are the metric and matter fields, including those of the standard model, and, in this case, there is no ``underlying theory'' (this will be discussed shortly \S\ref{sub:asygrav}). The point that I wish to emphasise here is that the bottom-up approach to GR as an EFT simply means remaining open-minded in regards to physics at high-energies, as \citet[][p. 218]{Donoghue1997} states,``We have no reason to suspect that the effects of our present theory are the whole story at the highest energies. Effective field theory allows us to make predictions at present energies without making unwarranted assumptions about what is going on at high energies.''
\paragraph*{}
The main problem with treating GR in the same way we treat other QFTs has been the non-renormalisability of gravity, that is, there is no renormalisable theory of the metric tensor that is invariant under general coordinate transformations.\footnote{The gravitational coupling constant (Newton's constant), $G$, has mass dimension -2 in units where $\hbar=c=1$, recalling from footnote (\ref{foot:renorm}) this means that we expect conventional perturbative QFT to be applicable only for energies $E^2\ll 1/G$.} However, as \citet{Burgess2004} and \citet{Donoghue1994, Donoghue1997} argue, the non-renormalisability is not actually a problem at the low-energies we are familiar with, thanks to the framework of EFT. Recall from \S\ref{subsub:EFTformalism}, that at low-energy the non-renormalisable interactions are highly suppressed. Hence, we are not prevented from making meaningful predictions; rather, predictions in this range are well-controlled due to the heavy mass in the low-energy expansion $E/M$. Choice of the heavy mass $M$ is dependent on the situation being studied. Although we might usually expect to use the Planck mass, $m_P = (\hbar c/G)^{1/2}$ (with $G$ being Newton's gravitational constant), in other cases it is more appropriate to choose a different scale (Burgess, 2004, p. 6).
\paragraph*{}
The low-energy Lagrangian, $\Lagr_{eff}$, consists of a sum of all possible interactions which are consistent with the symmetries of the GR (general covariance and local Lorentz invariance). The Einstein-Hilbert Lagrangian, $\sqrt{g}R$ (where $g = det(g_{\mu\nu})$ is the determinant of the metric tensor, and $R$ is the Ricci scalar) appears simply as the first (i.e. least suppressed) term in the expansion $\Lagr_{eff}$. Although a complete quantitative analysis of the size of quantum corrections remains a work in progress, the leading and next-to-leading quantum corrections can be calculated, and have been shown to be negligible. This is as we expect, given the success of GR in its familiar applications.

\subsection{Kinetic theory approach}
Another, distinct, bottom-up approach is Hu's kinetic theory \citep{Hu2002}. Although this theory is based on a strong analogy between condensed matter physics and spacetime, Hu explains his idea of GR emergent from a condensate by analogy with hydrodynamics as emergent from molecular dynamics: the metric and connection forms are, according to Hu, hydrodynamic variables. Rather than treating GR as an effective quantum field theory, the kinetic approach takes semiclassical gravity as its starting point (i.e. the coupling of the classical spacetime metric with the expectation value of the stress-energy tensor, where this tensor represents quantum matter fields).
\paragraph*{}
Hu’s stochastic gravity is the ``next level up'' from semiclassical gravity and involves the two-point function of the stress-energy tensor \citep{Hu1999, Hu2008}. The kinetic approach then builds on this, being a hierarchy, or ``staircase'', of equations which take into account the higher correlations of the stress-energy tensor and describe their effect on the higher-order induced fluctuations of the metric. \citet[][p. 393]{Mattingly2009} describes how this relates to the high-energy theory from which GR emerges, ``We know that any good [quantum gravity] theory will have to give correctly the correlations between the quantum fluctuations of matter at every order, and we hope that by accounting for these correlations by hand some new important insights will be found into the nature of that underlying theory.''

\subsection{Emergence}
The relation of emergence between GR and the high-energy theory beyond would mean that the only ``insights'' we can find using bottom-up approaches will be in the form of approximate, quantitative predictions. Using the EFT framework to work bottom-up, we are restricted by the availability of experimental input required to set the parameters of the theory. Also, there is the risk that perhaps the assumptions of the framework (e.g. the existence of well-separated heavy mass scales) are not fulfilled at high-energies. The bottom-up approach, in both examples, sustains no illusions, however: it is explicitly heuristic. It is not aimed at producing an elegant final theory, but is just a means of combining GR and QM to make predictions in the regimes where we are able to.
\paragraph*{}
The physical and conceptual aspects of the theories in the bottom-up approach are not supposed to resemble those of the underlying theory, so we should not be concerned if, for example, these theories neglect background independence. They are not themselves appropriate candidates for a quantum theory of gravity. As Mattingly (2009) and Burgess (2004) both express, the philosophy of the bottom-up approach entails a recognition that quantum gravity is not an immediate issue. Using the ideas of EFT, we are able to press on, slowly inching our way up until we hit crisis point.

\section{Asymptotic safety in quantum gravity}\label{sub:asygrav}
Treating GR in the bottom-up EFT framework, the effective action, $S_{eff}$, may be expressed as,\footnote{Following Weinberg (2009).}
\begin{multline}
S_{eff}=-\int d^4x\sqrt[]{-det g}[f_0(\Lambda)+f_1(\Lambda)R\\
+f_{2a}R^2+f_{2b}R^{(\mu\nu)}R_{(\mu\nu)}\\
+f_{3a}(\Lambda)R^3+\cdots],
\end{multline}
where $\Lambda$ is the ultraviolet cutoff, and the $f_n(\Lambda)$ are coupling parameters with a cutoff dependence chosen so that physical quantities are cutoff-independent. We can replace these couplings with dimensionless parameters $g_n(\Lambda)$,
\begin{equation}
g_0\equiv \Lambda^{-4}f_0; g_1\equiv\Lambda^{-2}f_1;g_{2a}\equiv f_{2a};
g_{2b}\equiv f_{2b}; g_{3a}\equiv\Lambda^2f_{3a};\cdots .
\end{equation}
Because these parameters are dimensionless, they must satisfy a RG equation of the form,
\begin{equation}
\Lambda \frac{d}{d\Lambda}g_n(\Lambda)=\beta_n(g(\Lambda))
\end{equation}
\paragraph*{}
In perturbation theory, all but a finite number of the $g_n(\Lambda)$ diverge as $\Lambda \rightarrow \infty$. Thus, we are apparently prevented from calculating anything at high-energy. As mentioned earlier (\S\ref{subsub:effQFT}), this proliferation of infinities at high-energies is typically taken to presage the ultimate failure of our theory in this regime. It is usually assumed that, when $\Lambda$ reaches some very high energy, that new physics will come into play: the appropriate high-energy degrees of freedom are not the metric and Standard Model fields. However, as \citet{Wuthrich2012} points out, ``these difficulties---at least in general relativity---simply result from insisting on forcing general relativity on the Procrustean bed of perturbation theory''.\footnote{This sentiment is also expressed by many proponents of so-called ``discrete'' approaches to quantum gravity, as discussed in the next chapter.} Weinberg (1979, 2009) has proposed that perhaps the couplings do not actually blow up at high-energy, but rather that they are attracted to a finite value $g_(n*)$, i.e. that they approach a UV fixed point. The suggestion is, thus, that gravity is \textit{asymptotically safe}, indicating that the physical quantities are ``safe'' from divergences as the cutoff is removed (taken to infinity).\footnote{This is similar to QCD, except that QCD is also asymptotically free, having a fixed point of zero; usually, in asymptotic safety, the fixed point is finite, but not zero.}
\paragraph*{}
The research programme focused on exploring this so-called ``asymptotic safety scenario in quantum gravity'' aims to place quantum gravity within the framework of known physics principles (this includes treating GR as a QFT, as above \S\ref{sub:GREFT}), so that we may use these familiar principles to explore the behaviour of the theory. Although doing this involves using our familiar low-energy degrees of freedom (metric and matter fields on a continuous. four-dimensional manifold), it is not (and needn't be) presupposed that these low-energy degrees of freedom will be appropriate at high-energy.\footnote{And so accords with the philosophy of ``effective EFT''.} 
\paragraph*{}
Instead, the strategy focuses on ``backtracking'', using the RG, toward the high-energy ``origin'' of these degrees of freedom \citep[][p. 5]{Niedermaier2006}. In particular, it is not supposed that the Einstein-Hilbert action (or a discretised form of the Einstein-Hilbert action) is the appropriate micro-action. However, unless the theory becomes purely topological at some scale, a metric will always be involved, and, from general covariance arguments, it will almost unavoidably contain an Einstein-Hilbert action. For this reason it might be thought that the Einstein-Hilbert action will play a role in the high-energy limit \citep{Percacci2009}.
\paragraph*{}
The arena on which the RG is applied is a space of actions; a typical action has the form $\sum_\alpha u_\alpha P_\alpha$, where $P_\alpha$ represent the interactions, and $u_\alpha$ are scale-dependent coefficients, or couplings.\footnote{This explanation is based on \citet{Niedermaier2006, Percacci2009}.} Using the Wilson-Kadanoff approach, the RG may be understood as a sequence of coarse-graining operations (as explained in \S\ref{sub:Blockspin}). It is stipulated that the dominant effect of the interactions in the extreme UV is \textit{antiscreening}, so that, by analogy with QCD (\S\ref{subsub:fixed points}) the interactions become weaker at high-energy. The RG thus flows toward a fixed point on the critical surface, and, recalling the description in \S\ref{subsect:criticalphen}, relevant couplings are repelled from the fixed point, while irrelevant ones flow, under the RG coarse-graining operation, towards it. The flow lines, known as renormalisation group \textit{trajectories}, emanating from the fixed point, sweep out a manifold that is defined as the \textit{unstable manifold}.
\paragraph*{}
The points on these flow lines correspond to actions from which we are able to obtain a \textit{continuum limit}: this is the limit in which we are able to calculate physical quantities that are strictly independent of the high-energy cutoff, independent of the form of the coarse-graining (RG) operation, and invariant under point transformations of the fields. That is, we have continuum properties even in the presence of a high-energy cutoff (which might otherwise be interpreted as ``discretising'' spacetime, as in \S\ref{subsub:effQFT}). The continuum limit represents a \textit{universality class} of scaling limits that give us continuum quantities, where a scaling limit is constructed by `backtracking' along an RG trajectory emanating from the fixed point. Any action on an RG trajectory describes identically the same physics on all energy scales lower than the one where it is defined. Because of this, if we follow the trajectory back (almost) into the fixed point, we can in principle extract unambiguous answers for physical quantities on all energy scales. Thus, the presence of the fixed point guarantees universality.
\paragraph*{}
As Niedermaier and Reuter (2006, p. 7) explain, the main drive for the asymptotic safety approach to quantum gravity is its potential for being able to ``propagate down'' the strongly-suppressed effects of the high-energy physics through many orders of magnitude toward experimentally accessible energies. Because of the \textit{universality} secured by the presence of the fixed point, the asymptotic safety approach to quantum gravity is not concerned with identifying the nature of the ``fundamental'' (i.e. high-energy) degrees of freedom: it is only the universality class that matters. Within this picture, Niedermaier and Reuter (2006, p. 7) state, even sets of fields or other variables that are non-locally and non-linearly related to one another may describe the same universality class, and, hence, the same physics.
\paragraph*{}
This leads us to the sense of \textit{emergence} appropriate to the asymptotic safety scenario. By direct comparison with the idea of emergence associated with universality and fixed points in \S\ref{sect:Univers}, we may tie our conception of emergence to the underdetermination of the high-energy degrees of freedom. The asymptotic safety scenario demonstrates is that the different choices of micro-action and dynamical micro-variables all lead to the same low-energy physics. In other words, the low-energy degrees of freedom, including the GR metric and Standard Model matter fields, are autonomous from their high-energy counterparts, dependent only on the universality class.
\paragraph{}
As in other examples of EFT, the novelty of the low-energy theory in the asymptotic safety scenario is expected to be synchronic. However, there may also be diachronic novelty, if the fixed point is associated with second-order phase transitions (recall the discussion in \S\ref{sect:Univers}, particularly the summary in \S\ref{sub:Uconcl}). In such a case the fixed point would represent a dynamical change of state of the universe, and the low-energy degrees of freedom could be seen as emergent compared to the universe before the phase transition. As will be discussed in the next chapter, there is some evidence, coming from the use of Regge calculus and causal dynamical triangulations (\S\ref{sub:CDT}), that the fixed point may indeed correspond to a second-order phase transition.
\paragraph*{}
Evidence for the existence of a UV fixed point in quantum gravity has come from calculations based on a number of different approximation techniques (for a brief overview, see \citet{Percacci2009}). These include the $2+\epsilon$ expansion \citep{Weinberg1979, Kawai1993, Kawai1996, Niedermaier2003, Niedermaier2010}, the $1/N$ approximation \citep{Smolin1982, Percacci2003, Percacci2006}, lattice methods \citep{Ambjorn2004, Ambjorn2005}, and the truncated exact renormalisation group equations (ERGE) \citep{Reuter2002, Reuter2009, Lauscher2002, Codello2006}.\footnote{For an extended list of references see \citet{Weinberg2009}.}

\section{Conclusion}
The analogue models of spacetime and the treatment of GR as an EFT represent two different directions in describing spacetime as an EFT. Both approaches exemplify the philosophy of EFT espoused in \S\ref{sub:effectiveEFT}: that EFT is itself an effective, pragmatic description of physics. The top-down approaches considered here serve to demonstrate the limitations on any approach to quantum gravity that conceives of GR as an EFT.\footnote{However, recall from discussion in the Introduction, \S\ref{sub:chal}, that there is more to recovering GR than simply an emergent metric and an effective dynamics for spacetime, and it is not clear that these models are capable of representing these extra features, see \citet{Carlip2012}.} I argued that these limitations are essentially tied to the conception of emergence (novelty and autonomy) appropriate to these models.
\paragraph*{}
Two types of analogue models were examined: the general case of an effective Lorentzian metric arising from the linearisation of a field theory around a non-trivial background, and Volovik's model in which a dynamical curved metric arises at low-energy in a condensed matter system of a particular universality class. The spacetime that arises at low-energy in both these models is strongly robust and autonomous from the high-energy physics, owing to the fact that the emergent spacetime depends only on the symmetries and general features (in the case of Volovik's model, the topology) of the condensed matter system rather than on any particular micro-details.
\paragraph*{}
Coming from the other direction, the bottom-up approaches to GR as an EFT treat gravity as if it were a QFT, and are valuable in that any and all testable predictions of quantum gravity can be calculated in this framework. These approaches tell us that quantum gravity is not an immediate concern, and also that new physics is not required at any energy scale accessible to experiment. The asymptotic safety scenario for gravity draws an analogy between GR and QFTs, and is the claim that we do not need new physics at \textit{any} energy scale in order to describe quantum gravity. If the fixed point postulated by the asymptotic safety scenario represents a second-order phase transition, then the situation is one in which spacetime as an EFT is diachronically novel as well as strongly autonomous from the high-energy physics underlying it. This suggestion is realised by several of the discrete approaches to quantum gravity, to be discussed in the next chapter.

%% file: discrete.tex
\chapter{Discrete approaches to quantum gravity}\label{sect:Discrete}
\section{Introduction}
The approaches considered in this chapter all describe discrete fundamental (i.e. theoretically basic) entities at high-energy and emphasise the importance of the notion of background independence. These \textit{discrete approaches} are radically different from the condensed matter physics approaches described in the previous chapter, and, indeed, many of the approaches that come under the heading of ``discrete'' are radically different from one another.

\paragraph*{}
There are various approaches to QG which might be called ``discrete''. The first of these grew out of  Regge calculus \citep{Regge1961}, which is a means of modelling spacetime in GR (i.e. solutions to the Einstein field equations), using discrete elements called \textit{simplices} (which will be described shortly, in \S\ref{sub:CDT}). \citet{Loll1998} provides a review of its application to quantum gravity. One of the several approaches which stemmed from Regge calculus is dynamical triangulations, problems with which led to the development of another theory, called \textit{causal dynamical triangulations} (discussed in \S\ref{sub:CDT}).\footnote{Some of the other discrete approaches are summarised in \citet{Williams2006}.} Most of the approaches that utilise, or stem from, Regge calculus involve taking a sum-over-histories in order to make the geometry quantum mechanical (this will also be explained in \S\ref{sub:CDT}). 

\paragraph*{}
As an example of a discrete approach which does not draw from Regge calculus, but still aims to implement the sum-over-histories is \textit{causal set theory} (\S\ref{sub:causet}). By contrast, and more recently, there are approaches which neither borrow from Regge calculus, nor implement a sum-over-histories, and two examples of these that I consider are \textit{quantum causal histories} and \textit{quantum graphity} (\S\ref{sub:pregeo}). These two approaches are examples of so-called ``pre-geometric'' quantum gravity, which claim to more fully exemplify the principle of \textit{background independence} than those theories that utilise a sum-over-histories. Because of this, the pre-geometric approaches purport to represent the most radical departure from spacetime at high-energy than any other present quantum gravity proposals.
\paragraph*{}
The discrete theories considered in this chapter not only share the spirit of the condensed matter approaches, but draw inspiration from techniques used in condensed matter physics. Accordingly, spacetime is conceived of as an effective, low-energy manifestation of very different high-energy degrees of freedom. Caravelli and Markopoulou, for instance, introduce their condensed matter models of quantum graphity by saying,
\begin{quotation}
Current research in the field is paying substantial attention to the numerous indications that gravity may only be emergent, meaning that it is a collective, or thermodynamical, description of microscopic physics in which we do not encounter geometric or gravitational degrees of freedom. An analogy to illustrate this point of view is  fluid dynamics and the transition from thermodynamics to the kinetic theory. What we currently know is the low energy theory, the analogue of fluid dynamics. We are looking for the microscopic theory, the analogue of the quantum molecular dynamics. Just as there are no waves in the molecular theory, we may not find geometric degrees of freedom in the fundamental theory. \citep[][p. 1]{Caravelli2011}. 
\end{quotation}
Another paper succinctly presents the question motivating these approaches:
\begin{quotation}
[O]ne can view the problem of quantum gravity as a problem in statistical physics or condensed matter theory: we know the low energy physics and are looking for the correct universality class of the microscopic quantum theory. By analogy to the Ising model for ferromagnetism, one can ask: What is the ``Ising model'' for gravity? \citep[][p. 1]{Hamma2010}
\end{quotation}
\paragraph*{}
The conceptions of emergent spacetime that these approaches suggest are thus very similar to those already considered in this thesis. Interestingly, many of them also provide evidence of a phase transition, and, by analogy with the conceptions of emergence explored in the previous chapter (\S\ref{sect:Univers}), may provide examples of diachronic novelty as well as autonomy and underdetermination. Before considering the details of these approaches and the bases for emergence they present, however, I look briefly at their main motivations and ``selling points'': discreteness (\S\ref{sub:discreteness}), non-perturbativity (\S\ref{sub:nonpert}) and background independence (\S\ref{sub:BI}).

\section{Discreteness}\label{sub:discreteness}
In the following, a theory is classified as ``discrete'' if it describes the micro-structure of spacetime as one consisting of discrete elements (points or simplices). This may be understood as a discrete geometry, in which case the system possesses a fundamental (shortest) length scale, or, as in the pre-geometric approaches, there may be no clear sense of geometry in the micro-theory. In all of the approaches considered in this chapter, the discreteness is \textit{assumed}---it is a basic postulate of the theory. This is in contrast with the theories in which the discreteness is \textit{derived} as a consequence (a commonly-cited example being loop quantum gravity, considered in \S\ref{sect:LQG}). In some approaches---for instance, causal dynamical triangulations---the discreteness is not taken as an ontological commitment of the theory, but can be understood as a calculational tool.
\paragraph*{}
Various motivations for the belief that spacetime is fundamentally discrete have been offered, although it is still unclear how much faith should be put in them. Essentially, these are all arguments for the existence of a fundamental length scale, which would serve as a natural cutoff and prevent the divergence of various physical quantities (i.e. render our theories finite in the UV). One of these arguments has already been discussed (\S\ref{subsub:axiom}): that spacetime must break down at some scale because it doing so provides the best explanation for why some particular QFTs (i.e. just those QFTs for which renormalisation is necessary) need be renormalised before they yield predictions. To simply re-state the answers to this: firstly, if we take our theories seriously as suggesting that something goes awry at high-energy, then the unknown, high-energy physics could be any number of things, not necessarily a physical breakdown of the continuum. The retort---that many quantum gravity proposals suggest that spacetime is discrete---is illegitimate, given, as stated, most of these approaches simply postulate discreteness at the outset, rather than derive it from accepted physics principles.
\paragraph*{}
Related to this is the claim that the non-renormalisability of gravity suggests that GR fails (or diverges) at high-energy, and so itself is in need of a physical cutoff provided by some fundamental length scale. As pointed out in the discussion of asymptotic safety (\S\ref{sub:asygrav}), and evidenced by the treatment of GR as an EFT (\S\ref{sub:GREFT}), however, it could be that the non-renormalisability of gravity is due simply to the misapplication of perturbation theory, rather than any physical problems with the theory at high-energy indicating its breakdown. In fact, as discussed in the next section (\S\ref{sub:nonpert}), at least one of the discrete approaches, causal dynamical triangulations, promotes the asymptotic safety scenario. This means that it does \textit{not} treat the non-renormalisability of gravity as evidence of a real physical cutoff (in the form of a shortest length scale), since theories that are asymptotically safe do not require a cutoff in order that they be well-defined (recall \S\ref{subsub:fixed points}).
\paragraph*{}
\citet{Wuthrich2012} considers two other arguments for discreteness; the first of these comes from \citet{Henson2009}, who states that we cannot obtain a finite value for black hole entropy unless we have a short-distance cutoff. That black hole entropy be finite is stated by Bekenstein's entropy formula, which may be arrived at by several different lines of reasoning. However, the ``entanglement entropy''---the entropy obtained when field values inside the black hole horizon are traced out---appears to be infinite (Henson admits this result is still in debate), and so if this is to be included in the black hole entropy, we need a finite cutoff in order to tame it. 

\paragraph*{}
W\"{u}thrich points out, however, acceptance of this argument means accepting that black hole entropy must be finite. He reminds us that Bekenstein's formula is derived using semiclassical approximations, ``i.e. by mixing and matching physical principles of which we cannot be certain will be licensed by a full quantum theory of gravity'' \citep[][p. 228]{Wuthrich2012}. While this is true, we should perhaps not be so hasty in dismissing the significance of the Bekenstein entropy, or the semiclassical approximation that it stems from, given that it is generally considered to be among the few ``predictions'' that a candidate quantum gravity theory is expected to reproduce.\footnote{Of course, this cries out for a more detailed philosophical exploration, which I am unable to provide here.} 

\paragraph*{}
The other argument for discreteness comes from \citet[][p. 6]{Reid2001}, who claims that time (and space, with it) being discrete is an effective way of explaining why photons of infinite energy do not exist. Of course, although it may be true that discreteness of spacetime would provide a possible explanation of why photons of infinite energy are not observed, the fact that we do not observe such photons is not, on its own, strong evidence for discreteness.
\paragraph*{}
In spite of there being no definite evidence for discreteness, these arguments do make a reasonable case in presenting some of the problems that would be solved if spacetime were discrete. On the other hand, if the discrete approaches to quantum gravity were successful in recovering spacetime, this would represent an important step toward arguing that spacetime is discrete. It is interesting in to note that there is a theorem in GR which states that given an idea of causal structure and volume information, almost all features of spacetime can be obtained, including dimension, topology, differential structure and metric structure.\footnote{Thanks to theorems presented by \citet{Hawking1976, Malament1977, Levichev1987}.} \citet{Dowker2005} argues that discreteness and causality go hand-in-hand, as two sides of the same coin, so, if this is the case, it seems the discrete approaches do have a good chance of being successful.

\section{Non-perturbative approaches}\label{sub:nonpert}
As we have seen, calculations in QFT are typically performed using perturbative approximations; particles are treated as excitations of fields that exist on a fixed spacetime background (usually the flat four-dimensional Minkowski space of special relativity). In most situations it is an excellent approximation, since at the small distance scales being considered, the gravitational force is much weaker than the other forces, so the dynamical nature of spacetime can be neglected. 

\paragraph*{}
However, the physical scenarios that quantum gravity is expected to treat---for instance, ``empty'' spacetime at distances of the order of the Planck scale ($10^{-35}$m), the extreme conditions in the vicinity of a black hole, and the ultra-dense state of the very early universe---are not in general able to be modelled in terms of linear fluctuations of the metric field around Minkowski space or any other fixed background. This fact is expressed as the statement that a quantum theory of gravity should be both \textit{background-independent} and \textit{non-perturbative} \citep{Ambjorn2006}. The former condition, background independence, will be discussed in the next sub-section, but the latter, of being non-perturbative, simply means that the interesting features of the theory will not show up in the weak field limit in which perturbation theory is applicable.

\paragraph*{}
As suggested above, the proponents of causal dynamical triangulations, interestingly, do not believe that spacetime will ``break down'' (i.e. reveal itself as comprised of discrete entities) at some point: instead, they maintain that the problems with perturbative quantum gravity which are commonly taken as motivation for new physics are simply problems with the perturbative approach to QFT applied in this context. More explicitly, proponents of causal dynamical triangulations claim that it is likely that the non-renormalisability of gravity is simply a consequence of using perturbation theory in a regime far beyond that where it is valid to do so. 
\paragraph*{}
This claim is typically tied to Weinberg's  \citeyearpar{Weinberg1979, Weinberg2009} asymptotic safety scenario (\S\ref{sub:asygrav}), where it is conjectured that there exists a non-Gaussian fixed point for gravity. The idea here is that there is a genuinely non-perturbative UV fixed point that governs the high-energy physics of quantum gravity, but the non-renormalisability described by perturbation theory only reflects the low-energy (IR) regime of an RG flow that originates from the fixed point \citep{Ambjorn2012}. Thus, proponents of causal dynamical triangulations maintain that the familiar framework of QFT is sufficient to construct quantum gravity, so long as the dynamical, causal\footnote{As we shall see, each approach has its own interpretation and means of implementing this notion.} and non-perturbative properties of such a theory are properly taken into account \citep{Loll2008}. 

\section{Background independence}\label{sub:BI}
As stated above, the discrete approaches to quantum gravity seek a theory that is both background independent as well as non-perturbative. It may seem that the desire for a the former condition is tied to the perceived need for the latter one, given that the perturbative approaches involve treating fluctuations around a fixed background metric (but note that an approach to QFT that is non-perturbative would not be classed as background independent simply on account of its non-perturbativity). There is a deeper motivation for background independence, however, stemming from the belief that the putative background independence of GR reflects an important insight into the nature of spacetime: one that should be preserved even at the micro-scale in a theory of quantum gravity.\footnote{We might even say that the force of this insight, and the perceived need to preserve it, is a driving factor in the desire for a non-perturbative fundamental theory.} In spite of this, it is unclear exactly what background independence amounts to, especially when it comes to the unique brand exemplified by GR, with many authors espousing different definitions of the term, or emphasising different aspects of it.
\paragraph*{}
A background dependent theory is taken as one which posits an absolute or fixed object (or objects), and typically relies on these objects in order to define the properties of other objects in the theory (for example, in Newtonian mechanics, absolute position is defined with respect to a fixed background space, and this fixed background space, together with an absolute time, is also used to define motion). Conversely, because of this, background independence has often been identified with \textit{relationalism}, under the observation that a theory that does not rely on a fixed background structure will have to define the properties of its objects purely in terms of the relations between these objects (see, e.g. Smolin, 2006). 
\paragraph*{}
The picture is more detailed than the one immediately suggested, however, since there are a number of different ways to understand what it means for an object to be \textit{fixed}. \citet{Butterfield1999} identify three different meanings of ``fixity'': 1. being classical (i.e. not quantised, no fluctuations), 2. being non-dynamical, and, 3. being given ``once and for all'' by theory (i.e. specified as being the same in all models of the theory). These three notions are mutually logically independent.\footnote{Cf. also \citet{Belot2011}.}

\paragraph*{}
In addition, if we are to understand a fixed background structure as an \textit{absolute} object, there is the connotation that this object be  somehow responsible for conferring properties to the dynamical objects in the theory, without itself being influenced by anything.\footnote{Cf. also Anderson's work on this: \citet{Anderson1964, Anderson1967, Anderson1971}.} Also, an absolute object is typically understood as being fundamental, in the sense that it not supervene on any further, underlying objects. So, when it is claimed that a background independent theory is relational, it means that the properties in the theory are defined only with respect to one another, and that these relationships are not fixed (in the sense of being non-dynamical), but evolve according to the equations of motion of the theory. While, for this reason, \citet[][p. 204]{Smolin2006} states that we may take ``relational'' and ``background independent'' as synonymous, \citet{Rickles2008} argues against the identification on the grounds that a substantival (as opposed to relational) view of spacetime is also compatible with background independence.
\paragraph*{}
At first glance, GR does possess background structure; recall that a spacetime is specified as $(M, g_{\mu\nu}, T_{\mu\nu})$, where
the manifold, $M$, encodes the dimension, topology, signature and differential structure; $g_{\mu\nu}$ is the metric, and $T_{\mu \nu}$, the energy-momentum tensor, denotes the matter fields. Here, the dimension, topology, differential structure and signature are fixed, in the sense of being non-dynamical (although we can alter the dimension and some other features, once specified they cannot vary), while the gravitational and matter fields are dynamical. GR is generally referred to as a background independent theory because of its \textit{diffeomorphism invariance}. 
\paragraph*{}
A diffeomorphism, $\phi$, is a smooth, invertible map from a manifold to itself:
\begin{equation}\label{eq:diffeo}
\phi (M, g_{\mu \nu}, T_{\mu \nu})\rightarrow (M, g'_{\mu \nu}, T'_{\mu \nu})
\end{equation}
This transformation takes a point $p$ to another point $\phi \ast p$ and drags the fields along with it by,
\begin{equation}
(\phi\ast f)(p)=f(\phi^{-1}\ast p)
\end{equation}
The diffeomorphisms of a manifold constitute a group, $Diff(M)$; a physical spacetime is defined to correspond not to a single $(M, g_{\mu\nu}, T_{\mu \nu})$, but to an \textit{equivalence class} of all manifolds, metrics and fields under all actions of $Diff(M)$, which may be denoted $\{M, g_{\mu\nu}, T_{\mu\nu}\}$. In other words, the physical content of the theory is wholly unaffected by the diffeomorphism transformation, which is interpreted as implying that the points of the manifold are not concrete things; they carry no physical meaning in themselves.\footnote{This is the basis of Einstein's famous ``hole argument'', see, e.g. \citet{Earman1987, Norton1988}.} 

\paragraph*{}
This is also the reason why GR is considered to be a relational theory: once we ``mod out'' by diffeomorphisms (remove the symmetry by taking spacetime to be the equivalence class $\{M, g_{\mu\nu}, T_{\mu\nu}\}$), all that is left is a system of relationships between events, where events are identifiable only by coincidences between the values of fields preserved by the actions of diffeomorphisms (Smolin, 2006).\footnote{Although \textit{sophisticated substantivalists} such as \citet{PooleyX} disagree, arguing that field values may be localised with respect to points if the points are taken to be defined by $M/Diff(M)$.} The observables in the theory do not make reference to the fixed elements of the manifold, and so the theory is background independent.\footnote{But cf. also \citet{Pitts2006} and \citet{Sus2010}.}
\paragraph*{}
As Rovelli puts it:
\begin{quotation}
In introducing the background stage, Newton introduced two structures: a spacetime manifold, and its non-dynamical metric structure. GR gets rid of the non-dynamical metric, by replacing it with the gravitational filed. More importantly, it gets rid of the manifold, by means of active diff invariance. In GR, the objects of which the world is made do not live over a stage and do not live on
spacetime: they live, so to say, over each other's shoulders. \citep[][p. 108]{Rovelli2001}
\end{quotation}
However, as \citet{Rickles2008} points out, it is not correct to describe the gravitational field as just one field among many, ``on par'' with the other fields, as Rovelli does. This owes to the existence of vacuum solutions to the Einstein field equations: the gravitational field may be distinguished by the fact that, while it is possible to have a dynamically possible world described by GR without any of the other fields being present, we cannot ever ``switch off'' the gravitational field. 

\paragraph*{}
However, if we are considering GR to be special because of, or characterised by, its background independence (even when understood as diffeomorphism invariance), we need to do some work in articulating exactly why this is the case.  The intuitive notion we have is that GR is special because we cannot define the physical observables of the theory without solving the dynamics, or, in other words, that ``there is no kinematics independent of dynamics'' \citep{Stachel2006}. In spite of this intuitive notion, it is quite difficult to define explicitly the idea of a dynamical object along the lines of ``being solved for''.
\paragraph*{}
The main problem, following \citet{Kretschmann1917}, is that we can turn \textit{any} background field into a dynamical object by making it satisfy some equations of motion, however physically vacuous they might happen to be. Even the metric in special relativity is able to be made dynamical in the sense that it satisfies some equations of motion---hence special relativity can be made background independent, which then conflicts with our basic intuitions of what background independence is supposed to be. As Rickles (2008, pp. 143--144 ) states, if diffeomorphism invariance is what underwrites background independence then the latter cannot be what makes general relativity \textit{special}.\footnote {There is a long-running debate on this point, and ``substantive general covariance''; for other recent contributions, see \citet{Earman2006, Pooley2010}.}

\paragraph*{}
Also, significantly, on this account, if we turn what were originally background fields into dynamical ones, the result is to introduce unobservable (``unphysical surplus'') content into the theory in the form of indistinguishable solutions (following Rickles' definition of ``genuine'' observables as those that are constant on gauge orbits). The solution suggested is to use the definition of \textit{physical observables} to ground the idea of background independence---if the observables are formed \textit{relationally}, then there is no dependence on anything ``external''. Taking this suggestion gives us not only a link between diffeomorphism invariance and background independence, but also an explanation that accords with our intuition that GR is special because of it.
\paragraph{}
The claim that quantum gravity should be background independent has perhaps been championed most prominently by Smolin (2006). He argues that the non-perturbative, background independent approaches to quantum gravity are more testable, on account of being more predictive and more falsifiable than background dependent theories (primarily referring to string theory), and are hence more scientific (apparently on a Popperian-type account of what it means for a theory to be scientific). The background independent approaches, claims Smolin (2006, p. 233),  are also more explanatory than the background dependent ones, not only by virtue of being more testable, but also because they do not require structures to be fixed in advance, i.e. put in ``by hand'', but instead have their structures follow from the dynamical laws and principles specified by the theory itself. 
\paragraph*{}
Finally, Smolin believes there is a correlation between a theory being ``more relational'' (i.e. having less background structure) and being more successful in solving the problems we expect a fundamental theory will solve. The arguments for this position come not only from Smolin's association of the background dependence/independence debate with the historical debates over substantivalism/relationalism, but also from the fact that theorists have so far been unsuccessful in discovering a fundamental formulation of string theory. Although Smolin's arguments for this last point are flawed\footnote{These arguments rely on Smolin's mistaken identification of background independence and relationalism, and the fact that there is no fundamental formulation of string theory. Although the latter fact might mean that background independent theories are more successful than string theory in regards to fundamental problems, it does not demonstrate that background independent theories---by virtue of their being background independent---are more successful than background dependent ones in solving fundamental problems.}, his claims that background independent theories are more testable and more explanatory than string theory seem correct; and, of course, it is interesting from the perspective of emergent spacetime to explore the meaning and implications of background independence.\footnote{For more on the definition of background independence, see \citet{Belot2011}.}
\paragraph*{}
The background independent approaches considered in this chapter do tend to postulate only minimal structure at high-energy, and aim to recover familiar spacetime structure at low-energy using only known physics principles, including the RG. Interestingly, given that each seeks to represent background independence, these approaches have different ideas about what it means for a theory to be background independent. Most generally, they appeal to the intuitive notion described above, captured in the slogan: ``no kinematics without dynamics''. The challenge, of course, lies in how this is to be implemented. 
\paragraph*{}
\citet{Markopoulou2009} takes her definition of a background independent theory as being one whose basic quantities and concepts do not presuppose the existence of a given background spacetime metric. As she points out, most of the background independent approaches seek to comply with this definition.\footnote{These include loop quantum gravity, causal set theory, spin foam models, causal dynamical triangulations and quantum Regge calculus.} Typically, there are two means by which these theories are made to describe quantum, rather than classical, geometry: one is that the theory be formulated in terms of quantum geometric degrees of freedom. The other, which is taken by the first two approaches considered here, is to have the geometry regularised at high-energy (by imposing a shortest length scale in the form of a finite-valued cutoff) and then implementing a sum-over-histories of the allowed geometries (i.e. causal histories).
\paragraph*{}
The second two discrete approaches considered in this chapter, being the so-called ``pre-geometric'' approaches, propound a similar line to Smolin's ``the more relational the better'': according to these approaches, even a sum-over-histories (also known as a sum-over-geometries) is too much like a geometry at the micro-level (although, as will be explained in \S \ref{sub:CDT}, the sum-over-histories in no way resembles the classical concept of geometry). Just as Smolin suggests that the problems with string theory stem from its employment of a background spacetime, the proponents of the pre-geometric theories suggest that perhaps the problems with the ``traditional'' discrete approaches stem from their use of some notion of quantum geometry at the basic level. 
\paragraph*{}
Instead, on this conception, a background independent theory is defined as one in which all observations are internal, i.e. made by observers inside the system (see, e.g. \citet{Markopoulou2000}). Such a theory cannot describe a global Hilbert space or wavefunction for the whole universe, but must instead feature a collection of local ones. Markopoulou (2009) describes how quantum causal histories as well as other pre-geometric approaches modelled on it (\S\ref{sub:pregeo}) are theories that satisfy the condition of all observations being internal also represent theories that are background independent in the sense of not describing anything resembling a geometry (whether classical or quantum) at the most basic level.
\paragraph*{}
Interestingly, Markopoulou (2009, p. 140) claims that what constitutes a background independent theory is ``a question that is currently being revisited'' and for which new suggestions are being offered. Given that this is the case, however, it seems an interesting question to ask why it is thought that these new conceptions of background independence at the fundamental level are expected to reproduce the conception of background independence demonstrated by GR at low-energies, and, indeed, what the motivation for background independence at the micro-level is (apart from the slow progress of string theory and the historical debates over relationalism and absolute space). In other words, if GR is an emergent theory, we might ask why its background independence is being taken as one feature that is to be preserved at high-energy, especially if it is allowed to take a radically different form. This great insight of GR---understood as the idea that observables should not refer to unphysical degrees of freedom (such as the labels of points)---is expected to survive GR itself.

\paragraph*{}
Aside from its association with GR, however, we might argue that background independence is a natural criterion for a fundamental theory just because the theory is supposed to be fundamental: having quantities dependent on an object that must be fixed (in the sense of being set ``by hand'') rather than following from the laws of the theory itself seems to suggest that there is something ``more fundamental'', not being captured by the theory. This point relates to Smolin's (2006) claim that background independent theories are more explanatory than background dependent ones.

\section{Causal set theory}\label{sub:causet}
Causal set theory owes much to its initial proponent, Rafael Sorkin, who continues to work on it \citep[see, e.g.][]{Sorkin1991a, Sorkin1991b, Sorkin2005}. The approach begins with some basic discrete elements (i.e. points, with no internal structure whatsoever) and takes them to be related only through a partial ordering that corresponds to a notion of \textit{causality}, i.e. a microscopic conception of ``before'' and ``after''. The continuum notions of time and distance intervals arise only as approximations to the theory on large scales---there are no corresponding features on the micro-scale described by causal set theory.\footnote{My presentation draws from \citet{Dowker2005} and \citet{Henson2009}.} 

\begin{figure}[h]
\centering
\includegraphics[height=5.36cm]{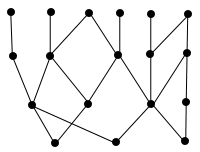}
\caption{Hasse diagram of a causal set.}
\label{fig:causet}
\end{figure}

\paragraph*{}
That causal structure is enough to recover metric relations in GR follows from a theorem due to \citet{Malament1977}.\footnote{\label{foot:theorems} Other powerful theorems to this effect are shown by \citet{Hawking1976, Levichev1987}.} Dowker (2005), inspired by this theorem, states that causality and discreteness go hand-in-hand, and claims that this is one of the strongest motivations for causal set theory as a description for the micro-structure of spacetime. A causal set is defined to be a set $C$ together with a relation, $\prec$, called ``precedes'', which satisfy the following axioms, \\ \\
\textbf{Axiom 1: Partial ordering}\\
\textit{Transitivity}: if $x\prec y$ and $y\prec z$ then $x\prec z, \forall x,y,z \in C$;\\
\textit{Antisymmetry}: if $x \prec y$ and $y \prec x$ then $x=y\, \forall x, y \in C$ \\ 
\textit{Reflexivity}: $x\prec x \forall x \in C$\\\\
\textbf{Axiom 2: Local finiteness} \\
For any pair of fixed elements $x$ and $z$ of $C$, the set $\{y|x \prec y \prec z  \}$ of elements lying between $x$ and $z$ is finite.
\paragraph*{}
One way to describe the structure of partial ordering is a genealogical terminology, where the causal set is thought of as a family tree; an element $x$ is said to be an ``ancestor'' of an element $y$ if $x \prec y$, and $y$ is then a ``descendent'' of $x$. Causal sets can be illustrated by \textit{Hasse diagrams}, which are graph-theoretic representations of finite partially ordered sets, as shown in Fig.\ref{fig:causet}. The vertices represent elements of $C$ and edges connecting the vertices represent their standing in the relation $\prec$. Since $\prec$ is transitive and reflexive, only a set's ``transitive reduction'' is drawn (i.e. points that are connected to each other via connections through other points are not shown as also being connected to each other directly).

\begin{figure}[h]
\centering
\includegraphics[height=10cm]{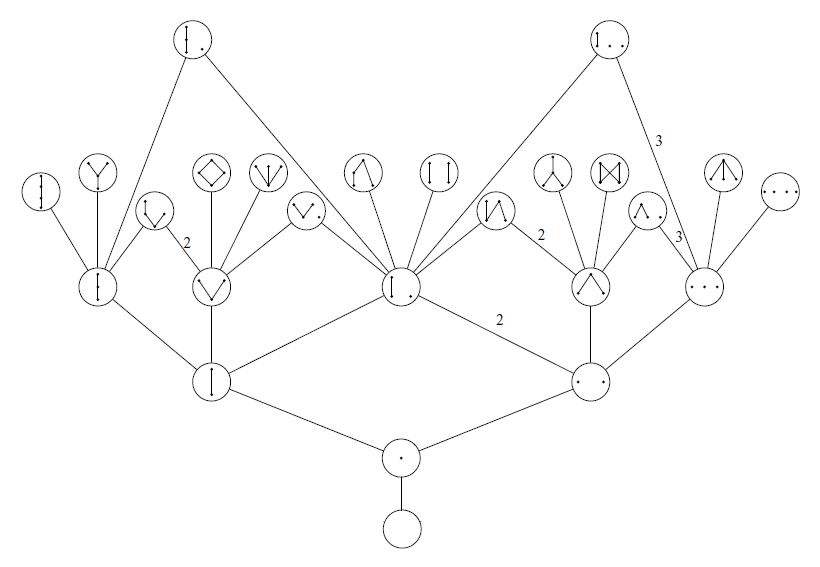}
\caption{Hasse diagram depicting the partially ordered set of finite causets (only causets up to size 4 are shown). The elements of this set are the finite causets. The numbers on some links indicate the number different possible ways in which new elements can be added, due to automorphisms of the ``parent'' causet. (Adapted from \citet{Henson2009}).}
\label{fig:bigset}
\end{figure}

\paragraph{}
In order to see how we can construct a spacetime from causal sets, it is useful to imagine ``working backwards'', i.e. first beginning with a continuum spacetime and then discretising it. Following the discretisation, the aim is to have the discrete structure be independent of the continuum from which it was derived. Then, in order to determine whether this discrete structure, known as a casual set, or \textit{causet}, really is a candidate for the micro-structure of spacetime, we need to recover the original continuum, or at least something similar to it. In the first step, of discretising the continuum, the discrete elements cannot be arranged in a lattice or grid formation, because the result will not be Lorentz invariant (i.e. there will be preferred directions).\footnote{The desire to preserve Lorentz invariance comes from the fact that there is no evidence of it being violated, notwithstanding the fact that there are approaches, most notably Ho\v{r}ava gravity \citep{Horava2009} according to which Lorentz invariance is violated at high-energy.}
\paragraph*{}
In order to ensure there is no preferred direction and Lorentz invariance is preserved, the points are distributed randomly on the manifold by a process called \textit{sprinkling}, which is the Poisson process of choosing countable subsets of the spacetime for which the expected number of points chosen from any given region of spacetime is equal to its volume in fundamental units. Having obtained a random distribution of points, we then remove the background prop of the continuum, to ``whisk away the tablecloth from under the crockery''. However, as Dowker (2005, p. 451) states, in the case of quantum gravity, the sprinkled elements are meant to \textit{be} the spacetime, and by whisking away the continuum, we remove the table, not just the tablecloth. Hence, the sprinkled elements need to be endowed with some extra structure if they aren't to collapse into a heap of unstructured dust.
\paragraph{}
What is required is the specification of a causal structure for the spacetime, being the information about which events can causally influence other events. For each point $p$ of the spacetime, we define the set $J^-(p)$ the causal past (and $J^+(p)$ the causal future of $p$), to be the set of points, $q$, in the spacetime for which there is a past (future) directed causal curve---a curve with an everywhere non-spacelike, past (future) pointing tangent vector---from $q$ to $p$. The collection of all these causal past and future sets is the causal structure of the spacetime. 
\paragraph*{}
Thus, the elements are sprinkled into the spacetime with the order given by the spacetime causal structure: elements $e_i$ and $e_j$ are sprinkled at points $p_i$ and $p_j$ respectively, satisfy $e_i \prec e_j$ if $p_i \in J^-(p_j)$. The set of sprinkled elements with this induced order is a causal set satisfying the axioms above. Now, to recover the spacetime: a spacetime $\Mani$ is a good approximation to a causal set $C$ if $C$ could have arisen from $\Mani$ by this discretisation process (i.e. sprinkling and endowing with causal order) with relatively high probability.
\paragraph{}
Of course, the aim of causal set theory is to explain how relativistic spacetime emerges at large-scales. This requires the theory  to have a dynamics: causal sets are supposed to be dynamical entities that grow by the creation of additional elements. The most popular of the proposed dynamics is the \textit{classical sequential growth dynamics} of \citet{Rideout1999}. This growth process is a discretised, stochastic Markov process\footnote{A stochastic process has the Markov property if the conditional probability distribution of future states of the process depends only upon the present state, not on the sequence of events that preceded it.}, which starts out from the empty set and adds new elements, one by one, to the future of the elements in the existing causal set. This is illustrated by the Hasse diagram in Fig.\ref{fig:bigset}, which shows all the possible routes that an evolution can take.

\paragraph*{}
The classical sequential growth model is just one suggestion for a possible dynamics for the theory. In an attempt to restrict the number of different possible dynamics, certain constraints are imposed, including \textit{discrete general covariance} and a \textit{Bell-type causal condition}. Discrete general covariance states that the probability of growing a particular finite partial causet does not depend on the order in which the elements are born. The Rideout-Sorkin dynamics produces a labelling on the elements, corresponding to the order in which they are born, but the principle of discrete general covariance requires that these labels be unphysical, i.e. meaningless: the only physical structure possessed by the causet is its partial order. This condition ensures that there is no ``fundamental time'' in the model, in spite of their being a dynamics defined by the ``birthing process''.\footnote{This is in contrast with quantum graphity (\S\ref{sub:pregeo}), for instance, where time is treated as fundamental while space is not, and a distinction is made between the time in the micro-dynamics and the emergent macro-time.} Bell causality is the closest possible analogue of the condition that gives rise to the Bell Inequalities, and is meant to imply that a birth taking place in one region of the causet cannot be influenced by any birth in a region spacelike to the first.
\paragraph*{}
Although these constraining principles certainly appear reasonable, and are both satisfied by the Rideout-Sorkin models, there is a serious problem in that they do not seem to result in causal sets that give rise to manifold-like spacetimes. This is an instance of a more general problem typical of all discrete approaches to quantum gravity: the sample space of causets (or discreta more generally) is always dominated in sheer number by non-manifold-like discreta, and a uniform distribution over the sample space will render these undesirable ones overwhelmingly more likely. In order to produce a measure over the sample space that is peaked on the manifold-like discreta, a dynamics must be specified that ``picks out'' the causal sets that give rise to manifold-like spacetimes. Research in this area is ongoing.
\paragraph{}
Causal set theory has been criticised because, as in the basic outline above, it is entirely classical; opponents emphasise that, in constructing a theory of quantum gravity, we cannot neglect quantum effects from the outset, as causal set theory does. While there has been recent progress in the attempt to formulate an entirely quantum dynamics \citep{Gudder2013}, for the most part, those who work on causal set theory have aimed to implement a ``sum-over-histories'' approach, where the sum is taken over a sample space, $\Omega$ of possible histories of the causal set, and a dynamics for the system expressed in terms of a quantum measure, $\mu$ on $\Omega$ \citep{Dowker2005, Sorkin2005}. Such an approach would ``build in'' the lessons and methods of quantum mechanics. The sum-over-histories approach is more fully implemented in a rival program, CDT, discussed below.

\subsection{Emergence in causal set theory}\label{subsub:ecauset}
Examining the idea of emergence in causal set theory might be considered gratuitous given that the approach on its own has so far been unable to recover relativistic spacetime. Without a description of how spacetime ``appears'' at low-energy, it is difficult to speak of emergence at all. Nevertheless, we can speculate about the nature of the relation between causal set theory and relativistic spacetime if we assume that it is possible that a well-behaved continuum limit of causal set theory resembling relativistic spacetime exists (although, in doing so, we can only speak in abstract regarding the dynamical details involved, assuming they resemble the Rideout-Sorkin models). 
\paragraph*{}
Without having implemented the quantum formulation of causal set theory, which is supposed to involve taking a sum over all possible (allowed) geometries, we may view the approach as purely classical at this stage. In this case, the discrete elements of causal set theory are supposed to literally represent ``atoms of spacetime'': the idea being that if we ``zoom in'' to small enough distance scales and the continuum described by our current theories of physics is revealed to be a low-energy approximation of a causal set of discrete elements.
\paragraph*{}
As with many of the other discrete approaches, causal set theory is able to utilise, at least in part, familiar physics techniques (including those from QFT). Dowker (2005, p. 460) contrasts this situation with suggestions from other approaches, where spacetime is ``replaced'' with some more radical entity at high-energy; in causal set theory, she says, spacetime is still considered to be ``real'', albeit discrete. It is not clear what Dowker means by ``real'', however. On the causal set picture, GR is taken as an effective theory valid at low-energies, and continuum spacetime is just a low-energy approximation. 
\paragraph*{}
The dramatic departure from familiar spacetime that causal set theory represents shouldn't be downplayed. \citet{Huggett2013} make a similar point, outlining the three ways in which the relations that govern the discrete landscape of causal set theory differ from ordinary spatiotemporal relations: firstly, there is nothing on the fundamental level corresponding to lengths and durations (i.e. spacetime intervals), secondly, the theory lacks the structure to identify ``space'' in the sense of a spacelike hypersurface, and thirdly, there is a tension between the discreteness of the causal set and the Lorentz invariance demanded of the emergent spacetime that threatens to render the intermediate physics non-local in a way unfamiliar to GR (this last suggestion is made in \citet{Sorkin2009}). In this sense, perhaps Dowker is a bit disingenuous in claiming a virtue of the theory is that it does not replace spacetime with a substance of ``completely different ilk''.
\paragraph*{}
Perhaps Dowker's claims here might be interpreted as suggesting that on the causal set picture, relativistic spacetime is ``resultant'' rather than ``emergent'' from causal set theory: at large scales the causal set ``turns into'' spacetime.\footnote{The idea of something being resultant rather than emergent, recall, is based on a conception of emergence as a failure of part-whole reductionism, as in \citet{Silberstein1999}, discussed in \S\ref{sub:Eemergence}.} I believe this suggestion is still misleading, however, since, again, the continuum is still phenomenological, just as it is in approaches where spacetime is ``replaced'' by some other entity at small-scales. 
\paragraph*{}
The best way to understand spacetime emergent from causal set theory is to take it as analogous to hydrodynamics emergent from the micro-physics of a particular system. In this analogy, spacetime may be viewed as a fluid whose low-energy effective field theory (GR) is distinct from---and describes very different degrees of freedom to---its micro-theory (causal set theory), which is a theory of the dynamics of particles (discrete elements). Recalling the results of \S\ref{sub:Uconcl}, we do not have diachronic novelty, because there is no dynamical change in the system akin to a phase transition (some of the other approaches I consider in this chapter, however, do feature such a transition).
\paragraph*{}
Just as in the case of hydrodynamics, we have quasi-autonomy of the macro- and micro-theories: although GR is distinct from causal set theory, the low-energy theory is not completely impervious to the details of the micro-theory. For instance, recall that improper sprinkling of the discrete elements will result in violations of Lorentz invariance at large distance-scales. In fact, of the discrete approaches considered in this section, spacetime emergent from causal set theory is apparently the most sensitive in regards to the details of its micro-constitution. Of course, this claim is speculative, however, since relativistic spacetime has not yet been recovered from causal set theory (though the reason for this lack of success perhaps has to do with the sensitivity of the macro-physics on the micro-physics). 
\paragraph*{}
We may speculate, too, how the ideas of \textit{underdetermination} and \textit{universality} that were discussed earlier in regards to hydrodynamics feature in causal set theory: different causal sets may correspond to the same spacetime (as is seen by considering the random nature of the sprinkling technique), and so spacetime might be considered emergent with respect to the causal sets.\footnote{Although obvious, it seems important to again emphasise that the analogy between causal set theory and hydrodynamics cannot be taken as seriously as this discussion perhaps suggests, given that the discrete elements are non-spatiotemporal. For instance, we cannot use the RG to ``zoom out'' from the causal set to larger distance scales (owing to the fact that the causal set has no concept of distance associated with it).} This suggestion, though, may be threatened by the idea of superpositions of causets that would feature if the sum-over-histories approach were properly implemented.
\paragraph*{}
Butterfield's \citep{Butterfield2011b, ButterfieldForthcoming} conception of emergence as tied to limiting relations seems as though it could provide an alternative basis for understanding emergence in causal set theory, although it suffers the same lack of support as the conception of emergence based on underdetermination and universality. In this case, continuum spacetime is supposed to be recovered as the number, $N$, of discrete elements goes to infinity (though the $N \rightarrow\infty$ needn't be considered physically real, as sufficiently novel behaviour appears before the limit is reached (\S\ref{subsub:varied})). Because causal sets are constructed by sprinkling the elements into a spacetime and endowing them with causal order, the idea of the continuum limit is the same as taking the elements to be sprinkled ``densely''.

\section{Causal dynamical triangulations (CDT)}\label{sub:CDT}
The sum-over-histories approach is central to the theory of CDT; it is analogous to the familiar Feynman path integral approach to quantum mechanics, the basic idea of which is to obtain a solution to the quantum dynamics of a system by taking a superposition of all possible configurations of the system. Each configuration contributes a complex weight $e^{iS}$ where the classical action is $S=\int dt \Lagr(t)$, with $\Lagr$ being the system's Lagrangian. In the case of a non-relativistic particle moving in a potential, the configurations are paths in space between two fixed points. 

\paragraph*{}
However, the individual paths being superposed are not themselves physical trajectories, nor even solutions to the particle's classical equations of motion. Rather, they are \textit{virtual} paths: any curves that one can draw between the fixed initial and end-points. The magic of the path integral is that the true quantum physics of the particle is encoded precisely in the superposition of all these virtual paths. In order to extract the physical properties, suitable quantum operators are evaluated on the ensemble of paths contributing to the integral (for instance, the expectation values of position or momentum can be calculated).
\paragraph*{}
Analogously, a path integral for gravity might be thought of as a superposition of all the virtual ``paths'' the universe can follow ``in time''---though, this is certainly difficult to conceptualise! More concretely, the virtual paths are different configurations of the metric field variables $g_{\mu \nu}$, so we have a superposition of all possible ways empty spacetime can be curved.\footnote{This is a simple path integral for ``pure gravity'', although inclusion of the matter degrees of freedom is not thought problematic.} These paths are also known as \textit{spacetime histories}. Again, it is important to note that the individual geometries contributing to the integral may be arbitrarily remote from classical spacetime---in fact, most of them are. 

\paragraph*{}
Spacetime histories that have any geometric resemblance to a classical spacetime are so rare that their contribution to the path integral is effectively negligible \citep{Ambjorn2006} .The superposition itself is nothing like a geometry: as Markopolou puts it,
\begin{quotation}
The monstrosity we just created does not even have a sensible notion of here and there, the most basic aspect of geometry. It also does not have a notion of dimension. It's only the fact that we call it ``quantum geometry'', a combination of two words we understand, that fools us into thinking we comprehend it. \citep[][p. 4]{Markopoulou2008}
\end{quotation}
Nevertheless, analogously to the Feynman path integral for a particle, we expect to be able to retrieve the full quantum dynamics of spacetime from the gravitational path integral, by evaluating suitable quantum operators on the ensemble of geometries contributing to it. 

\paragraph*{}
The gravitational path integral is formulated as,
\begin{equation}\label{eq:sumhistories}
Z(G_N, \Lambda_C)=\int\mathcal{D}de^{iS[g]}
\end{equation}
where the integral is taken over all spacetime geometries $g\in \mathcal{G}$, and $S$ is the four-dimensional Einstein-Hilbert action:
\begin{equation}
S=\frac{1}{G_N}\int d^4 x \sqrt{det g}(R-2\Lambda_C)
\end{equation}
where $G_N$ is the gravitational coupling (Newton's constant) and $\Lambda_C$ the cosmological constant, and the integral is to be taken over all spacetimes, subject to specified boundary conditions. As Loll (2008) points out, the path integral \ref{eq:sumhistories} is better regarded as a statement of intent rather than a well-defined prescription, insofar as the boundary conditions, integration space $\mathcal{G}$ and integration measure, $\mathcal{D}$ remain unspecified. CDT provides a non-perturbative means of defining and evaluating the path integral \ref{eq:sumhistories} given a positive value of $\Lambda_C$, in particular, it gives a definite prescription of how the contributing virtual paths should be chosen.\footnote{The discussion here draws from \citet{Ambjorn2006, Ambjorn2012, Loll2008}.}
\paragraph*{}
The integral is approximated\footnote{Although, as Loll (2008) points out, calling it an approximation is misleading, since it supposes that we have prior knowledge of what it is we are approximating, when in this case, we do not.} by constructing the space of all spacetimes $\mathcal{G}$ using four-dimensional triangular ``building blocks'', known as 4-simplices, illustrated below in Fig.\ref{fig:simplices}.
\begin{figure}[h]
\centering
\includegraphics[height=6cm]{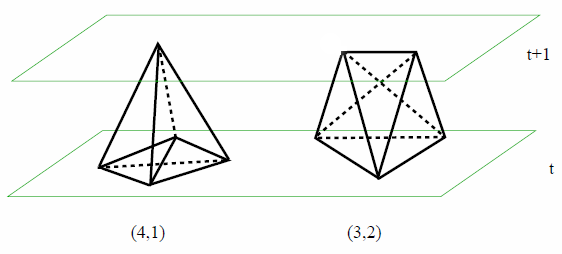}
\caption{The two fundamental building blocks of CDT. The four-simplex (4,1) has four of its vertices at time $t$ and one at time $t+1$, and analogously for the (3,2) type of simplex. (From \citet[][p. 347]{Ambjorn2009}).}
\label{fig:simplices}
\end{figure}
Just as we can imagine constructing a two-dimensional curved surface by gluing together flat triangular pieces of cardboard, we imagine the 4-simplices as the four-dimensional analogues.\footnote{Although, surprisingly, simply gluing together the 4-simplices generically does \textit{not} result in a four-dimensional manifold: more is required for the gluing-rules, as discussed below.} The 4-simplices have side-length $a$, which serves as a cutoff for the integral, so we obtain a regularised version $\mathcal{G}_{a, N}$ of the space of all spacetimes $\mathcal{G}$, by gluing together $N$ 4-simplices. However, it is important to note that the side-length $a$ is \textit{not} to be regarded as a minimal fundamental length in the theory; rather, the path integral is studied in the limit as $a \rightarrow 0$ and $N \rightarrow \infty$, so that the individual building blocks are completely shrunk away. The path integral, $Z$ according to CDT is thus,
\begin{equation}
Z=\lim_{N \rightarrow \infty, a \rightarrow 0} \sum \frac{1}{C_g}e^{iS^{Regge[g]}}
\end{equation}
where $C_g$ denotes the order of the automorphism group of the geometry $g$, and $S^{Regge}$ is a Regge-version of the Einstein-Hilbert action.\footnote{See \citet{Regge1961}.}
\paragraph*{}
Taking the limit $a \rightarrow 0$ while renormalising the coupling constants featured in the theory as a function of $a$, enables the study of the scaling behaviour of the physical parameters. This is the same method as used when studying critical phenomena, and it ensures the continuum theory that results---\textit{provided} a well-defined continuum limit actually exists---will not depend on many of the regularisation details (for instance, the precise geometry, including side-length, of the building blocks and the details of the gluing rules). 

\begin{figure}[h]
\centering
\includegraphics[height=10cm]{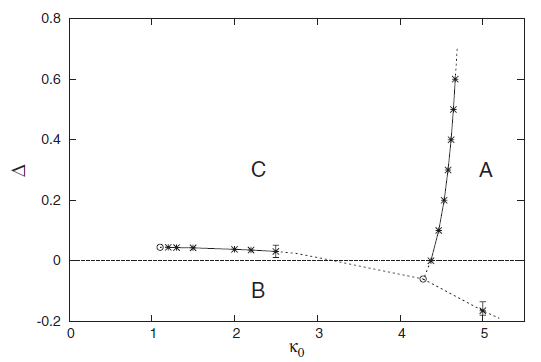}
\caption{Phase diagram of CDT. Crosses represent actual measurements, dashed lines are extrapolations. (From Ambj\o{}rn, Jordan, et al., 2012, p. 124044-3).}
\label{fig:phase}
\end{figure}

\paragraph*{}
Just as in studying critical phenomena, the system in the continuum limit exhibits \textit{universality}: the Planck-scale physics is \textit{robust}, or largely independent of the regularisation details. Proponents of CDT emphasise that the features of the 4-simplices are arbitrarily chosen and not to be regarded as fundamental: they are tools for constructing the continuum theory. Considered as a theory of quantum gravity, CDT is supposed to apply at the Planck scale, but, as should be clear, it does not feature a fundamental length scale; in fact, it predicts a fractal-like picture of spacetime at high-energies, which means that the theory is scale-invariant in the extreme UV.
\paragraph{}
The gluing rules for the 4-simplices must be such that they lead to a well-defined path integral and a four-dimensional continuum spacetime. In a manifestation of the standard problem with discrete approaches to quantum gravity, most of the apparently reasonable models of 4-simplices do not generically lead to a good classical limit---they do not give rise to spacetimes that are macroscopically extended and four-dimensional. However, this is rectified by imposing the condition of \textit{causality}, in the sense that the histories allowed in the sum must be those with a well-behaved causal structure. This is implemented in the gluing rules by giving the geometries a globally layered structure, labelled by a global, geometrically defined integer-valued proper time $t$. 
\paragraph*{}
The resulting strict subset of possible spacetime histories leads to a sum that has a stable extended geometry as its continuum limit, with an effective dimension of four at large scales. The recovery of the correct spacetime dimension at large scales is the primary achievement of the CDT approach. These results come from numerical (Monte-Carlo) simulations of the gravitational path integral defined by CDT. At very small length-scales, spacetime according to CDT is effectively two-dimensional, with a fractal-like structure.
\paragraph*{}
Recently, Monte-Carlo simulations, as well as other numerical techniques, have shown that CDT predicts a second-order phase transition \citep{Ambjorn2012a, Ambjorn2011}. The Regge-action used in the simulations is,
\begin{equation}
S^{Regge}= -\kappa_0N_0+\tilde{\kappa}_4N_4+\Delta(N^{(4,1)}_4-6N_0)
\end{equation}
where $N_0$, $N_4$ and $N^{(4,1)}_4$ denote the number of vertices, 4-simplices and 4-simplices of type (4, 1) (having four vertices on one hypersurface and the fifth on a neighbouring hypersurface, as shown in Fig. \ref{fig:simplices}). The three couplings $\kappa_0$, $\tilde{\kappa}_4$ and $\Delta$ depend on the gravitational coupling, the bare cosmological coupling, and the edge-length, $a$. The phase diagram that results from the simulations is Fig. \ref{fig:phase}, below.

\paragraph*{}
The average large-scale geometry in phase $C$ shows the scaling behaviour of a four-dimensional universe, as described above, with the average volume profile matching that of a Euclidean de-Sitter spacetime. The situation in the other phases is completely different: the typical volume profile of configurations in phase $A$ shows an almost uncorrelated sequence of spatial slices, while the configurations in phase $B$ are characterised by an almost vanishing time extension. This is illustrated below in Fig. \ref{fig:phasecolour}.

\begin{figure}[h]
\centering
\includegraphics[height=8cm]{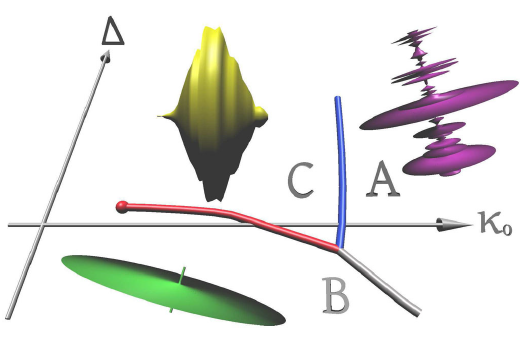}
\caption{Visualisation of the CDT phase diagram, based on actual measurements, illustrating the distinct volume profiles characterising the three observed phases A, B and C. (From Ambj\o{}rn, Jordan, et al., 2012, p. 124044-3).}
\label{fig:phasecolour}
\end{figure}

\paragraph*{}
Evidence suggests that the $A-C$ phase transition line is of first-order. There is strong evidence, however that the $B-C$ transition line is of second-order. The triple point where all three phases meet has so far not been closely examined (the dashed part of the $B-C$ transition line in Fig. \ref{fig:phase} represents a region where the conventional methods have been insufficient to measure the location of the phase transition with acceptable accuracy). The presence of the second-order phase transition may be further evidence for the existence of a UV-fixed point, but research in this area is ongoing.\footnote{For details on how the CDT second-order phase transition might make contact with Weinberg's asymptotic safety scenario, and the consequences of assuming there is a UV-fixed point somewhere along the transition line, see \citet{Ambjorn2012}.}

\subsection{Emergence in CDT}\label{subsub:ecdt}
In the case of CDT, the building blocks (4-simplices) are not considered real---although they are the basic elements described by the theory, they are removed from the theory before any claims about spacetime are made. In other words, they are mathematical tools, used simply to approximate the spacetime integral. This is in contrast with causal set theory, where the elements of the causal set are taken as real and fundamental. Thus, in CDT, spacetime is not considered a low-energy manifestation of something ``more fundamental'' at high-energy: the high energy degrees of freedom are still spatiotemporal. As described above, CDT aims to describe relativistic spacetime at the quantum (presumed to be Planck) scale, and the physics at this scale is robust and autonomous compared to the details of the building blocks used to define and calculate the integral.
\paragraph*{}
There is an interesting notion of \textit{universality}, where spacetime is universal (multiply-realised) with respect to the details of the mathematical tools used to model it. Because of the similarity of the mathematical techniques used, proponents of CDT describe this idea of universality as akin to that featured in condensed matter theories of critical phenomena. However, it should be clear that the ontological implications are very different: in condensed matter theory, the discrete elements that serve to regularise the integral have physical meaning, being the particles of the micro-system that become ``visible'' when the system studied is probed with high enough energy. In the case of CDT, the discrete elements introduced to regularise and define the integral do not have physical meaning---hence, the universality in this sense is not of direct relevance to the emergence of spacetime, insofar as spacetime is not thought to physically ``emerge'' from these elements.
\paragraph*{}
CDT recovers four-dimensional spacetime at low-energies, while at high-energies, spacetime has a fractal structure and is effectively two-dimensional.\footnote{The prediction that spacetime is two-dimensional at small scales is not one unique to CDT: \citet{Carlip2010} examines several different approaches that suggest this phenomenon, which he terms ``spontaneous dimensional reduction''.} The suggestion of a fractal geometry is intriguing as it means that the theory would be \textit{scale invariant} for all high-energies beyond the point at which the ``spontaneous dimensional reduction'' occurs. The suggestion of scale invariance means that there is a fundamental (smallest) length scale, in the sense that we would be unable to define anything smaller beyond the scale at which the system reveals its fractal structure. If  we are willing to consider the two-dimensional high-energy manifestation of spacetime to indeed \textit{be} spacetime, then spacetime cannot be said to break down at any point. On the other hand, it may seem plausible, perhaps, in comparison with causal set theory, to consider this two-dimensional manifestation of spacetime as a sort of ``more fundamental'' precursor to familiar relativistic spacetime, in which case it may be thought that the latter does emerge from this two-dimensional version.
\paragraph*{}
On the other hand, there may be an interesting conception of emergence related to the second-order phase transition of the B-C line in Fig. \ref{fig:phase}, especially if it is indeed associated with a fixed point. In this case, there would be \textit{underdetermination} and \textit{universality} just as in considering critical phenomena. Recalling the discussion in \S\ref{subsect:criticalphen}, we would expect some parameter of the system (analogous to the correlation length) to diverge at the critical point, but then ``zooming out'' using the RG would not reveal a change in the coupling constants of the theory, as is characteristic of a system at a fixed point. The universality class would then be defined as those different systems (i.e. with different micro-physics) that lie in the basin of attraction of the same fixed point, i.e. having the same scaling behaviour, characterised by the same values of their critical exponents.\footnote{The critical exponent for the B-C transition is cited in  \citet{Ambjorn2012a} as $\tilde{v}=2.51(3)$.} 
\paragraph*{}
In \S\ref{sect:Univers}, I argued that the phenomenon of universality means that we may consider the critical behaviour as emergent in the sense that it is independent of the system's micro-constitution. It seems strange to make such a claim in the case of CDT, as the theory does not suggest that there is any micro-structure to spacetime. Here, however, it is worth noting that the couplings involved in the theory (i.e. as plotted in Fig.\ref{fig:phase}) do involve the side-length, $a$, of the simplices. Perhaps, then, the conception of universality associated with the idea of critical phenomena here may not be physical, but again related to the mathematical models used by the theory.

\paragraph*{}
Finally, it is worth briefly commenting on how both CDT and causal set theory rely on the notion of \textit{causality}. In the case of CDT, causality was discovered to be the means by which to select those classes of geometries that were to be included in the path integral; while in causal set theory, causality is implemented at the fundamental level and is supposed, together with the discrete elements which realise the partial ordering (of ``precedes''), to be responsible for the appearance of spacetime at large-scales.

\section{Pre-geometric approaches}\label{sub:pregeo}
Most of the ``traditional'' discrete/background independent theories, which utilise a sum-over-histories-type approach, encounter significant difficulties with their main aim of recovering relativistic spacetime as a low-energy limit (apart from CDT, which has been more successful than most in this respect). In response, some physicists have suggested that perhaps, rather than utilising a sum-over-histories-type approach, which presupposes some sort of quantum geometry, we need implement a stronger sense of background independence, where there is no underlying micro-geometry.\footnote{Even recalling the significant differences between ordinary geometry and ``quantum geometry'' outlined above (\pageref{sub:CDT}).} These approaches are thus known as \textit{pre-geometric} approaches to quantum gravity, and I will consider the two main ones only, being \textit{quantum graphity}\footnote{For details, see: \citet{Quach2012,  Caravelli2011, Hamma2011, Konopka2008}.} and \textit{quantum causal histories}.\footnote{For details see:  \citet{Markopoulou2009, Livine2007, Hawkins2003, Markopoulou2000}.} 

\paragraph*{}
The promise of these approaches is ``truly emergent'' geometry, as well as gravity. One of the core principles of the pre-geometric approaches is that the geometry be defined \textit{intrinsically} via the interactions of the micro-elements described by the theory, and is dependent on the dynamics of these micro-elements; because of this, these approaches are set to recover geometry that is dynamical, as well as being emergent.
\paragraph*{}
Recall that causal set theory is discrete, but not quantum-mechanical at its heart (the quantum effects are supposed to be taken into account later when the sum-over-histories is implemented in the approach). Proponents of quantum causal histories (QCH) view this as a mistake, and attempt to rectify it by quantising the discrete causal structure. One way to do this is to attach Hilbert spaces to the events (points) of a causal set. Markopoulou (2000, p. 309) suggests that, in doing this, we are interpreting the events as elementary quantum-mechanical systems that (are assumed to) exist at the Planck scale, and that these elementary systems interact and evolve by rules that give rise to a discrete causal history. 
\paragraph*{}
Because the theory is fundamentally discrete, and thus possesses a natural cutoff-energy scale, it is assumed that these Hilbert spaces are finite-dimensional. Therefore, in attaching them to the nodes of the causal set, we have constructed a causal network of finite-dimensional Hilbert spaces. There is a problem, however, in that this network would not in general respect local causality (as explained in Markopoulou, 2000). The solution is to attach the Hilbert spaces to the causal relations (i.e. the edges of the graph) rather than to the events. The events, instead, are promoted to quantum evolution operators, which, Markopoulou (2000) says, is consistent with the intuition that an event in the causal set denotes change, and so is most naturally represented by an operator. In QCH with the Hilbert spaces on the causal relations and the events as operators, the quantum evolution strictly respects the underlying causal set.
\paragraph*{}
QCH is capable of modelling other pre-geometric approaches (including quantum graphity, to be discussed below), and it makes some claims that hold generally for these approaches. One such general claim regards the importance of there only being \textit{internal} observers describing the universe and defining spatial and temporal intervals. This constraint is implemented (or manifests itself) in several different ways in the different approaches; for instance, in QCH, defining the causal relations and operators as above results in there being no single Hilbert space, or wave function, for the entire universe. 
\paragraph*{}
Internal observations contain only partial information about the universe, that which is in the causal past of an observer at the corresponding spacetime The notions of observers and spacetime intervals, of course, are only supposed to arise at the emergent, low-energy level. In the pre-geometric approaches, the emergent, low-energy level is not simply the system as viewed at long-distances, however. Instead, there is a \textit{phase transition} in which geometry is supposed to appear. This is called \textit{geometrogenesis}.
\paragraph*{}
Markopoulou (2009, p. 145) states, ``A typical feature of a phase transition is that the degrees of freedom that characterize each of the two phases are distinct [...], with the emergent degrees of freedom being collective excitations of the microscopic ones''. This is not the complete story, however---as we have seen, the idea of a phase transition involves the micro-degrees of freedom somehow arranging themselves such that there is long-range coherence. In this regime the long-wavelength fluctuations dominate, and this is indicated by the correlation length becoming large (diverging). 

\begin{figure}[ht]
\begin{minipage}[b]{0.45\linewidth}
\centering
\includegraphics[width=6cm]{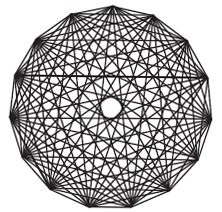}
\caption{High-energy (pre-geometric) phase of quantum graphity. \citep{Markopoulou2008}.}
\label{fig:qgraphhi}
\end{minipage}
\hspace{0.5cm}
\begin{minipage}[b]{0.45\linewidth}
\centering
\includegraphics[width=6cm]{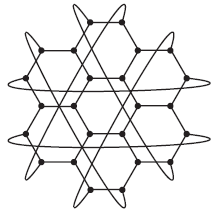}
\caption{Low-energy (post-geometrogenesis) phase of quantum graphity. (Markopoulou, 2008).}
\label{fig:qgraphlo}
\end{minipage}
\end{figure}

\paragraph*{}
Hence, the strategy of QCH is to identify the long-range, coherent collective excitations, and to determine whether or not they behave as though there is a spacetime. Of course, as with the condensed-matter approaches discussed earlier, this is not an easy task, since we do not expect many (if any) of the properties described by the micro-theory to be present in the effective theory. The task is made more complicated by the approach's premise that spacetime is to be defined internally, using only operations that are accessible to parts of the system itself.
\paragraph*{}
A main advantage touted by the pre-geometric approaches compared with the ``traditional'' background-independent theories (again, excluding CDT, which has been particularly successful) is its ability to deal naturally with \textit{dynamics}. It seems that the dynamics of the micro-theory is one aspect that does manifest itself in the effective theory; \citet{Dreyer2004} states that it is the same low-energy excitations (and the interactions of these excitations) of the micro-system that will be used to define both the geometry as well as the energy-momentum tensor, $T_{\mu\nu}$ in the effective theory that results from the geometrogenesis phase transition. If this is the case, then the geometry is necessarily dynamical, owing to the micro-dynamics, and tied to $T_{\mu\nu}$. Markopoulou (2009) thus states that the pre-geometric approaches properly take into account the fact that it is not possible to cleanly separate matter and gravity.
\paragraph*{}
Quantum graphity is another example of the pre-geometric approaches, and also draws explicitly from techniques and theorems of condensed matter physics. What distinguishes this approach is its dynamics: the dynamics is not a movement or ``birthing'' of points, but rather a change in the connections between the points. The connections, represented by the edges of the graph, are able to be in two states ``on'' or ``off'', and, being quantum-mechanical, are able to exist in superpositions of both ``on'' and ``off'' states. 
\paragraph*{}
Basically, we have a finite, quantum system of $N$ points, $a, b, \dots = 1, 2, \dots, N$, with a Hilbert space, $\mathcal{H}_{ab}$ attached to each link $ab$ (as in QCH above), so that a $|1 \rangle$ state on link $ab$ means that the link is ``on'' and that $a$ and $b$ are \textit{local}, and a $|0\rangle$ state means that the link is ``off''.
\begin{figure}[h]
\centering
\includegraphics[height=2.5cm]{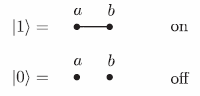}
\end{figure}
$K_N$ is the complete graph on $N$ vertices, i.e. the graph in which there is one edge connecting every pair of points, so that there is a total of $N(N-1)/2$ edges, and each vertex has degree $N-1$. The total state space, $\mathcal{H}$ of the system is,
\begin{equation}
\mathcal{H}=\bigotimes \frac{N(N-1)}{2} \mathcal{H}_{ab}
\end{equation}
A generic state of $\mathcal{H}$ is a superposition of subgraphs of $K_N$. A Hamiltonian operator $H$ assigns energy $E(G)=\langle \Psi_G|H|\Psi_G\rangle$ to a graph, $G$.
\paragraph*{}
The micro-description of the early (pre-geometric) universe is understood as a \textit{complete} graph (as shown in Fig. \ref{fig:qgraphhi}, below), this is a high-energy, maximally-connected state. In such a state, the dynamics is invariant under permutation of the vertices, and, because the entire universe is one-edge adjacent to any vertex, there is no notion of locality. Also, this means there is no conception of a subsystem or local neighbourhood, since the neighbourhood of any point is the entire $K_N$. The micro-degrees of freedom are the states of the links, and these evolve in time under the Hamiltonian. The system at low-energy (i.e. at its ground state) is a graph with far fewer edges than $K_N$: the permutation invariance breaks, and instead translation invariance arises. At this stage, because of the presence of subsystems, locality is able to be defined and we gain a meaningful sense of relational geometry. This is the picture of geometrogenesis, illustrated in Fig.\ref{fig:qgraphlo}, below.

\paragraph*{}
The low-energy graph appears to be low-dimensional, and consideration of the free energy associated with the dominant terms in the dynamics shows that this low-energy state is thermodynamically stable under local perturbations. The model has also succeeded in producing a U(1) gauge symmetry. This is done by utilising the string-net condensation mechanism of Levin and Wen (2005), a theory borrowed from condensed matter physics.\footnote{Detailed in \citet{Konopka2008}.} 
\paragraph*{}
The geometrogenesis phase transition occurs as the universe cools and condenses. This suggests the presence of a reservoir at a tunable temperature, which is a problem since the graph is supposed to be interpreted as the entire universe. There is a question, then, of whether this external temperature is indeed a physical temperature or some other renormalisation parameter.\footnote{In \citet{Hamma2010}, this problem is addressed by introducing additional degrees of freedom (bosonic particles on the vertices) in order to have the system thermalise and reach an equilibrium distribution.}
\paragraph*{}
Finally, since the geometry is to be defined internally in these approaches, it means that time is also internal: defined by the observer, using only operations accessible to parts of the system, and identified with the $g_{00}$ component of the metric tensor that emerges at low-energy. Hence, there is a distinction between the \textit{fundamental} time that appears in the dynamics of the micro (high-energy) system, known as \textit{external time}, and the additional \textit{geometric} time that appears at low-energy, post-geometrogenesis, known as the \textit{internal time}.

\subsection{Emergence in quantum graphity and QCH}\label{subsub:epregeo}
The pre-geometric approaches boast, perhaps, the most radical departure from spacetime. What differentiates them from the other background-independent approaches, such as causal set theory and CDT, is that they do not appeal to a sum-over-histories, but instead build-in the quantum nature of the universe at the fundamental level without any mention of geometry. The importance of the micro-dynamics is emphasised in these theories, given the difficulties in implementing it in many of the other approaches. 

\paragraph*{}
It is the dynamics that is responsible for the emergence of spacetime, with the micro-degrees of freedom producing long-range coherent fluctuations that survive to low-energy and produce an effective spacetime geometry that is itself dynamical. The dynamical nature of the emergent geometry is, of course, desirable, and, since the energy-momentum tensor also emerges from the same coherent fluctuations, the pre-geometric approaches claim to go some way toward \textit{explaining} gravity, rather than just quantising it.
\paragraph*{}
For all this discussion, however, it is unfortunate that it is left unclear exactly what the pre-geometric approaches to quantum gravity have managed to recover in terms of geometry. It is certainly not the metric tensor of GR. Rather, what we know of the low-energy structure is simply that it is a regular lattice structure of low-dimensionality, which has a notion of locality (i.e. that subsystems are able to be defined), translation symmetry, and effective U(1) matter. 

\paragraph*{}
Nevertheless, it is interesting to consider the conception of emergence---or, more precisely, of \textit{novelty} and \textit{autonomy}---associated with the geometrogenesis phase transition, since, if it is to be understood analogously to familiar phase transitions (as discussed in \S\ref{sect:Univers}), we might expect it to feature diachronic novelty. The idea of diachronic novelty in the case of a phase transition, recall, comes from considering the comparison class not as the micro-system, but the system before the phase transition (keeping in mind that prior to the phase transition there is no distinction between ``micro'' and ``macro'').
\paragraph*{}
Geometrogenesis is not supposed to be a second-order phase transition, and so it is not characterised by a critical exponent; hence, the system does not exhibit universality in the sense associated with critical phenomena discussed earlier (\S\ref{sub:Uconcl}). However, there is still a strong sense in which the low-energy physics is autonomous from the high-energy physics, thanks to the symmetry breaking involved in the geometrogenesis phase transition. Pre-geometrogenesis, the system features a permutation symmetry of its graph nodes. This symmetry is broken to a translation symmetry once geometrogenesis occurs. Caravelli and Markopoulou (2011), model the quantum graphity system with an Ising Hamiltonian, and find, using mean field theory that an order parameter for the model is the average valence of the graph.\footnote{Recall from \S\ref{subsect:criticalphen} that the order parameter indicates when a phase transition has occurred.} 
\paragraph*{}
\citet{Quach2012} describe how the phase transition produces domain structures that are analogous to the crystallographic defects that occur when familiar condensed matter systems cool and crystallise. As argued in \S\ref{subsub:symmetrybreaking}, \ref{subsub:SBHOP}, the symmetry-breaking of the phase transition means that there is underdetermination of the high-energy physics given the low-energy physics, since the latter is dependent on the symmetry-breaking pattern \textit{only} (and the symmetry-breaking pattern is able to be ``realised'' by a number of high-energy systems). Hence, there is a strong sense in which the geometry is autonomous from the details of the quantum graphity system. We may treat the symmetry-breaking as a higher-organising principle, being itself responsible for the emergence of geometry.

\paragraph*{}
This account is not completely accurate, however: the involvement of a prior ``external time'' related to the dynamics of the graph complicates the picture. Contrary to the speculation (typically stemming from the problem of time) that time does not exist at the fundamental level (\S\ref{subsub:canon}), the pre-geometric approaches hold that time is fundamental---indeed, they hold that time is fundamental whereas geometry is not. 

\paragraph*{}
This raises the question of how time is to be defined in the absence of geometry: a question that is not addressed in these approaches (the fundamental time appears simply in order that there be fundamental dynamics, where the dynamics is understood as the changes in the states of the vertices). There is the worry, then, that perhaps the existence of a ``background'' spacetime is in conflict with the diffeomorphism invariance of GR. This is similar to the concern raised by the condensed matter models of GR (\S\ref{sub:analogue}), although in the condensed matter approaches the problem is perhaps more severe because there is an underlying spacetime.
\paragraph*{}
Of course, just as it is a possibility in the condensed matter approaches, the diffeomorphism invariance of GR may just be a property of the effective theory. This seems to be the suggestion of the pre-geometric approaches with their fundamental time. The micro-theory has its own time, tied to its dynamics, and undergoes a phase transition in which geometry appears. Once this occurs, and we move to low-energy, we see spacetime described by an effective theory that is effectively diffeomorphism invariant. 

\paragraph*{}
The micro-dynamics is responsible for the appearance of spacetime, and, on this picture, is responsible for the dynamical nature of spacetime: it is still present at low-energies. Nevertheless, it must be that the way in which the micro-time manifests itself at low-energy is such that it is not perceptible as being in violation of the diffeomorphism invariance of GR. Indeed, the suggestion seems to be that the micro-dynamics is somehow responsible for the diffeomorphism invariance of GR at low-energies.
\paragraph*{}
In spite of this, it may still appear conceptually problematic to have spacetime depend on some sort of fundamental time: after all, the point of the pre-geometric approaches is to take background independence as the basic guiding principle. If these approaches are motivated by the need to uphold what is perceived as the most important aspect of GR, then it might seem strange to have the background independence of GR somehow dependent on a background time provided by a background independent theory. 

\paragraph*{}
However, we might instead understand the approaches' pre-geometric starting point as reflecting the idea, suggested in \S\ref{sub:BI}, that background independence is a deeper physical principle than GR and even spacetime itself. If this is the case, then the existence of a fundamental time, from which GR (with its effective diffeomorphism invariance) emerges, is not so distasteful; although the diffeomorphism invariance of GR is only effective (not preserved at high-energy), quantum gravity it still background independent in a different sense.

\section{Conclusion}
The discrete approaches to quantum gravity explicitly appeal to the idea of emergent spacetime, and, in many of the approaches, we may conceive of it in a similar manner as the condensed matter approaches---as a low-energy collective manifestation of very different high-energy degrees of freedom. However, while the condensed matter approaches are background dependent, in the sense of having an underlying spacetime metric, the discrete approaches aim for background independence. The most important and yet most problematic aspect of these approaches is the dynamics: background independence is realised by having a dynamical lattice structure at high-energy, and notions of distance (spatial and temporal) are to be recovered at low-energy by the dynamics of the particular theory. 

\paragraph*{}
The fact that the low-energy degrees of freedom are expected to be very different from the high-energy ones in any of these theories, together with the difficulties in implementing a dynamics for them, means that it is tempting to draw conclusions about the physical content of the theory before we have taken dynamics into account.\footnote{This is also a difficulty that arises in LQG, considered in the next chapter (\S\ref{sect:LQG}).} As Markopoulou (2009, p. 141) states, ``This is analogous to considering a spin system in condensed matter physics and inferring properties of its continuum limit by looking at the spins, independently of the Hamiltonian''.
\paragraph*{}
The problem with dynamics is most severe in causal set theory, where, because of the absence of a dynamics that results in the emergence of a manifold-like spacetime at low-energies, the nature of the high-energy degrees of freedom remains unclear. Nevertheless, some general claims were made about the nature of spacetime that could perhaps emerge from causal set theory, assuming that a Rideout-Sorkin model, plus the two standard conditions---Bell-type causality and discrete general covariance---could in fact produce a good notion of spacetime. The picture suggested is analogous to the idea of treating spacetime as hydrodynamics; at high-energy it is composed of discrete elements, the interactions between which result in an effective spacetime when viewed at low-energy. In other words, at low-energy scales, the system is described by an effective theory framed in terms of very different degrees of freedom. 

\paragraph*{}
Recalling \S\ref{sub:Uconcl}, we would say that there is no diachronic novelty, since there is no phase transition involved: the system changes only at a level of description as we move from high- to low-energies. The degree to which the low-energy description of spacetime is autonomous (or decoupled) from the high-energy physics is unclear, but the fact that different ``sprinklings'' of the elements produce effects that survive at low-energy suggests that spacetime is quite sensitive to the high-energy physics. Indeed, the fact that the approach has so far been unable to recover spacetime might also be taken as evidence for sensitivity. For these reasons, one would be reluctant to use the term ``emergence'' to describe the relationship between causal sets and spacetime.
\paragraph*{}
The sense in which spacetime might be considered emergent from CDT is also unclear. Its indication of a second-order phase transition is exciting because it could potentially be associated with a UV fixed point, and hence provide evidence for Weinberg's asymptotic safety scenario. If this were the case, we would have diachronic novelty and a strong sense of autonomy (good bases for a conception of emergence). However, the involvement of the side-length, $a$, as a parameter upon which the transition depends suggests that the correct interpretation is that spacetime is strongly autonomous with respect to the (unphysical) parameters involved in the mathematical models used to calculate the path integral, rather than autonomous from its micro-structure. This is consistent with the fact that CDT aims to describe spacetime at high-energies, and the high-energy degrees of freedom it refers to are not different from the low-energy ones of GR (although at higher energies the theory suggests a two-dimensional spacetime).
\paragraph*{}
On the other hand, the pre-geometric approaches are premised on there being a physical phase transition in which spacetime geometry comes into being. These approaches do not calculate quantum spacetime using a path integral over classical histories, but attempt to ``build in'' the quantum nature of spacetime at the outset, encoding it in the dynamics of the theory. It is hoped that by doing this, the pre-geometric approaches might avoid the problems in regards to dynamics that many of the other discrete approaches suffer. These approaches feature diachronic novelty, associated with the emergent post-geometrogenesis macro-physics compared with the system before it undergoes the phase transition. 

\paragraph*{}
There is also a strong sense in which spacetime might be said to be \textit{autonomous} from the micro (high-temperature) system, that comes from considering the symmetry-breaking featured in the phase transition as a higher-organising principle. In this case, the geometry might be said to depend only on the symmetry-breaking pattern, rather than the details of the high-energy physics. This interpretation is complicated, however, by the involvement of an ``external time'' tied to the dynamics of the micro-theory, upon which the dynamics of spacetime at low-energies is supposed to depend.

%% file: moo.tex
\chapter{Loop quantum gravity}\label{sect:LQG}

\section{Introduction}\label{sub:LQGintro}
LQG is one of the most well-established quantum gravity programs (along with string theory). Proponents of LQG hold that the most important lesson of GR is the diffeomorphism invariance of the gravitational field, and thus seek to preserve diffeomorphism invariance at the high-energy level of quantum gravity. Like the discrete approaches to quantum gravity discussed in the previous chapter, LQG is non-perturbative and researchers in LQG suggest that the problems with perturbative approaches (i.e. the problems associated with the non-renormalisability of gravity) may be a consequence of the failure of perturbation theory when applied at the scales being considered. And, like the discrete approaches, LQG describes the small-scale structure of spacetime as being discrete. The difference, though, is that some proponents of LQG claim that the discrete nature of spacetime is not postulated from the outset, as it is in the discrete approaches, but rather follows from the theory itself, as a prediction. However, it is not clear that this is indeed the case, since, as they stand, the discrete operators described by LQG are not physical observables.

\paragraph*{}
This chapter is concerned with the conception of spacetime described by LQG. The most well-developed formulation of the theory is based on canonical quantum gravity and uses the Hamiltonian formalism, which, as described in (\S\ref{subsub:canon}), means that 4-dimensional spacetime is split into (3+1)-dimensions; the kinematics of the theory concerns primarily the microstructure of \textit{space}, which is introduced in this section, and discussed in \S \ref{subsub:microspace}. The microstructure of spacetime will be discussed in \S\ref{subsub:microst}. In \S \ref{subsub:semiclas}, I consider the semiclassical limit and the recovery of (large-scale) spacetime. Finally, the idea of emergent spacetime in LQG is discussed in \S\ref{sub:LQGemergence}.
\paragraph*{}
The birth of LQG is generally acknowledged as having occurred in 1987, and began when Ted Jacobson and Lee Smolin rewrote the Wheeler de-Witt equation (\ref{eq:WdW}) using Ashtekar variables, which Abhay Ashtekar \citeyearpar{Ashtekar1986, Ashtekar1987}  had used to construct a novel formulation of GR the year before, building upon the work of Amitabha Sen.\footnote{This discussion draws upon \citet{Ashtekar2004, Carlip2001, Nicolai2007, Rovelli2003, Rovelli2004, Rovelli2008a, Rovelli2011, WuthrichForthcoming}.} Ashtekar variables are \textit{connection} variables rather than metric ones and allow GR to be cast in a form similar to a Yang-Mills theory, and thus in a way that more closely resembles the standard model than it does otherwise.\footnote{Since the standard model is a quantum Yang-Mills theory, meaning it has local (non-Abelian) gauge symmetry.} Jacobson and Smolin \citeyearpar{Jacobson1988} discovered that, when rewritten using the Ashketar variables, the Wheeler-DeWitt equation has solutions that seem to describe loop excitations of the gravitational field. 
\paragraph*{}
As Rovelli (2004, p. 15-16) points out, there is a natural old idea that a Yang-Mills theory is really a theory of loops: recalling Faraday's intuition that there are ``lines of force'' that connect two electric charges and which form closed loops in the absence of charges (the direction of the electric field at any point along such a line is given by the tangent vector at that point). More technically: the relevant mathematical quantity is the holonomy of the gauge potential along the line, and in LQG the holonomy is a quantum operator that creates ``loop states''. A loop state is one in which the field vanishes everywhere except along a single Faraday line.\footnote{The idea of loops and loop states will be discussed again below, and will hopefully become clearer by the end of this section.} In 1987, Carlo Rovelli visited Smolin, and together they defined a theory of (canonical) quantum gravity in terms of loop variables. Doing so, they discovered that not only did the formerly intractable Wheeler-DeWitt equation become manageable and admit a large class of exact solutions, but that there were solutions to all the quantum constraint equations in terms of knot states (loop functionals that depend only on the knotting of the loops). In other words, knot states were proven to be exact physical states of quantum gravity \citep{Rovelli1988a, Rovelli1990}.
\paragraph*{}
Although the idea that loops are the appropriate variables for describing Yang-Mills fields is perhaps a natural one, Rovelli (2004, 2008) explains that it was never able to be properly implemented except within lattice theories (QFT on a lattice).\footnote{For example, Wilson loops, as a gauge-invariant observables obtained from the holonomy of the gauge connection around a given loop, were developed in the 1970s to study the strong interaction in QCD \citep[after][]{Wilson1974a}. They now play an important role in lattice QCD.} One of the problems with using loops in a continuum theory is that loop states on a continuous background are over-abundant; a loop situated at one position on the background spacetime must be considered a different loop state from one that is positioned only an infinitesimal distance away, and so there are an infinite number of loop states on the continuum. Thus, the space spanned by the loop states is non-separable and therefore unsuitable for providing a basis of the Hilbert space of a QFT.\footnote{This will be discussed in more detail in \S\ref{subsub:microspace}.}
\paragraph*{}
Although it is not an obvious matter, \citet{Fairbairn2004} argue that this problem does not arise for a background independent (diffeomorphism invariant) theory, such as GR. The argument is that, if we treat spacetime itself as made up of loops, then the position of a loop state of a QFT is relevant only with respect to other loops, rather than a continuum background spacetime, and so there is no sense in saying that two loops are separated in spacetime. An infinitesimal (coordinate) displacement will not produce a distinct physical state, but only a gauge equivalent representation of the same physical state. Therefore, the size of the state space is dramatically reduced by diffeomorphism invariance; only a finite displacement, which involves a loop being moved across another loop, will represent a physically different state. In the context of GR, the, loop states are thus (arguably) able to provide a basis of the Hilbert space, and the state space of LQG is a separable Hilbert space, $\mathcal{H}_K$ spanned by loop states (Rovelli, 2004, p. 18).\footnote{The subscript $K$ is used in order to signal that this is the \textit{kinematical} Hilbert space of the theory.} Quantum states are represented in terms of their expansion on the loop basis, that is: as functions on a space of loops. 
\paragraph*{}
In a quantum theory, the discrete values of a physical quantity can be found by calculating the eigenvalues of its corresponding operator. In a theory of quantum gravity (where the gravitational field is identified with the geometry of spacetime), any quantity that depends on the metric becomes and operator, and it is by studying the spectral properties of these operators that we can learn about the quantum structure of spacetime. Most significant in LQG is the operator, \textbf{\^{A}}, associated with the area, \textbf{A}, of a given surface, $\mathcal{S}$, and the operator, \textbf{\^{V}}, associated with the volume, \textbf{V}, of a given spatial region, $\mathcal{R}$. 

\begin{figure}[ht]
\begin{minipage}[b]{0.45\linewidth}
\centering
\includegraphics[width=6cm]{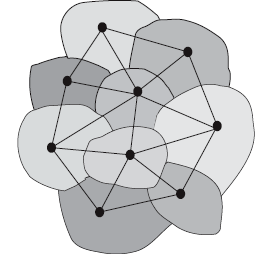}
\caption{Quanta of volume (grey blobs). Adjacent chunks are separated by a surface $\mathcal{S}$ of quantised area. The corresponding spin network graph is overlaid. Each link ``cuts'' one quantised surface $\mathcal{S}$. }
\label{fig:chunksgrey}
\end{minipage}
\hspace{0.5cm}
\begin{minipage}[b]{0.45\linewidth}
\centering
\includegraphics[width=6cm]{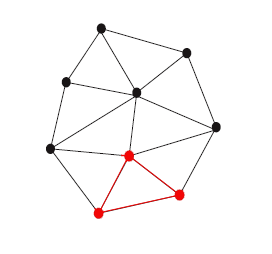}
\caption{Spin network: Nodes represent quanta of volume, which are adjacent if there is a link between them. Connected links form loops, like the one highlighted in red. (Adapted from Rovelli (2004, p. 20)).}
\label{fig:chunksspin}
\end{minipage}
\end{figure}

\paragraph*{}
The area operator \textbf{\^{A}} can be calculated by taking the standard expression for the area of a surface, replacing the metric with the appropriate function of the loop variables, and then promoting these loop variables to operators. An essentially similar procedure can be followed in order to construct \textbf{\^{V}}.\footnote{More precisely: the construction of the area operator first requires the classical expression be regularised, then the limit of a sequence of operators, in a suitable operator topology, be taken. Both \textbf{\^{A}}  and \textbf{\^{V}} have been derived several times using different regularisation techniques \citep[e.g.][]{Ashtekar1997, Ashtekar1997a, Frittelli1996, Loll1995, Loll1996, Rovelli1995}. } 
Both \textbf{\^{A}} and \textbf{\^{V}} are mathematically well defined self-adjoint operators in the kinematical Hilbert space $\mathcal{H}_K$; their spectra, first derived in 1994, were found to be discrete \citep{Rovelli1995}. For instance, the spectrum for \textbf{\^{A}} is given by:
\begin{equation}\label{eq:Aspectra}
\textnormal{\textbf{A}} =8\pi\gamma\hbar G \sum_i \sqrt{j_i(j_i+1)}
\end{equation} 
Where $i=1,\dots, n$, so that $j$ is an $n$-tuplet of half-integers, labelling the eigenvalues, and $\gamma$ is the Immirzi parameter, which is a free dimensionless constant (i.e. not determined by the theory).
\paragraph*{}
The discrete spectra of the area and volume operators implies that the gravitational field is quantised. These quanta of space may intuitively be thought of as ``chunks'' of space, of definite volume given by the eigenvalues of \textbf{\^{V}}. Each chunk (or region, $\mathcal{R}$) can be thought of as bounded by a surface: if two chunks are adjacent to one another (i.e. direct neighbours), then the part of the surface that separates them (i.e. the fence that lies between the two neighbours) is $\mathcal{S}$, of area given by the eigenvalues of \textbf{\^{A}}. This idea is illustrated in Fig. \ref{fig:chunksgrey}, where the grey blobs represent the chunks of space. 
\paragraph*{}
In LQG, this intuitive picture takes the form of abstract graphs called \textit{spin networks}, in which each volume chunk is represented by a node (the black dots in each grey blob in Fig. \ref{fig:chunksgrey}), and each $\mathcal{S}$ separating two adjacent chunks is represented by a link (the lines joining the nodes). The  spin network without the heuristic background illustration is shown in Fig. \ref{fig:chunksspin}. This diagram also enables us to visualise the loops of LQG: they are the links that meet up to enclose white space, for instance the red loop highlighted.

\begin{figure}[h]
\centering
\includegraphics[height=5.36cm]{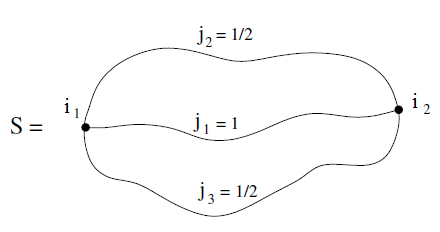}
\caption{A simple spin network with colouring. Labels $i_n$ indicate the quantised volume of the corresponding node, and $j_l$ give the quantised area represented by the corresponding link. (Rovelli, 2004, p. 19).}
\label{fig:spinnet}
\end{figure}

\paragraph*{}
An  spin network graph, $\Gamma$ with $N$ nodes represents a quantum state of space, $|s\rangle$ formed by $N$ quanta of space. The graph has each node $n$ labelled $i_n$, which is the quantum number of the volume (i.e. the volume of the corresponding quanta, or chunk, of space), and each link $l$ labelled $j_l$, which is the quantum number of the area (i.e. the quantised value of the area of $\mathcal{S}$ separating the two adjacent chunks of space being represented by those nodes being linked). The choice of labels is called the \textit{colouring} of the graph. The area of a surface cutting $n$ links of the  spin network with labels $j_i$ $(i = 1,\dots, n)$ is given by the spectrum \ref{eq:Aspectra}.  The spin network $s$ may thus be designated $s=(\Gamma , i_n, j_l)$ as shown in Fig. \ref{fig:spinnet}: these quantum numbers completely characterise and uniquely identify an spin network state. It is worth briefly mentioning (because it will be important in \S\ref{subsub:microspace}) that the labels $j_l$ attached to the links are called \textit{spins}, while the labels $i_n$ are \textit{intertwiners} associated to the nodes.\footnote{The meaning of these terms will not be discussed here, except to say that these quantum numbers are determined by the representation theory of the local gauge group, $SU(2)$. See \citet[][pp. 234--236]{Rovelli2004}.} 

\paragraph*{}
The spin network states $|s\rangle$ provide a basis for $\mathcal{H}$, and represent the general (unmeasured) quantum states of the gravitational field. Since (a region of) physical space is a state in $\Hilb_K$, it is a quantum superposition of abstract spin network states. A loop state is a spin network state in which the graph $\Gamma$ has no nodes, i.e. it is a single loop, and in such a state, the gravitational field has support only on the loop itself, with the direction of the field at any point along the loop given by the tangent vector at that point (again, recalling Faraday's intuition mentioned above). It is important to emphasise that spin networks are \textit{abstract} graphs: spin network states are not quantum states of a physical system \textit{in} space, rather they are the quantum states of physical space itself; only abstract combinatorial relations defining the graph are (physically) significant, not its shape or position in space.\footnote{When referring to ``spin networks'' I mean only the \textit{abstract} graphs. Embedded (i.e. non-abstract) spin networks are of significance, and will be discussed in \ref{subsub:microspace}, where they will be explicitly referred to as embedded spin networks. The abstract spin network states $|s\rangle$ are equivalence classes under diffeomorphism invariance of the embedded spin networks, and are also known as \textit{s}-knots.}
\paragraph*{}
The significance of the spin network states as providing a basis for the Hilbert space of LQG was only realised in 1995 \citep{Rovelli1995a}. However, it turns out that spin networks were only ``rediscovered'' rather than invented in the context of LQG: the spin networks themselves had been created independently many years earlier, by Roger Penrose, based simply on what he imagined quantum space could look like (e.g. Penrose, 1971). It was Penrose who named these graphs ``spin networks'', since their quantum numbers and their algebra resembled the spin angular momentum quantum numbers of elementary particles.

\section{Spacetime in LQG}
Recall that the canonical quantisation program (on which LQG is based) begins with canonical GR, which casts GR as a Hamiltonian system with constraints. The goal of the quantisation procedure is to find the Hilbert space corresponding to the physical state space of theory, and to define operators on the Hilbert space that represent the relevant physical quantities. In LQG, the procedure begins with a classical phase space coordinatised by the ``holonomy'' and its conjugate ``flux'' variable, which are constructed from the Ashtekar connection $A^i_a$ and its conjugate, $E^a_i$ a densitised triad ``electric field''.\footnote{Where $i=1,2,3$ are ``internal'' indices that label the three axes of a local triad, and $a=1,2,3$ are spatial indices. A densitised electric field has $\rho(E)=1$.} 
\paragraph*{}
The geometrical structure of the classical phase space is encoded by the canonical algebra given by the Poisson brackets among these basic variables; in the quantisation, an initial functional Hilbert space of quantum states is defined, and the basic canonical variables are turned into operators whose algebra is determined by their commutation relations, which come from the classical Poisson brackets. These are then used in the construction of the constraints, which, in turn, serve to select a subset of states that correspond to the physical states of the theory.
\paragraph*{}
In LQG, there are three types of constraint: the $SU(2)$ Gauss gauge constraints, which come about from expressing GR as an $SU(2)$ Yang-Mills theory, and are comparatively easy to solve; the spatial diffeomorphism constraints, which stem from diffeomorphism invariance and are hard to solve; and the Hamiltonian constraint (the general form of which is the Wheeler-DeWitt equation (\ref{eq:WdW})), which has not yet been solved. In fact, there are many different forms of the Wheeler-DeWitt equation in LQG, and it is not clear which---if any---is correct. Solving the Gauss constraints and the diffeomorphism constraints gives us the kinematical Hilbert space $\Hilb_K$ (i.e. this is the Hilbert space we obtain from the states which get annihilated by the Gauss and diffeomorphism constraints). The Hamiltonian constraint (its general form being like the Schr\"{o}dinger equation), represents the dynamics of the theory. 
\paragraph*{}
Because of the technical and conceptual difficulties with the Hamiltonian constraint equation, proponents of LQG have sought alternative ways of understanding the dynamics of LQG. Here I will focus on the idea of treating spin networks as ``initial'' and ``final'' states, and the dynamics of the theory being determined by the transition probability amplitudes $W(s)$ between them, i.e. by taking the sum-over-histories approach. This represents a \textit{covariant} formulation of LQG as opposed to the canonical one (although, I should point out that the covariant representation can be approached from different starting points, and the one presented here stems from the canonical formulation). 
\paragraph*{}
The covariant formulation of LQG (also called ``spin foam theory'') is a relatively new area of research and is less-developed than the canonical formulation of LQG. The aim of the formalism is to provide a means of calculating the transition amplitudes in LQG: it does this as a sum-over-histories, where the ``histories'' being summed-over are known as \textit{spin foams}. A spin foam can be thought of as a world-history of a spin network, and represents a ``spacetime'' in the way that a spin network represents a ``space''. These ideas will be discussed in  \S \ref{subsub:microst}.

\subsection{Micro-structure of space: Spin networks}\label{subsub:microspace}
\begin{figure}[h]
\centering
\includegraphics[height=6 cm]{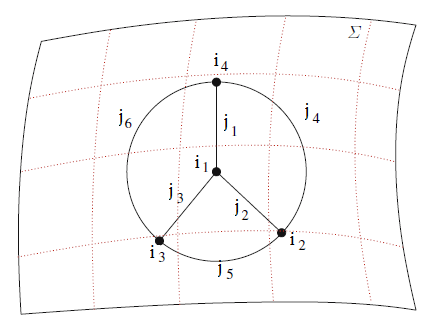}
\caption{A spin network embedded on the spatial hypersurface $\Sigma$ (adapted from \citet[][p. 156]{Nicolai2007}).}
\label{fig:embedsn}
\end{figure}

Spin networks do not start out as abstract graphs: rather, LQG begins with a three-dimensional spatial manifold, $\Sigma$, on which the holonomies and spin networks are defined. The manifold is used to label the  positions of the vertices and edges with coordinates; embedded spin network states are designated $|S\rangle$ (i.e. with a capital $S$ rather than the lowercase $s$ of the abstract spin network states). An embedded spin network is shown in Fig. \ref{fig:embedsn}.

\paragraph*{}
The spin network wave functions only ``probe'' the geometry on one-dimensional sub-manifolds (i.e. along the one-dimensional edges), and are insensitive to the geometry elsewhere on $\Sigma$ \citep[][p. 154]{Nicolai2007}. For any two embedded spin network states, the scalar product is defined as,

\begin{equation}\label{eq:scalar}
\langle \Psi_{\Gamma, j_l, i_n}| \Psi^\prime_{\Gamma^\prime, j^\prime_l, i^\prime_n}\rangle=\begin{cases}0, & \mbox{if }\Gamma\neq\Gamma^\prime \\ \int\prod_{j_l\in\Gamma}\mathrm{d}h_{j_l}\; \bar{\psi}_{\Gamma, j_l, i_n}\psi^\prime_{\Gamma^\prime, j^\prime_l, i^\prime_n}, & \mbox{if } \Gamma=\Gamma^\prime \end{cases}
\end{equation}
where $\Psi$ are the spin network wave functions, $\Gamma$ the spin network graphs, $j_l$ are the spins attached to the edges (links) and $i_n$ the intertwiners associated to the nodes of the graph.

\paragraph*{}
The \textit{pre-kinematical Hilbert space}, $\Hilb_{K^*}$ is defined using the scalar product (\ref{eq:scalar}), which induces a peculiar discretisation (one entirely different from the discreteness of a lattice or the naive discretisation of space).\footnote{The full kinematical Hilbert space, $\Hilb_K$ (which is separable) is defined once the diffeomorphism constraint has been implemented, as will be explained below.} The resulting topology is similar to the discrete topology of the real line with countable unions of points as the open sets. Because the only notion of ``closeness'' between two points in this topology is whether or not they are coincident, \textit{any} function is continuous in it \citep[][p. 156]{Nicolai2007}. Thus, it is already difficult to see how it would be possible to recover any conventional notion of continuity in LQG. The effect of the scalar product (\ref{eq:scalar}) means that non-coincident states are orthogonal, and the expectation values of operators that depend on some parameter do not vary continuously as the parameters upon which they depend are continuously varied. 

\paragraph*{}
This leads to the traditional problem mentioned above, of the spin network basis being ``too large'': any operation which moves the graphs around continuously corresponds to an uncountable sequence of mutually orthogonal states in $\Hilb_{K^*}$. The Hilbert space does not admit a countable basis and so is non-separable. No matter how ``small'' the deformation of the graph in $\Sigma$, the associated elements of $\Hilb_{K^*}$ always remain a finite distance apart---this means that continuous motion in ``real'' space gets mapped to highly discontinuous motion in $\Hilb_{K^*}$.

\paragraph*{}
The separable kinematical state space, $\Hilb_K$, of the theory comes about once the diffeomorphism constraint has been implemented. This involves factoring out the gauge equivalent loop representations according to diffeomorphism invariance (i.e. any two graphs can be deformed into one another): the uncountable, ``too large'' basis is reduced once gauge redundancy is taken into account, and the non-separable Hilbert space becomes separable (Rovelli, 2008).  As mentioned in the introduction to this chapter, the loop representation of GR (as a Yang-Mills field theory) is only of value because of the theory's diffeomorphism invariance. Implementing diffeomorphism invariance removes the significance of the manifold $\Sigma$, and the natural basis states are abstract spin network states  (also called \textit{s}-knots), which are equivalence classes (under diffeomorphism invariance) of the embedded spin networks. Construction of the constraints makes use of the operators corresponding to the relevant physical observables. 

\paragraph*{}
The important operators are \textbf{\^{A}}, which measures the area of a two-dimensional surface, $S\subset\Sigma$, and \textbf{\^{V}}, which measures the volume of a three-dimensional subset of $\Sigma$. These operators, however, \textit{cannot} be classed as physical observables, since they do not commute with the constraints. In particular, being defined for surfaces and regions on $\Sigma$, the area and volume operators are not invariant under the transformations generated by the diffeomorphism constraint. Because \textbf{\^{A}} and \textbf{\^{V}} are not diffeomorphism invariant, \citet[][p. 423]{Rickles2005} argues that these quantities are not measurable unless they are \textit{gauge fixed} and taken to correspond to the area and volume of some \textit{physically defined} surface or region. \citet[][p. 157]{Nicolai2007} make a similar point, though emphasising the failure of the operators to commute with the Hamiltonian, and the necessity of the inclusion of matter in defining a physical surface or region.

\paragraph*{}
\citet{Dittrich2009} argue that if the discrete spectra of the area and volume operators are to be taken as representing physical discreteness of geometry, then we need to investigate whether the kinematical discreteness of the spectra survives at the gauge-invariant level once the operators have been ``turned into'' gauge-invariant quantities. This process depends on the mechanism used in order to turn the gauge-dependent operators into gauge-invariant ones, as well as on the interpretation of generally covariant quantum theory---a point that \citet{Rovelli2007} emphasises. In standard quantum theory, a quantity is predicted to have discrete values if the corresponding quantum operator has a discrete spectrum. Carrying over this idea to generally-covariant quantum field theory---a framework that is not fully developed or understood, but which LQG is modelled on---is not unambiguous because the distinction between the kinematics and the dynamics of such a theory is not clear cut \citep[][p. 1]{Rovelli2007}. 

\paragraph*{}
The debate between \citet{Dittrich2009} and \citet{Rovelli2007} demonstrates that the discrete spectra of \textbf{\^{A}} and \textbf{\^{V}} do not necessarily represent physical discreteness of Planck-scale geometry. \citet{Dittrich2009} present a simplified, non-LQG (not quantum gravitational) case-study in which the discreteness of the kinematical operators' spectra does not survive once the operators have been turned into gauge-invariant ones. \citet{Rovelli2007} argues that not only is this example not analogous to the case in LQG, but that Dittrich and Thiemann utilise a particular interpretation of generally covariant quantum theory---one that Rovelli does not endorse. 

\paragraph*{}
On this interpretation, it is possible in principle for a ``partial observable'' \textit{\^{f}}, representing the area of a coordinate surface, to have a discrete spectrum, while the corresponding ``complete observable'', \textit{\^{F}} representing the area of a physically defined surface, have continuous spectrum. Thus, on this interpretation, there is no ``guarantee'' that the discrete spectra of the operators of LQG, understood as partial observables, indicates physical discreteness of spacetime. On a different interpretation (the one that Rovelli recommends), the physical quantisation depends on the spectra of the kinematical operators in $\Hilb_{K^*}$, and so it ``follows immediately'' that a physical measurement of the area or volume operators of LQG would yield a discrete value.

\paragraph*{}
What is needed now, if we follow Rovelli's line of reasoning, is an examination of the motivations for preferring one interpretation over the other. Unfortunately, I cannot undertake this here. It is enough to say that it is at least plausible that the discrete spectra of the operators represent physical predictions of LQG.\footnote{Even granting this, though, we should consider the meaning of a prediction that cannot be tested---if the discrete spectra are physical predictions, then LQG states that \textit{if} we were able to probe extremely high energy scales \textit{and} had a means of detecting the discrete structure of spacetime itself at these energies, then we would find spacetime to be discrete. Needless to say, this is a long shot. Also, we must remember that, because LQG is not based on any physical data, the prediction is a consequence of a particular combination of principles and assumptions \citep{CrowtherForthcoming}. Because of this, the distinction between \textit{postulated} versus \textit{predicted} discreteness is perhaps not an interesting one to push (as we might be tempted to do in comparing LQG with the ``discrete'' approaches, where discreteness is explicitly acknowledged as an assumption or principle of the theory).} If the spectra of \textbf{\^{V}} and \textbf{\^{A}} are physical predictions, then, according to LQG, space itself is discrete and combinatorial, and, because the theory has a natural cutoff at the Planck scale, there are no ultraviolet divergences. On the other hand, if the discrete spectra of the kinematical operators is not physical, then they might be understood in a similar way to the discrete elements in CDT: part of the formalism of the theory, but not themselves evidence of spacetime discreteness. In any case, the structures described by LQG are very different from our familiar conception of space.

\paragraph*{}
Additionally, there is another, perhaps even stronger, way in which the spin networks differ from the emergent spacetime they are supposed to underlie: the fundamental relation of \textit{adjacency} which is meant to correspond to the notion of two objects being ``nearby'' to one another in LQG does not, typically, translate into this notion in the emergent spacetime. Recall that two nodes being linked by an edge in a graph represents two adjacent quanta of space. The idea of spacetime emerging from the more basic spin networks means that there is a ``mapping'' of spin network nodes onto events in the emergent spacetime. 

\paragraph*{}
However, two nodes that are adjacent in the basic (high-energy) description can be arbitrarily large distances away from one another as measured in the emergent metric---in other words, they will, in general, not be mapped to ``nearby points'' in the emergent spacetime.\footnote{\citet{Huggett2013, WuthrichForthcoming} both emphasise this point and the associated problems for our understanding of locality.} The fact that the adjacency relations described by the fundamental spin networks do not typically feature in the emergent spacetime (i.e. are not ``translated'' into the corresponding spatiotemporal relations at low-energy) means that many of them (i.e. all those adjacencies which do not get translated into Planck-sized neighbourhoods in the spacetime) are suppressed at low-energy.\footnote{Again, \citet{Huggett2013, WuthrichForthcoming} also make this point, calling the suppressed effects ``non-localities''.}

\subsection{Semiclassical limit: Weaves}\label{subsub:semiclas}
Finding the low-energy limit of LQG has proven very difficult, and all attempts to recover GR from LQG have so far been unsuccessful. One obvious handicap is the fact that all such attempts have been confined to working with the kinematical Hilbert space $\Hilb_K$, rather than the physical Hilbert space of the theory. Thus, there are questions regarding both the viability and the meaningfulness of relating the kinematical states to corresponding classical spacetimes (or spaces), as \citet{WuthrichForthcoming} notes. 

\paragraph*{}
The most prominent of the attempts to construct semiclassical states (i.e. states in which the quantum fluctuations are minimal and the gravitational field behaves almost classically) is based on \textit{weave states}, which were first introduced by \citet{Ashtekar1992}.\footnote{Alternative methods aiming to overcome the shortcomings of weave states have, and are still, being explored; for a discussion, see \citet[][\S II.3]{Thiemann2001} or \citet[][\S 11]{Thiemann2007}.} The intuitive idea is captured by analogy: at familiar scales, the fabric of a t-shirt is a smooth, two-dimensional curved surface, but when we examine it more closely, we see that the fabric is composed of one-dimensional threads woven together.\footnote{The analogy comes from \citet{Ashtekar1992}.} The suggestion is that LQG presents a similar picture: while the geometry of space at large-scales is a three-dimensional continuum, at high-energy it is revealed to be a very large lattice of Planck-sized spacing (i.e. a spin network with a very large number of nodes and links).

\paragraph*{}
Consider a classical three-dimensional gravitational field $e$, which determines a three-dimensional metric $g_{ab}(\vec{x})=e^i_a(\vec{x})e_{ib}(\vec{x})$, and a macroscopic three-dimensional region $\mathcal{R}$ of spacetime with this metric, bounded by the two-dimensional surface $\mathcal{S}$ (the values of area and volume being large compared to the Planck scale). It is possible to construct an (embedded) spin network state $|S\rangle$ that approximates this metric at a length scale $\Delta\gg l_P$, where $l_P$ is the Planck length. To do this involves selecting spin network states that are eigenstates of the volume and area operators for the region $\mathcal{R}$ and the surface $\mathcal{S}$ with eigenvalues that approximate the corresponding classical values for the volume of $\mathcal{R}$ and area of $\mathcal{S}$ as given by $e$. The classical value for the area \textbf{A} of a surface $\mathcal{S}\subset \Mani$ and the classical value for the volume of a region $\mathcal{R}\subset \Mani$ with respect to a fiducial gravitational field $c^i_a$ are given by,\footnote{This presentation is based on \citet[][p. 26]{WuthrichForthcoming}.}

\begin{equation}\label{eq:A}
\textnormal{\textbf{A}}[e,\mathcal{S}]=\int|d^2\mathcal{S}|
\end{equation}

\begin{equation}\label{eq:Vol}
\textnormal{\textbf{V}}[e,\mathcal{R}]=\int|d^3\mathcal{R}|
\end{equation} 
where the relevant measures for the integrals are determined by $c^i_a$. 

\paragraph*{}
Now, we require that  $|S\rangle$ is an eigenstate of \textbf{\^{A}} and \textbf{\^{V}}, with eigenvalues given by (\ref{eq:A}) and (\ref{eq:Vol}), respectively, up to small corrections of the of  $l_P/\Delta$,

\begin{equation}\label{eq:AS}
\textnormal{\textbf{\^{A}}}(\mathcal{S})|S\rangle=(\textnormal{\textbf{A}}[e,\mathcal{S}]+\mathcal{O}(l^2_P/\Delta^2))|S\rangle
\end{equation}
\begin{equation}\label{eq:VR}
\textnormal{\textbf{\^{V}}}(\mathcal{R})|S\rangle=(\textnormal{\textbf{V}}[e,\mathcal{S}]+\mathcal{O}(l^3_P/\Delta^3))|S\rangle
\end{equation}

If an embedded spin network state $|S\rangle$  satisfies (\ref{eq:AS}) and (\ref{eq:VR}), then it is a \textit{weave state} of the metric $g_{ab}$.\footnote{Although embedded, rather than abstract, spin network graphs are used, this definition of weave states is able to be carried over to the diffeomorphism-invariant level of abstract spin network states ($s$-knots) without issue. If we introduce a map $P_{Diff}: \Hilb_(K^*)\rightarrow\Hilb_K$, which projects states in the pre-kinematical Hilbert space into the same elements of the kinematical Hilbert space, then the state $|s\rangle = P_{Diff}|S\rangle$ is a weave state of the classical three-geometry $[g_{ab}]$, i.e. the equivalence class of three-metrics $g_{ab}$, just in case $|S\rangle$ is a weave state of the classical three-metric $g_{ab}$ \citep{WuthrichForthcoming}.} At length scales of order $\Delta$ or larger, the weave state is a good approximation  to the corresponding classical geometry, as $|S\rangle$ determines the same volumes and areas as $g_{ab}$. At length scales much smaller than  $\Delta$, however, the quantum fluctuations become relevant, and the weave state can no longer be considered a valid semiclassical approximation. 

\paragraph*{}
Finally, it is worth pointing out that (\ref{eq:AS}) and (\ref{eq:VR}) do not determine the state $|S\rangle$ uniquely for a given three-metric $g_{ab}$. This is because (\ref{eq:AS}) and (\ref{eq:VR}) involve quantities that are averaged over the macroscopic surface $\mathcal{S}$ and region $\mathcal{R}$. There are many different spin network states that can represent these averaged values, whereas there is only one classical metric that corresponds to these values. In this way, the actual microstate of a given macroscopic geometry is \textit{underdetermined} by the equations (\ref{eq:AS}) and (\ref{eq:VR}). However, as Rovelli (2004, p. 270) points out, the generic quantum state of the macroscopic spacetime is not a weave state, but a superposition of weave states--- in other words, what LQG describes is not a lattice structure for spacetime, but a ``cloud of lattices''. 

\subsection{Micro-structure of spacetime: Spin foams}\label{subsub:microst}
The preceding two sub-sections have described only the kinematics of the theory: since the spin networks are based in the kinematical Hilbert space rather than the physical Hilbert space, they represent microstates of \textit{space} rather than \textit{spacetime}. There is a covariant version of LQG which aims to discover the dynamics of the theory without engaging with the Hamiltonian constraint of canonical LQG. This formulation, which is known as \textit{spin foam theory} describes the micro-structure of spacetime as a \textit{spin foam}, which is a history of spin networks. Presently, this theory is also incomplete and its relation to canonical LQG not known. It should also be noted that the sum-over-histories approach described in this sub-section is not representative of the full covariant LQG program, in that its starting-point draws from the concepts and results of canonical LQG.

\begin{figure}[h]
\centering
\includegraphics[height=3 cm]{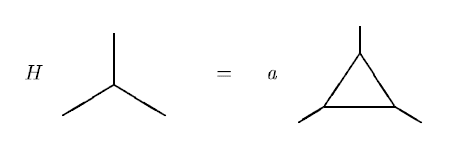}
\caption{Scheme of the action of $H$ on a node of a spin network. (Rovelli, 2004, p. 25).}
\label{fig:Haction}
\end{figure}

\paragraph*{}
Recall that in quantum mechanics, a complete description of the dynamics of a particle is provided by the transition probability amplitudes, $A$, defined as,
\begin{equation}\label{eq:amp}
A=\langle \psi^\prime |e^{\frac{i}{\hbar}H_0(t-t^\prime)}| \psi \rangle
\end{equation}
where $|\psi\rangle$ is the initial quantum state prepared at $t$, and $|\psi^\prime \rangle$ is the final state of the system, measured at $t^\prime$, and $H_0$ is the Hamiltonian operator. Following Feynman, this amplitude can be calculated as a sum-over-paths between the ``initial'' and ``final'' states. The same is true in LQG, where the dynamics of the theory may be described entirely by the spin network transition amplitudes $W(s^\prime, s)$, governed by the Hamiltonian operator $H$, which is defined on the space of the spin networks. The space of solutions of the Wheeler-DeWitt equation is  the \textit{physical Hilbert space}, denoted $\Hilb$. There is an operator $P:\Hilb_K \rightarrow \Hilb$ that projects $\Hilb_K$ on the space of solutions of the Wheeler-DeWitt equation.\footnote{(Rovelli, 2004, \S 1.2.3)} The transition amplitudes between an ``initial'' spin network state $|s^\prime \rangle$ and the ``final'' spin network state $|s\rangle$ (recalling that there is no external time variable in the theory) are the matrix elements of the operator $P$,
\begin{equation}\label{eq:LQGamp}
W(s, s^\prime)=\langle s|P|s^\prime\rangle_{Hilb_K} = \langle s|s^\prime\rangle_{\Hilb}
\end{equation}
The Hamiltonian operator $H$ (in all its different versions in LQG) acts only on the nodes of the spin network graph; in the vicinity of a node, the action of $H$ upon a generic spin network state $|s\rangle$, is to change the topology and labels of the graph. Typically, $H$ splits a node into three nodes and multiplies the state by a number $a$ that depends on the labels of the spin network around the node. This is illustrated in Fig. \ref{fig:Haction}.

\begin{figure}[h]
\centering
\includegraphics[height=6 cm]{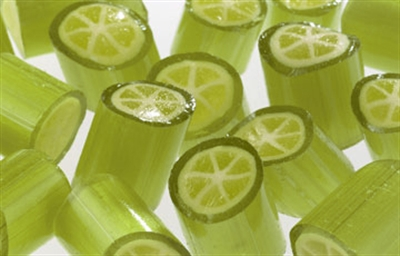}
\caption{``Lime'' rock candy: A slice width-wise reveals a cross-section with a lime-shaped pattern, which represents a spin network state $s_1$. The face at the ``bottom'' of the stick is $s^\prime$, and the one at the top is $s$. The whole length of the stick represents a history of spin networks, i.e. a spin foam, $\sigma = (s, s_N, \dots, s_1, s^\prime)$. (Of course, every slice of rock candy will reveal a cross-section of essentially the same pattern, whereas ``slices'' of a spin foam would reveal different shaped spin networks, since the spin networks are transformed under the action of $H$).}
\label{fig:rockcandy}
\end{figure}

\paragraph*{}
The transition amplitude $W(s, s^\prime)$ can be represented as a sum-over-histories: a representation that follows from summing over different histories of sequences of actions of $H$ that send $s^\prime$ to $s$.\footnote{Although Rovelli (2004, p. 26) states that this is just one of several ways to arrive at the sum-over-histories representation of $W$.} The histories (of spin networks) being summed over are spin foams; a history of going from $s^\prime$ to $s$ is a spin foam, $\sigma$, bounded by $s^\prime$ and $s$. The heuristic way to picture a spin foam is to imagine a 4-d spacetime in which the graph of a spin network $s$ is embedded. If this graph moves along ``upwards'' through the ``time'' coordinate of the 4-d spacetime, then it ``sweeps out'' a ``worldvolume'' (a 3-dimensional version of a worldline). Actually, it's sort of like rock candy, shown in the picture, Fig. \ref{fig:rockcandy}.

\begin{figure}[h]
\centering
\includegraphics[height=6 cm]{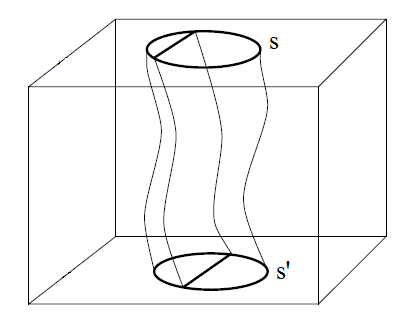}
\caption{A simple spinfoam: the worldsheet of a spin network (``colouring'' of the faces and edges not indicated). (Adapted from Rovelli, 2004, p. 325).}
\label{fig:spinfoam}
\end{figure}

\paragraph*{}
More schematically, Fig. \ref{fig:spinfoam} illustrates the worldsheet of a spin network that's shaped like $\theta$. The surfaces traced out by the links of the spin network graph are called \textit{faces}; the worldlines traced out by the nodes of the spin network graph are called \textit{edges}. A spin foam, $\sigma$ also includes a colouring, where faces are labelled by the area quantum numbers $j_l$ and edges are labelled by the volume quantum numbers $i_n$.

\paragraph*{}

\begin{figure}[h]
\centering
\includegraphics[height=5 cm]{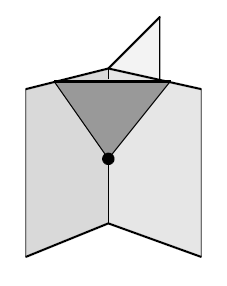}
\caption{Vertex of a spin foam. (Rovelli, 2004, p. 325).}
\label{fig:vertex}
\end{figure}

\begin{figure}[h]
\centering
\includegraphics[height=6 cm]{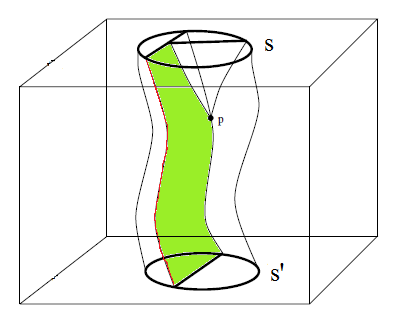}
\caption{Spinfoam with one vertex. One \textit{face} has been tinted green, and one \textit{edge} is shown in red.  (Adapted from Rovelli, 2004, p. 326).}
\label{fig:spinvert}
\end{figure}

When $H$ acts on a node, it results in the corresponding edge of the spin foam to branch off into three edges, in the ``3-dimensional'' version of the action shown in Fig. \ref{fig:Haction}. The point where the edges branch is called a \textit{vertex}, $v$, as shown in Fig. \ref{fig:vertex}. A spin foam with a vertex is shown in Fig. \ref{fig:spinvert}.

\paragraph*{}
A spin foam is a Feynman graph of spin networks, however it differs from usual Feynman graphs in that it has one additional structure: while Feynman graphs have edges and vertices, spin foams have edges, vertices and faces. In the perturbative expansion of $W(s, s^\prime)$ , each spin foam $\sigma$, bounded by $s^\prime$ and $s$, is weighted by an amplitude which is given by (a measure term $\mu(\sigma)$-times) the product over the vertices, $v$ of a vertex amplitude, $A_v(\sigma)$. The vertex amplitude is determined by the matrix elements of $H$ between the incoming and outgoing spin networks and depends on the labels of the faces and the edges adjacent to the vertex (Rovelli, 2004, pp. 26--27). This is analogous to the amplitude of a standard Feynman vertex, which is determined by the matrix element of the Hamiltonian between the incoming and outgoing states.
\paragraph*{}
The sum-over-histories is thus,
\begin{equation}\label{eq:spinsum}
W(s, s^\prime)=\sum_{\sigma}\mu(\sigma)\prod_vA_v(\sigma)
\end{equation}
Just as Feynman's sum-over-histories can be interpreted as a sum over different possible classical paths that a particle might take between two points, so too the spin foam sum (\ref{eq:spinsum}) can be interpreted as a sum over spacetimes. And yet, although Feynman's sum-over-histories can be interpreted as a sum over different possible particle trajectories, we know (in quantum theory) that there are no classical trajectories. The sum-over-histories is not itself a history. Although a single spin foam can be thought of as representing a spacetime, the theory states that spacetime is not a single spin foam, but a sum over spin foams. For this reason, Rovelli (2004, p. 31) states that the notion of spacetime disappears in quantum gravity in the same way that the notion of a particle trajectory disappears in the quantum theory of a particle. This interpretation of spin foam theory may also be applicable to the discrete quantum gravity approaches which utilise a sum-over-histories. 

\paragraph*{}
\citet[][p. 175]{Nicolai2007} distinguish between two types of discrete quantum gravity approaches which utilise a sum-over-histories (i.e. approaches which attempt to define a discretised path integral in quantum gravity). The first group represents those approaches where spacetime is approximated by a fixed number of simplices and the integration is performed over all edge lengths: quantum Regge calculus is an example. The second group represents those approaches, including causal dynamical triangulations, where the simplices are assigned fixed edge lengths, and the sum is taken over different triangulations while keeping the number of simplices fixed (thus changing the ``shape'' of the triangulation but not its ``volume''). Spin foam theory falls into the former category, along with quantum Regge calculus, since, in the first step of the procedure---which is calculating the partition function for a given spin foam---all spins are summed over (for the given spin foam), but there is no addition, removal or replacement of edges, vertices or faces. 

\paragraph*{}
In the second step (which aims to recover the continuum limit), the sum is taken over all spin foams, as in (\ref{eq:spinsum}). The method by which to perform this step is not formally agreed upon; \citet{Nicolai2007} suggests that one way of doing it would be to weight each spin foam term in the sum according to its ``shape'', in order to achieve formal independence of the triangulations. This would thus resonate with CDT (\S\ref{sub:CDT}), where, recall, the spacetime obtained was autonomous from the 4-simplices used to approximate the path integral, and so \citet{Nicolai2007} state that we might interpret such a spin foam model as a hybrid of the two classes just distinguished.\footnote{Note that, in spin foam theory, this method of constructing the sum would mean a sum over spin foams with different numbers of simplices and different edge lengths, which is not how it is done in CDT.} 

\paragraph*{}
A key difference between CDT and spin foam theory is the ontological interpretation of the two approaches. In spin foam theory, following LQG, the discrete elements described by the theory are interpreted realistically, as the ultimate constituents of spacetime.\footnote{Although, as described above, there may be reason to question this interpretation of LQG.} Recall that in CDT, however, the 4-simplices are taken simply to be mathematical tools that aid in the regularisation procedure used to define the integral---at the end of calculations, the aim is to remove the discretisation and recover a continuum theory (as in lattice QFT). Hence, spacetime according to CDT (and quantum Regge calculus) is not fundamentally discrete.\footnote{\citet{OritiForthcoming} describes some further differences between LQG, Regge calculus and CDT, and explores what they might teach us about the fundamental nature of space and time, in combination with the \textit{group field theory} (GFT) approach. GFT is essentially similiar to LQG and spin foam models, but with a key advantage in its definition of the dynamics---in particular, it prescribes a strict means by which to calculate the weights for the terms in the path integral.} The fact that the discrete elements of space are interpreted realistically in LQG is the reason why the continuum limit cannot be recovered as it is in the other theories: the ``lattice spacing'' cannot be taken to zero (as described above in \S\ref{subsub:semiclas}).

\section{Emergence}\label{sub:LQGemergence}
Although LQG is incomplete and its physical Hilbert space undefined, there are still some potential bases for a conception of emergence in the theory. Interestingly, there does not appear to be a conception of emergence that might be based on the idea of a limiting relation in the theory (or, at least, not based on a limiting relation alone). This is because macroscopic geometry is not recovered in the limit as the density of the weave (lattice) of loops goes to infinity. Intuitively, of course, it seems as though it would be the case that the continuum could be approximated in this way---indeed, originally in LQG, before the spectra of the area and volume operators had been derived (and, hence, before the significance of the spin network basis had been revealed), this was believed to be the case. The limiting procedure was thought to run analogously to that in conventional QFT, where a continuum theory is defined by taking the limit of a lattice theory, as the lattice spacing $a$ goes to zero. However, when the limit of the corresponding ``loop constant'' (analogous to the lattice spacing) in LQG was taken to zero, surprisingly, there was no increase in the accuracy of the LQG approximation to macroscopic geometry (Rovelli, 2004, p. 269).  

\paragraph*{}
The reason the theory fails to approximate a smooth geometry in the limit as the lattice spacing goes to zero is that the physical density of the loops does not increase in this limit. Instead, what occurs is that the eigenvalues of the area and volume operators increase, meaning that the areas and volumes in the region being studied grow larger. In other words, the loop density remains constant because we look at greater volumes. If we believe that area and volume are quantised, then this result can be readily interpreted: there is a minimum size for the loops, and thus, a minimal physical scale.  The theory refuses to approximate a smooth geometry as the loop constant is taken below $l_P$ because there is no physical length scale below $l_P$. Thus, such a limit is unable to serve as the means by which to recover spacetime from its fundamental spin network structure. This seems to suggest that a conception of emergence based on the idea of a limiting relation (e.g. that of \citet{Butterfield2011a, Butterfield2011b}) will not be applicable in LQG.

\paragraph*{}
\citet{WuthrichForthcoming}, however, suggests that perhaps the failure of the limiting procedure to recover spacetime is due to the fact that it is only capable of representing one of the two necessary transitions involved in the recovery process. The process \citet{WuthrichForthcoming} proposes is comprised of two steps: firstly, there is an approximating procedure which turns the quantum states into semiclassical ones, and, secondly, there is the limiting procedure which relates the semiclassical states to the phase space of the classical (i.e. the ``emergent'') theory. The associated notion of emergence comes from \citet{Butterfield1999, Butterfield2000}, where a theory $T_1$ emerges from another theory, $T_2$, \textit{iff} $T_1$ can be arrived at from $T_2$ by either a limiting procedure or an approximation procedure, or both. As discussed above (\S\ref{sub:emerphysics}), this is the physicists' sense of emergence, and is not concerned with capturing the ideas of novelty and autonomy that are important in defining a conception of emergence. 

\paragraph*{}
A \textit{limiting procedure} is defined as taking the mathematical limit of some physically relevant parameter(s) in the underlying theory $T_2$ in order to recover the emergent theory $T_1$. An \textit{approximation procedure} is defined as the process of either neglecting some physical magnitudes, and justifying such neglect, or selecting a proper subset of states in the state space of the approximating theory, and justifying such selection, or both, in order to arrive at a theory whose values of physical quantities remain sufficiently close to those of the theory to be approximated \citep{Butterfield1999}. In LQG, \citet{WuthrichForthcoming} imagines the limiting procedure to be something akin to the one mentioned above, where the semiclassical weave states are mapped to classical spacetimes. The weave states themselves are supposed to be arrived at via an approximation procedure. 

\paragraph*{}
The method of constructing weave states, described above, fits the definition of an approximation procedure, since they must be carefully selected to include only those states which are peaked around the geometrical values (of area and volume) determined by the fiducial metric $e^i_a$. \citet[][p. 26]{WuthrichForthcoming} says that this can be achieved either by neglecting all those operators constructed from connection operators (since the ``geometrical'' eigenstates are maximally spread in these operators), or, if this cannot be justified (as the approximation procedure requires), then only the seimiclassical states that are peaked in both the connection and the triad basis, and peaked in such a way that they approximate classical states, should be considered. This approximation procedure is taken to represent whatever the physical mechanism is that drives the quantum states to semiclassical ones. Although LQG makes no reference to any such mechanism, we can suppose (as \citet{WuthrichForthcoming} does), that the physical mechanism justifying this approximation procedure is \textit{decoherence}. \citet[][p. 23]{WuthrichForthcoming} claims, based on arguments by \citet{Landsman2006}, that this approximation procedure is necessary in addition to the limiting procedure because not even the $\hbar\rightarrow 0$ limit can resolve a quantum superposition into a classical state.

\paragraph*{}
The $\hbar\rightarrow 0$  limit is the ``textbook'' means of justifying the use (or explaining the success) of ``older'', classical, theories for large orbits and low energies. The fact that it is paired with the idea of decoherence suggests that the limiting procedure could be interpreted as representing the micro/macro transition (a rescaling of the theory) while decoherence represents the quantum/classical transition---both being necessary for an account of emergence. This interpretation seems to accord with the claim that the  $\hbar\rightarrow 0$  limit has the $N\rightarrow\infty$ limit as a special case; while the $N\rightarrow\infty$ limit implies moving to a large system of many particles, the $\hbar\rightarrow 0$  limit means moving to a description scales where $\hbar$ can be treated as negligible. This is certainly what is being indicated by the weave analogy in LQG. 

\paragraph*{}
It is worth emphasising again that both the idea of decoherence and the $\hbar\rightarrow\infty$ limit are \textit{external} to the theory itself: they have been imposed in an attempt to have LQG match up with low-energy classical physics, including GR. Decoherence and the $\hbar\rightarrow\infty$ limit are among the traditional means by which quantum theories are shown to ``reduce'' to classical physics (although, again, the former represents a physical mechanism, and its interpretation more controversial than the mathematical limit in this case), and it would be distressing if they did not work in the context of LQG,  given that the theory itself does not offer any ``natural'' means by which to recover spacetime: LQG does not (on its own, without the additional assumption of decoherence) explain how or why some states (those that are able to be ``mapped'' to classical states) are ``selected''.\footnote{As explained in the Introduction (\S\ref{subsub:deco}), there is a strong possibility that quantum gravity will provide some insight into the physical means (mechanism) by which quantum superpositions are resolved into classical states (e.g. decoherence), even in familiar quantum mechanics. LQG does not, as it stands, offer such insight, even though this physical mechanism is likely to play a role in the emergence of spacetime from LQG.} 

\paragraph*{}
However, this might be expected, given that LQG is a quantisation of GR; perhaps LQG may be conceived as a sort of ``stepping stone'', offering us access to the information contained in quantised GR, without (uniquely) capturing the micro-dynamics. On such an interpretation, the recovery of large-scale physics might be less important than making predictions---an interpretation along the same lines as treating GR as an EFT, as in (\S\ref{sub:GREFT}).

\paragraph*{}
Leaving these concerns aside, though, we can begin to sketch how a conception of emergence based in the \textit{underdetermination} of the underlying physics given the emergent theory could apply in LQG. In fact, even more generally, it seems as though the conception of emergence advocated by \citet{WuthrichForthcoming}, based in the ideas of an approximation procedure plus a limiting procedure, leads to an underdetermination of the ``more basic'' quantum description by the emergent classical one. 

\paragraph*{}
The idea of emergence that W\"{u}thrich utilises is inspired by \citet{Landsman2006}, who argues that while neither the limiting procedure nor decoherence is sufficient, on its own, for understanding how the classical picture emerges from the quantum world, together these procedures indicate that it comes from ignoring certain states and observables in the quantum theory. ``Thus the classical world is not created by observation (as Heisenberg once claimed), but rather by the lack of it'' \citep[][p. 417]{Landsman2006}. On this account, the classical realm is correlated with certain ``classical'' states and observables, those which are robust against coupling to the environment, and which ``survive'' the approximation procedure that eliminates the non-``classical'' states and observables. 

\paragraph*{}
Hence, although W\"{u}thrich's conception of emergence is a physicist's sense of emergence---aimed at demonstrating the posssibility of recovering the older theory from the newer one in the former's domain of applicability---we can see that it also leads to the more philosophically-oriented conception of emergence in physics that has been expounded in this thesis. More specifically, it embodies the idea of underdetermination which provides the basis for the \textit{autonomy} of the emergent theory from the one it emerges from. If this conception of emergence does apply in LQG, as W\"{u}thrich indicates, then the classical spacetime that emerges from LQG will be independent of many of the micro-states described by the theory. 

\paragraph*{}
Unfortunately, this claim is a very vague one since the relation between the spin network states, weave states and continuum spacetime being utilised is, at this stage, only a crude sketch. Nevertheless, the idea of \textit{novelty} that accompanies that of autonomy in characterising the conception of emergence advocated in this thesis can also be found in LQG. The requisite novelty seems clearly embodied, given the substantial differences between the fundamental structures of LQG and those of GR---in particular, the failure of the relation of adjacency to map onto the corresponding notion of ``closeness'' in the classical geometry (as described in \S\ref{subsub:microspace}).

\paragraph*{}
The failure of this relation to translate properly into the emergent spacetime suggests another possible conception of emergence associated with the idea of an approximation procedure. Because the spin networks generically give rise to geometries in which the notion of adjacency is not respected (from the point of view of the spin network), all those spin networks which correspond to geometries in which the spatial counterparts of two adjacent nodes are separated by more than a Planck length must be suppressed. In other words, we must select ``classical'' spin networks and ignore the rest, as accords with the definition of an approximating procedure, so long as some physical justification is provided for the neglect. Thus, it seems as though understanding this procedure and the justification for it could potentially lead to another conception of emergence in LQG.\footnote{Recall that the idea of emergence associated with neglecting certain states has also been proposed by \citet{Bain2013a}, in the context of EFT. This is dicussed in \S\ref{sub:emergence}, where I also tie it to the idea of underdetermination. } 

\paragraph*{}
Finally, the weave states furnish yet another possible basis for emergence, where the relevant conception of emergence resembles that associated with hydrodynamics, and is again related to the idea of underdetermination. The underdetermination comes about because the equations (\ref{eq:AS}) and (\ref{eq:VR}) do not uniquely determine a weave state for a given metric. The reason for this, recall (\S\ref{subsub:semiclas}), is that the equations only utilise \textit{averaged} properties, which could be represented by a number of different microstates. As \citet[][p. 26]{WuthrichForthcoming} points out, this is similar to the case of thermodynamics, where an averaged, macroscopic property, such as temperature, will correspond to many different microstates in a system with a large number of micro-level degrees of freedom. 

\paragraph*{}
Hence, this account seems to accord with the picture of emergence associated with hydrodynamics, presented in \S\ref{subsub:hydro}, although there are two differences of potential relevance: firstly, emergence in hydrodynamics can be connected to the idea of universality, the definition of universality, however refers to the RG, and, given the absence of a physical Hilbert space in LQG, it is not clear to me that we can utilise the RG within LQG at this stage. Secondly, macroscopic geometry is supposed to correspond to a superposition of weave states (\S\ref{subsub:semiclas}), whereas in hydrodynamics this is typically not the case.

\paragraph*{}
Most of the potential bases for a conception of emergence in LQG that I have presented here utilise the idea of underdetermination (as providing an explanation for spacetime being largely autonomous of its micro-structure). The suggestion that the micro-structure of spacetime is a superposition of microstates---which is made not only in regards to the weave states in LQG, but also in spinfoam theory (\S\ref{subsub:microst})---raises some interesting questions in regards to how the classical idea of underdetermination corresponds to quantum indeterminacy. This is another point where the quantum/classical transition intersects with the micro/macro transition, and perhaps the idea of decoherence will be of some help. As already stated above, however, I will not engage with these questions here.

\paragraph*{}
Given the substantial differences between LQG and GR, and the absence of any limiting procedure linking the two theories, it might seem pointless to attempt to frame a conception of emergence with GR taken to be emergent from LQG. The conception of emergence related to underdetermination that has been presented here perhaps fuels this worry---on it, \textit{any} low-energy theory might be said to be emergent from LQG in the same way that GR is supposed to be. 

\paragraph*{}
Of course, the only reply is that LQG is a direct quantisation of GR.  It is hoped (or assumed) that GR must somehow be emergent from LQG because we are able to ``go the other way'' and arrive at LQG from GR. The aim of the project is not to recover some other spacetime theory from LQG, but to approximate GR in the regime where the accuracy of the latter theory has been proven. Here, it is worth pointing out, however, that a theory of quantum gravity need not be a quantisation of GR (as \citet{Butterfield2000} remind us, it could be a quantisation of a theory other than GR, or it might not be a quantisation of any classical theory): conversely, a quantisation of GR does not necessarily produce a theory of quantum gravity.

\section{Conclusion}
LQG and spin foam theory are incomplete, with no definite means by which to describe the dynamics in either theory: there are several different options for a Hamiltonian operator in LQG, and several different options for calculating the measures in the spin foam sum, but no indication that any choice is correct in either of the cases. This incompleteness means that we are unable to develop a concrete picture of how spacetime could emerge from LQG. It seems as though such a picture would involve both the micro/macro transition as well as the quantum/classical transition, where the latter perhaps will need to be understood before the former can be implemented---nevertheless, I have purposefully avoided engaging with questions related to the quantum/classical transition and the idea of decoherence here. \citet{WuthrichForthcoming} suggests that one consequence of not yet understanding the role of decoherence is the failure of the continuum limit of LQG. 
 
\paragraph*{}
Because the limit in which the density of the weave states goes to infinity (or the ``lattice spacing'' goes to zero) fails to approximate continuum spacetime, it is unclear how a conception of emergence based on the idea of a limiting relation, such as described by \citet{Butterfield2011a, Butterfield2011b}, could apply in LQG. This could potentially be problematic, since, as described in \S\ref{sub:recovery} and \S\ref{sub:emerphysics}, a limiting relation is the typical means by which a newer theory is shown to relate to the older theory it is supposed to supplant, and the demonstration of the recovery of GR (which may be done through the use of a continuum limit) is generally taken as necessary for a theory of quantum gravity. An RG scaling procedure is also unable to be implemented as a means of recovering spacetime at large-distances, since the physical Hilbert space of LQG is undefined.
 
\paragraph*{}
Although the area and volume operators of LQG have discrete spectra, the fact that they are not gauge invariant, only existing at the kinematical level, means we cannot say definitely that LQG predicts spacetime discreteness. Additionally, LQG is faced with the problem of time and the problem of space, which are also related to difficulties of interpreting gauge invariance (amplified by tensions between quantum theory and GR, \S\ref{subsub:canon}). It may be that the problems with LQG have to do with the fact that it is a quantisation of GR. This fact also makes the lack of a low-energy limit of the theory particularly worrisome---given that LQG is a quantisation of GR, we would expect it to be relatively easy\footnote{Well, compared to other approaches!} to recover GR through decoherence plus a semiclassical limit. This having not been done is perhaps further motivation for considering an approach to quantum gravity that does not have a quantisation of GR as its starting point.

 \paragraph*{}
In spite of these difficulties, a number of potential bases for emergence can be identified in LQG. The requisite criterion of \textit{novelty} is fulfilled, since the macro-structures of GR differ in several major ways from the micro-structures described by LQG. Not only is the discreteness of the spin networks (and spin foams) of LQG a departure from the structures of GR, but the generic micro-state of spacetime is not supposed to even be a single spin network (spin foam)---rather a quantum superposition of such states. Furthermore, the fundamental relation of adjacency in a spin network, which indicates that two ``quanta of space'' are next to one another, is not typically preserved in the macroscopic geometry that the spin network is supposed to underlie: two adjacent quanta of space in a spin network may be arbitrarily far away from one another in the emergent space.
 
\paragraph*{}
The idea of \textit{autonomy} that forms the second part of the account of emergence presented in this thesis is furnished primarily by the underdetermination of the micro-states given the macro-states. For instance, while a weave state is constructed so as to represent the micro-state of a given (macroscopic, three-dimensional) metric, the construction of weave states means that, for a given metric there is a multitude of potential micro-states.\footnote{More correctly: a weave state is constructed so as to demonstrate the existence of a micro-state for a given macro-state.} In other words, the emergent spacetime is able to be represented by a number of different spin networks, and so is independent of the details of the micro-theory. The reason for this is the fact that the weave states depend only on average values, and so this idea of emergence resembles that associated with hydrodynamics/thermodynamics, where a number of micro-states correspond to the same macro-state of a system. The emergent structures depend only on \textit{collective properties} of the micro-degrees of freedom.

%% file: concl.tex
\chapter{Conclusion: Now here from nowhere}
The search for a theory of quantum gravity represents the quest for a more unified picture of the world. Yet, this search also reveals the utility and significance of a divided picture: one in which different levels emerge at different energy scales. The emergent theories are novel compared to the higher energy theory from which they each emerge, describing features and behaviour that only exists at the lower energy scales. Additionally, the emergent theories are importantly autonomous from their bases. This comes about because not many of the details of the high-energy physics get ``carried down'' to lower energies. Even so, there is much that the different levels have in common. In fact, it is the relationships that tie these theories together, and the features that they share that are responsible for the novelty and autonomy that characterise emergence. 

\paragraph*{}
It is thanks to these shared features---for instance, symmetries, phase transitions and scaling behaviour---that we are led to draw analogies between the different domains of physics: between high-energy particle physics and (comparatively) low-energy condensed matter physics, and even between the universe at large, as in cosmology, and the universe at its very smallest, as in quantum gravity. The existence of these shared features is demonstrated by the effectiveness of techniques such as the RG and the framework of EFT; and it is interesting that these techniques, used to highlight which features of a theory will survive to lower-energy scales, are themselves robust ``across the levels''. The techniques and principles, like the physical features they rely on, transcend the different theories and disciplines of physics.

\paragraph*{}
But, because emergence means that the link between the micro- and macro-levels is minimal, with the small-scale physics being underdetermined by the large-scale theory, we are restricted in what we can legitimately take from the analogy and how much we can learn about the small-scale theory of quantum gravity. The idea that the details of the high-energy theory are underdetermined by the emergent physics forms a large part of the conception of emergence (related to autonomy) in this thesis, but the underdetermination is only significant because of its origin: it comes from the way we move between theories, via the RG and limiting relations, and dynamical mechanisms such as symmetry-breaking. It is these relationships between the theories that give us emergence.

\paragraph*{}
Of course, the aim of this thesis has not been just to better understand emergence in physics, but the emergence of spacetime from quantum gravity. In doing this, I have looked at the relation of emergence with an interest in seeing it as a means of recovering GR (or some of the structures of GR, or an approximation to GR) from the high-energy theory that describes the micro-degrees of freedom of spacetime. I have found that, even though several of the theories do not have spacetime or GR as a low-energy limit (some not even having a well-defined low energy limit at all), there are at least plausible bases for understanding emergence in all of the approaches considered here. In many of the cases, especially those utilising the RG or describing phase transitions, it is relatively straightforward to see how the relationship between continuum spacetime and its discrete micro-degrees of freedom could be understood as emergence. 

\paragraph*{}
While the idea of emergence here is not exactly in line with the traditional philosophical conceptions which focus on a failure of reduction or derivation (in some sense), it is not incompatible with such accounts: instead it is meant as a more general and inclusive conception, able to encompass different types of novelty. Although some philosophers may view this generality as a weakness of the account, it is, in fact, the best we can do---since there is no theory of quantum gravity and all the approaches are incomplete, it is not sensible to speculate on the ontological or epistemological implications of a derivation that does not (yet) exist.

\paragraph*{}
Nevertheless, adherence to the GCP (as defined on p. \pageref{def:GCP}) as a principle of quantum gravity means maintaining that something akin to reduction or derivation must hold between GR and quantum gravity. In exploring quantum gravity, essentially, we are experimenting with combinations of different principles---we choose the ingredients and see whether the results accord with our expectations or desires of the theory. Lessons about quantum gravity are essentially lessons about the combination of principles we have chosen.

\paragraph*{}
Although physics without spacetime would be a conceptual revolution, the techniques and ideas we have seen adopted, including the RG and symmetry-breaking, are familiar. If these are indeed transcendent in their applicability, and something of the analogy ``between levels'' holds true, then we may expect them to serve us even in the inaccessible high-energy realm. Yet these ideas, which may themselves be understood as guiding principles, have employed in order to fulfil the GCP: to demonstrate the link between old and new, known and unknown. 

\paragraph*{}
In this thesis, the GCP has been kept constant as a principle, itself otherwise unevaluated. Its aim is to enable us to make contact with current physics, to ``save the phenomena'', but there is a question of how much we need to recover: if not full GR, then what approximation to it? Do we really need enough to explain the success of GR, or just enough to make us feel as though we could explain its success (if we were to ``push through'' with further derivations and approximations)? On its own, the GCP is not enough to lead us to quantum gravity; it says nothing, for instance, about the motivation, domain of applicability, or the structure of the theory. What has been learnt, from its combination with the other principles of interest in this thesis, is that we potentially require only minimal structures to recover spacetime, and that different theories may be able to fulfil, or represent, these. It is a question, now, of finding the structures from which we can travel enough of the way back to GR that we believe the principle to be satisfied. Once this has been done, we will have arrived back from nowhere.